\documentclass[%
onecolumn,
superscriptaddress,
amsmath,amssymb,aps,
pra,
]{revtex4-1}

\usepackage{graphicx}
\usepackage{dcolumn}
\usepackage{bm}
\usepackage{xcolor}
\usepackage{array}
\usepackage{tikz}
\usepackage{lineno}
\usepackage{amsmath}
\usepackage{comment} 

\newcommand{\unpolarizedSymbol}{%
  \mathord{%
    \vcenter{%
      \hbox{%
        \begin{tikzpicture}[baseline=0pt]
          \draw[line width=0.4pt] (0,0) circle[radius=0.18cm];
          \draw[<->, >=latex, line width=0.1pt] (-0.18,0) -- (0.18,0);
          \draw[<->, >=latex, line width=0.1pt] (0,-0.18) -- (0,0.18);
        \end{tikzpicture}%
      }%
    }%
  }%
}

\begin{document}

\preprint{APS/123-QED}

\title{Structured coherence: A modern perspective on optical coherence as a resource}

\author{Ayman F. Abouraddy}
\affiliation{CREOL, The College of Optics \& Photonics, University of Central Florida, Orlando, Florida 32816, USA}

\author{Bahaa E. A. Saleh}
\affiliation{CREOL, The College of Optics \& Photonics, University of Central Florida, Orlando, Florida 32816, USA}


\begin{abstract} 
Optical coherence is a well-established branch of physical optics in which the statistical properties of fluctuating optical fields are described in terms of correlation functions over continuous spatial and temporal degrees of freedom (DoFs). Nevertheless, in any practical setting, only discrete DoFs are ever accessible experimentally (e.g., sampling the field with a detector array), and there are many settings in which the DoFs are intrinsically discrete (polarization or spatial modes). In these scenarios, the modes themselves are fixed, stable, and deterministic, and partial coherence arises solely from random relative complex amplitudes, a field configuration we refer to as `structured coherence'. Advances in structured coherence have recently helped unveil new conceptual ground and surprising capabilities in optical communications and information processing in which partial coherence may be preferable to full coherence, which we call a `coherence advantage'. In this Tutorial, we present a discretized formulation of structured coherence in terms of coherence matrices to facilitate the investigation of these recent theoretical and experimental breakthroughs. We first review the partial coherence of an optical field characterized by a binary DoF, which is described mathematically by a $2\times2$~Hermitian, unity trace, positive semi-definite coherence matrix. This allows us to introduce key concepts that take on new significance for DoFs characterized by a large-dimensional modal set. Next, we examine the structured coherence of two binary DoFs, which can be described by $4\times4$ coherence matrices whose structure reflects coupling between the two DoFs. We introduce the concept of coherence rank (the number of non-zero eigenvalues of the coherence matrix), entropy swapping (reversibly transferring entropy between the two DoFs), and optical cross-purity (the interplay between separability and symmetry of the coherence matrix). In the perspective outlined here, coherence is viewed as a `resource', which can be exchanged between DoFs, concentrated into a DoF or into particular modes, or spread over the DoFs. We then briefly examine larger-dimensional modal sets, which allow for more versatile applications in optical information processing. The formulation presented here lends itself particularly to the manipulation of partial optical coherence in integrated photonic platforms, thereby opening myriad avenues for novel fundamental investigations of structured coherence and potentially exploiting the coherence advantage in optical communications and information processing.
\end{abstract}


\maketitle

\tableofcontents
\clearpage

\section{Introduction}

\subsection{Historical overview}

There will always be a need for studying the partial coherence of light. Indeed, all natural sources of light are partially coherent \cite{SalehBook07} -- whether solar  \cite{Mashaal12OL,Divitt15Optica,Dellieu17MTP,Ricketti22SR} and stellar radiation \cite{Michelson1890PM,Michelson21AJ,Zagury12OC,Hanbury-Brown56Nature,Hanbury-Brown74Book,Tan14AJL}, luminescence (electro-, chemo-, or bioluminescence) \cite{Tang87APL,Ono95Book,Vij98Book,Wilson98ARCDB,Brenny14JAP}, fluorescence \cite{Lakowicz06book,Diaspro11book}, or scintillation \cite{Anger58RSI,Codona87RS}. Moreover, partially coherent light is at the center of the study of vision \cite{Wesemann87JOSAA,Sahin25AO}, lighting, and viewing systems (e.g., virtual-reality and augmented-reality displays \cite{Lu24npjN}). Furthermore, optical fields that are initially coherent can be rendered partially coherent upon traversing a variety of media; e.g., a turbulent atmosphere \cite{Beran70JOSA,Fante74JOSA,Fante75ProcIEEE,Gbur02JOSAA,Ponomarenko02OL,Dogariu03OL,Shirai03JOSAA,Korotkova04OE,Berman06PRA,Berman07PRE,Wang15PER,Cox20IEEEJSTQE}, a turbid medium \cite{Alfano15InBook}, biological tissue \cite{Nolte24RPP}, or scattering surfaces such as painted walls \cite{Orchard68JOCCA,Vargas00SE,Auger12JCTR}. Crucially, there are applications in optics for which partial coherence provides salutary benefits. For example, imaging using partially coherent light eliminates the unwanted and deleterious speckle that accompany coherent light \cite{Fujii75NRO,Goodman07Book,Peng21SA,Evered25OLT}. More generally, finite coherence can be used in metrology, sensing \cite{Islam23arXiv}, imaging \cite{Deng17SR,Akcay02AO}, among other applications. In this tutorial, we are interested in partially coherent optical fields as they pertain to possible applications in optical communications and information processing, which requires considering what we call `structured coherence'.

The study of optical coherence has roots extending back to the demonstration of optical interference by Young \cite{Young1804PTRS} and Fresnel \cite{Young1804PTRS}, which was an epochal moment in the development of optical physics by confirming the wave nature of light \cite{Kipnis91Book}, thereby overturning a century-long dominance of Newton's particulate `emission theory' of light \cite{Sabra81Book}. Michelson's introduction of the formula for visibility in 1891 helped extract a \textit{quantitative} feature from optical interference \cite{Michelson1891PM1} (see also Refs.~\cite{Michelson1890PM,Michelson1891PM2,Michelson82PM}). The investigations of Zernike \cite{Zernicke} and others \cite{Laue07AdP,Berek26ZP,Wiener30AM,vanCittert39Physica,Hopkins51PRS} in the early twentieth century established the connection between interference visibility and a `degree of coherence'. Emil Wolf then firmly established optical coherence in the 1950s on solid mathematical and physical foundations \cite{Wolf53Nature,Wolf54NC,Wolf54PRSA,Wolf55PRSA,Wolf57PM,Wolf59INC,Mandel61JOSA2,Mandel65RMP} (see also \cite{Forrester56AJP,Gamo56OB,Hopkins57JOSA,Parrent59JOSA,Parrent59OA,ONeill61PhysToday,Bracewell62ProcIRE}). Starting from the premise that  electromagnetic fields cannot be observed directly at optical frequencies, he established a description of the optical field based on physically observable, continuous correlation functions in space and time (in addition to polarization) that capture the impact of the random fluctuations undergirding partially coherent light. Since then, our understanding of the properties of coherence functions has deepened, a variety of insights have been gained \cite{Martienssen64AJP,McCutchen66JOSA,Wolf96RPP,Devaney97OL,Dorrer04JOSA}, and new applications continue to emerge \cite{Wolf07Book,Friberg16JOSAA,Korotkova20PO,Chen20PO,Yu23PQE}. Nevertheless, it is probably fair to say that progress in the study of partially coherent fields has not kept apace with that for their coherent (laser) counterparts. Specifically, the major recent trends in optics and photonics, such as reliance on integrated photonics platforms, have left almost no impact on partially coherent light to date, thereby practically precluding its use in modern optical information processing technologies. 

\subsection{Why reconsider the matrix treatment of optical coherence?}

In this tutorial we focus on a matrix formulation of optical coherence. This approach will be central to the potential applications of partially coherent light that are emerging in the area of optical information processing. Optical polarization being inherently a discrete degree-of-freedom (DoF) of the optical field \cite{Wolf59NC,Wolf60PRS} has always been treated using matrix algebra (Jones vectors, Jones matrices, and Mueller matrices) \cite{Parrent60NC,Shurcliff66book,Azzam77book,Brosseau98Book}. Applying this matrix approach to spatial, temporal, or spectral modes is less common -- but not without precedent. An early salient effort by Hideo Gamo \cite{Gamo64PO} used the sampling theorem to formulate optical coherence in terms of matrices to accommodate the spatial discretization of the intensity profile associated with detector arrays (see also Ref.~\cite{ONeill61JPSJ}). Unfortunately, this approach was largely ignored in subsequent years.

Recent developments point towards a need for revisiting this matrix formulation of partial coherence and further extending this methodology to what we call `structured coherence'. Specifically, there are many scenarios where stable, fixed, deterministic modes are maintained by a class of optical systems, whether guided modes in multimode fibers \cite{Li14AOP} or on-chip waveguides \cite{Bogaerts20Nature}, or orbital angular momentum (OAM) modes in free space \cite{Willner15AOP}, among many other possibilities \cite{Levy16PO}. In these cases, it is most beneficial to exploit such a favored modal basis to analyze the optical field -- especially when the field is partially coherent. We delineate here some salient motivating factors driving current interest into matrix formulations of structured coherence.

\subsubsection{Only a finite number of measurements are possible in practice}

Although optical coherence is nominally cast in terms of continuous functions, only a finite number of measurements can be acquired in practice. For example, the spatial intensity is typically sampled at a finite number of discrete points (e.g., the pixels of a CCD). The ubiquity of optical detector arrays therefore motivates describing optical coherence with matrices -- represented in a modal basis of sampled positions -- rather than continuous functions \cite{Gamo64PO}.

\subsubsection{Formulating partially coherent light in terms of new modal sets}

The development of lasers has brought to the fore several families of optical modes, some of which arise naturally in the context of optical resonators \cite{Siegman86book}; examples include Hermite-Gaussian (HG) \cite{Zhou18OL} and  Laguerre-Gaussian (LG) modes \cite{Fu18OE}, Bessel beams \cite{Durnin87PRL}, OAM modes \cite{Allen92PRA,Willner15AOP}, among many others. Modern beam-shaping technologies (e.g., spatial light modulators \cite{Neff90ProcIEEE,Maurer11LPR}, micro-mirror arrays \cite{Jang04OE,Brennesholtz08Book,Hellman19OE}, diffractive optics \cite{Buralli91AO,Gil03JVSTB,Banerji19Optica}, and metasurfaces \cite{Yu11Science}) have simplified the precise sculpting of spatial field profiles, thereby enabling the unique characteristics of such modal sets to be exploited in applications ranging from particle trapping \cite{Dholakia08AAMOP} and manipulation \cite{Grier03Nature} to optical microscopy \cite{Maurer11LPR} and laser control \cite{Ngcobo13NC,Forbes24NPR}, a topic known under the umbrella term `structured light' \cite{Forbes21NP}.

Moreover, optical fibers and waveguides make clear the importance of thinking in terms of `modes'. Although the field emerging from a multimode fiber or waveguide may appear random, it is nevertheless constrained to be a superposition of the guided modes, which are predetermined by the guiding structure -- only the modal weights can vary, and may take on deterministic or random values. The study of all these novel modal sets in the context of partial coherence has been limited \cite{Saleh81AO}.

\subsubsection{Exploiting multiple DoFs}

Studying multiple DoFs of the optical field can be facilitated by a discretization of the associated modal basis for each DoF. One pertinent example is that of Young's double-slit interference in a vector field. The polarization at each slit is described by a $2\times2$ \textit{polarization} coherence matrix, and spatial coherence at the two points (when ignoring polarization) can itself be captured by a $2\times2$ \textit{spatial} coherence matrix \cite{Abouraddy14OL,Abouraddy19Optica,Halder21OL}. Because the spatial and polarization DoFs are physically independent, encompassing both DoFs requires a $4\times4$ coherence matrix over the direct product of the two subspaces for the spatial and polarization DoFs \cite{Kagalwala13NP,Abouraddy14OL,Kagalwala15SR}. This $4\times4$ spatial-polarization coherence matrix has been studied by Gori and others in the context of investigating the impact of polarization transformations on the visibility of Young's interference \cite{Gori06OL} (see also \cite{Abouraddy17OE}). The mathematical structure of this $4\times4$ coherence matrix is isomorphic to that of the $4\times4$ density matrix of a pair of qubits (two-level systems) in quantum mechanics \cite{Peres93Book}. This mathematical correspondence allows the tremendous progress in quantum information processing to be harnessed in the study of optical coherence. For example, the non-separability of a classical optical field with respect to its DoFs has given rise to the concept of `classical entanglement' \cite{Spreeuw98FP,Kagalwala13NP}, in mathematical analogy to the counterpart concept of quantum entanglement for bipartite quantum systems \cite{Pan12RMP}.

\subsubsection{Exploiting advances in photonic technologies: On-chip structured coherence}

Many well-known results in optical coherence are restricted to freely propagating optical fields; e.g., double-slit interference, the van~Cittert-Zernike theorem \cite{Born99Book}, cross-spectral purity \cite{Mandel61JOSA}, among other thoroughly studied phenomena. In contrast, the new applications for partially coherent light in information processing motivate studying the propagation of optical coherence in new systems that are described in terms of a discretized transfer function. Examples include (1) multimode optical fibers and waveguides \cite{Saleh81AO}; (2) on-chip platforms, photonic integrated circuits (PICs), and programmable photonics \cite{Bogaerts20Nature}; and (3) coupled waveguides in discrete optics \cite{Christodoulides03Nature}. All of these platforms demand a matrix formulation of structured coherence. Experimental demonstrations of on-chip manipulation of optical coherence are just now emerging \cite{Hashemi26TwoModes,Hashemi26FourModes,Mor26arxiv}.

\subsubsection{Utilizing partial coherence in optical information processing}

Optical coherence is central to the study of imaging, radiometry, lighting, solar energy, atmospheric turbulence, among other areas. However, new applications for partially coherent light are emerging in the areas of optical information processing and communications. Recent examples include proposals for increasing the information capacity in optical fiber communications \cite{Nardi22OL}, communicating across strongly scattering optical channels \cite{Harling25APLP}, optical cryptography \cite{Peng21P,Liu25LPR}, and even for improving on-chip optical computing \cite{Dong24Nature}, all of which require a discretized matrix formulation of optical coherence. Moreover, novel perspectives and insights regarding aspects of partially coherent light that have traditionally attracted only limited attention are emerging from this matrix formulation, especially with regards to the entropy-carrying capacity, leading to such novel concepts as entropy swapping \cite{Okoro17Optica,Harling22OE,Harling23JO}, coherence rank \cite{Harling24PRA2}, and locked entropy \cite{Harling24PRA}.

\subsection{What is structured coherence?}

\begin{figure}[t!]
\centering
\includegraphics[width=10cm]{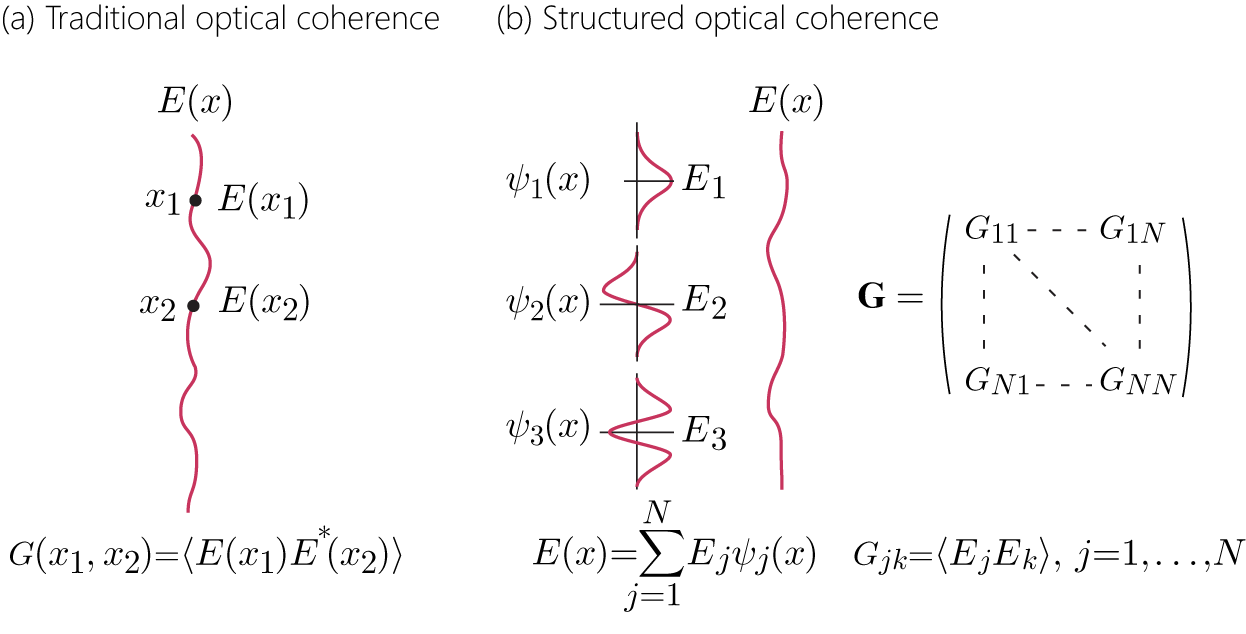}
\caption{(a) In conventional optical coherence, a continuous correlation function $G(x_{1},x_{2})=\langle E(x_{1})E_{2}^{*}(x_{2})\rangle$ describes the spatial coherence of the field. (b) In structured coherence, the field is viewed in terms of a finite set of modes $\{\psi_{j}(x)\}_{j=1}^{N}$, so that we can express the field as a superposition $E(x)=\sum_{j=1}^{N}E_{j}\psi_{j}(x)$. The modes themselves are fixed, and only the modal coefficients $\{E_{j}\}$ can vary. When the modal coefficients are random variables, the structured coherence of the field is captured with an $N\times N$ coherence matrix, whose elements are correlations between the modal coefficients, $G_{jk}=\langle E_{j}E_{k}^{*}\rangle$.}
\label{fig:StructuredCoherence}
\end{figure}

In traditional optical coherence, the statistical nature of the field is encoded in correlation functions. In the case of spatial coherence, for example, the coherence function at positions $x_{1}$ and $x_{2}$ is given by $G(x_{1},x_{2})=\langle E(x_{1})E^{*}(x_{2})\rangle$, where $\langle\cdot\rangle$ denotes a statistical average over an ensemble [Fig.~\ref{fig:StructuredCoherence}(a)]. The field at each point can be considered a random variable, with the function $G$ representing the correlation between pairs of such points. This formulation can reveal, for example, the transverse coherence width: the separation after which the correlation between a pair of points drops below some threshold. The development of such correlation functions is at the heart of traditional coherence theory \cite{Wolf54NCC,Wolf55PRSA,Wolf59INC,Karczewski63INC,Mandel65RMP,Perina72Book,Born99Book,Wolf07Book,Goodman15Book,Agarwal20PO}.

In \textit{structured coherence}, on the other hand, underpinning the field is a set of modes $\{\psi_{j}(x)\}_{j=1}^{N}$, which are stable, fixed, deterministic field distributions that are maintained by the relevant optical systems. Because these modes are deterministic, they do not themselves display random features. If these modes form a complete basis of dimension $N$, the optical field can be expressed as a superposition of these modes, $E(x)=\sum_{j=1}^{N}E_{n}\psi_{j}(x)$, with complex coefficients $\{E_{j}\}$. In the context of structured coherence, only these relative complex coefficients are random variables, while the modes themselves remain fixed [Fig.~\ref{fig:StructuredCoherence}(b)]. Consequently, any statistical correlations would involve only the modal coefficients $\{E_{j}\}$ but \textit{not} the modes themselves. Studying the correlations between pairs of points, while of course remaining a valid approach, does not describe the statistical features of the field in an efficient manner, and does not capture the essential features of the field structure. Instead, it is more beneficial to consider the correlations between pairs of the $N$ modal coefficients $\{E_{j}\}$, which can be tabulated in an $N\times N$ matrix that we call the \textit{coherence matrix} $\mathbf{G}$, whose elements are $G_{jk}=\langle E_{j}E_{k}^{*}\rangle$, $j,k=1,\cdots N$. Capturing such correlations presume the ability to measure the modal weights and to explicitly manipulate the optical field in the selected basis $\{\psi_{j}(x)\}$, rather than in the more common positional basis. Section~\ref{sec:ModalDetection} describes the various approaches currently available for measuring modal weights in different modal bases. Much more work is needed along these lines. This Tutorial offers basic tools for tackling the nascent research area of structured coherence.

\subsection{The `coherence advantage': What can partially coherent light do that coherent light cannot?}

The modern developments regarding coherent light were launched by the invention of the laser. Although light produced from natural sources is partially coherent, optical information processing has relied almost solely on coherent light, especially for optical communications. Of course, a partially coherent source (e.g., LEDs in Li-Fi \cite{Tsonev14ProcSPIE}) can be used for optical communications over short distances, but this is typically only a matter of convenience. The partial coherence of the source itself plays no role in the communications scheme. The study of optical coherence is currently experiencing something of a renaissance driven by investigations related to the question of a `coherence advantage' in optical information processing. The motivating question that underpins the study of `structured coherence' is the following: are there settings in optical information processing in which partially coherent light (described in terms of coherence matrices) can outperform coherent light? An affirmative answer to this question would then correspond to a `coherence advantage'.

One well-known `coherence advantage' is the following. A coherent field spanned by $N$ modes requires $2N-2$ real parameters for its identification (removing a normalized length and an overall phase), whereas a partially coherent field supported by the same $N$ modes requires $N^{2}-1$ real parameters to identify the associated $N\times N$ Hermitian coherence matrix (removing a normalization factor) \cite{Waller12NP}. The much larger set of real parameters $\mathcal{O}(N^{2})$ for a partially coherent field, compared to the $\mathcal{O}(N)$ required for a coherent field, makes the synthesis and characterization of partially coherent fields significantly more challenging than their coherent counterparts \cite{Waller12NP}. This feature suggests the richer information-carrying capacity of partially coherent light. Nevertheless, no experimental realizations have taken advantage of this opportunity to date. Other recent examples of the coherence advantage demonstrated experimentally include improving the parallelization of optical computing \cite{Dong24Nature}, new schemes of optical cryptography \cite{Peng21P,Liu25LPR}, and scattering-free optical communications using the coherence rank \cite{Harling25APLP}. Many more such examples of the coherence advantage are expected to be discovered over the next few years.

\subsection{Structure of this Tutorial}

The goal of this Tutorial is to present the basic concepts and mathematics for describing partially coherent fields in terms of discrete modal bases, and thus working with coherence matrices instead of continuous correlation functions. This formulation is particularly geared to on-chip implementations for the processing of partially coherent light. This Tutorial is structured as follows. We first consider the concepts of optical modes and modal bases, which permit us to provide a discrete formulation of optical fields using the mathematics of linear vector spaces as is common in quantum mechanics. As a first step, we describe current approaches to detecting the modal weights in a given discrete modal basis, which we classify as `modal projectors' and `modal analyzers' (Section~\ref{sec:ModalDetection}). We next establish in Section~\ref{sec:OneBinaryDoF} the $2\times2$ coherence matrices characterizing an optical field in which a single DoF characterized by two modes is pertinent (a binary DoF). Within this context we define several key concepts: the degree of coherence (Section~\ref{sec:DegreeOfCoherenceOfSingleDoF}) and its measurement (Section~\ref{sec:MeasuringTheDegreeOfCoherence}), the coherence entropy \cite{Peres93Book} (Section~\ref{sec:EntropyOfOneDoF}), modal Stokes parameters (Section~\ref{sec:StokesSingleDoF}) for reconstructing the coherence matrix (Section~\ref{sec:ReconstructingGSingleDoF}), unitary (Section~\ref{sec:UnitarySingleDoF}) and non-unitary (Section~\ref{sec:NonUnitaryOperators}) transformations. As concrete examples, we apply this mathematical framework to the polarization DoF (Section~\ref{sec:polarizationDoF}), where it is commonly used, and to a spatial DoF comprising two spatial modes (Section~\ref{sec:SpatialDoF}), where this formalism is less common, but which is the building block for large-scale on-chip systems for processing structured coherence. We end this Section by describing how partial coherence can enable optical communications through a scattering channel (Section~\ref{sec:Communications1DoF}) before a brief comparison with corresponding concepts in quantum mechanics (Section~\ref{sec:QuantumSingleDoF}). Next, we extend this formulation to partially coherent optical fields characterized by two binary DoFs utilizing $4\times4$ coherence matrices (Section~\ref{sec:2DoFs}). Extending the dimensionality of the coherence matrix allows the introduction of several novel concepts, such as the coherence rank (the number of non-zero eigenvalues of the coherence matrix, Section~\ref{sec:CoherenceRank}), reduced and restricted coherence matrices (Section~\ref{sec:ReducedRestricted}), unitaries encompassing two DoFs (Section~\ref{sec:Unitary2DoFs}), the non-uniqueness of entropy to identify the fields that can be inter-converted into each other via unitary transformations (Section~\ref{sec:EntropyNonUniqueness}), entropy concentration and swapping (Section~\ref{sec:EntropySwapping}), in addition to describing the tomographic reconstruction of the coherence matrix via modal Stokes parameters associated with Kronecker-Pauli matrices as an intermediary (Section~\ref{sec:Stokes2DoFs}), maximizing double-slit interference visibility in a vector field (Section~\ref{sec:MaximizingDoubleSlits}), optical cross-purity (Section~\ref{sec:CrossPurity}), coherence-rank communications (Section~\ref{sec:Communications2DoFs}) and the correspondences and differences to bipartite quantum states (Section~\ref{sec:Quantum2DoFs}). Finally, we provide a brief description of various extensions for structured coherence (Section~\ref{sec:Discussion}), especially with regard to larger-dimensional modal sets and other physical DoFs. We end the tutorial with a roadmap for expected developments in the area of structured coherence over the course of the next few years (Section~\ref{sec:Roadmap}), with emphasis on burgeoning efforts for on-chip processing of structured coherence.

\section{Modal detection for structured coherence}\label{sec:ModalDetection}

\subsection{Modes, modal bases, and the Dirac notation}

Although optical fields are described mathematically using continuous functions in space and time, in addition to polarization \cite{Wolf07Book}, the finite, discrete set of measurements that are made constitute a \textit{de facto} discretization of the field. We consider here `structured coherence': partially coherent optical fields supported by a discrete, finite modal basis. Taking into account an underlying modal structure can have profound implications for efficient representation, manipulation, and characterization of the field. This requires that we first define `modes' and `modal bases'.

By `mode' we denote a stable, fixed, deterministic field structure that is well-defined and is orthogonal to the other modes (i.e., the modes can be unambiguously distinguished from each other).  Selection of the modes is usually determined by a pertinent optical system. For example, in the case of propagation along a multimode fiber, the modes selected will naturally be the guided modes of the fiber; alternatively, in a photonic integrated circuit, the modes are those of the on-chip waveguides. Such modes can in principle traverse the optical channel of interest in a stable manner. By `modal basis' we denote a collection of modes that is `closed'. Any optical field of interest can be constructed out of these modes. Upon traversing any system, the modes may couple to each other, but no new modes outside the basis can contribute to the field. Strictly speaking, an infinite modal basis is typically required. Nevertheless, it is usual practice to set an upper limit on the dimension of the selected modal basis in order to exclude modes whose contribution to the field dynamics is negligible. This upper limit usually emerges from the finiteness of experimental resources and practical limits of the source and detectors (e.g., finite pixel resolution, finite aperture sizes, etc.).

We adopt the Dirac notation for our formulation. A mode is represented by a vector $|\psi_{j}\rangle$ with integer index $j$. For the $N$ modes, from $|\psi_{1}\rangle$ to $|\psi_{N}\rangle$, we define each mode on the basis by a $N\times1$ vector having~0 for all elements except the $j^{\mathrm{th}}$ element that is set to~1 when considering $|\psi_{j}\rangle$. The optical field vector is thus defined in terms of a field vector $|E\rangle$, which is in general a weighted superposition of the modes:
\begin{equation}\label{eq:GeneralFieldVector}
|E\rangle=\sum_{j=1}^{N}E_{j}|\psi_{j}\rangle=E_{1}\left(\begin{array}{c}1\\0\\\vdots\\0\end{array}\right)+E_{2}\left(\begin{array}{c}0\\1\\\vdots\\0\end{array}\right)+\cdots E_{N}\left(\begin{array}{c}0\\0\\\vdots\\1\end{array}\right)=\left(\begin{array}{c}E_{1}\\E_{2}\\\vdots\\E_{N}\end{array}\right);
\end{equation}
here the complex coefficients $\{E_{j}\}$ are referred to as modal amplitudes, and we refer to their squared magnitudes $\{|E_{j}|^{2}\}$ as modal weights. We define the Hermitian conjugate (complex conjugate plus the transpose operation) of a vector as $(|\psi\rangle)^{\dagger}=\langle\psi|$, which is a $1\times N$ row vector. Consequently, the inner product of any mode with itself (a measure of its length) is unity; more generally, $\langle\psi_{j}|\psi_{k}\rangle=\delta_{jk}$, so that the modes are orthonormal. This entails that the length of the field vector is $\langle E|E\rangle=\sum_{j=1}^{N}|E_{j}|^{2}$, which we typically normalize in turn to unity $\langle E|E\rangle=1$. Each modal weight thus corresponds to the fraction of power associated with that mode. We also define an outer product $|\psi_{j}\rangle\langle\psi_{k}|$ which is an $N\times N$ matrix with all the elements zero, except for the element in row $j$ and column $k$ that is~1. Any $N\times N$ matrix $\mathbf{M}$ can then be written as a sum of outer products, $\mathbf{M}=\sum_{jk}m_{jk}|\psi_{j}\rangle\langle\psi_{k}|$, where $m_{jk}$ is the matrix entry at the $j^{\mathrm{th}}$ row and $k^{\mathrm{th}}$ column.

Our goal here is to elucidate how to measure the modal weights for an optical field in a prescribed basis. Most recent progress along these lines has been confined to coherent fields. However, the schemes described here are equally applicable to partially coherent light, and indeed constitute a major component of future progress towards structured coherence.

\subsection{Examples of modal bases in optics}

\subsubsection{Polarization}

The polarization DoF is naturally formed of a `discrete' modal basis of finite dimension. In general, the polarization at a point in the field is represented by a 3D complex vector with the field components taken along three Cartesian axes. In the paraxial regime, polarization is approximately confined to a transverse plane orthogonal to the propagation axis (the $z$-axis). The polarization vector is then a 2D complex vector with components along the $x$ and $y$ axes, which we denote $|\mathrm{H}\rangle$ and $|V\rangle$ (the horizontal and vertical polarization modes, respectively): 
\begin{equation}
|E\rangle=E_{\mathrm{H}}|\mathrm{H}\rangle+E_{\mathrm{V}}|\mathrm{V}\rangle=E_{\mathrm{H}}\left(\begin{array}{c}1\\0\end{array}\right)+E_{\mathrm{V}}\left(\begin{array}{c}0\\1\end{array}\right)=\left(\begin{array}{c}E_{\mathrm{H}}\\E_{\mathrm{V}}\end{array}\right).
\end{equation}

\begin{figure}[t!]
\centering
\includegraphics[width=13.3cm]{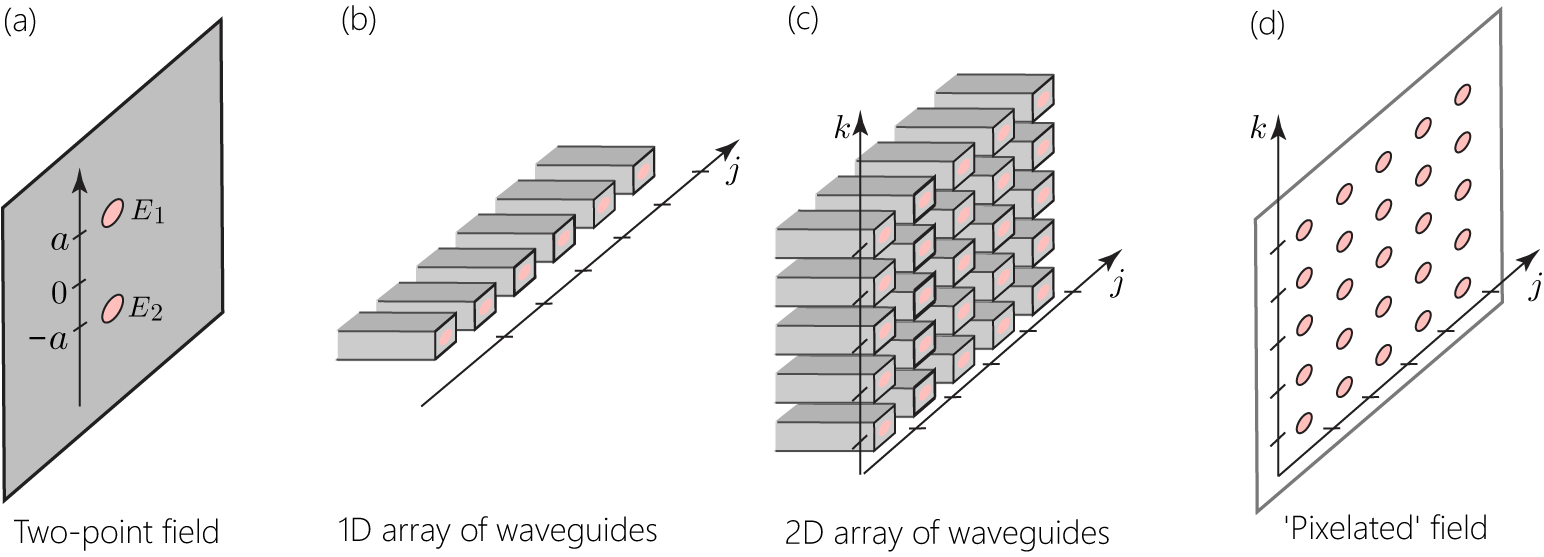} 
\caption{Modal bases for the spatial DoF with \textit{non-overlapping} modes. (a) A two-point field. (b) A 1D array of waveguides. (c) A 2D array of waveguides. (d) A 2D optical field in an axial plane.}
\label{fig:NonOverlappingSpatialModes}
\end{figure}

\subsubsection{Spatial modal bases}

We will focus in this Tutorial on spatial modes in addition to polarization (see the Discussion in Section~\ref{sec:Discussion} for other DoFs). We classify such modes as spatially `non-overlapping' or `overlapping'. We depict in Fig.~\ref{fig:NonOverlappingSpatialModes} prominent scenarios in optics in which the field is restricted to a discrete superposition of non-overlapping spatial modes. The simplest example is a two-point field [Fig.~\ref{fig:NonOverlappingSpatialModes}(a)], as occurs in Young's double-slit experiment. If the continuous field before the slits is $E_{\mathrm{o}}(x)$, then after a pair of identical slits of width $W$ located at $x=\pm a$ the field can be written as $E(x)=E_{1}(x)+E_{2}(x)$, where the field from the first slit is $E_{1}(x)=f(\tfrac{x-a}{W})E_{\mathrm{o}}(x)$ and from the second $E_{2}(x)=f(\tfrac{x+a}{W})E_{\mathrm{o}}(x)$; here $f(x)$ is a function of width unity (representing the slit). We can model this scenario as a two-mode field: the modes are the fields at the slits that do not overlap in space, and are thus `orthogonal', $\int\!dx\;E_{2}^{*}(x)E_{1}(x)=0$. Assuming that the slits are identical and are sufficiently narrow that the spatial variation of $E(x)$ over $W$ is negligible, only the complex amplitudes of the fields at $x=\pm a$ are relevant, $E_{\mathrm{o}}(a)$ and $E_{\mathrm{o}}(-a)$. This scenario can thus be described with two spatial modes $|\psi_{1}\rangle$ and $|\psi_{2}\rangle$,
\begin{equation}\label{eq:BinarySpatialModes}
|E\rangle=E_{1}|\psi_{1}\rangle+E_{2}|\psi_{2}\rangle=E_{1}\left(\begin{array}{c}1\\0\end{array}\right)+E_{2}\left(\begin{array}{c}0\\1\end{array}\right)=\left(\begin{array}{c}E_{1}\\E_{2}\end{array}\right).
\end{equation}
We have thus abstracted from the continuous fields at the slits to a discrete bimodal basis.

A second example depicted in Fig.~\ref{fig:NonOverlappingSpatialModes}(b) is a 1D array of $N$ identical single-mode waveguides. Here the set of displaced copies $|\psi_{j}\rangle$ of the fundamental spatial mode of the waveguide, which are non-overlapping spatially, form the modal basis, and we write the field as in Eq.~\ref{eq:GeneralFieldVector}, where $E_{j}$ are the complex field amplitudes of the spatial modes at each waveguide. This example can be extended to a 2D array of identical waveguides as illustrated in Fig.~\ref{fig:NonOverlappingSpatialModes}(c), so that $|E\rangle=\sum_{j,k=1}^{N}E_{jk}|\psi_{jk}\rangle$. Finally, we can consider the optical field $E(x,y)$ in a transverse plane at a fixed axial position, which is `pixellated' [Fig.~\ref{fig:NonOverlappingSpatialModes}(d)] into identical bins centered at coordinates $(x_{j},y_{k})$, whereupon the field can be expressed in terms of modes $|\psi_{jk}\rangle$, each corresponding to a pixel at position  $(x_{j},y_{k})$.

\begin{figure}[t!]
\centering
\includegraphics[width=13.3cm]{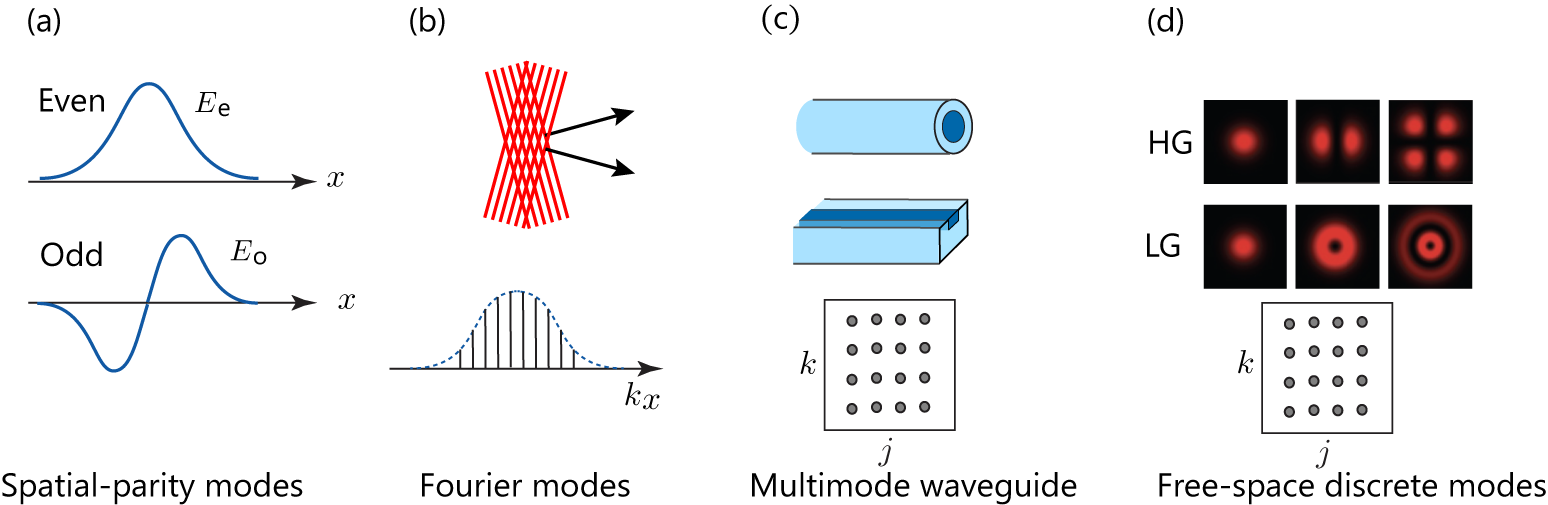} 
\caption{Modal bases for the spatial DoF with spatially \textit{overlapping} modes. (a) Parity modes along 1D. (b) Fourier modes comprising tilted planes waves at the same wavelength. (c) Modes in a multimode optical fiber or array. (d) A free field formed of a superposition of any of a variety of available modal bases (Laguerre-Gaussian, Hermite-Gaussian, etc.).}
\label{fig:OverlappingSpatialModes}
\end{figure}

The common feature shared by the examples depicted in Fig.~\ref{fig:NonOverlappingSpatialModes} is that the spatial modes do \textit{not} overlap spatially. However, in many common scenarios, the field is a superposition of \textit{overlapping} spatial modes. A simple example is depicted in Fig.~\ref{fig:OverlappingSpatialModes}(a) in which the field is formed of a superposition of two functions $\psi_{\mathrm{e}}(x)$ and $\psi_{\mathrm{o}}(x)$ having different spatial `parity': one is even and the other odd, which we denote in vector form as $|\psi_{\mathrm{e}}\rangle$ and $|\psi_{\mathrm{o}}\rangle$, respectively \cite{Abouraddy07PRA,Yarnall07PRL1,Yarnall07PRL2,Yarnall08OE}. Consequently, the two modes are still `orthogonal': $\int\!dx\;\psi_{\mathrm{o}}^{*}(x)\psi_{\mathrm{e}}(x)=\langle\psi_{\mathrm{o}}|\psi_{\mathrm{e}}\rangle=0$. The field can be written in the same form as Eq.~\ref{eq:BinarySpatialModes}, with $|E\rangle=E_{\mathrm{e}}|\psi_{\mathrm{e}}\rangle+E_{\mathrm{o}}|\psi_{\mathrm{o}}\rangle$. Another familiar example is that of Fourier optics with a monochromatic field in which the field is viewed as a superposition of overlapping planes waves each tilted at a different angle with the propagation axis [Fig.~\ref{fig:OverlappingSpatialModes}(b)]. In one transverse dimension $E(x)=\int\!dk_{x}\;e^{ik_{x}x}\widetilde{E}(k_{x})$ \cite{GoodmanBook05}, so that discretizing the continuous span of $k_{x}$ to bins centered on the values $k_{j}$ yields $|E\rangle=\sum_{j}E_{j}|\psi_{j}\rangle$, where $|\psi_{j}\rangle\rightarrow\widetilde{E}(k_{j})$. 

\begin{figure}[b!]
\centering
\includegraphics[width=13.3cm] {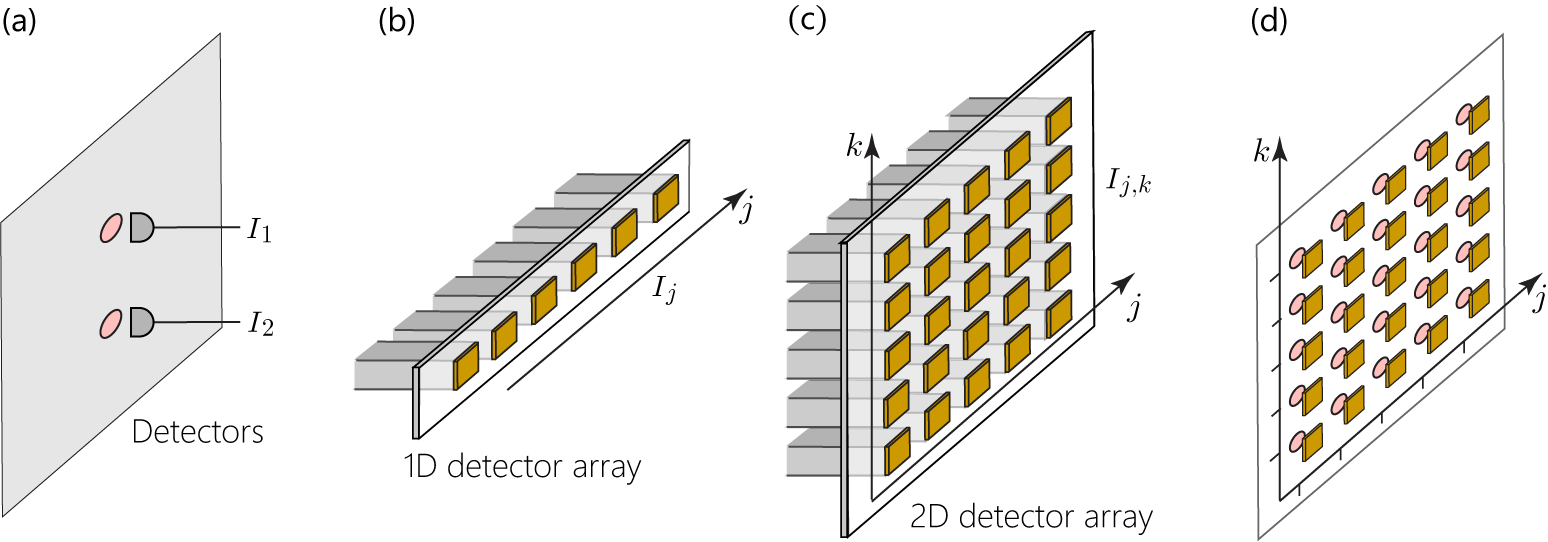} 
\caption{Projections onto a modal bases for non-overlapping spatial DoFs. (a) A two-point field. (b) A 1D array of waveguides. (c) A 2D array of waveguides. (d) A 2D optical field in an axial plane.}
\label{fig:DetectionNonOverlappingSpatialModes}
\end{figure}

Another example is that of a multimode optical fiber or waveguide. The boundary conditions guarantee that the guided field is constrained to be a superposition of the discrete set of guided modes identified by independent radial and azimuthal indices [Fig.~\ref{fig:OverlappingSpatialModes}(c)]. Finally, a freely propagating field can be viewed as a discrete superposition of propagation modes \cite{Levy16PO} overlapping in space [Fig.~\ref{fig:OverlappingSpatialModes}(d)].

\subsection{Finding the modal weights}

A central task in structured coherence is to measure the modal weights: the relative fractions of power contributed by each of the underlying modes to the field. This process is straightforward for non-overlapping spatial modes by utilizing an appropriate detector array, whether a pair of detectors for two-point fields [Fig.~\ref{fig:DetectionNonOverlappingSpatialModes}(a)], 1D or 2D detector arrays for 1D and 2D waveguide arrays [Fig.~\ref{fig:DetectionNonOverlappingSpatialModes}(b,c)], a CCD array (or other pixellated detector arrays) to spatially sample the field at positions $(x_{i},y_{j})$ in the pixel basis [Fig.~\ref{fig:DetectionNonOverlappingSpatialModes}(d)]. 

\begin{figure}[t!]
\centering
\includegraphics[width=10.3cm] {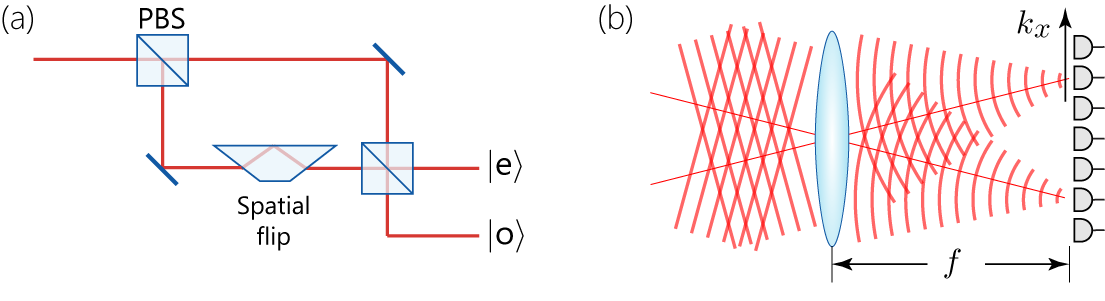}
\caption{(a) A parity projector formed of a balanced MZI in which a spatial flip (here a dove prism) is inserted in one arm. The two outputs of the parity projector correspond to the modal basis $\{|\mathrm{e}\rangle,|\mathrm{o}\rangle\}$. (b) A Fourier-mode projector that separates the plane-wave components of the optical fields. Each plane wave is focused to a point in the the focal plane where a detector array is placed.}
\label{fig:DetectionParityFourier}
\end{figure}

\begin{figure}[b!]
\centering
\includegraphics[width=8.5cm]{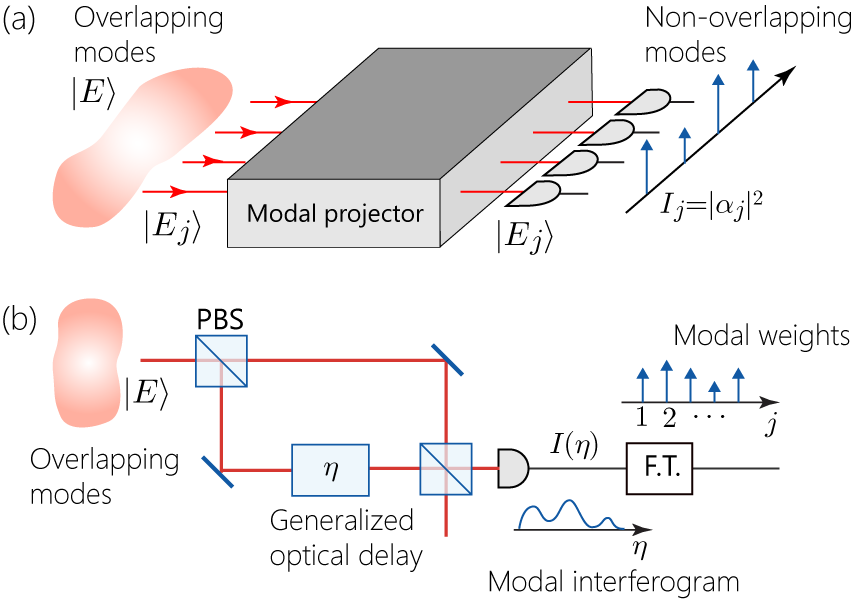}
\caption{(a) A modal projector separates spatially overlapping modes into spatially non-overlapping modes, so that a detector array captures the modal weights $|E_{j}|^{2}$. (b) A modal analyzer contains a real parameter $\eta$ that is scanned and a signal $I(\eta)$ is recorded with a single detector. Harmonic analysis of $I(\eta)$ with respect to $\eta$ reveals the modal weights $|E_{j}|^{2}$.}
\label{fig:ProjectionAnalyzer}
\end{figure}

However, such detector arrays cannot measure the modal weights directly in the case of overlapping spatial modes, and a different procedure must be followed. Procedures for obtaining the modal weights for different sets of modes have been developed. For example, in the case of optical parity modes $\{|\psi_{\mathrm{e}}\rangle,|\psi_{\mathrm{o}}\rangle\}$ [Fig.~\ref{fig:OverlappingSpatialModes}(a)], a parity projector [Fig.~\ref{fig:DetectionParityFourier}(a)] comprises a spatial flip (e.g., via a Dove prism \cite{Abouraddy07PRA}) in one arm of a balanced Mach-Zehnder interferometer (MZI). The spatial flipper $\psi(x)\rightarrow\psi(-x)$ implements the transformation $|\psi_{\mathrm{e}}\rangle\rightarrow|\psi_{\mathrm{e}}\rangle$ and $|\psi_{\mathrm{o}}\rangle\rightarrow-|\psi_{\mathrm{o}}\rangle$, thus separating an input field $|E\rangle=E_{\mathrm{e}}|\psi_{\mathrm{e}}\rangle+E_{\mathrm{o}}|\psi_{\mathrm{o}}\rangle$ into the components $E_{\mathrm{e}}|\psi_{\mathrm{e}}\rangle$ and $E_{\mathrm{o}}|\psi_{\mathrm{o}}\rangle$ at the MZI output ports \cite{Yarnall07PRL1,Yarnall07PRL2,Yarnall08OE}. For the plane-wave basis [Fig.~\ref{fig:OverlappingSpatialModes}(b)], the Fourier-mode projector formed of a lens in a $2f$ configuration [Fig.~\ref{fig:DetectionParityFourier}(b)] separates the Fourier modes and directs each to a point in the focal plane. In both of these scenarios, the overlapping spatial modes are converted into spatially non-overlapping modes, whereupon conventional detectors or detector arrays measure the modal weights. The question remains as how to systematically construct optical systems that acquire the modal weights $|E_{j}|^{2}$ for non-overlapping spatial modes. One of two general strategies can be followed, which we refer to as `modal projectors' [Fig.~\ref{fig:ProjectionAnalyzer}(a)] and `modal analyzers' [Fig.~\ref{fig:ProjectionAnalyzer}(b)].

\subsubsection{Modal projectors}\label{sec:ModalProjectors}

This strategy relies on constructing a unitary transformation that converts the underlying spatially overlapping modes $\{|\psi_{j}\rangle\}$ into spatially \textit{non}-overlapping modes $\{|\psi_{j}'\rangle\}$ in a one-to-one manner, in which case the complex modal amplitudes are preserved: $|E\rangle=\sum_{j}E_{j}|\psi_{j}\rangle\rightarrow|E'\rangle=\sum_{j}E_{j}|\psi'_{j}\rangle$. Once this conversion is accomplished, 1D or 2D detector arrays can then register the modal weights $I_{j}=|E_{j}|^{2}$ [Fig.~\ref{fig:ProjectionAnalyzer}(a)]. An elementary example of a modal projector is a polarizing beam splitter (PBS) that separates the $|\mathrm{H}\rangle$ and $|\mathrm{V}\rangle$ polarization modes into separate spatial paths. Other examples are the parity projector [Fig.~\ref{fig:DetectionParityFourier}(a)] and Fourier-mode projector [Fig.~\ref{fig:DetectionParityFourier}(b)].

\begin{figure}[b!]
\centering
\includegraphics[width=3.6in] {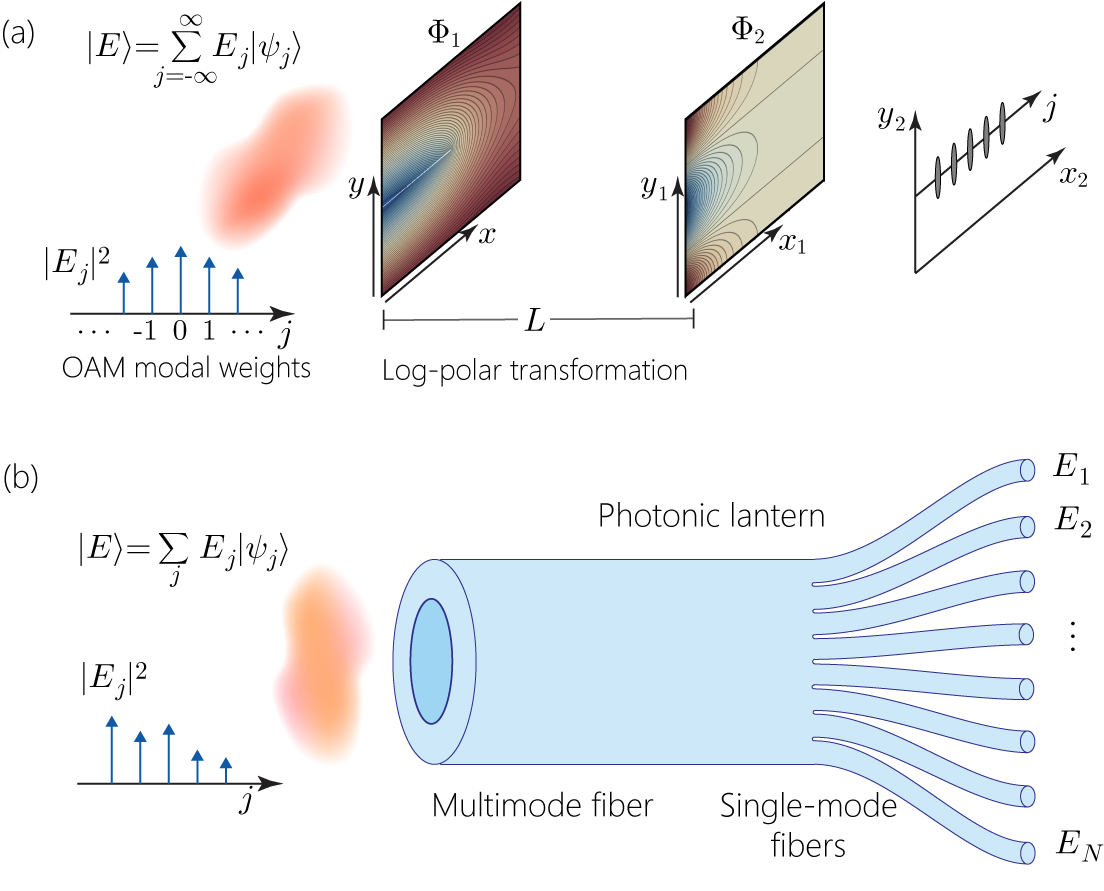}
\caption{(a) Modal projector in the OAM basis via a log-polar coordinate transformation. The device comprises two phase plates $\Phi_{1}$ and $\Phi_{2}$ whose phase distributions are depicted. Initially, the OAM modes are spatially overlapping, but are subsequently separated into spatially non-overlapping modes. (b) An optical lantern that separates the field in a multimode fiber supporting $N$ guided fiber modes into $N$ single-mode fibers (SMFs). Each of the fiber modes shown on the left (constituting the modal basis) is coupled to one of the SMFs on the right.}
\label{fig:LogPolarLantern}
\end{figure}

We consider three examples here. The first example is that of the log-polar coordinate transformation between Cartesian and polar coordinate systems [Fig.~\ref{fig:LogPolarLantern}(a)]. Such a transformation can unravel spatially overlapping OAM modes in a polar coordinate system into transversely displaced spatially non-overlapping patches in a Cartesian coordinate system [Fig.~\ref{fig:LogPolarLantern}(a)] implemented using two phase plates $\Phi_{1}$ and $\Phi_{2}$ separated by a lens in a $2f$ configuration (or alternatively adding the appropriate additional phase distributions to the phase plates to replace the lens). This concept was reported initially in 1974 by Bryngdal \cite{Bryngdahl74JOSA}, and developed and realized subsequently in \cite{Hossack87JOMO,Berkhout10PRL,Lavery12OE}. This approach has been utilized as a modal projector for OAM modes \cite{Li19OE,Yaraghi25NC}, and has also been operated in the opposite direction to produce spatiotemporallly structured optical fields known as space-time wave packets localized in all dimensions \cite{Yessenov22NC,Yessenov22AOP,Yessenov25JOSAA,Yessenov25NC}.

A second example is fiber lanterns \cite{BlandHawthorn09OE,LeonSaval10OE,Davenport21OME} that convert the field confined to a multimode fiber, which comprises spatially overlapping modes, into a set of spatially non-overlapping modes delivered into separate single-mode fibers (SMFs). In Fig.~\ref{fig:LogPolarLantern}(b), a multimode fiber that supports $N$~modes constrains the guided field to be a superposition of these spatially overlapping fiber modes. The modes then couple to $N$~SMFs, one mode to each SMF. The power delivered in each SMF is the modal weight of one mode from the multimode fiber.

\begin{figure}[t!]
\centering
\includegraphics[width=3.8in] {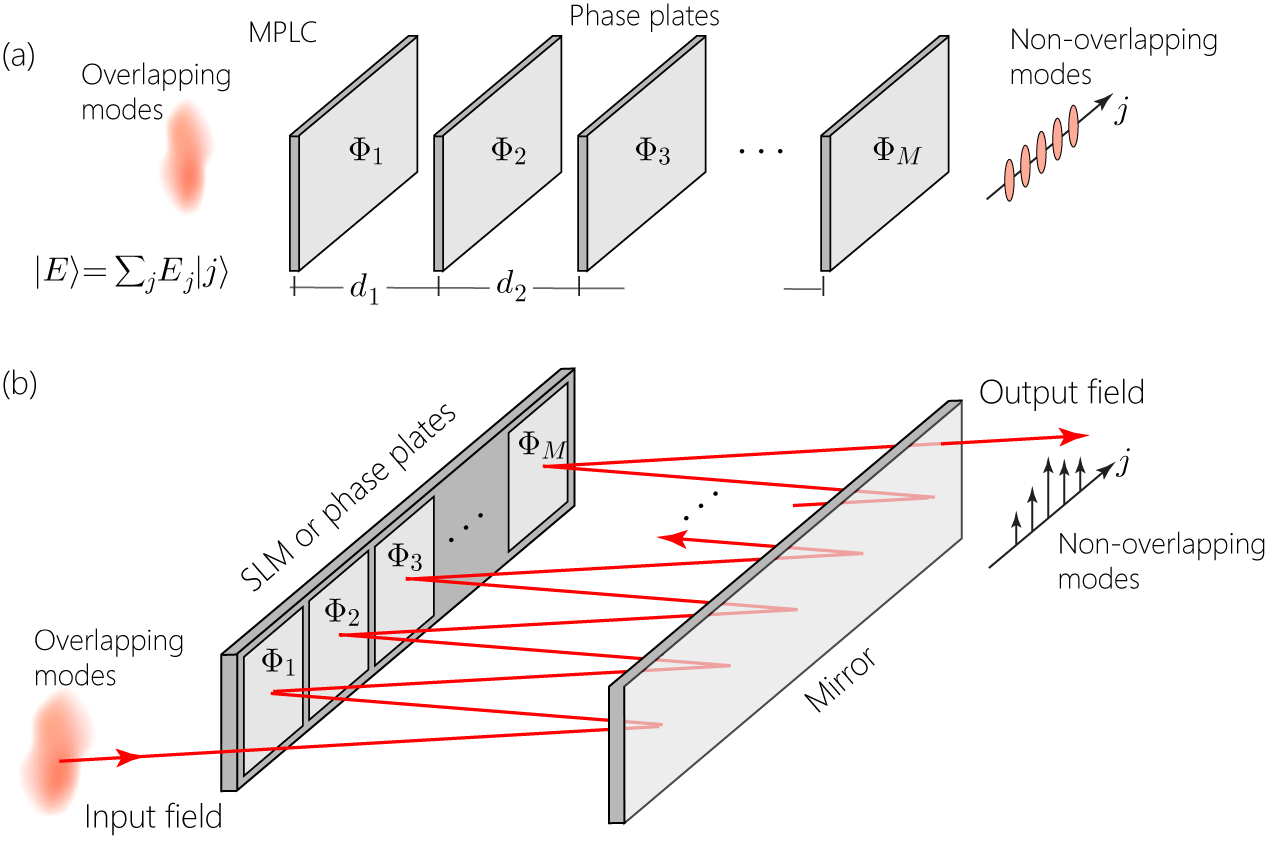}
\caption{(a) Conceptual scheme for an MPLC formed of a sequence of phase plates with phase distributions $\Phi_{j}$ separated by free-space propagation. (b) Realization of an MPLC by multiple reflections between a segmented reflective spatial light modulator (SLM) and a mirror. The input field bounces back and firth between the SLM and the mirror. At the device output, the spatially overlapping modes of the input field are converted to spatially non-overlapping modes.}
\label{fig:MPLC}
\end{figure}

Each of these two examples of modal projectors is designed for a particular modal basis. They are nevertheless extremely useful because they address modal bases that are heavily used in optics (OAM modes and multimode fiber modes). The third example of a modal projector we describe here is the most versatile developed to date. Known as a multi-plane light converter (MPLC) \cite{Mounaix20NC,CruzDelgado22NP,Martinez-Becerril24OE}, this device can map between any pair of modal bases (among other tasks it can perform). For our purpose here, the MPLC is useful in mapping a set of spatially overlapping modes to another set of spatially non-overlapping modes. A conceptual implementation of an MPLC is sketched in Fig.~\ref{fig:MPLC}(a) comprising a set of $M$ thin phase plates imparting phase distribution $\Phi_{1}$ through $\Phi_{M}$ separated by lengths $d_{j}$ ($j=1,\cdots,M$) of free-space propagation to introduce diffraction (lenses in a $2f$ configuration can be used instead). As such, an MPLC has more degrees of freedom than the log-polar coordinate transformation in Fig.~\ref{fig:LogPolarLantern}(a), which is the reason for its versatility.

Rather than implementing a sequence of phase plates [Fig.~\ref{fig:MPLC}(a)], it is more common in practice to use a single reflecting phase plate or SLM that is segmented, with each segment displaying a particular phase distribution. This reflecting phase device is used in conjunction with a mirror as depicted in Fig.~\ref{fig:MPLC}(b). The input field comprising spatially overlapping modes bounces back and forth between the SLM (or reflecting phase plate) and the mirror. The field thus encounters $M$ different phase distributions separated by fixed distances of free-space propagation. At the end of the device, the field is converted into a set of spatially non-overlapping modes that can be delivered to a detector array, thereby obtaining the modal weights.


\subsubsection{Modal analyzers}\label{sec:ModalAnalyzers}

Instead of the detector array required in a modal projector to record the modal weights [Fig.~\ref{fig:ProjectionAnalyzer}(a)], a \textit{modal analyzer} requires a single detector [Fig.~\ref{fig:ProjectionAnalyzer}(b)]. This strategy generalizes to the spatial DoF the well-known approach for obtaining the optical spectrum via an MZI. In a conventional MZI, sweeping a temporal delay inserted in one arm yields a temporal interferogram, followed by a Fourier transform (FT) to obtain the spectral intensity. The conventional temporal delay is replaced for our purposes by a `generalized delay' (GD) that operates in the modal basis of interest. Sweeping this GD yields a \textit{modal} interferogram, whose FT yields the modal weights [Fig.~\ref{fig:ProjectionAnalyzer}(b)]. Because this general strategy is applicable in principle to an arbitrary modal basis, it has been termed a `Hilbert-space analyzer' \cite{Martin17SR}.

\begin{figure}[t!]
\centering
\includegraphics[width=11.3cm]{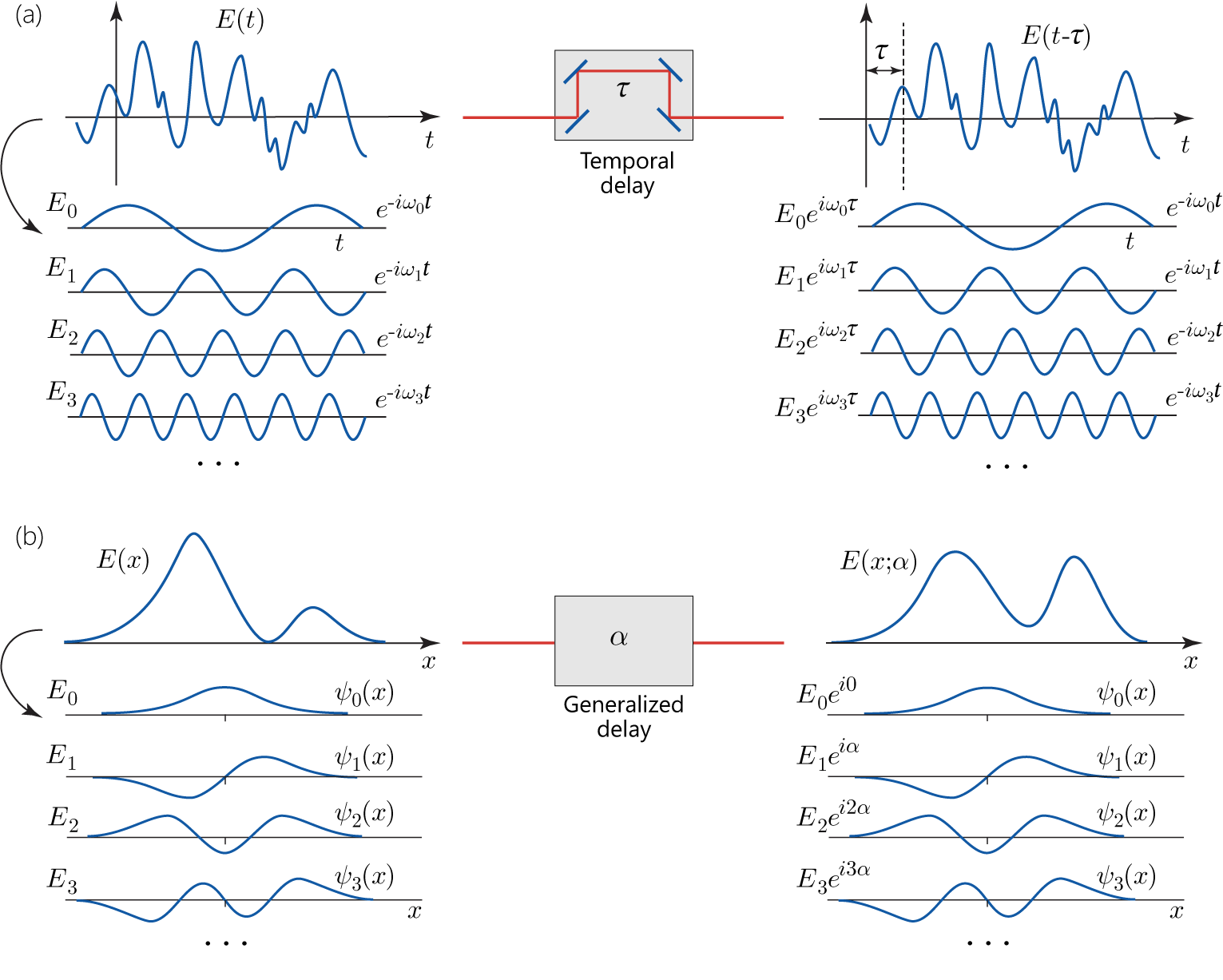}
\caption{(a) Conventional optical delay $\tau$ viewed from the perspective of a modal basis of temporal harmonics $\{e^{-i\omega t}\}$. The field is a superposition of these harmonics, $E(t)=\int\!d\omega\widetilde{E}(\omega)e^{-i\omega t}$, and the delay introduces phases of the form $e^{i\omega\tau}$ to each harmonic without changing the magnitudes, resulting in a displacement operation $\delta(t-t'-\tau)$ on the input field to produce the output $E'(t)=\int\!dt'\delta(t-t'-\tau)E(t')=\int\!d\omega\widetilde{E}(\omega)e^{-i\omega(t-\tau)}=E(t-\tau)$. (b) A generalized optical delay in an arbitrary modal basis operates analogously with the conventional delay in (a). The input field $E(x)=\sum_{n}c_{n}\psi_{n}(x)$ traverses a generalized delay $\alpha$ that introduces relative phases $e^{in\alpha}$ between the modes, yielding $E'(x)=\sum_{n}c_{n}e^{in\alpha}\psi_{n}(x)$.}
\label{fig:GenDelayConcept}
\end{figure}

To set the stage for introducing the concept of a GD, we first examine an MZI containing a temporal optical delay $\tau$ \cite{SalehBook07}. Consider a plane-wave pulse $E(t)=e^{-i\omega_{\mathrm{o}}t}\psi(t)$, where $\omega_{\mathrm{o}}$ is the central frequency, and the slowly varying envelope is: $\psi(t)=\int\!d\Omega\;\widetilde{\psi}(\Omega)e^{-i\Omega t}$, where the spectrum $\widetilde{\psi}(\Omega)$ is the Fourier transform of $\psi(t)$. Taking the modal basis to be the spectral harmonics $\{e^{-i\omega t}\}_{\omega}$, indexed by the temporal frequency $\omega$, the optical delay $\tau$ implements the transformation $E(t)\rightarrow E(t-\tau)$ in the time domain, corresponding in the frequency domain to a phase factor $e^{i\omega\tau}$, $\widetilde{E}(\omega)\rightarrow e^{i\omega\tau}\widetilde{E}(\omega)$, which is linear in the delay $\tau$ and also in the frequency $\omega$ [Fig.~\ref{fig:GenDelayConcept}(a)]. The transformation produced by the delay represented by the operator $\Lambda(t,t';\tau)$ is:
\begin{equation}
\Lambda(t,t';\tau)=\delta(t-t'-\tau)=\int\!d\omega\;e^{-i\omega(t-t'-\tau)}=\int\!d\omega\;e^{i\omega\tau}(e^{-i\omega t})(e^{-i\omega t'})^{*}.
\end{equation}
The spectral harmonics are thus the \textit{eigenfunctions} of the delay operator with eigenvalues $e^{i\omega\tau}$. The time averaged intensity at the MZI output is:
\begin{equation}
I(\tau)=\int\!dt|E(t)+E(t-\tau)|^{2}=1+\mathrm{Re}\int\!dt\;E(t)E^{*}(t-\tau)=1+\int\!d\omega\;|\widetilde{E}(\omega)|^{2}\cos\omega\tau,
\end{equation}
whose FT with respect to $\tau$ reveals the modal weights $|\widetilde{E}(\omega)|^{2}$.

Guided by this picture we generalize these familiar concepts to an arbitrary modal basis $\{\psi_{n}(x)\}$, where $n$ is an integer index, so that the field is expressed as $E(x)=\sum_{n}c_{n}\psi_{n}(x)$, and the modal coefficients are $c_{n}=\int\!dx\;\psi_{n}^{*}(x)E(x)$. The generalized phase operator (GPO) is a linear transformation $\Lambda(x,x';\alpha)$ defined as:
\begin{equation}
E(x)=\sum_{n}c_{n}\psi_{n}(x)\rightarrow E'(x;\alpha)=\int\!dx'\;\Lambda(x,x';\alpha)E(x')=\sum_{n}e^{in\alpha}c_{n}\psi_{n}(x),
\end{equation}
where $\alpha$ is a real continuous parameter. Therefore, $\Lambda$ has the modal basis $\{\psi_{n}(x)\}$ as its eigenfunctions with eigenvalues $e^{in\alpha}$, $\int\!dx'\;\Lambda(x,x';\alpha)\psi_{n}(x')=e^{in\alpha}\psi_{n}(x)$, and $\alpha$ is a generalized delay (GD) in analogy to the conventional temporal optical delay [Fig.~\ref{fig:GenDelayConcept}(b)]. The transformation $\Lambda(x,x';\alpha)$ can thus be expressed as $\Lambda(x,x';\alpha)=\sum_{n}e^{in\alpha}\psi_{n}(x)\psi_{n}^{*}(x')$.

In general, a GPO has the mathematical structure of a group, with the composition rule $\int\!dx'\Lambda(x,x';\alpha)\Lambda(x',x'';\beta)=\Lambda(x,x'';\alpha+\beta)$, $\Lambda(x,x';0)=\delta(x-x')$ is the identity element, and the inverse of $\Lambda(x,x';\alpha)$ is $\Lambda(x,x';-\alpha)$. The definition of $\Lambda$ implies that $\Lambda(\alpha+2m\pi)=\Lambda(\alpha)$ for $m$ integer. The inverse of $\Lambda(x,x';\alpha)$ is $\Lambda(x,x';-\alpha)=\Lambda^{*}(x',x;\alpha)$, so that $\Lambda$ is unitary, $\int\!dx\;\Lambda(x,x';\alpha)\Lambda^{*}(x,x'';\alpha)=\delta(x'-x'')$. In other words, the operator $\Lambda^{*}(x'',x';\alpha)$ undoes the operator $\Lambda(x,x';\alpha)$.

\begin{figure}[t!]
\centering
\includegraphics[width=13.3cm]{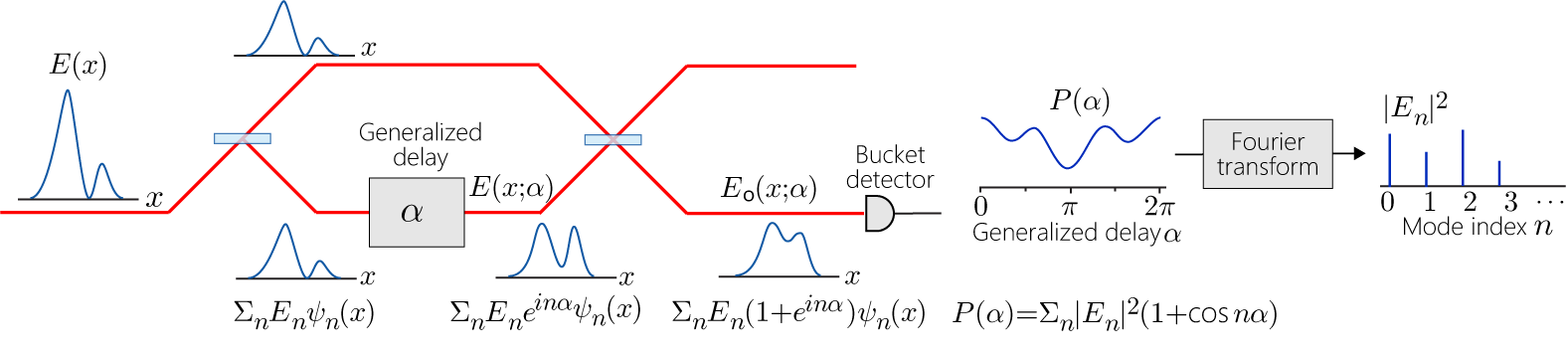}
\caption{Concept of a Hilbert-space modal analyzer. The field is incident from the left, and two copies are produced at the first beam splitter. One copy traverses a generalized optical delay represented by a GPO with parameter $\alpha$, which introduces relative phases between the modes. The delayed and reference fields are superposed, and the resulting field is detected and harmonic analysis performed on the detected signal. This reveals the squared magnitudes of the modal weights.}
\label{fig:GenDelayInterferometer}
\end{figure}

Consider an MZI [Fig.~\ref{fig:GenDelayInterferometer}] with a GD in one arm. The output after integrating over $x$ is:
\begin{equation}
I(\alpha)=\int\!dx\;\big|E(x)+E'(x;\alpha)\big|^{2}=\int\!dx\;\biggr|\sum_{n}E_{n}(1+e^{in\alpha)})\psi_{n}(x)\biggr|^{2}=1+\sum_{n}|E_{n}|^{2}\cos n\alpha.
\end{equation}
The FT of $I(\alpha)$ yields the modal weights $|E_{n}|^{2}$.

We list in Table~\ref{TableSummary} GPOs corresponding to a broad range of DoFs. In addition to conventional spectral analysis, the analogous system for a Fourier modal basis of plane waves $\{e^{ikx}\}$ relies on a `spatial delay' corresponding to a transverse displacement, $\psi(x)\rightarrow\psi(x-\Delta)$, yielding the GPO $\Lambda_{2}(x,x';\Delta)=\delta(x-x'-\Delta)$. Placing this GD in one arm of a balanced MZI yields the modal weights $|\widetilde{E}(k_{x})|^{2}$ at the output. Of course, the lens-based Fourier-analysis system in Fig.~\ref{fig:DetectionParityFourier}(b) is simpler to implement, but this example serves to show that the GPO approach is amenable to systematic generalization.

As an example of a discrete modal basis, consider a field in the radial and azimuthal dimensions written in a modal basis of LG functions $E(\rho,\varphi)=\sum_{\ell,p}c_{\ell p}u_{\ell p}(\rho,\varphi)$, where $u_{\ell p}(\rho,\varphi)=A_{\ell p}\rho^{|\ell|}L_{p}^{|\ell|}(\rho^{2})e^{-\rho^{2}/2}e^{i\ell\varphi}$, $\rho=\sqrt{2}r/w$ is a normalized radial coordinate, $r$ is the radial coordinate, $w$ is the beam waist, $A_{\ell p}=\sqrt{\tfrac{p!}{\pi(p+|\ell|)!}}$ is a normalization constant, and $L_{p}^{|\ell|}$ is the associated Laguerre polynomial. First, consider the OAM modal basis $\{e^{i\ell\varphi}\}$ after setting $p=0$, where the GPO is:
\begin{equation}
\Lambda_{3}(\varphi,\varphi';\alpha)=\frac{1}{2\pi}\sum_{\ell}e^{i\ell\alpha}e^{i\ell\varphi}(e^{i\ell\varphi'})^{*}=\frac{1}{2\pi}\sum_{\ell}e^{i(\varphi-\varphi'+\alpha)}=\delta(\varphi-\varphi'+\alpha),
\end{equation}
which corresponds to a rotation by an angle $\alpha$. Therefore, placing a spatial rotator in one arm of a balanced MZI yields at its output an interferogram $I(\alpha)=1+\sum_{\ell}|c_{\ell}|^{2}\cos\ell\alpha$ when the input field is $E(\varphi)=\sum_{\ell}c_{\ell}e^{i\ell\varphi}$ and $c_{\ell}$ are the expansion coefficients of the field in the OAM basis.

Consider now the radial modes associated with OAM modes, after setting $\ell=0$, whereupon $u_{0p}=A_{0p}L_{p}^{0}(\rho^{2})e^{-\rho^{2}/2}$ and $A_{0p}=\tfrac{1}{\sqrt{\pi}}$. The GPO associated with this basis set would introduce to each mode the phase $e^{ip\alpha}$, so that the GPO takes the form \cite{Abouraddy11OL}:
\begin{equation}
\Lambda_{4}(r,r';\alpha)=\sum_{p=0}^{\infty}e^{ip\alpha}L_{p}^{|\ell|}(r)\left(L_{p}^{|\ell|}(r')\right)^{*}=i\frac{e^{-i\alpha}}{sin\alpha}\exp\left\{-i\frac{\rho^{2}+\rho'^{2}}{2\tan\alpha}\right\}J_{0}\left(\frac{\rho\rho'}{\sin\alpha}\right).
\end{equation}
This is nothing but the \textit{fractional Hankel} transform \cite{Namias80JIMA2} (studied in optics as the limiting form of the 2D fractional Fourier transform in systems with cylindrical symmetry \cite{Namias80JIMA1,Yu98OL,Alieva99OL,Ozaktas01Book}), which can be realized using lenses \cite{Martin17SR}. Combining the two GPOs associated with the radial and azimuthal coordinates allows for finding the modal weights $|c_{\ell p}|^{2}$.

Finally, consider the modal basis comprising HG modes $\{H_{n}(x)\}$ along one transverse dimension, where the field is written as $E(x)=\sum_{n}c_{n}H_{n}(x)$. The GPO $\Lambda_{6}(x,x';\alpha)$ has HG functions as eigenfunctions: $\int\!\Lambda_{5}(x,x';\alpha)H_{n}(x')=e^{in\alpha}H_{n}(x)$, so that it can be expressed as $\Lambda_{6}(x,x';\alpha)=\sum_{n}e^{in\alpha}H_{n}(x)H_{n}^{*}(x')$. It was shown by Namias~\cite{Namias80JIMA1} that this operator corresponds to a fractional Fourier transform:
\begin{equation}
\Lambda_{6}(x,x';\alpha)=\sqrt{1-i\cot\alpha}\exp\left\{i\pi\left(\cot\alpha x^{2}-2\csc\alpha xx'+\cot\alpha x'^{2}\right)\right\},
\end{equation}
which can be realized via combination of lenses \cite{Martin17SR}. Combining fractional Fourier transforms along $x$ and $y$ allows for analysis of any field in terms of a basis of HG functions $\{H_{n}(x)H_{m}(y)\}$.


\begin{table}[t!]
\caption{Realizations of GPO's for generalized interferometry.} \label{TableSummary}

\begin{tabular}{ccccc}
\\ 
\hline 
\textbf{Degree of Freedom} & \textbf{Modes} & \textbf{GPO} & \textbf{Realization} & \textbf{Implementation}
\vspace{0mm}
\\
\hline
\vspace{2mm}

Temporal Spectrum  & $\quad\{e^{-i\omega t}\}_\omega\quad$ &  $\quad\Lambda_{1}(t,t';\tau)\!=\!\delta(t-t'-\tau)\quad$ & delay $\tau$ & \includegraphics[width=2cm]{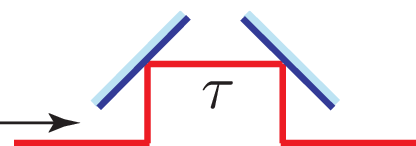}
\\ [0pt]

Spatial Spectrum &  \vspace{-15pt} 
$\{e^{ikx}\}_k$ & \vspace{-15pt}  
$\Lambda_{2}(x,x';\Delta)\!=\!\delta(x-x'+\Delta)$ & \vspace{-15pt} 
transverse shift $\Delta$ & \includegraphics[width=2cm]{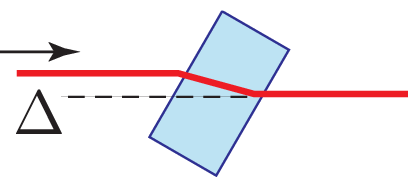}
\\ [.7in]

Angular Momentum & \vspace{-15pt}  $\{e^{i\ell\varphi}\}_\ell$ & \vspace{-15pt}  $\Lambda_{3}(\varphi,\varphi';\theta)\!=\!\delta(\varphi-\varphi'+\theta) $& \vspace{-15pt}  rotation $\theta$ & \includegraphics[width=2cm]{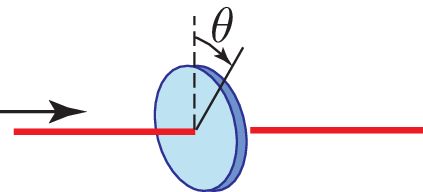}
\\ [.8in]

Radial Modes & \vspace{-15pt} $\{L_p^{|\ell|}(r)\}$ &\vspace{-15pt} $\Lambda_{4}(r,r';\alpha)\!=\!K_p(r,r';\alpha)$ &\vspace{-15pt} fHT of order $\alpha$ & \includegraphics[width=2cm]{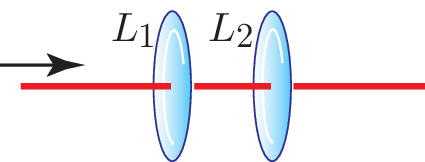}
\\ [.8in]

Transverse Modes & \vspace{-15pt} $\{H_{n}(x)\}_n$ & \vspace{-15pt} $\Lambda_{5}(x,x';\alpha)\!=\!F(x,x';\alpha)$ &\vspace{-15pt} fFT of order $\alpha$ & \includegraphics[width=2cm]{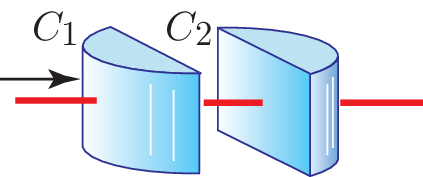}
\\ [.8in]
\hline

\end{tabular}

\end{table}


\subsection{Outlook}

The modal projectors and analyzers described here have all been demonstrated experimentally -- albeit solely with coherent laser light to date. Nevertheless, both approaches are equally applicable to partially coherent light. In other words, the modal projectors and analyzers described above will also reveal the modal weights in a modal basis of interest even when the initial field is partially coherent. So far, only spatially \textit{non}-overlapping modal bases have been utilized in the study of partially coherent light. We anticipate that much work will be done in the years to come on structured coherence utilizing spatially \textit{overlapping} modes.

We proceed to examine in detail structured coherence with a limited number of generic modes assuming the possibility of measuring the modal weights, using either modal projectors or analyzers.


\section{Optical fields in a single binary DoF}\label{sec:OneBinaryDoF}

We now proceed to study in detail partially coherent optical fields described by \textit{one binary DoF}. We consider a generic DoF spanned by two modes $|\psi_{1}\rangle$ and $|\psi_{2}\rangle$ that are normalized so that $\langle\psi_{1}|\psi_{1}\rangle=\langle\psi_{2}|\psi_{2}\rangle=1$, and are orthogonal to each other $\langle\psi_{1}|\psi_{2}\rangle=0$. We will consider first the representation of coherent and partially coherent fields using a generic pair of modes before giving concrete examples using polarization (Section~\ref{sec:polarizationDoF}) and spatial (Section~\ref{sec:SpatialDoF}) modes.

\subsection{Coherent fields}

\begin{figure}[t!]
\centering
\includegraphics[width=4.5cm]{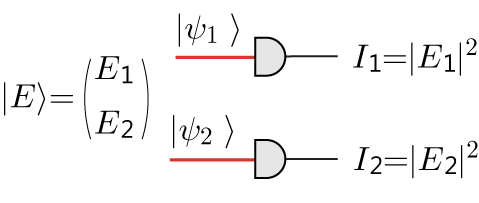}
\caption{The field vector $|E\rangle$ is expressed with complex modal amplitudes $E_{1}$ and $E_{2}$ in a generic modal basis $\{|\psi_{1}\rangle,|\psi_{2}\rangle\}$ represented by two spatial paths (red lines). The modal weights are obtained by the detectors. When the field is coherent, $I_{1}=|E_{1}|^{2}$ and $I_{2}=|E_{2}|^{2}$; and when the field is partially coherent, $I_{1}=G_{11}$ and $I_{2}=G_{22}$.}
\label{fig:BasicBinaryDoF}
\end{figure}

We write the field vector $|E\rangle$ for a \textit{coherent} binary-DoF field as a superposition of the two modes $|E\rangle=E_{1}|\psi_{1}\rangle+E_{2}|\psi_{2}\rangle$, where $E_{1}$ and $E_{2}$ are complex modal amplitudes or coefficients that are normalized to $|E_{1}|^{2}+|E_{2}|^{2}=1$, thereby maintaining $\langle E|E\rangle=1$. Using the representation $|\psi_{1}\rangle=\left(\begin{array}{c}1\\0\end{array}\right)$ and $|\psi_{2}\rangle=\left(\begin{array}{c}0\\1\end{array}\right)$, we have $|E\rangle=\left(\begin{array}{c}E_{1}\\E_{2}\end{array}\right)$. The two modal weights (the squared magnitudes of the complex modal amplitudes) are recorded via the two detectors in Fig.~\ref{fig:BasicBinaryDoF}. These detectors may correspond to a modal projector in the basis of interest (Section~\ref{sec:ModalProjectors}) or to a modal analyzer (Section~\ref{sec:ModalAnalyzers}). In either case, the detectors acquire the modal weights: $I_{1}=|\langle\psi_{1}|E\rangle|^{2}=|E_{1}|^{2}$ and $I_{2}=|\langle\psi_{2}|E\rangle|^{2}=|E_{2}|^{2}$, where $I_{1}$ and $I_{2}$ are thus the fractions of power in the modes $|\psi_{1}\rangle$ and $|\psi_{2}\rangle$, respectively. The field vector may thus be expressed as:
\begin{equation}\label{eq:GeneralStateCoherent1DoF}
|E\rangle=\left(\begin{array}{c}\cos\tfrac{\theta}{2}\\e^{i\varphi}\sin\tfrac{\theta}{2}\end{array}\right),
\end{equation}
where $\varphi$ is the relative phase between the two modes (we ignore the overall phase), and $\theta$ determines the relative modal weights: $I_{1}=\cos^{2}\tfrac{\theta}{2}=\tfrac{1}{2}\{1+\cos\theta\}$ and $I_{2}=\sin^{2}\theta=\tfrac{1}{2}\{1-\cos\theta\}$.

Mathematically, extracting the modal weights can be modeled via projection operators expressed as outer products \cite{Gamo64PO}:
\begin{equation}
\hat{P}_{1}=\left(\begin{array}{cc}1&0\\0&0\end{array}\right)=|\psi_{1}\rangle\langle\psi_{1}|,\;\;\hat{P}_{2}=\left(\begin{array}{cc}0&0\\0&1\end{array}\right)=|\psi_{2}\rangle\langle\psi_{2}|,
\end{equation}
which allow us to write the modal weights as follows:
\begin{eqnarray}\label{eq:ModalWeights}
I_{1}&=&|E_{1}|^{2}=|\langle\psi_{1}|E\rangle|^{2}=\langle E|\psi_{1}\rangle\langle\psi_{1}|E\rangle=\langle E|\hat{P}_{1}|E\rangle=\mathrm{Tr}\left\{\hat{P}_{1}|E\rangle\langle E|\right\},\nonumber\\
I_{2}&=&|E_{2}|^{2}=|\langle\psi_{2}|E\rangle|^{2}=\langle E|\psi_{2}\rangle\langle\psi_{2}|E\rangle=\langle E|\hat{P}_{2}|E\rangle=\mathrm{Tr}\left\{\hat{P}_{2}|E\rangle\langle E|\right\},
\end{eqnarray}
where $\mathrm{Tr}\{\cdot\}$ refers to the trace of a matrix. In Eq.~\ref{eq:ModalWeights} we made use of several general mathematical properties: (1) $(\langle E|\psi_{1}\rangle)^{*}=\langle\psi_{1}|E\rangle$; (2)
$\mathrm{Tr}\{\hat{A}\hat{B}\}=\mathrm{Tr}\{\hat{B}\hat{A}\}$ for matrices $\hat{A}$ and $\hat{B}$; and
(3) for a scalar $\eta$, $\mathrm{Tr}\{\eta\}=\eta$. Because $\hat{P}_{1}+\hat{P}_{2}=\hat{\mathbb{I}}_{2}$, where $\hat{\mathbb{I}}_{2}$ is the $2\times2$ identity matrix, the measurement in Fig.~\ref{fig:BasicBinaryDoF} captures the full power of the field by adding the detector outputs:
\begin{equation}
I_{1}+I_{2}=\mathrm{Tr}\{(\hat{P}_{1}+\hat{P}_{2})|E\rangle\langle E|\}=\mathrm{Tr}\{|E\rangle\langle E|\}=\mathrm{Tr}\{\langle E|E\rangle\}=\langle E|E\rangle=1.    
\end{equation}

This overall formalism is familiar from optical polarization, where $|E\rangle$ corresponds to the so-called Jones vector \cite{Jones41JOSA} for a purely polarized field (see \cite{Brosseau98Book}). However, this formalism is applicable to any binary DoF \cite{Eberly17Optica,Abouraddy19Optica,Halder21OL}.

\subsection{Unitary transformations}\label{sec:UnitarySingleDoF}

We are interested in the coherence dynamics of optical fields traversing a broad range of passive optical systems; e.g., unitary, filtering, and decohering. We consider the first family here and deal with the latter two below in Section~\ref{sec:NonUnitaryOperators} (we do not consider systems with optical gain).

A unitary system operating on a binary DoF is represented mathematically by a $2\times2$  unitary matrix $\hat{U}$ whose defining property is that $\hat{U}\hat{U}^{\dagger}=\hat{U}^{\dagger}\hat{U}=\hat{\mathbb{I}}_{2}$, which implies that the inverse of $\hat{U}$ is its Hermitian conjugate $\hat{U}^{-1}=\hat{U}^{\dagger}$. Conceptually, a \textit{unitary} optical system can be thought of as lossless and reversible. After traversing $\hat{U}$, the field vector $|E\rangle$ becomes $|E'\rangle=\hat{U}|E\rangle$ [Fig.~\ref{fig:PolarizationSpatialUnitaries}(a)]. By cascading after $\hat{U}$ its Hermitian conjugate $\hat{U}^{\dagger}$, the field vector regains its original configuration, $|E''\rangle=\hat{U}^{\dagger}|E'\rangle=\hat{U}^{\dagger}\hat{U}|E\rangle=|E\rangle$ [Fig.~\ref{fig:PolarizationSpatialUnitaries}(b)].

\subsubsection{Properties of unitaries}

\begin{figure}[t!]
\centering
\includegraphics[width=11.3cm]{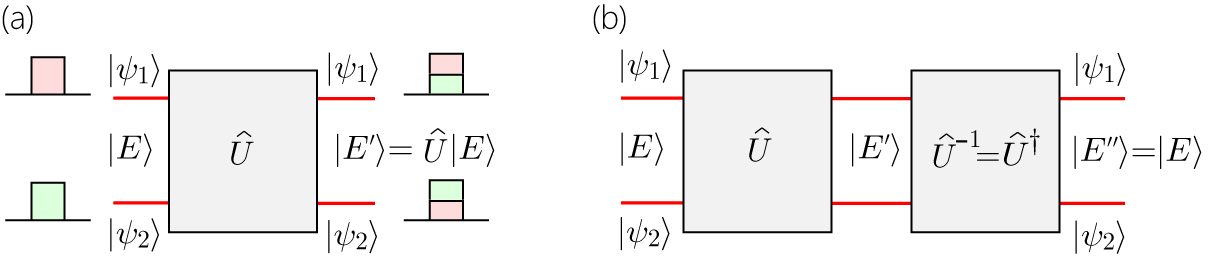}
\caption{(a) A unitary transformation for a binary DoF with a modal basis $\{|\psi_{1}\rangle,\psi_{2}\rangle\}$. The unitary $\hat{U}$ converts the field vector $|E\rangle$ into $|E'\rangle=\hat{U}|E\rangle$. In general, each complex modal amplitude at the output $|E'\rangle$ receives contributions from both input complex modal amplitudes. (b) Cascading the unitary $\hat{U}^{\dagger}$ after $\hat{U}$ corresponds to the identity operator. The final field vector is $|E''\rangle=\hat{U}^{\dagger}|E'\rangle=\hat{U}^{\dagger}\hat{U}|E\rangle=|E\rangle$.}
\label{fig:PolarizationSpatialUnitaries}
\end{figure}

A unitary transformation (or `unitary' henceforth for brevity) converts the field vector $|E\rangle=\left(\begin{array}{c}E_{1}\\E_{2}\end{array}\right)$ into a new field $|E'\rangle=\hat{U}|E\rangle=\left(\begin{array}{c}E_{1}'\\E_{2}'\end{array}\right)$, which is also a superposition of the modes $|\psi_{1}\rangle$ and $|\psi_{2}\rangle$ but with new modal amplitudes [Fig.~\ref{fig:PolarizationSpatialUnitaries}(a)]. The definition of a unitary entails several relevant mathematical properties:
\begin{enumerate}
\item A unitary does not change the length of $|E\rangle$. If $|E'\rangle=\hat{U}|E\rangle$, then $\langle E'|E'\rangle=\langle E|\hat{U}^{\dagger}\hat{U}|E\rangle=\langle E|E\rangle=1$; i.e., unitaries change the components of a field vector while retaining its length.
\item A unitary does not change the `angle' between field vectors: for $|E_{1}'\rangle=\hat{U}|E_{1}\rangle$ and $|E_{2}'\rangle=\hat{U}|E_{2}\rangle$, $\langle E_{1}'|E_{2}'\rangle=\langle E_{1}|\hat{U}^{\dagger}\hat{U}|E_{2}\rangle=\langle E_{1}|E_{2}\rangle$. Orthogonal field vectors remain orthogonal after a unitary.
\item Because $\hat{U}^{\dagger}\hat{U}=\hat{\mathbb{I}}_{2}$, then $\mathrm{det}\{\hat{U}^{\dagger}\hat{U}\}=\mathrm{det}\{\hat{U}^{\dagger}\}\mathrm{det}\{\hat{U}\}=\mathrm{det}\{\hat{U}\}^{*}\mathrm{det}\{\hat{U}\}=|\mathrm{det}\{\hat{U}\}|^{2}$, but $\mathrm{det}\{\hat{U}^{\dagger}\hat{U}\}=\mathrm{det}\{\hat{\mathbb{I}}_{2}\}=1$, so that $|\mathrm{det}\{\hat{U}\}|=1$.
\end{enumerate}

\subsubsection{Representation of a unitary matrix}

The definition of a $2\times2$ unitary $\hat{U}\hat{U}^{\dagger}=\hat{\mathbb{I}}_{2}$ determines its general form when $\mathrm{det}\{\hat{U}\}=1$:
\begin{equation}\label{eq:General2x2unitary}
\hat{U}=\left(\begin{array}{cc}u_{1}&-u_{2}\\u_{2}^{*}&u_{1}^{*}\end{array}\right),
\end{equation}
where $\mathrm{det}\{\hat{U}\}=|u_{1}|^{2}+|u_{2}|^{2}=1$, so the rows and columns of $\hat{U}$ have unit length, the two rows are orthogonal to each other, and the two columns are orthogonal to each other. The two columns of $\hat{U}$ are the field vectors produced after operating on the modes $|\psi_{1}\rangle$ and $|\psi_{2}\rangle$, respectively: $\hat{U}|\psi_{1}\rangle=u_{1}|\psi_{1}\rangle+u_{2}^{*}|\psi_{2}\rangle$ and $\hat{U}|\psi_{2}\rangle=-u_{2}|\psi_{1}\rangle+u_{1}^{*}|\psi_{2}\rangle$, or $\hat{U}\left(\begin{array}{c}1\\0\end{array}\right)=\left(\begin{array}{c}u_{1}\\u_{2}^{*}\end{array}\right)$ and $\hat{U}\left(\begin{array}{c}0\\1\end{array}\right)=\left(\begin{array}{c}-u_{2}\\u_{1}^{*}\end{array}\right)$. The unitary $\hat{U}$ in Eq.~\ref{eq:General2x2unitary} can be expressed in the useful form: 
\begin{equation}\label{eq:2x2U}
\hat{U}=\left(\begin{array}{cc}e^{i\varphi_{1}}\cos\tfrac{\theta}{2}&-e^{-i\varphi_{2}}\sin\tfrac{\theta}{2}\\e^{i\varphi_{2}}\sin\tfrac{\theta}{2}&e^{-i\varphi_{1}}\cos\tfrac{\theta}{2}\end{array}\right),
\end{equation}
where it can be easily checked that $\hat{U}\hat{U}^{\dagger}=\hat{\mathbb{I}}_{2}$. The unitary is thus expressed in terms of two phases $\varphi_{1}$ and $\varphi_{2}$ and a mixing ratio of the two modes determined by $\theta$ (a total of 3~real parameters). It many cases, it is sufficient to utilize a restricted unitary of the form:
\begin{equation}\label{eq:UnitaryRestricted}
\hat{U}=\left(\begin{array}{cc}\cos\tfrac{\theta}{2}&-e^{-i\varphi}\sin\tfrac{\theta}{2}\\e^{i\varphi}\sin\tfrac{\theta}{2}&\cos\tfrac{\theta}{2}\end{array}\right),
\end{equation}
which is characterized by only two real parameters.

\begin{figure}[t!]
\centering
\includegraphics[width=6cm]{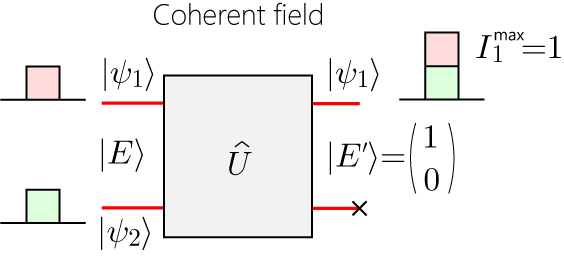}
\caption{One of the modes, $|\psi_{2}\rangle$, is blocked after the field traverses a unitary $\hat{U}$, or we may want to avoid delivering power to $|\psi_{2}\rangle$. For a coherent field, it is guaranteed that a unitary exists such that all the input power, initially distributed between $|\psi_{1}\rangle$ and $|\psi_{2}\rangle$, can be directed to the mode $|\psi_{1}\rangle$. In other words, for any field vector $|E\rangle$, one can always find a unitary $\hat{U}$ such that $\hat{U}|E\rangle=|\psi_{1}\rangle$.}
\label{fig:ConcentratingTheField}
\end{figure}

Consider two orthogonal field vectors $|E\rangle=\left(\begin{array}{c}\cos\tfrac{\theta}{2}\\e^{i\varphi}\sin\tfrac{\theta}{2}\end{array}\right)$ and $|E_{\perp}\rangle=\left(\begin{array}{c}-e^{-i\varphi}\sin\tfrac{\theta}{2}\\\cos\tfrac{\theta}{2}\end{array}\right)$, where $\langle E_{\perp}|E\rangle=0$, which result from the restricted unitary $\hat{U}$ operating on the basis modes: $|E\rangle=\hat{U}|\psi_{1}\rangle$ and $|E_{\perp}\rangle=\hat{U}|\psi_{2}\rangle$. The unitary $\hat{U}'=\hat{U}^{\dagger}$ implements the transformation $\hat{U}'|E\rangle=|\psi_{1}\rangle$ and $\hat{U}'|E_{\perp}\rangle=|\psi_{2}\rangle$. This result leads to a critical observation: a coherent field -- in which the power is distributed among a set of modes -- can always be converted into a single mode via a unitary. In other words, the power of a coherent field can always be concentrated into a single mode, even when initially distributed among the modes. In some scenarios, one of the modes may be `blocked' or is inaccessible, or one may want to avoid delivering power to a particular mode [Fig.~\ref{fig:ConcentratingTheField}]. The question is: how much of the total input power can be delivered to the other mode? For a binary DoF, \textit{coherence} guarantees that if one mode is blocked, \textit{all} the input power can still be directed to the other mode by utilizing the unitary that yields $\hat{U}|E\rangle=|\psi_{1}\rangle$. We will see in Section~\ref{sec:ExtenstioToN} that this statement extends to DoFs with higher dimensionality $N>2$.

\subsubsection{Construction of a general $2\times2$ unitary}

\begin{figure}[t!]
\centering
\includegraphics[width=13.3cm]{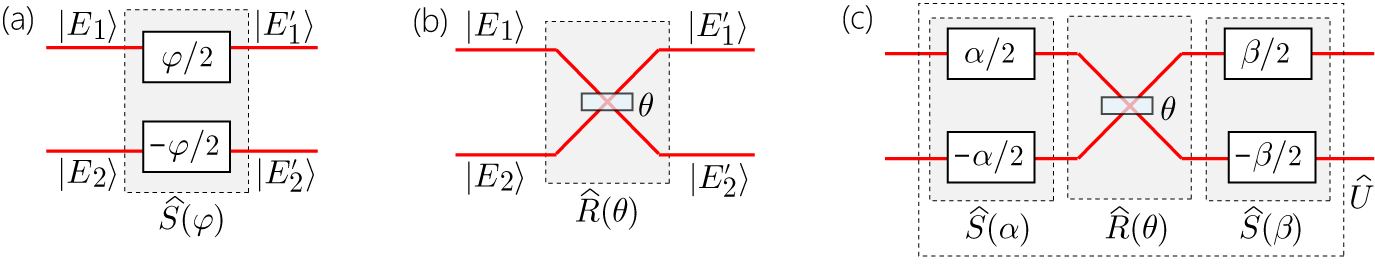} 
\caption{(a) The unitary $\hat{S}(\varphi)$ introduces a phase $\varphi$ between the mode $|\psi_{1}\rangle$ and $|\psi_{2}\rangle$ (Eq.~\ref{eq:PhaseAndRotationUnitaries}). (b) The unitary $\hat{R}(\theta)$ `rotates' the field by an angle $\theta$ (Eq.~\ref{eq:PhaseAndRotationUnitaries} and Eq.~\ref{eq:RotationUnitaries}). (c) A general unitary $\hat{U}=\hat{S}(\beta)\hat{R}(\theta)\hat{S}(\alpha)$ is formed of a sequence of three simpler unitaries (Eq.~\ref{eq:GeneralUnitaryDecomp}).}
\label{fig:2x2Unitary}
\end{figure}

We first consider two basic unitaries depicted in Fig.~\ref{fig:2x2Unitary}(a,b): a phase unitary $\hat{S}(\varphi)$ that introduces a relative phase $\varphi$ between the two modes, and a rotation unitary $\hat{R}(\theta)$ that `rotates' the field by forming superpositions of the two modes, which are given in matrix form as follows:
\begin{equation}\label{eq:PhaseAndRotationUnitaries}
\hat{S}(\varphi)=\left(\begin{array}{cc}e^{i\varphi/2}&0\\0&e^{-i\varphi/2}\end{array}\right),\;\;\;\;\hat{R}(\theta)=\left(\begin{array}{cc}\cos\tfrac{\theta}{2}&-\sin\tfrac{\theta}{2}\\\sin\tfrac{\theta}{2}&\cos\tfrac{\theta}{2}\end{array}\right).
\end{equation}
The phase operator $\hat{S}(\varphi)$ only introduces phase factors: $\hat{S}(\varphi)|\psi_{1}\rangle=e^{i\varphi/2}|\psi_{1}\rangle$ and $\hat{S}(\varphi)|\psi_{2}\rangle=e^{-i\varphi/2}|\psi_{2}\rangle$, whereas the rotation operator $\hat{R}(\theta)$ forms superpositions of the two modes,
\begin{eqnarray}\label{eq:RotationUnitaries}
\hat{R}(\theta)\left(\begin{array}{c}1\\0\end{array}\right)&=&\left(\begin{array}{c}\cos\tfrac{\theta}{2}\\\sin\tfrac{\theta}{2}\end{array}\right),\;\;\;|\psi_{1}\rangle\rightarrow\cos\tfrac{\theta}{2}|\psi_{1}\rangle+\sin\tfrac{\theta}{2}|\psi_{2}\rangle,\nonumber\\
\hat{R}(\theta)\left(\begin{array}{c}0\\1\end{array}\right)&=&\left(\begin{array}{c}-\sin\tfrac{\theta}{2}\\\cos\tfrac{\theta}{2}\end{array}\right),\;\;\;|\psi_{2}\rangle\rightarrow-\sin\tfrac{\theta}{2}|\psi_{1}+\cos\tfrac{\theta}{2}|\psi_{2}\rangle.
\end{eqnarray}
When the components of $|E\rangle=\left(\begin{array}{c}\cos\tfrac{\theta}{2}\\\sin\tfrac{\theta}{2}\end{array}\right)$ are real, the impact of $\hat{R}(\theta')$ can be interpreted as a rotation of a vector in the Cartesian plane, $\hat{R}(\theta')|E\rangle=\left(\begin{array}{c}\cos\tfrac{\theta+\theta'}{2}\\\sin\tfrac{\theta+\theta'}{2}\end{array}\right)$.

The general unitary in Eq.~\ref{eq:2x2U} can be decomposed into a sequence of phase and rotation operators, as depicted in Fig.~\ref{fig:2x2Unitary}(c): from left to right, the field traverses a phase operator that introduces a phase $\alpha$ between the two modes, a `rotation' in the space of the two modes by $\theta$, followed by a second operator that introduces a phase $\beta$ between the two modes:
\begin{equation}\label{eq:GeneralUnitaryDecomp}
\hat{U}=\hat{P}(\beta)\hat{R}(\theta)\hat{P}(\alpha)=
\left(\begin{array}{cc}e^{i\beta/2}&0\\0&e^{-i\beta/2}\end{array}\right)
\left(\begin{array}{cc}\cos\frac{\theta}{2}&-\sin\frac{\theta}{2}\\\sin\frac{\theta}{2}&\cos\frac{\theta}{2}\end{array}\right)
\left(\begin{array}{cc}e^{i\alpha/2}&0\\0&e^{-i\alpha/2}\end{array}\right),
\end{equation}
which corresponds to Eq.~\ref{eq:2x2U} with $\varphi_{1}=\tfrac{\alpha+\beta}{2}$ and $\varphi_{2}=\tfrac{\alpha-\beta}{2}$, or $\alpha=\varphi_{1}+\varphi_{2}$ and $\beta=\varphi_{1}-\varphi_{2}$.

When $\theta=\alpha=\beta=0$, we have $\hat{U}=\hat{\mathbb{I}}_{2}$ and the field vector is invariant: $\hat{U}|E\rangle=|E\rangle$; i.e., the unitary acts as a `relay'. When $\theta=\pi$ and $\alpha=\beta=0$, we have $\hat{U}=\left(\begin{array}{cc}0&-1\\1&0\end{array}\right)$, so that $\hat{U}|\psi_{1}\rangle=|\psi_{2}\rangle$ and $\hat{U}|\psi_{2}\rangle=-|\psi_{1}\rangle$; i.e., the unitary acts as a `switch'. When $\theta=\tfrac{\pi}{2}$ and $\alpha=\beta=0$, each mode is transformed into an equal-weight superposition of the two modes; i.e., a balanced beam splitter.

\subsection{The coherence matrix}\label{sec:CoherenceMatrix1DoF}

\subsubsection{Definition of the coherence matrix}

Whereas the field vector $|E\rangle$ suffices to describe coherent fields associated with a binary DoF, a \textit{partially coherent} field associated with the same binary DoF -- in presence of random fluctuations in the modal amplitudes -- cannot be described by such a vector. Rather, a partially coherent field is described by a $2\times2$ coherence matrix $\mathbf{G}$:
\begin{equation}
\mathbf{G}=\left(\begin{array}{cc}G_{11}&G_{12}\\G_{21}&G_{22}\end{array}\right).
\end{equation}
In lieu of the field vector $|E\rangle$ representing a deterministic field, $\mathbf{G}$ is a mathematical object that describes the impact of statistical fluctuations in the field on measurements of intensity.

The significance of the elements of $\mathbf{G}$ can be understood as follows. We have shown for a coherent field (Eq.~\ref{eq:ModalWeights}) that $I_{1}=\mathrm{Tr}(\hat{P}_{1}|E\rangle\langle E|)$ and $I_{2}=\mathrm{Tr}(\hat{P}_{2}|E\rangle\langle E|)$. Our goal is to find a description of the field that yields equivalent definitions for the measurements of the modal weights after replacing $|E\rangle\langle E|$ in these equations with a new mathematical object that captures the statistical nature of the field. A partially coherent field can be viewed as a statistical ensemble of fields $\{|E(\xi)\rangle\}$, where $\xi$ represents a set of random variables, which are described collectively in terms of a probability distribution $P(\xi)$, with $\int\!d\xi\;P(\xi)=1$. The members of this ensemble are normalized $\langle E(\xi)|E(\xi)\rangle=1$, but need not be orthogonal to each other. Applying Eq.~\ref{eq:ModalWeights} to each member $E(\xi)$ of the ensemble, statistical averaging over the ensemble yields:
\begin{equation}
I_{1}=\int\!d\xi\;P(\xi)\mathrm{Tr}\left\{\hat{P}_{1}|E(\xi)\rangle\langle E(\xi)|\right\}=\mathrm{Tr}\left\{\hat{P}_{1}\mathbf{G}\right\},
\end{equation}
and similarly $I_{2}=\mathrm{Tr}(\hat{P}_{2}\mathbf{G})$, where we now have a definition for the coherence matrix: 
\begin{equation}\label{eq:BasisDefinitionOfG}
\mathbf{G}=\int\!d\xi\;P(\xi)|E(\xi)\rangle\langle E(\xi)|.
\end{equation}
Expressing each field vector $|E(\xi)\rangle$ in the modal basis $\{|\psi_{1},|\psi_{2}\rangle\}$, the matrix elements of $\mathbf{G}$ are:
\begin{eqnarray}
G_{11}\!\!&=&\!\!\langle\psi_{1}|\mathbf{G}|\psi_{1}\rangle=\int\!d\xi\;P(\xi)\langle\psi_{1}|E(\xi)\rangle\langle E(\xi)|\psi_{1}\rangle=\langle E_{1}E_{1}^{*}\rangle=I_{1},\nonumber\\
G_{22}\!\!&=&\!\!\langle\psi_{2}|\mathbf{G}|\psi_{2}\rangle=\int\!d\xi\;P(\xi)\langle\psi_{2}|E(\xi)\rangle\langle E(\xi)|\psi_{2}\rangle=\langle E_{2}E_{2}^{*}\rangle=I_{2},\nonumber\\
G_{12}\!\!&=&\!\!\langle\psi_{1}|\mathbf{G}|\psi_{2}\rangle=\int\!d\xi\;P(\xi)\langle\psi_{1}|E(\xi)\rangle\langle E(\xi)|\psi_{2}\rangle=\langle E_{1}E_{2}^{*}\rangle=G_{21}^{*},
\end{eqnarray}
where $\langle E_{j}E_{k}^{*}\rangle=\int\!d\xi\;P(\xi)E_{j}(\xi)E_{k}^{*}(\xi)$, $j,k=1,2$, and $\langle\cdot\rangle$ is the ensemble average. The matrix elements $G_{11}$ and $G_{22}$ represent the average power in each of the modes $|\psi_{1}\rangle$ and $|\psi_{2}\rangle$, respectively, which are thus detected via the setup in Fig.~\ref{fig:BasicBinaryDoF} ($I_{1}=G_{11}$ and $I_{2}=G_{22}$), and $\mathrm{Tr}\{\mathbf{G}\}=G_{11}+G_{22}$ corresponds to the total power of the field. The normalization $\langle E(\xi)|E(\xi)\rangle=1$ of each member of the ensemble guarantees that the trace of $\mathbf{G}$ is unity: $\mathrm{Tr}\{\mathbf{G}\}=1$. Therefore, $G_{11}$ and $G_{22}$ represent the fractions of power in each mode (normalized to unity), or modal weights.

\begin{figure}[t!]
\centering
\includegraphics[width=13.3cm]{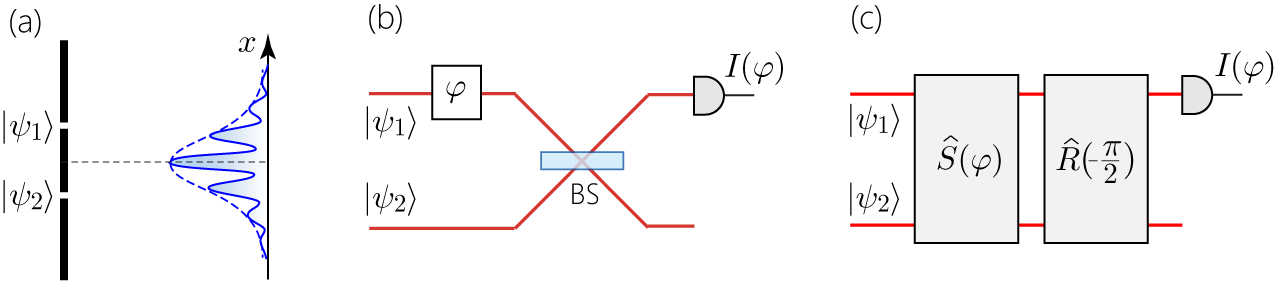}
\caption{(a) Conventional double-slit interference. (b) The field in two paths interfere after introducing a relative phase $\varphi$ and superposition at a beam splitter (BS). (c) In general, for modes $|\psi_{1}\rangle$ and $|\psi_{2}\rangle$, a phase operator $\hat{S}(\varphi)$ followed by a rotation $\hat{R}(-\tfrac{\pi}{2})$ yields an interferogram in a recorded modal weight $I(\varphi)$.}
\label{fig:Interference}
\end{figure}

The off-diagonal coherence-matrix elements $G_{12}$ and $G_{21}=G_{12}^{*}$ are not directly measurable via modal projectors or analyzers as in Fig.~\ref{fig:BasicBinaryDoF}. Rather, they represent quantitatively the statistical correlations between the two field components, and thus indicate the possibility that the fields associated with the two modes can display optical interference when superposed. Specifically, if $G_{12}=0$, no interference is produced when the two modes are superposed. An example is traditional double-slit interference [Fig.~\ref{fig:Interference}(a)], where the visibility is related to $G_{12}$ at the two slit positions. Using a generic pair of modes $|\psi_{1}\rangle$ and $|\psi_{2}\rangle$, the simplest interference setting involves superposing the two modes with equal amplitudes after introducing a phase $\varphi$ between the two modes [Fig.~\ref{fig:Interference}(b)]. The detected intensity is $I=\tfrac{1}{2}\langle|e^{i\varphi}E_{1}+E_{2}|^{2}\rangle=\tfrac{1}{2}\{I_{1}+I_{2}+2\mathrm{Re}\langle E_{1}E_{2}^{*}\rangle e^{i\varphi}\}=\tfrac{1}{2}+|G_{12}|\cos\left(\chi+\varphi\right)$, where $I_{1}=\langle E_{1}E_{1}^{*}\rangle=G_{11}$, $I_{2}=\langle E_{2}E_{2}^{*}\rangle=G_{22}$, $I_{1}+I_{2}=\mathrm{Tr}\{\mathbf{G}\}=1$, and $G_{12}=|G_{12}|e^{i\chi}$. As the relative phase $\varphi$ is swept, interference fringes with alternating maxima and minima are observed. The minimum and maximum values of $I(\varphi)$ are $I_{\mathrm{min}}=\tfrac{1}{2}-|G_{12}|$ and $I_{\mathrm{max}}=\tfrac{1}{2}+|G_{12}|$, respectively. Michelson defined the visibility $V$ as \cite{Michelson1891PM1}:
\begin{equation}
V=\frac{I_{\mathrm{max}}-I_{\mathrm{min}}}{I_{\mathrm{max}}+I_{\mathrm{min}}},
\end{equation}
whereupon $V=2|G_{12}|$. The phase $\chi$ of $G_{12}$ shifts the interference fringes: $I_{\mathrm{max}}$ is achieved when $\varphi=-\mathrm{arg}\{G_{12}\}$ rather than at $\varphi=0$.

\subsubsection{Unitaries operating on the coherence matrix}

As described above, a unitary $\hat{U}$ transforms a coherent field according to $|E\rangle\rightarrow|E'\rangle=\hat{U}|E\rangle$. The definition of $\mathbf{G}$ in Eq.~\ref{eq:BasisDefinitionOfG} involves an outer product of each member of the ensemble, which implies that $\hat{U}$ converts a coherence matrix as follows:
\begin{equation}
\mathbf{G}\rightarrow\mathbf{G}'=\hat{U}\mathbf{G}\hat{U}^{\dagger}.    
\end{equation}
Interferometers are unitaries that convert $\mathbf{G}$ in such a way that the off-diagonal elements that quantify the correlations between the two modes are brought onto the diagonal of $\mathbf{G}$, so that they can be detected directly by measuring the modal weights (or via a power measurement). For example, consider when $\hat{U}=\hat{S}(-\varphi)\hat{R}(-\tfrac{\pi}{2})\hat{S}(\varphi)=\tfrac{1}{\sqrt{2}}\left(\begin{array}{cc}1&e^{-i\varphi}\\-e^{i\varphi}&1\end{array}\right)$, whereupon $\mathbf{G}'=\hat{U}\mathbf{G}\hat{U}^{\dagger}$, so that $I_{1}'=G_{11}'=\tfrac{1}{2}(1+2\mathrm{Re}\{G_{12}e^{i\varphi}\})$ and $I_{2}'=G_{22}'=\tfrac{1}{2}(1-2\mathrm{Re}\{G_{12}e^{i\varphi}\})$. We obtain at each output of the unitary an interferogram, where the visibility of the interference fringes is $V=2|G_{12}|$ in both cases [Fig.~\ref{fig:Interference}(c)]. The two interferograms are complementary, $I_{1}'+I_{2}'=1$, so that the maxima and minima of $I_{1}'$ coincide with the minima and maxima of $I_{2}'$.

\subsubsection{Properties of the coherence matrix}

In addition to the trace of $\mathbf{G}$, $\mathrm{Tr}\{\mathbf{G}\}=G_{11}+G_{22}=1$, we also define its determinant, $\mathrm{det}\{\mathbf{G}\}=G_{11}G_{22}-G_{12}G_{21}=G_{11}G_{22}-|G_{12}|^{2}$. Both the trace and the determinant are unitary invariants, $\mathrm{Tr}\{\hat{U}\mathbf{G}\hat{U}^{\dagger}\}=\mathrm{Tr}\{\mathbf{G}\}$ and $\mathrm{det}\{\hat{U}\mathbf{G}\hat{U}^{\dagger}\}=\mathrm{det}\{\mathbf{G}\}$, which follow from the fact that $\mathrm{Tr}\{\hat{A}\hat{B}\}=\mathrm{Tr}\{\hat{B}\hat{A}\}$ and $\mathrm{det}\{\hat{A}\hat{B}\}=\mathrm{det}\{\hat{B}\hat{A}\}$ for any two matrices $\hat{A}$ and $\hat{B}$.

The coherence matrix $\mathbf{G}$ has the following relevant mathematical properties:
\begin{enumerate}
\item $\mathbf{G}$ is Hermitian, $\mathbf{G}^{\dagger}=\mathbf{G}$, which entails the following characteristics:
    \begin{enumerate}
    \item $G_{11}$ and $G_{22}$ are real and $G_{12}=G_{21}^{*}$.
    \item Both $\mathrm{Tr}\{\mathbf{G}\}$ and $\mathrm{det}\{\mathbf{G}\}$ are real.
    \item The eigenvalues $\lambda_{1}$ and $\lambda_{2}$ of $\mathbf{G}$ are real, and are given by:
    \begin{equation}\label{eq:EigenvalesNoNormalization}
    \lambda_{1,2}=\frac{\mathrm{Tr}\{\mathbf{G}\}}{2}\pm\sqrt{\left(\frac{\mathrm{Tr}\{\mathbf{G}\}}{2}\right)^{2}-\mathrm{det}\{\mathbf{G}\}}. 
    \end{equation}
    \item The eigenvectors of $\mathbf{G}$ corresponding to $\lambda_{1}$ and $\lambda_{2}$ are orthogonal when $\lambda_{1}\neq\lambda_{2}$.
    \item $\mathbf{G}$ can be diagonalized via a $2\times2$ unitary $\hat{U}$, whereby $\mathbf{G}$ is converted into the diagonal form $\mathbf{G}^{\mathrm{D}}$:
        \begin{equation}        \mathbf{G}^{\mathrm{D}}=\hat{U}\mathbf{G}\hat{U}^{\dagger}=\left(\begin{array}{cc}\lambda_{1}&0\\0&\lambda_{2}\end{array}\right).
        \end{equation}
    \end{enumerate}
Unless otherwise stated, the eigenvalues are arranged in descending value, $\lambda_{1}\geq\lambda_{2}$.
\item In our normalization scheme, we have $\mathrm{Tr}\{\mathbf{G}\}=1$, in which case:
\begin{equation}
\mathrm{Tr}\{\mathbf{G}\}=\mathrm{Tr}\{\mathbf{G}^{\mathrm{D}}\}=\lambda_{1}+\lambda_{2}=1,
\end{equation}
with $0\leq|\lambda_{1}-\lambda_{2}|\leq1$, and the eigenvalues given explicitly by:
\begin{equation}\label{eq:EigenvaluesUnityTrace}
\lambda_{1,2}=\frac{1}{2}\pm\frac{1}{2}\sqrt{1-4\mathrm{det}\{\mathbf{G}\}}=\frac{1}{2}\pm\sqrt{\left(\frac{G_{11}-G_{22}}{2}\right)^{2}+|G_{12}|^{2}}.
\end{equation}
\item $\mathbf{G}$ is positive semi-definite: $\lambda_{1},\lambda_{2}\geq0$. This can be confirmed directly by inspecting Eq.~\ref{eq:EigenvaluesUnityTrace}. Consequently, the diagonal elements $G_{11},G_{22}\geq0$ are always positive (see Eq.~\ref{eq:General2x2CoherenceMatrix} below).
\end{enumerate}

\subsubsection{General form of the coherence matrix}

A general form of the coherence matrix can be obtained starting from the diagonal form $\mathbf{G}^{\mathrm{D}}=\left(\begin{array}{cc}\lambda_{1}&0\\0&\lambda_{2}\end{array}\right)$, where $\lambda_{1}+\lambda_{2}=1$ and $1\geq\lambda_{1}\geq\lambda_{2}\geq0$. Utilizing the general unitary $\hat{U}$ from Eq.~\ref{eq:2x2U}, we have:
\begin{equation}\label{eq:General2x2CoherenceMatrix}
\mathbf{G}=\hat{U}\mathbf{G}^{\mathrm{D}}\hat{U}^{\dagger}=\frac{1}{2}\left(\begin{array}{cc}1+(\lambda_{1}-\lambda_{2})\cos\theta&(\lambda_{1}-\lambda_{2})e^{-i\varphi}\sin\theta\\(\lambda_{1}-\lambda_{2})e^{i\varphi}\sin\theta&1-(\lambda_{1}-\lambda_{2})\cos\theta\end{array}\right),
\end{equation}
where only a single phase $\varphi=\varphi_{1}-\varphi_{2}$ remains. Note that the same general coherence matrix is obtained when the restricted unitary in Eq.~\ref{eq:UnitaryRestricted} operates on $\mathbf{G}^{\mathrm{D}}$. The coherence matrix can be diagonalized by the unitary $\hat{U}^{\dagger}$, $\mathbf{G}^{\mathrm{D}}=\hat{U}^{\dagger}\mathbf{G}\hat{U}$. The eigenvectors of $\mathbf{G}$ are:
\begin{equation}
|u_{1}\rangle=\left(\begin{array}{c}\cos\tfrac{\theta}{2}\\e^{i\varphi}\sin\tfrac{\theta}{2}\end{array}\right),\;\;|u_{2}\rangle=\left(\begin{array}{c}-e^{-i\varphi}\sin\tfrac{\theta}{2}\\\cos\tfrac{\theta}{2}\end{array}\right),
\end{equation}
where $\mathbf{G}|u_{1}\rangle=\lambda_{1}|u_{1}\rangle$ and $\mathbf{G}|u_{2}\rangle=\lambda_{2}|u_{2}\rangle$.

\begin{figure}[t!]
\centering
\includegraphics[width=6cm]{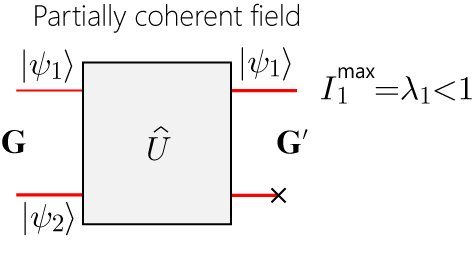}
\caption{The mode $|\psi_{2}\rangle$ is blocked after the field traverses a unitary $\hat{U}$, or we may want to avoid delivering power to $|\psi_{2}\rangle$. For a partially coherent field, a unitary does \textit{not} exist that can direct all the input power to the unblocked mode $|\psi_{1}\rangle$ -- in contrast to the case for a coherent field [Fig.~\ref{fig:ConcentratingTheField}]. For any coherence matrix $\mathbf{G}$, the maximum output at mode $|\psi_{1}\rangle$ is $\lambda_{1}<1$ (the larger of the two eigenvalues of $\mathbf{G}$). There always remains power in $|\psi_{2}\rangle$.}
\label{fig:PartiallyCoherentBlocked}
\end{figure}

We now pose the same question for a partially coherent field that we tackled above for a coherent field: when a partially coherent field described by a coherence matrix $\mathbf{G}$ traverses a unitary $\hat{U}$, how much of the input power -- initially distributed between $|\psi_{1}\rangle$ and $|\psi_{2}\rangle$ ($G_{11}\neq0$ and $G_{22}\neq0$) -- can be delivered to the mode $|\psi_{1}\rangle$? From the general form of $\mathbf{G}$ in Eq.~\ref{eq:General2x2CoherenceMatrix}, $I_{1}=G_{11}=\tfrac{1}{2}\{1+(\lambda_{1}-\lambda_{2})\cos\theta\}$ reaches a maximum value of $I_{1}^{\mathrm{max}}=\tfrac{1}{2}\{1+(\lambda_{1}-\lambda_{2})\}=\lambda_{1}<1$ at $\theta=0$. Because $\lambda_{1}<1$ for partially coherent fields, it is no longer possible to deliver all the input power to the single mode $|\psi_{1}\rangle$, and consequently the mode $|\psi_{2}\rangle$ can\textit{not} be fully avoided [Fig.~\ref{fig:PartiallyCoherentBlocked}]. This optimal power concentration corresponds to diagonalizing $\mathbf{G}$, which is achieved via the Hermitian conjugate of the unitary $\hat{U}$ in Eq.~\ref{eq:UnitaryRestricted}.

Note that $G_{11}G_{22}=\tfrac{1}{4}\{1-(\lambda_{1}-\lambda_{2})^{2}\cos^{2}\theta\}$ and $|G_{12}|^{2}=\tfrac{1}{4}(\lambda_{1}-\lambda_{2})^{2}\sin^{2}\theta$, so that $\mathrm{det}\{\mathbf{G}\}=\tfrac{1}{4}\{1-(\lambda_{1}-\lambda_{2})^{2}\}=\lambda_{1}\lambda_{2}\geq0$, so that $G_{11}G_{22}\geq|G_{12}|^{2}$ (which is also an expression of the Schwartz inequality). For future reference, we define the difference between the diagonal elements:
\begin{equation}\label{eq:2X2Delta}
\Delta=G_{11}-G_{22}=(\lambda_{1}-\lambda_{2})\cos\theta,
\end{equation}
which is \textit{not} a unitary invariant because it contains $\theta$. The maximum value attained by $\Delta$ under unitaries is $\Delta_{\mathrm{max}}=\lambda_{1}-\lambda_{2}$ when $\theta=0$, at which point $\mathbf{G}$ is diagonalized $\mathbf{G}\rightarrow\mathbf{G}^{\mathrm{D}}=\left(\begin{array}{cc}\lambda_{1}&0\\0&\lambda_{2}\end{array}\right)$. Alternatively, when $\theta=\tfrac{\pi}{2}$, then $\Delta=0$, $G_{11}=G_{22}=\tfrac{1}{2}$, and $|G_{12}|=\tfrac{1}{2}(\lambda_{1}-\lambda_{2})$.

\subsubsection{Degree of coherence for the coherence matrix}\label{sec:DegreeOfCoherenceOfSingleDoF}

Two limit cases for partially coherent optical fields stand out with reference to the eigenvalues of the associated coherence matrix. The first limit is when $\lambda_{1}=1$ and $\lambda_{2}=0$, whereupon Eq.~\ref{eq:General2x2CoherenceMatrix} becomes:
\begin{equation}
\mathbf{G}=\tfrac{1}{2}\left(\begin{array}{cc}1+\cos\theta&e^{-i\varphi}\sin\theta\\e^{i\varphi}\sin\theta&1-\cos\theta\end{array}\right)=|E\rangle\langle E|,\;\;\;|E\rangle=\left(\begin{array}{c}\cos\tfrac{\theta}{2}\\e^{i\varphi}\sin\tfrac{\theta}{2}\end{array}\right).  
\end{equation}
The coherence matrix in this case is an outer product (or projection operator) of a field vector, which is the hallmark of a \textit{fully coherent field} \cite{Gamo64PO}. Any unitary conserves $\lambda_{1}$ and $\lambda_{2}$, and  thus does not change the coherence of the field. The second limit occurs when $\lambda_{1}=\lambda_{2}=\tfrac{1}{2}$, whereupon Eq.~\ref{eq:General2x2CoherenceMatrix} becomes $\mathbf{G}=\tfrac{1}{2}\left(\begin{array}{cc}1&0\\0&1\end{array}\right)$. We take this limit to correspond to a \textit{maximally incoherent field}. The off-diagonal elements are zero in this diagonal form, and they remain non-zero after \textit{any} unitary, $\hat{U}\mathbf{G}\hat{U}^{\dagger}=\mathbf{G}$, so no interference can be observed. The general case where $\lambda_{1}>\lambda_{2}>0$ corresponds to \textit{partially coherent} fields, in which case $\mathbf{G}$ cannot be expressed as a projection, but interference can be nevertheless observed upon an appropriate unitary.

Based on these considerations, we define a quantitative measure to characterize the coherence of the field, which we call the `degree of coherence' $D$:
\begin{equation}
D=\lambda_{1}-\lambda_{2},\;\;\lambda_{1}\geq\lambda_{2}.
\end{equation}
Using the expressions for $\lambda_{1}$ and $\lambda_{2}$ in Eq.~\ref{eq:EigenvaluesUnityTrace}, we can define $D$ is terms of the elements of $\mathbf{G}$:
\begin{equation}
D=\sqrt{1-4\mathrm{det}\{\mathbf{G}\}}=\sqrt{(G_{11}-G_{22})^{2}+4|G_{12}|^{2}},
\end{equation}
the eigenvalues can be expressed as $\lambda_{1}=\tfrac{1}{2}(1+D)$ and $\lambda_{2}=\tfrac{1}{2}(1-D)$, and the coherence matrix can then be written as:
\begin{equation}\label{eq:GeneralGwithD}
\mathbf{G}=\frac{1}{2}\left(\begin{array}{cc}1+D\cos\theta&e^{-i\phi}D\sin\theta\\e^{i\phi}D\sin\theta&1-D\cos\theta\end{array}\right).
\end{equation}

\begin{figure}[t!]
\centering
\includegraphics[width=5.25in]{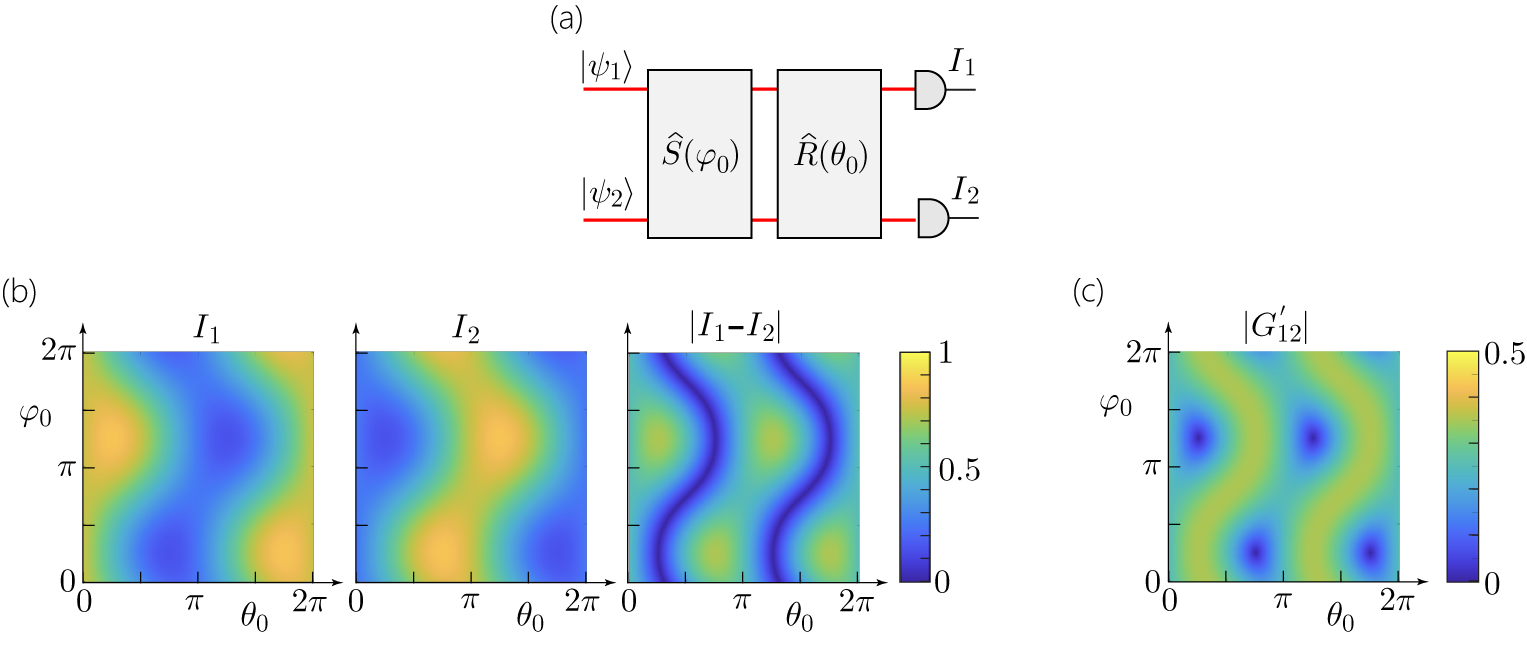}
\caption{(a) General two-mode interference, whereby the input field traverses a unitary $\hat{U}(\theta_{\mathrm{o}},\varphi_{\mathrm{o}})$ in which two angular parameters $\theta_{\mathrm{o}}$ and $\varphi_{\mathrm{o}}$ are scanned while recording the modal weights $I_{1}$ and $I_{2}$. (b) Plots of $I_{1}(\theta_{\mathrm{o}},\varphi_{\mathrm{o}})$, $I_{2}(\theta_{\mathrm{o}},\varphi_{\mathrm{o}})$, and $\Delta(\theta_{\mathrm{o}},\varphi_{\mathrm{o}})=I_{1}(\theta_{\mathrm{o}},\varphi_{\mathrm{o}})-I_{2}(\theta_{\mathrm{o}},\varphi_{\mathrm{o}})$
for an input $\mathbf{G}$ (Eq.~\ref{eq:GeneralGwithD}) with $\theta=\tfrac{\pi}{4}$, $\varphi=\tfrac{\pi}{4}$, and $D=0.5$. (c) Plot of $|G_{12}'(\theta_{\mathrm{o}},\varphi_{\mathrm{o}})|$ corresponding to (b).}
\label{fig:GeneralInterference}
\end{figure}

The degree of coherence is a unitary invariant and is limited to the range $0\leq D\leq1$, with $D=1$ corresponding to a completely \textit{coherent} field ($\lambda_{1}=1$ and $\lambda_{2}=0$), $D=0$ corresponding to a completely \textit{incoherent} field ($\lambda_{1}=\lambda_{2}=\tfrac{1}{2}$), and partially coherent otherwise. In general, $\Delta=|G_{11}-G_{22}|=(\lambda_{1}-\lambda_{2})\cos\theta\leq D$ and $|G_{12}|=\tfrac{1}{2}(\lambda_{1}-\lambda_{2})|\sin\theta|$. The visibility of the interference fringes observed after overlapping the fields associated with the two modes and varying a relative phase is $V=2|G_{12}|=(\lambda_{1}-\lambda_{2})|\sin\theta|\leq D$. In general, the visibility does not reveal $D$. This requires first implementing a unitary that results in $G_{11}=G_{22}=\tfrac{1}{2}$, $\Delta=0$, $|G_{12}|=\tfrac{1}{2}(\lambda_{1}-\lambda_{2})$, and $V=2|G_{12}|=\lambda_{1}-\lambda_{2}=D$. In other words, the visibility $V=2|G_{12}|$ reaches a maximum value of $V=D$ when $\Delta=0$, corresponding to $\theta=\tfrac{\pi}{2}$ in Eq.~\ref{eq:GeneralGwithD}.

This necessitates a unitary that changes the modal weights and not only introduces a relative phase [Fig.~\ref{fig:GeneralInterference}(a)]. The input coherence matrix is given by the general form in Eq.~\ref{eq:GeneralGwithD}, which traverses a phase operator $\hat{S}(\varphi_{\mathrm{o}})$ followed by a rotator $\hat{R}(\theta_{\mathrm{o}})$ and another phase operator $\hat{S}(-\varphi_{\mathrm{o}})$, which combine to form the unitary $\hat{U}(\theta_{\mathrm{o}},\varphi_{\mathrm{o}})=\left(\begin{array}{cc}\cos\tfrac{\theta_{\mathrm{o}}}{2}&
-e^{-i\varphi_{\mathrm{o}}}\sin\tfrac{\theta_{\mathrm{o}}}{2}\\
e^{i\varphi_{\mathrm{o}}}\sin\tfrac{\theta_{\mathrm{o}}}{2}&
\cos\tfrac{\theta_{\mathrm{o}}}{2}\end{array}\right)$. The new coherence matrix is $\mathbf{G}'=\hat{U}\mathbf{G}\hat{U}^{\dagger}$, and the detectors record the modal weights and their difference $\Delta$:
\begin{eqnarray}
I_{1}(\theta_{\mathrm{o}},\varphi_{\mathrm{o}})&=&G_{11}'=\frac{1}{2}\left\{1+D\cos\theta\cos\theta_{\mathrm{o}}-D\sin\theta\sin\theta_{\mathrm{o}}\cos(\varphi-\varphi_{\mathrm{o}})\right\},\nonumber\\
I_{2}(\theta_{\mathrm{o}},\varphi_{\mathrm{o}})&=&G_{22}'=\frac{1}{2}\left\{1-D\cos\theta\cos\theta_{\mathrm{o}}+D\sin\theta\sin\theta_{\mathrm{o}}\cos(\varphi-\varphi_{\mathrm{o}})\right\},\nonumber\\
\Delta(\theta_{\mathrm{o}},\varphi_{\mathrm{o}})&=&I_{1}(\theta_{\mathrm{o}},\varphi_{\mathrm{o}})-I_{2}(\theta_{\mathrm{o}},\varphi_{\mathrm{o}})=D\cos\theta\cos\theta_{\mathrm{o}}-D\sin\theta\sin\theta_{\mathrm{o}}\cos(\varphi-\varphi_{\mathrm{o}}),\nonumber\\
G_{12}'(\theta_{\mathrm{o}},\varphi_{\mathrm{o}})&=&\frac{1}{2}De^{-i\varphi_{\mathrm{o}}}\left\{\cos\theta\sin\theta_{\mathrm{o}}+\sin\theta\left[\cos\theta_{\mathrm{o}}\cos(\varphi_{\mathrm{o}}-\varphi)+i\sin(\varphi_{\mathrm{o}}-\varphi)\right]\right\}.
\end{eqnarray}
We plot in Fig.~\ref{fig:GeneralInterference}(b) $I_{1}(\theta_{\mathrm{o}},\varphi_{\mathrm{o}})$ and $I_{2}(\theta_{\mathrm{o}},\varphi_{\mathrm{o}})$ for fixed values of $\theta$, $\varphi$, and $D$ in the input coherence matrix $\mathbf{G}$. The maxima and minima of $I_{1}$ and $I_{2}$ occur at $\varphi_{\mathrm{o}}=\varphi$. Setting $\varphi_{\mathrm{o}}=\varphi$ yields the simplified equations $I_{1}(\theta_{\mathrm{o}})=\tfrac{1}{2}\{1+D\cos(\theta_{\mathrm{o}}+\theta)\}$, $I_{2}(\theta_{\mathrm{o}})=\tfrac{1}{2}\{1-\cos(\theta_{\mathrm{o}}+\theta)\}$, $\Delta(\theta_{\mathrm{o}})=D\cos(\theta_{\mathrm{o}}+\theta)$, and $G_{12}'(\theta_{\mathrm{o}})=\tfrac{1}{2}e^{-i\varphi}D\sin(\theta_{\mathrm{o}}+\theta)$. We then have $I_{1}^{\mathrm{max}}=I_{2}^{\mathrm{max}}=\tfrac{1}{2}\{1+D\}$ and $I_{1}^{\mathrm{min}}=I_{2}^{\mathrm{min}}=\tfrac{1}{2}\{1-D\}$, but $I_{1}^{\mathrm{max}}$ and $I_{2}^{\mathrm{min}}$ occur at $\theta_{\mathrm{o}}=-\theta$, while $I_{1}^{\mathrm{min}}$ and $I_{2}^{\mathrm{max}}$ occur at $\theta_{\mathrm{o}}=-\theta+\pi$. Both interferograms observed in $I_{1}$ and $I_{2}$ have a visibility $V=D$. We also plot in Fig.~\ref{fig:GeneralInterference}(b) $\Delta(\theta_{\mathrm{o}},\varphi_{\mathrm{o}})$ and $|G_{12}(\theta_{\mathrm{o}},\varphi_{\mathrm{o}})|$ in Fig.~\ref{fig:GeneralInterference}(c), with $\Delta_{\mathrm{max}}=D=\lambda_{1}-\lambda_{2}$ occuring when $\theta_{\mathrm{o}}=-\theta$ and $\varphi_{\mathrm{o}}=\varphi$, accompanied by the minimization of $|G_{12}'|=0$. On the other hand, $\Delta=0$ occurs when $\varphi_{\mathrm{o}}=\varphi$ and $\theta_{\mathrm{o}}=-\theta+\tfrac{\pi}{2}$ is accompanied by a maximization of $|G_{12}'|=\tfrac{1}{2}D$.

\subsubsection{Entropy of the coherence matrix}\label{sec:EntropyOfOneDoF}

\begin{figure}[t!]
\centering
\includegraphics[width=2.5in]{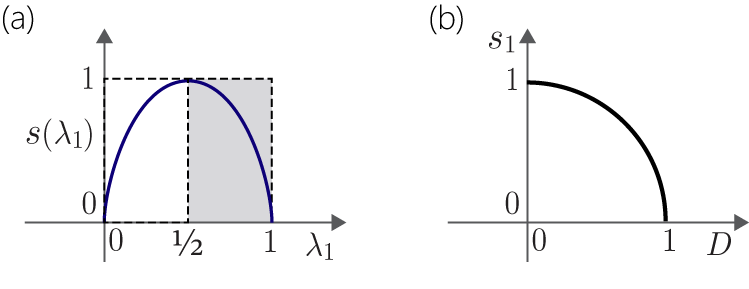}
\caption{(a) Plot of the entropy $S(\lambda_{1})$ for a binary DoF as a function of the eigenvalue $\lambda_{1}$. The shaded portion corresponds to the range $\lambda_{1}\in[\tfrac{1}{2},1]$, in which $S$ is uniquely determined. (b) A plot of the relationship between $S$ and $D$.}
\label{fig:EntropyOneDoF}
\end{figure}

We define the (von~Neumann) entropy $S$ \cite{Gamo64PO,Peres93Book} for the coherence matrix $\mathbf{G}$ as follows:
\begin{equation}
S=-\mathrm{Tr}\{\mathbf{G}\log_{2}\mathbf{G}\}=-\mathrm{Tr}\{\mathbf{G}^{\mathrm{D}}\log_{2}\mathbf{G}^{\mathrm{D}}\}=-\lambda_{1}\log_{2}\lambda_{1}-\lambda_{2}\log_{2}\lambda_{2},
\end{equation}
where $0\leq S\leq1$. The entropy of the coherent and incoherent limits identified above are as follows: $S=0$ when $\lambda_{1}=1$ and $\lambda_{2}=0$ (coherent field free of random fluctuations), and $S=1$ when $\lambda_{1}=\lambda_{2}=\tfrac{1}{2}$ (incoherent field). A binary DoF can thus carry at most $S=1$~bit of entropy; see the plot of $S(\lambda_{1})$ in Fig.~\ref{fig:EntropyOneDoF}(a). Because $S(\lambda_{1})=S(1-\lambda_{1})$, $S(\lambda_{1})$ is symmetric around $\lambda_{1}=\tfrac{1}{2}$, $S(\lambda_{1})$ is uniquely defined in the shaded area in Fig.~\ref{fig:EntropyOneDoF}(a) where $\lambda_{1}\in[\tfrac{1}{2},1]$. Both $S$ and $D$ have a one-to-one relationship with $\lambda_{1}$ when restricted to $\tfrac{1}{2}\leq\lambda_{1}<1$. Consequently, there is a one-to-one relationship between $S$ and $D$, so that $S$ uniquely defines $D$,
\begin{equation}
S=-\left(\frac{1+D}{2}\right)\log_{2}\left(\frac{1+D}{2}\right)-\left(\frac{1-D}{2}\right)\log_{2}\left(\frac{1-D}{2}\right),
\end{equation}
which we plot in Fig.~\ref{fig:EntropyOneDoF}(b).

Note that $S$ is a unitary invariant of $\mathbf{G}$; that is, any two coherence matrices $\mathbf{G}$ and $\mathbf{G}'=\hat{U}\mathbf{G}\hat{U}^{\dagger}$ that are related through a unitary have the same entropy $S$. Conversely, any two coherence matrices $\mathbf{G}$ and $\mathbf{G}'$ that have the same entropy can be inter-converted via a unitary \cite{Brosseau06PO}. Each value of $S$ therefore defines a family of coherence matrices that are related unitarily. We shall see below that this conclusion does \textit{not} extend to multiple DoFs or higher-dimensional coherence matrices.

\subsection{Measuring the degree of coherence for a binary DoF}\label{sec:MeasuringTheDegreeOfCoherence}

Traditionally, measuring the degree of coherence has followed physical procedures adapted to each DoF. For example, the degree of spatial coherence at two points $|a\rangle$ and $|b\rangle$ in the field is measured by recording the interference pattern formed when placing two slits at $|a\rangle$ and $|b\rangle$ and allowing the fields to propagate and overlap in the far field, producing spatial interference fringes with visibility $V=\tfrac{I_{\mathrm{max}}-I_{\mathrm{min}}}{I_{\mathrm{max}}+I_{\mathrm{min}}}$, where $I_{\mathrm{max}}$ and $I_{\mathrm{min}}$ are the maximum and minimum intensity values in the interferogram, respectively \cite{Michelson1891PM1}. In contrast, measuring the degree of polarization coherence relies on rotating the appropriate arrangement of wave plates in the path of the field. The interference visibility $V$ for any DoF may correspond to $D$ in some cases; in general $V\leq D$. Of course, $V=1$ requires that $I_{\mathrm{min}}=0$, which indicates that $D=1$. In other words, not extinguishing the field at some point indicates that the field may be partially coherent.

We aim here to systematize the measurement of $D$ for \textit{any} binary DoF by restricting ourselves to measurements of the modal weights for the two modes $|\psi_{1}\rangle$ and $|\psi_{2}\rangle$. In general, the field is transformed unitarily via $\hat{U}$, and $I_{1}$ and $I_{2}$ are measured as the parameters of $\hat{U}$ are varied. Two general strategies are of particular interest, which we refer to as Procedure~I (diagonalization) and Procedure~II (equalization).

\subsubsection{Procedure~I: Diagonalization (variational processing)}

\begin{figure}[t!]
\centering
\includegraphics[width=5in] {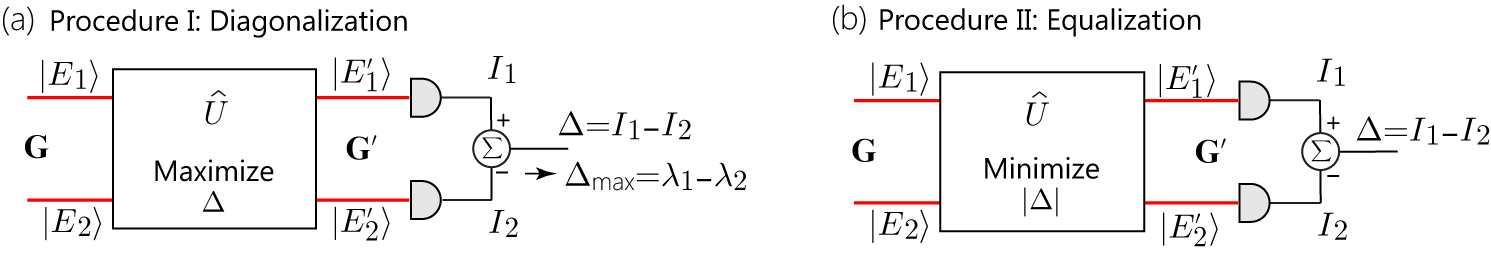} 
\caption{(a) Procedure~I, diagonalization. A unitary $\hat{U}$ is placed in the field path, $I_{1}$ and $I_{2}$ are measured, and their difference $\Delta=I_{1}-I_{2}$ is obtained. The parameters in $\hat{U}$ are varied to maximize $\Delta\rightarrow\Delta_{\mathrm{max}}$. Once achieved, $\Delta_{\mathrm{max}}=D$, and $\mathbf{G}$ is diagonalized. (b) Procedure~II, equalization. The overall approach matches procedure~I in Fig.~\ref{fig:ProcedureI}, except we aim to reach $\Delta=0$ (or minimize $|\Delta|$). Once achieved, $D=2|G_{12}'|$.}
\label{fig:ProcedureI}
\end{figure}

In Procedure~I, the field is first transformed via a unitary to $\mathbf{G}'=\hat{U}\mathbf{G}\hat{U}^{\dagger}=\left(\begin{array}{cc}G_{11}'&G_{12}'\\G_{21}'&G_{22}'\end{array}\right)$, and the power $I_{1}=G_{11}'$ and $I_{2}=G_{22}'$ are recorded. We monitor the difference $\Delta=I_{1}-I_{2}$ while varying the parameters of the unitary $\hat{U}$ so as to maximize $\Delta$: $\Delta\rightarrow\Delta_{\mathrm{max}}$ [Fig.~\ref{fig:GeneralInterference} and Fig.~\ref{fig:ProcedureI}(a)]. From Eq.~\ref{eq:2X2Delta}, we have:
\begin{equation}
\Delta_{\mathrm{max}}=\lambda_{1}-\lambda_{2}=D,
\end{equation}
which occurs when $G_{11}'=\lambda_{1}$, $G_{22}'=\lambda_{2}$, and $G_{12}'=0$, corresponding to $\theta\rightarrow0$ in Eq.~\ref{eq:General2x2CoherenceMatrix}. In other words, \textit{maximizing $\Delta=I_{1}-I_{2}$ is equivalent to diagonalizing $\mathbf{G}$}, which then reveals the degree of coherence $\Delta_{\mathrm{max}}=D$.

When the field is coherent ($\lambda_{1}=1$ and $\lambda_{2}=0$), we reach $\Delta_{\mathrm{max}}=1$. In other words, extinguishing one of the measured outputs is the hallmark of complete coherence. Equivalently, the ability to transfer all the field power to one mode signifies full coherence [Fig.~\ref{fig:ConcentratingTheField}]. When the field is incoherent ($\lambda_{1}=\lambda_{2}=\tfrac{1}{2}$), then $\Delta=\Delta_{\mathrm{max}}=0$. In this case, changing the parameters of $\hat{U}$ has no impact on $\Delta$, and the field power is always distributed equally among the modes (this feature is also independent of the dimensionality of the modal basis). When the field is partially coherent, $\Delta_{\mathrm{max}}=\lambda_{1}-\lambda_{2}<1$, and neither output can be extinguished [Fig.~\ref{fig:PartiallyCoherentBlocked}].

As described in Fig.~\ref{fig:GeneralInterference}, finding $\Delta_{\mathrm{max}}$ involves sweeping through a 2D parameter space spanned by the parameters $\theta_{\mathrm{o}}$ and $\varphi_{\mathrm{o}}$ of the unitary $\hat{U}$. This strategy has been called `variational processing'. One may speed up the search process by using optimization or machine-learning algorithms \cite{roques2024measuring}.

\subsubsection{Procedure~II: Equalization}

In Procedure~II we follow the same approach of Procedure~I except that the parameters of the unitary $\hat{U}$ are varied so as to reach $\Delta=0$ [Fig.~\ref{fig:ProcedureI}(b)]; that is we aim to equalize $I_{1}$ and $I_{2}$, or $G_{11}'=G_{22}'=\tfrac{1}{2}$. This occurs when $\theta=\tfrac{\pi}{2}$ in Eq.~\ref{eq:General2x2CoherenceMatrix}, whereupon $G_{12}'=\tfrac{1}{2}(\lambda_{1}-\lambda_{2})e^{-i\varphi}$ and:
\begin{equation}
D=2|G_{12}'|.
\end{equation}
In other words, whenever the power is equally divided between the two modes $|\psi_{1}\rangle$ and $|\psi_{2}\rangle$, the magnitude of the off-diagonal element of $\mathbf{G}$ is $\tfrac{1}{2}D$; i.e., the off-diagonal element indicates the degree of coherence. When the field is fully coherent ($\lambda_{1}=1$, $\lambda_{2}=0$, and $D=1$), equalizing $I_{1}$ and $I_{2}$ always yields $|G_{12}|=\tfrac{1}{2}$. When the field is completely incoherent, $G_{12}=0$ in all cases. When the field is partially coherent, equalizing the modal weights $I_{1}=I_{2}=\tfrac{1}{2}$ corresponds to $0<|G_{12}|=\tfrac{1}{2}(\lambda_{1}-\lambda_{2})<1$.


Using the configuration shown in Fig.~\ref{fig:GeneralInterference}(a), the equalization condition $\Delta=0$ requires that $I_{1}=I_{2}=\tfrac{1}{2}$, which is reached when $\varphi_{\mathrm{o}}=\varphi$ and $\theta_{\mathrm{o}}=-\theta+\pi$, whereupon $G_{12}'=\tfrac{1}{2}De^{-i\varphi}$. Once again, a search in the 2D parameters space spanned by $\theta_{\mathrm{o}}$ and $\varphi_{\mathrm{o}}$ is required.

Of course, extracting $D$ via this procedure requires an extra step to estimate $|G_{12}|$ after equalization. The fields associated with the two modes that may be superposed with a relative phase to record an interferogram whose visibility $V$ corresponds to the degree of coherence $D$ (as done in double-slit interference, for example [Fig.~\ref{fig:Interference}]). Alternatively, the value of $G_{12}$ can be estimated directly by reconstructing $\mathbf{G}$, which we proceed to elucidate.


\subsection{The Poincar{\'e} sphere and the Stokes parameters}\label{sec:PSandSP}

\subsubsection{The Poincar{\'e} sphere}

\begin{figure}[t!]
\centering
\includegraphics[width=13.3cm]{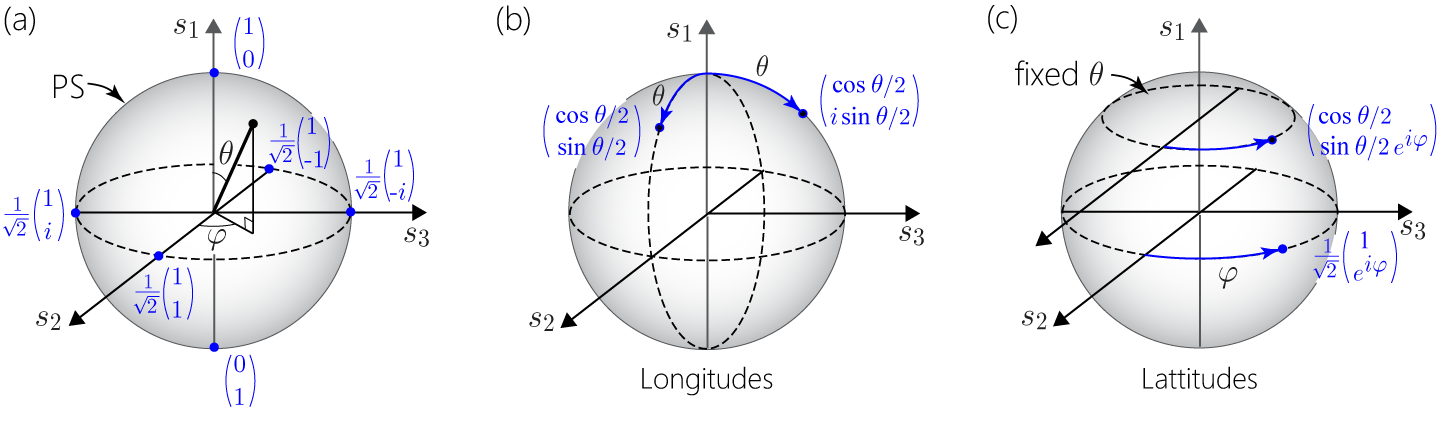}
\caption{Representation of a coherent field with a binary DoF as a point on the surface of the Poincar{\'e} sphere (PS). (a) Representative points defined on the PS surface: a general point with angular coordinates $(\theta,\varphi)$, the north and south poles, and points on the equator. (b) The longitudes on the PS surface correspond to families of coherent field vectors in which $\theta$ is changed at fixed $\varphi$. (c) The latitudes on the PS surface correspond to families of coherent field vectors in which $\varphi$ is changed at fixed $\theta$.}
\label{fig:PScoherent}
\end{figure}

The Poincar{\'e} sphere (PS) is a unit-radius sphere in a 3D space that provides a useful visual representation of the field vector $|E\rangle$ for binary DoFs. The PS is widely utilized in polarization optics, but it is equally useful for any binary DoF \cite{Abouraddy14OL,Abouraddy19Optica,Halder21OL}. Its usefulness stems from the fact that any coherent field with a binary DoF is represented by a unique point on its surface. When the field vector is defined by two angular parameters $\theta$ and $\varphi$, $|E\rangle=\left(\begin{array}{c}\cos\tfrac{\theta}{2}\\e^{i\varphi}\sin\tfrac{\theta}{2}\end{array}\right)$, it can be represented directly by a point of angular coordinates $(\theta,\varphi)$ on the PS surface in a spherical coordinate system; where $\theta$ is the angle measured with the $z$-axis and $\varphi$ is the angle measured with the $x$-axis in the $(x,y)$-plane [Fig.~\ref{fig:PScoherent}(a)]. The PS north and south poles are identified with the $|\psi_{1}\rangle=\left(\begin{array}{c}1\\0\end{array}\right)$ and $|\psi_{2}\rangle=\left(\begin{array}{c}0\\1\end{array}\right)$, respectively; the equator corresponds to the family of field vectors $|E\rangle=\tfrac{1}{\sqrt{2}}\left(\begin{array}{c}1\\e^{i\varphi}\end{array}\right)$, where the power is equally divided between the two modes; and the four points at the intersection of the equator with the $x$ and $y$ axes correspond to $\tfrac{1}{\sqrt{2}}\left(\begin{array}{c}1\\1\end{array}\right)$, $\tfrac{1}{\sqrt{2}}\left(\begin{array}{c}1\\i\end{array}\right)$, $\tfrac{1}{\sqrt{2}}\left(\begin{array}{c}1\\-1\end{array}\right)$, and $\tfrac{1}{\sqrt{2}}\left(\begin{array}{c}1\\-i\end{array}\right)$. Each longitude, or great circle incorporating the north and south poles, corresponds to the family of field vectors in which $\theta$ is varied while $\varphi$ is held fixed [Fig.~\ref{fig:PScoherent}(b)]. Each latitude, a circle parallel to the equator, corresponds to a family of field vectors in which $\varphi$ is varied while $\theta$ is held fixed [Fig.~\ref{fig:PScoherent}(c)]. Consequently, the PS is useful in visualizing unitary dynamics of the field vector. Unitaries, which conserve the length of $|E\rangle$, transport the point representing the field vector along trajectories bound to the PS surface.

In the case of partially coherent fields represented by a $2\times2$ coherence matrix $\mathbf{G}$ ($D<1$), the field can\textit{not} be represented by a point \textit{on} the PS surface, but rather a point \textit{inside} the PS volume. To elucidate this, we first introduce the Stokes parameters.

\subsubsection{The Stokes parameters}\label{sec:StokesSingleDoF}

The Stokes parameters are four real parameters $\{s_{0},s_{1},s_{2},s_{3}\}$ that correspond to the coefficients of the expansion of $\mathbf{G}=\tfrac{1}{2}\sum_{j=0}^{3}s_{j}\hat{\sigma}_{j}$ in terms of the Pauli matrices $\{\hat{\sigma}_{0},\hat{\sigma}_{1},\hat{\sigma}_{2},\hat{\sigma}_{3}\}$:
\begin{equation}
\hat{\sigma}_{0}=\left(\begin{array}{cc}1&0\\0&1\end{array}\right),\;
\hat{\sigma}_{1}=\left(\begin{array}{cc}1&0\\0&-1\end{array}\right),\;
\hat{\sigma}_{2}=\left(\begin{array}{cc}0&1\\1&0\end{array}\right),\;
\hat{\sigma}_{3}=\left(\begin{array}{cc}0&-i\\i&0\end{array}\right).
\end{equation}
Although the Stokes parameters have been traditionally associated with polarization, they can nevertheless be adapted for \textit{any} binary DoF \cite{Abouraddy14OL,Halder21OL,Abouraddy19Optica}. 

We can write $\hat{\sigma}_{0}$ and $\hat{\sigma}_{1}$ directly in terms of the projection operators $\hat{P}_{0}$ and $\hat{P}_{1}$: $\hat{\sigma}_{0}=\hat{P}_{1}+\hat{P}_{2}$ and $\hat{\sigma}_{1}=\hat{P}_{1}-\hat{P}_{2}$. The Pauli matrices $\{\hat{\sigma}_{j}\}$ for $j\neq0$ have the following properties:
\begin{enumerate}
\item They are zero-trace matrices: $\mathrm{Tr}\{\hat{\sigma}_{j}\}=0$.
\item They are Hermitian $\hat{\sigma}_{j}^{\dagger}=\hat{\sigma}_{j}$, all sharing the same eigenvalues 1 and -1.
\item The Pauli matrices $\hat{\sigma}_{2}$ and $\hat{\sigma}_{3}$ can be obtained from $\hat{\sigma}_{1}$ via unitaries: $\hat{\sigma}_{2}=\hat{U}_{2}^{\dagger}\hat{\sigma}_{1}\hat{U}_{2}$ and $\hat{\sigma}_{3}=\hat{U}_{3}^{\dagger}\hat{\sigma}_{1}\hat{U}_{3}$, where
\begin{equation}\label{eq:U2U3}
\hat{U}_{2}=\frac{1}{\sqrt{2}}\left(\begin{array}{cc}1&1\\-1&1\end{array}\right),\;\hat{U}_{3}=\frac{1}{\sqrt{2}}\left(\begin{array}{cc}1&-i\\-i&1\end{array}\right).
\end{equation}
\item They are unitary matrices $\hat{\sigma}_{j}^{\dagger}\hat{\sigma}_{j}=\hat{\sigma}_{j}^{2}=\hat{\mathbb{I}}_{2}$, with $\mathrm{det}\{\hat{\sigma}_{j}\}=-1$.
\item $\hat{\sigma}_{j}\hat{\sigma}_{k}=i\epsilon_{jk\ell}\hat{\sigma}_{\ell}$ ($j,k\neq0$), where $\epsilon_{ijk}$ is the anti-symmetric tensor coefficient (the Levi-Civita symbol).
\end{enumerate}

The coherence matrix $\mathbf{G}$ is thus expressed in terms of the Stokes parameters as follows:
\begin{equation}\label{eq:CoherenceMatrixStokesDecomp}
\mathbf{G}=\frac{1}{2}\sum_{j=0}^{3}s_{j}\hat{\sigma}_{j}=\frac{1}{2}\left(\begin{array}{cc}
s_{0}+s_{1}&s_{2}-is_{3}\\s_{2}+is_{3}&s_{0}-s_{1}
\end{array}\right).
\end{equation}
The last property of the Pauli matrices listed above allows us to obtain the Stokes parameters via projections onto the Pauli matrices: $s_{j}=\mathrm{Tr}\{\hat{\sigma}_{j}\mathbf{G}\}$, which are related to the elements of $\mathbf{G}$ as follows:
\begin{eqnarray}
s_{0}&=&G_{11}+G_{22},\nonumber\\
s_{1}&=&G_{11}-G_{22},\nonumber\\
s_{2}&=&G_{12}+G_{21}=2\mathrm{Re}\{G_{12}\},\nonumber\\
s_{3}&=&i(G_{12}-G_{21})=-2\mathrm{Im}\{G_{12}\}.
\end{eqnarray}

The Stokes parameters have the following properties:
\begin{enumerate}
\item For unity-trace $\mathbf{G}$, we have the normalization $s_{0}=1$. 
\item The Stokes parameters are real because $\mathbf{G}$ is Hermitian.
\item $|s_{j}|\leq1$.
\item The Stokes parameters are \textit{not} unitary invariants. 
\item However, the sum of the squared Stokes parameters is a unitary invariantd:
\begin{equation}
s_{1}^{2}+s_{2}^{2}+s_{3}^{2}=(\mathrm{Tr}\{\mathbf{G}\})^{2}-4\mathrm{det}\{\mathbf{G}\}=D^{2}.
\end{equation}
\end{enumerate}

\begin{figure}[t!]
\centering
\includegraphics[width=8.9cm]{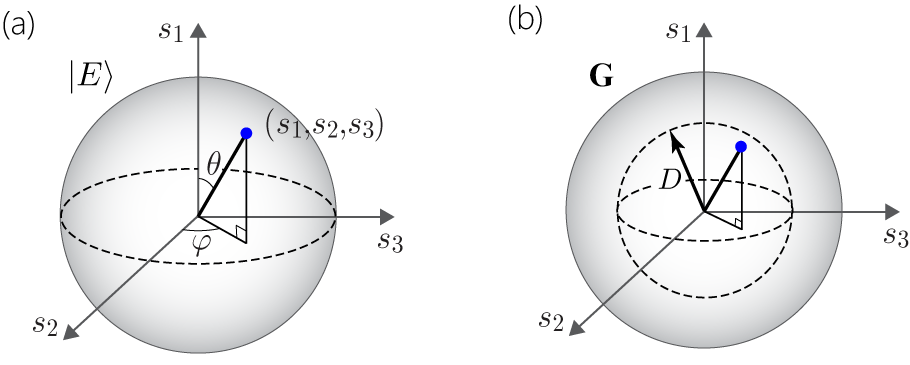}
\caption{Representation of a field with a binary DoF as a point on the PS surface in terms of the Stokes parameters. The coordinates of a point representing a field in this space are $(s_{1},s_{2},s_{3})$. The distance of a point with these coordinates from the origin is thus $\sqrt{s_{1}^{2}+s_{2}^{2}+s_{3}^{2}}$. (a) For a coherent field, $s_{1}^{2}+s_{2}^{2}+s_{3}^{2}=D^{2}=1$, and the point representing the field vector lies on the PS surface. (b) For a partially coherent field, $s_{1}^{2}+s_{2}^{2}+s_{3}^{2}=D^{2}<1$, and the point representing the coherence matrix $\mathbf{G}$ lies within the PS rather than on its surface. A spherical surface of radius $D$ represents all the iso-entropy fields; that is, fields of equal entropy $S$ or, equivalently, equal $D$.}
\label{fig:SokesPoincare}
\end{figure}

For a coherent field expressed as $|E\rangle=\left(\begin{array}{c}\cos\frac{\theta}{2}\\e^{i\varphi}\sin\tfrac{\theta}{2}\end{array}\right)$, we have
\begin{eqnarray}
s_{1}&=&\cos\theta,\nonumber\\
s_{2}&=&\sin\theta\cos\varphi,\nonumber\\ s_{3}&=&\sin\theta\sin\varphi,
\end{eqnarray}
so that $s_{1}^{2}+s_{2}^{2}+s_{3}^{2}=1$. The Stokes parameters for a coherent field therefore correspond to the coordinates of the point on the PS representing the field vector. In other words, the 3D space in which the PS is represented is that spanned by $\{s_{1},s_{2},s_{3}\}$ [Fig.~\ref{fig:SokesPoincare}(a)].

We make use of this observation to establish a PS-representation for partially coherent fields. We first cast the Stokes parameters for a partially coherent field in the following form:
\begin{eqnarray}
s_{1}&=&D\cos\theta,\nonumber\\
s_{2}&=&D\sin\theta\cos\varphi,\nonumber\\ s_{3}&=&D\sin\theta\sin\varphi,
\end{eqnarray}
where the angles $\theta$ and $\varphi$ are once again those in a spherical coordinate system, and $s_{1}^{2}+s_{2}^{2}+s_{3}^{2}=D^{2}$. The coherence matrix is represented by a point in this spherical coordinate system $(\theta,\varphi)$, but the degree of coherence $D<1$ is now the radius of the sphere on which the point is located [Fig.~\ref{fig:SokesPoincare}(b)]. We can thus represent partially coherent fields \textit{inside} the unit-radius PS. Unitary transformations move this point representing the coherence matrix on a spherical surface of fixed radius.

The points on the PS surface, or the points on any other spherical surface of radius $D<1$, represent all the iso-entropy fields. Any two points on the same PS surface have the same entropy. Consequently, these two points can be unitarily inter-converted into each other. Conversely, the family of optical fields that can be inter-converted into each other are all represented on a PS surface of fixed radius.

\subsubsection{Radial and angular parameters}

We distinguish between `radial' and `angular' parameters of the coherence matrix $\mathbf{G}$. In general, $\mathbf{G}$ is identified by 3 independent real parameters ($D,\theta$, and $\varphi$ in Eq.~\ref{eq:GeneralGwithD}), which are reduced to 2 for a coherent field ($\theta$ and $\varphi$ in Eq.~\ref{eq:GeneralStateCoherent1DoF}). We refer to $\theta$ and $\varphi$ as the `angular parameters' and to $D$ as a `radial parameter'. A unitary transformation changes the angular parameters $\theta$ and $\varphi$, but the point representing the field vector or coherence matrix retains its radius (whether on the PS surface or within its volume). The distinction between these two classes of parameters is clear: the angular parameters are those that change upon implementing a unitary, whereas the radial parameters are those that are invariant with unitaries. This distinction is relatively obvious for a binary DoF, but it becomes more significant for larger-dimensional coherence matrices.

\subsubsection{Reconstructing the coherence matrix}\label{sec:ReconstructingGSingleDoF}

\begin{figure}[t!]
\centering
\includegraphics[width=13.3cm]{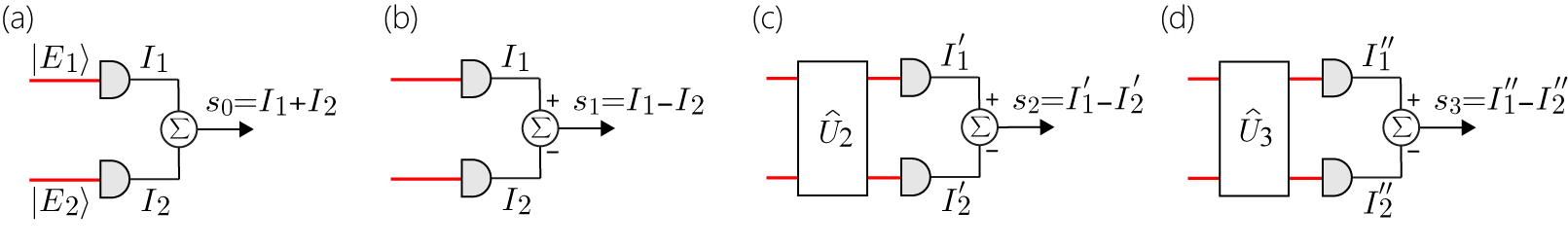} 
\caption{Configurations for measuring the Stokes parameters: (a) $s_{0}$, (b) $s_{1}$, (c) $s_{2}$, and (d) $s_{3}$. Note that $s_{0}$ and $s_{1}$ are obtained in the same setting, so that only three different configurations suffice to extract all the Stokes parameters.}
\label{fig:StokesGeneral}
\end{figure}

We are now in a position to describe a systematic approach to reconstruct the coherence matrix $\mathbf{G}$. Because a $2\times2$ coherence matrix is identified by 3 real parameters, in addition to one more parameter (the total power) to ensure normalization, we expect that 4 independent measurements will be needed to reconstruct $\mathbf{G}$. One useful parameterization of these measurements makes use of the Stokes parameters $\{s_{0},s_{1},s_{2},s_{3}\}$, and the four measurement configurations required to acquire them are illustrated in Fig.~\ref{fig:StokesGeneral}.

Obtaining $s_{0}$ corresponds to measuring the total power [Fig.~\ref{fig:StokesGeneral}(a)]:
\begin{equation}
s_{0}=\mathrm{Tr}\{\mathbf{G}\}=\mathrm{Tr}\{(\hat{P}_{1}+\hat{P}_{2})\mathbf{G}\}=\mathrm{Tr}\{\hat{P}_{1}\mathbf{G}\}+\mathrm{Tr}\{\hat{P}_{2}\mathbf{G}\}=I_{1}+I_{2}.
\end{equation}
The Stokes parameter $s_{0}$ therefore provides the normalization. Dividing all the Stokes parameters by this value yields $s_{0}=1$, $|s_{j}|\leq1$, and $s_{1}^{2}+s_{2}^{2}+s_{3}^{2}\leq1$. The Stokes parameter $s_{1}$ is the difference between the modal weights [Fig.~\ref{fig:StokesGeneral}(b)]:
\begin{equation}
s_{1}=\mathrm{Tr}\{\hat{\sigma}_{1}\mathbf{G}\}=\mathrm{Tr}\{\hat{P}_{1}\mathbf{G}\}-\mathrm{Tr}\{\hat{P}_{2}\mathbf{G}\}=I_{1}-I_{2}.
\end{equation}

Writing the Pauli matrix $\hat{\sigma}_{2}$ as $\hat{\sigma}_{2}=\hat{U}_{2}^{\dagger}\hat{\sigma}_{1}\hat{U}_{2}$, where $\hat{U}_{2}$ is given in Eq.~\ref{eq:U2U3}, the Stokes parameter $s_{2}$ is:
\begin{equation}
s_{2}=\mathrm{Tr}\{\hat{\sigma}_{2}\mathbf{G}\}=\mathrm{Tr}(\hat{U}_{2}\hat{\sigma}_{1}\hat{U}_{2}^{\dagger}\mathbf{G})=\mathrm{Tr}(\hat{\sigma}_{1}\mathbf{G}')=I_{1}'-I_{2}',
\end{equation}
where $\mathbf{G}'=\hat{U}_{2}\mathbf{G}\hat{U}_{2}^{\dagger}$, $I_{1}'=\mathrm{Tr}(\hat{P}_{1}\mathbf{G}')$ and $I_{2}'=\mathrm{Tr}(\hat{P}_{2}\mathbf{G}')$, with $I_{1}'+I_{2}'=1$; i.e., $s_{2}$ is the difference between the modal weights after implementing $\hat{U}_{2}$ [Fig.~\ref{fig:StokesGeneral}(c)]. Similarly, the Stokes parameter $s_{3}$ is:
\begin{equation}
s_{3}=\mathrm{Tr}(\hat{\sigma}_{3}\mathbf{G})=\mathrm{Tr}(\hat{U}_{3}\hat{\sigma}_{1}\hat{U}_{3}^{\dagger}\mathbf{G})=\mathrm{Tr}(\hat{\sigma}_{1}\mathbf{G}'')=I_{1}''-I_{2}'',
\end{equation}
where $\hat{U}_{3}$ is given in Eq.~\ref{eq:U2U3}, $\mathbf{G}''=\hat{U}_{3}\mathbf{G}\hat{U}_{3}^{\dagger}$, $I_{1}''=\mathrm{Tr}(\hat{P}_{1}\mathbf{G}'')$ and $I_{2}''=\mathrm{Tr}(\hat{P}_{2}\mathbf{G}'')$; once again, $I_{1}''+I_{2}''=1$. In other words, $s_{3}$ is the difference between the modal weights after implementing the unitary $\hat{U}_{3}$ [Fig.~\ref{fig:StokesGeneral}(d)]. The four measurements illustrated in Fig.~\ref{fig:StokesGeneral} therefore extract the Stokes parameters $\{s_{0},s_{1},s_{2},s_{3}\}$, from which we reconstruct the coherence matrix $\mathbf{G}$ (Eq.~\ref{eq:CoherenceMatrixStokesDecomp}).

In the measurement settings in Fig.~\ref{fig:StokesGeneral}, $I_{1}+I_{2}=I_{1}'+I_{2}'=I_{1}''+I_{2}''=1$ after $\hat{U}_{2}$ and $\hat{U}_{3}$ (after the normalization with respect to $s_{0}=I_{1}+I_{2}=1$). Consequently,
\begin{eqnarray}
s_{1}&=&I_{1}-I_{2}=2I_{1}-1,\nonumber\\
s_{2}&=&I_{1}'-I_{2}'=2I_{1}'-1,\nonumber\\
s_{3}&=&I_{1}''-I_{2}''=2I_{1}''-1,
\end{eqnarray}
so that only one detector in principle is needed in any setting; for example, the detector that projects onto $|\psi_{1}\rangle$. The coherence matrix $\mathbf{G}$ can then be reconstructed in terms of the measurements:
\begin{equation}
\mathbf{G}=\left(\begin{array}{cc}
I_{1}&(I_{1}'-\tfrac{1}{2})-i(I_{1}''-\tfrac{1}{2})\\(I_{1}'-\tfrac{1}{2})+i(I_{1}''-\tfrac{1}{2})&1-I_{1}
\end{array}\right).
\end{equation}
From a practical perspective, however, measuring both modal weights for $|\psi_{1}\rangle$ and $|\psi_{2}\rangle$ is preferable to guarantee that correct normalization is maintained throughout.
 
\subsection{Non-unitary transformations}\label{sec:NonUnitaryOperators}

Unitaries conserve the degree of coherence $D(\hat{U}\mathbf{G}\hat{U}^{\dagger})=D(\mathbf{G})$ and the entropy $S(\hat{U}\mathbf{G}\hat{U}^{\dagger})=S(\mathbf{G})$. In contrast, non-unitary optical systems can increase or decrease $D$ and $S$. We consider here two classes of non-unitary operators: filtering and decohering.

\subsubsection{Filtering operators}\label{sec:Filtering}

Filtering operations usually \textit{increase} $D$, but may nevertheless \textit{decrease} $D$ and even yield a fully incoherent field. Writing the filtering operator as $\hat{F}=\hat{U}F^{\mathrm{D}}\hat{U}^{\dagger}$, where $\hat{F}^{\mathrm{D}}=\left(\begin{array}{cc}\sqrt{d_{1}}&0\\0&\sqrt{d_{2}}\end{array}\right)$, $d_{1}$ and $d_{2}$ are real and nonnegative, $0\leq d_{1},d_{2}\leq1$ and $\hat{U}$ is a $2\times2$ unitary. It is clear that $\hat{F}$ reduces the two modal weights by the factors $d_{1}$ and $d_{2}$ in a basis rotated via $\hat{U}$. Of course, this non-unitary filter does \textit{not} conserve the normalization of $|E\rangle$ or $\mathbf{G}$; that is, a filtering operation is \textit{not} trace-preserving. For a diagonalized filter $\hat{F}^{\mathrm{D}}$ operating on a coherent field:
\begin{equation}
\hat{F}^{\mathrm{D}}|E\rangle=\left(\begin{array}{cc}\sqrt{d_{1}}&0\\0&\sqrt{d_{2}}\end{array}\right)\left(\begin{array}{c}E_{1}\\E_{2}\end{array}\right)\rightarrow|E'\rangle=\frac{1}{\sqrt{d_{1}|E_{1}|^{2}+d_{2}|E_{2}|^{2}}}\left(\begin{array}{c}\sqrt{d_{1}}E_{1}\\\sqrt{d_{2}}E_{2}\end{array}\right),
\end{equation}
so that $\langle E'|E'\rangle=1$ when $d_{1}|E_{1}|^{2}+d_{2}|E_{2}|^{2}\neq0$. The filter $\hat{F}^{\mathrm{D}}$ modifies the field vector and moves the point representing it on the PS surface along a longitude, but does \textit{not} change $D=1$; i.e., the field remains coherent ($S=0$).

\begin{figure}[t!]
\centering
\includegraphics[width=9cm]{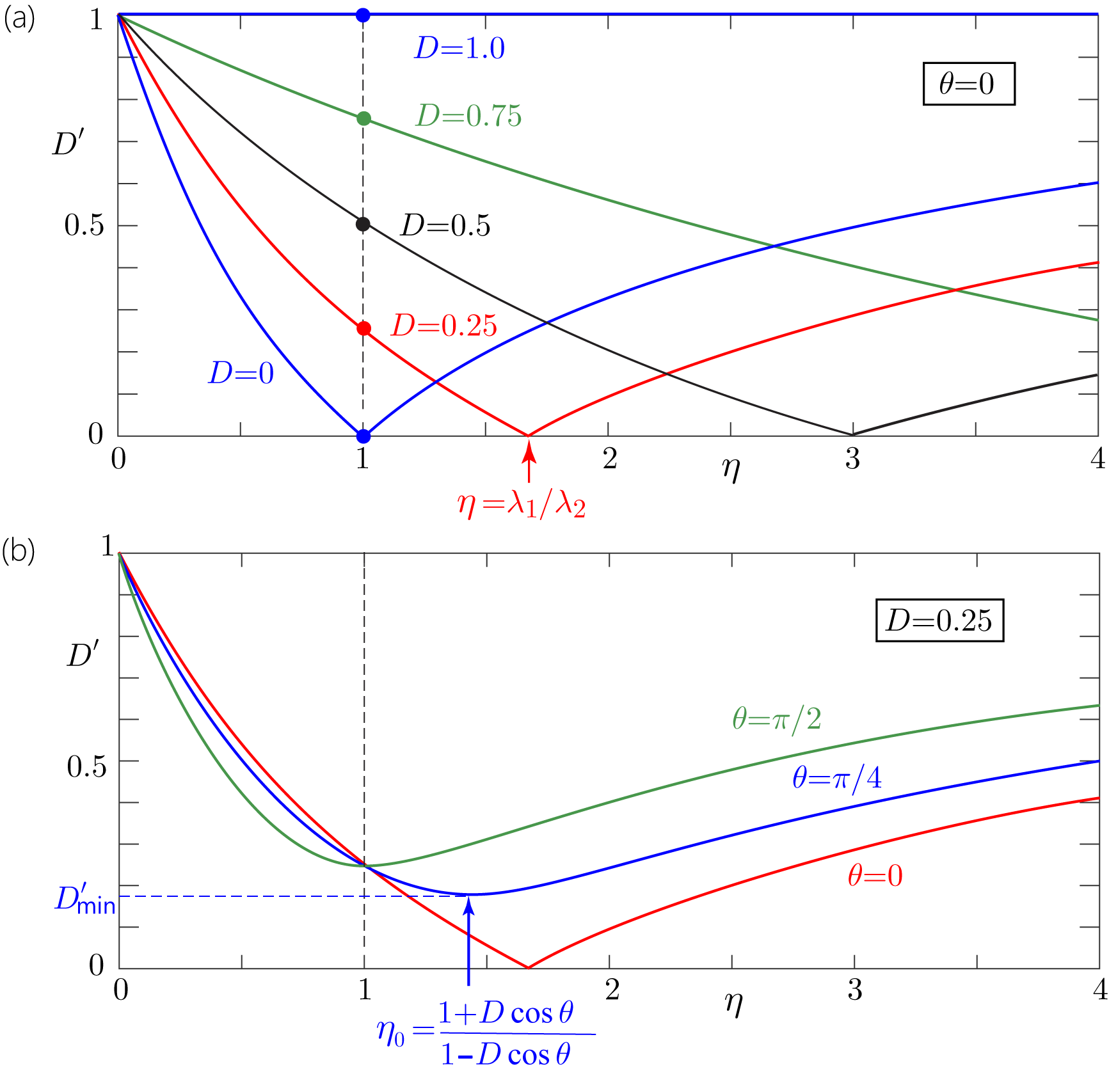}
\caption{(a) The tunability of the degree of coherence $D'$ after a filtering operator $\hat{F}^{\mathrm{D}}$ when $\mathbf{G}=\mathbf{G}^{\mathrm{D}}$ is diagonal ($\theta=0$ in Eq.~\ref{eq:GeneralGwithD}) as we vary $\eta$ (Eq.~\ref{eq:DFilteringDiagonal}). The curves correspond to different initial values of $D$. Varying $\eta=\tfrac{d_{2}}{d_{1}}$ allows tuning $D'$ from~0 (incoherent, reached when $\eta=\tfrac{\lambda_{1}}{\lambda_{2}}$) to~1 (coherent, reached when $\eta=0$ or $\eta\rightarrow\infty$, corresponding to $d_{2}=0$ or $d_{1}=0$, respectively). When $\eta=1$ (i.e., $d_{1}=d_{2}$ and $\hat{F}^{\mathrm{D}}=d_{1}\hat{\mathbb{I}}_{2}$ is an attenuator), we have $D'=D$. (b) The tunability of $D'$ after a filtering operator $\hat{F}^{\mathrm{D}}$ when $\mathbf{G}$ is not diagonal. The curves correspond to different values of $\theta$, but $D$ is held fixed ($D=0.25$ here). When $\theta=0$, the tuning curve corresponds to that curve in (a) having $D=0.25$, in which case $D'$ can be tuned from~0 ($\eta=\tfrac{\lambda_{1}}{\lambda_{2}}$) to~1 ($\eta=0$ or~$\infty$). When $\theta=\tfrac{\pi}{2}$, varying $\eta$ tunes $D'$ from a lower limit of $D'=D$ at $\eta=1$ to $D'=1$ when $\eta=0$ or~$\infty$. We cannot lower $D'$ below the initial value $D$ when $\theta=\tfrac{\pi}{2}$; the filtering operator always increases $D'$ in this case. For $0\leq\theta<\tfrac{\pi}{2}$, we have $D'=D$ when $\eta=1$, $D'=0$ at $\eta=0$ or~$\infty$, and the filter reduces $D'$ to a minimum value of $D_{\mathrm{min}}'=\tfrac{D\sin\theta}{\sqrt{1-D^{2}\cos^{2}\theta}}<D$ which is reached at $\eta_{\mathrm{o}}=\tfrac{1+D\cos\theta}{1-D\cos\theta}$.}
\label{fig:Filtering}
\end{figure}

In the more general case of a partially coherent field ($\lambda_{1}\geq\lambda_{2}\neq0$, and $0<D<1$), a filtering operator may increase \textit{or} decrease $D$. First, consider a diagonal coherence matrix,
\begin{equation}
\mathbf{G}^{\mathrm{D}}=\left(\begin{array}{cc}\lambda_{1}&0\\0&\lambda_{2}\end{array}\right)\rightarrow \hat{F}^{\mathrm{D}}\mathbf{G}\hat{F}^{\mathrm{D}\dagger}\rightarrow\mathbf{G}'=\frac{1}{d_{1}\lambda_{1}+d_{2}\lambda_{2}}\left(\begin{array}{cc}d_{1}\lambda_{1}&0\\0&d_{2}\lambda_{2}\end{array}\right),
\end{equation}
in which case the degree of coherence changes from $D=\lambda_{1}-\lambda_{2}$ to:
\begin{equation}\label{eq:DFilteringDiagonal}
D'=\frac{|d_{1}\lambda_{1}-d_{2}\lambda_{2}|}{d_{1}\lambda_{1}+d_{2}\lambda_{2}}=\frac{|\lambda_{1}-\lambda_{2}\eta|}{\lambda_{1}+\lambda_{2}\eta},
\end{equation}
where $\eta=\tfrac{d_{2}}{d_{1}}$ and $0\leq\eta<\infty$. We can tune $D'$ across the entire range $0\leq D'\leq1$ independently of the initial value of $D$ (as long as $D\neq1$) by varying $\eta$ [Fig.~\ref{fig:Filtering}(a)]. The field can be rendered fully \textit{coherent} $D'=1$ by setting $d_{1}=0$ or $d_{2}=0$ (eliminating one of the modes). Alternatively, to render the field fully \textit{incoherent} $D'=0$, we need $\eta=\tfrac{\lambda_{1}}{\lambda_{2}}$ or $d_{1}\lambda_{1}=d_{2}\lambda_{2}$. To vary $D'$, one can thus start with an incoherent field with $D=0$ ($\lambda_{1}=\lambda_{2}=\tfrac{1}{2}$), so that Eq.~\ref{eq:DFilteringDiagonal} yields $D'=\tfrac{|d_{1}-d_{2}|}{d_{1}+d_{2}}=\tfrac{|1-\eta|}{1+\eta}$. By varying the parameter $\eta$ from $\eta=1$ ($d_{1}=d_{2}$) to $\eta=0$ ($d_{2}=0$) we can tune $D'$ continuously from~0 to~1 [Fig.~\ref{fig:Filtering}(a)]. Likewise, varying $\eta$ from~1 to~$\infty$ tunes $D'$ from~0 to~1.

Less control is afforded over $D'$ when $\mathbf{G}$ is non-diagonal. We need not consider both $\mathbf{G}$ and $\hat{F}$ to be non-diagonal; rather we can operate in a basis where only one of them is diagonal. We maintain a diagonal matrix $\hat{F}^{\mathrm{D}}$ and adopt a non-diagonal $\mathbf{G}$ initially having $D=\sqrt{1-4\mathrm{det}\{\mathbf{G}\}}$. After traversing $\hat{F}^{\mathrm{D}}$, the coherence matrix becomes:
\begin{equation}\label{eq:NondiagonalFiltering}
\hat{F}^{\mathrm{D}}\mathbf{G}\hat{F}^{\mathrm{D}\dagger}\rightarrow\mathbf{G}'=\frac{1}{d_{1}G_{11}+d_{2}G_{22}}\left(\begin{array}{cc}d_{1}G_{11}&\sqrt{d_{1}d_{2}}G_{12}\\\sqrt{d_{1}d_{2}}G_{21}&d_{2}G_{22}\end{array}\right),
\end{equation}
and the associated degree of coherence $D'$ is:
\begin{equation}
D'=\sqrt{1-\frac{d_{1}d_{2}(1-D^{2})}{(d_{1}G_{11}+d_{2}G_{22})^{2}}}=\sqrt{1-\frac{4d_{1}d_{2}(1-D^{2})}{[(d_{1}+d_{2})+(d_{1}-d_{2})D\cos\theta]^{2}}},
\end{equation}
where we substituted for $G_{11}$, $G_{22}$, and $G_{12}$ from Eq.~\ref{eq:GeneralGwithD}. When the field is initially coherent ($D=1$), then the field remains coherent $D'=1$ independently of $d_{1}$ and $d_{2}$. To achieve $D'=1$ when initially $D\neq1$, we require $d_{1}=0$ or $d_{2}=0$. With such a filtering operator, we can reach $D'=1$ independently of the values of $D$ or $\theta$ [Fig.~\ref{fig:Filtering}(a,b)].

However, we can\textit{not} reach $D'=0$ as we did in the above scenario when considering a field represented by a diagonal coherence matrix. This can be seen from Eq.~\ref{eq:NondiagonalFiltering} where the off-diagonal element $G_{12}'=\sqrt{d_{1}d_{2}}G_{12}$ cannot be eliminated (as needed when $D'=0$) when initially $G_{12}\neq0$ unless we set $d_{1}=0$ or $d_{2}=0$, which instead yields $D'=1$. For a coherence matrix with given values of $D$ and $\theta$, varying $\eta$ tunes $D'$ from~1 to a minimum value $D_{\mathrm{min}}'$ given by:
\begin{equation}
D_{\mathrm{min}}'=\frac{D\sin\theta}{\sqrt{1-D^{2}\cos^{2}\theta}},
\end{equation}
which is reached when:
\begin{equation}
\eta=\eta_{\mathrm{o}}=\frac{1+D\cos\theta}{1-D\cos\theta}.
\end{equation}
In the special case when $\theta=0$ ($\mathbf{G}$ is initially diagonal), $D_{\mathrm{min}}'=0$ at $\eta_{\mathrm{o}}=\tfrac{1+D}{1-D}=\tfrac{\lambda_{1}}{\lambda_{2}}$, and we retrieve the previous result; see Fig.~\ref{fig:Filtering}(b).

\subsubsection{Decohering operators}\label{sec:Decohering}

In contrast to filtering operators that change $D$ at the cost of reducing the power, a `decohering' system reduces $D$ by introducing randomness or stochastic fluctuations into the optical field without loss of energy. We model such a system as an ensemble of unitaries $\{\hat{U}(\chi)\}$ containing random parameters $\chi$ characterized by a probability distribution $P(\chi)$, where $\chi$ can refer to a single or multiple stochastic variables, and $\int\!d\chi P(\chi)=1$. When traversing a unitary $\hat{U}(\chi)$ selected from this ensemble, the coherence matrix $\mathbf{G}$ is transformed according to $\mathbf{G}\rightarrow\mathbf{G}'=\hat{U}(\chi)\mathbf{G}\hat{U}^{\dagger}(\chi)$. When averaged over the ensemble, the coherence matrix is given by:
\begin{equation}
\mathbf{G}'=\int d\chi\;\left\{\hat{U}(\chi)\mathbf{G}\hat{U}^{\dagger}(\chi)\right\}P(\chi).
\end{equation}
In contrast to filtering operators, this definition of a decohering operator is \textit{trace-preserving}. Note that such an operation always \textit{increases} the entropy $S$ (or \textit{reduces} the degree of coherence $D$).

Consider a unitary comprising a phase operator $\hat{S}(\varphi)$ followed by a rotator $\hat{R}(\theta)$, which is described by the unitary $\hat{U}_{\mathrm{d}}(\theta,\varphi)$ given by:
\begin{equation}
\hat{U}_{\mathrm{d}}(\theta,\varphi)=\hat{R}(\theta)\hat{S}(\varphi)=\left(\begin{array}{cc}
e^{i\varphi/2}\cos\tfrac{\theta}{2}&
-e^{-i\varphi/2}\sin\tfrac{\theta}{2}\\
e^{i\varphi/2}\sin\tfrac{\theta}{2}&
e^{-i\varphi/2}\cos\tfrac{\theta}{2}\end{array}\right).
\end{equation}
The elements of a coherence matrix $\mathbf{G}$ with Stokes parameters $\{s_{0},s_{1},s_{2},s_{3}\}$ are transformed by $\hat{U}_{\mathrm{d}}(\theta,\varphi)$ to:
\begin{eqnarray}
G_{11}'&=&\frac{1}{2}\left\{1+s_{1}\cos\theta-(s_{2}\cos\varphi+s_{3}\sin\varphi)\sin\theta\right\},\nonumber\\
G_{22}'&=&\frac{1}{2}\left\{1-s_{1}\cos\theta+(s_{2}\cos\varphi+s_{3}\sin\varphi)\sin\theta\right\},\nonumber\\
G_{12}'&=&\frac{1}{2}\left\{s_{1}\sin\theta+s_{2}(\cos\theta\cos\varphi+i\sin\varphi)-is_{3}(\cos\varphi+i\cos\theta\sin\varphi)\right\}.
\end{eqnarray}
Consequently, the Stokes parameters $\{s_{0}',s_{1}',s_{2}',s_{3}'\}$ for $\mathbf{G}'$ are given by:
\begin{eqnarray}
s_{1}'&=&s_{1}\cos\theta-(s_{2}\cos\varphi+s_{3}\sin\varphi)\sin\theta,\nonumber\\
s_{2}'&=&s_{1}\sin\theta+(s_{2}\cos\varphi+s_{3}\sin\varphi)\cos\theta,\nonumber\\
s_{3}'&=&-s_{2}\sin\varphi+s_{3}\cos\varphi.
\end{eqnarray}
If the rotation angle $\theta$ is varied randomly over the range $[0,\pi]$ with a uniform probability, then the elements of the coherence matrix become $G_{11}'=G_{22}'=\tfrac{1}{2}$ and $G_{12}'=\tfrac{i}{2}(s_{2}\sin\varphi-s_{3}\cos\varphi)$; or, equivalently, $s_{1}'=s_{2}'=0$ and $s_{3}'=-s_{2}\sin\varphi+s_{3}\cos\varphi$. In this case, setting $\varphi=0$ (corresponding to utilizing $\hat{R}(\theta)$ alone) yields $G_{11}'=G_{22}'=\tfrac{1}{2}$ and $G_{12}'=-\tfrac{i}{2}s_{3}$ (or $s_{1}'=s_{2}'=0$ and $s_{3}'=s_{3}$). Consequently, $\mathbf{G}'=\tfrac{1}{2}\hat{\mathbb{I}}_{2}$ (fully decohered) for any input field in which $s_{3}=0$. However, to guarantee that $\mathbf{G}'=\tfrac{1}{2}\hat{\mathbb{I}}_{2}$ for \textit{any} input field, rather than setting $\varphi=0$ we instead randomly vary $\varphi$ over the range $[0,2\pi)$ with a uniform probability (in addition to the random variation in $\theta$).

Instead of the random variation in $\theta$ and/or $\varphi$ over a continuous range, one may consider a discrete ensemble of only 4 unitaries:
\begin{eqnarray}
\hat{U}_{1}&=&\hat{U}_{\mathrm{d}}(0,0)=\left(\begin{array}{cc}1&0\\0&1\end{array}\right),\;\;
\hat{U}_{2}=\hat{U}_{\mathrm{d}}(\pi,0)=\left(\begin{array}{cc}0&-1\\1&0\end{array}\right),\nonumber\\
\hat{U}_{3}&=&\hat{U}_{\mathrm{d}}(0,\pi)=i\left(\begin{array}{cc}1&0\\0&-1\end{array}\right),\;\;
\hat{U}_{4}=\hat{U}_{\mathrm{d}}(\pi,\pi)=i\left(\begin{array}{cc}0&1\\1&0\end{array}\right). 
\end{eqnarray}
By randomly selecting a unitary from this ensemble with equal probabilities $P_{j}=\tfrac{1}{4}$ ($j=1,\cdots,4$), we have $\mathbf{G}'=\sum_{j=1}^{4}P_{j}\hat{U}_{j}\mathbf{G}\hat{U}_{j}^{\dagger}=\tfrac{1}{2}\hat{\mathbb{I}}_{2}$ independently of the initial coherence matrix $\mathbf{G}$.

Such decohering systems always \textit{increase} the entropy $S$ of the field. Consequently, starting with an incoherent field $\mathbf{G}=\tfrac{1}{2}\hat{\mathbb{I}}_{2}$ with maximum entropy $S=1$, a decohering matrix has no impact because the entropy cannot be increased further. Nevertheless, there exist decohering operators that are \textit{trace-preserving} and can \textit{increase or decrease the entropy}, but requires introducing a second DoF for the field (Section~\ref{sec:ReducedRestricted}).

\subsection{Example~1: Polarization DoF}\label{sec:polarizationDoF}

\begin{figure}[t!]
\centering
\includegraphics[width=11.7cm]{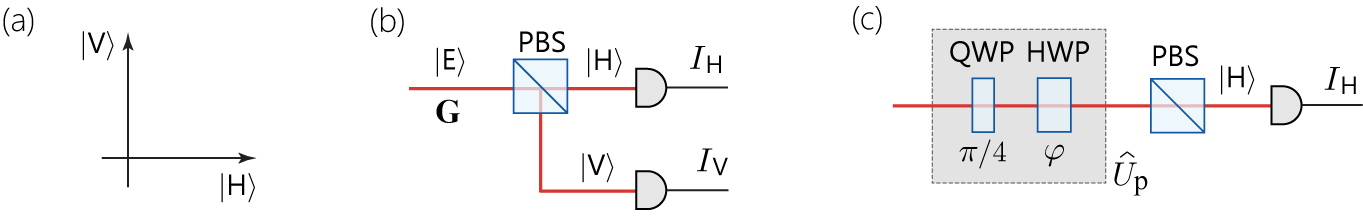}
\caption{(a) The polarization DoF is spanned by the modes $|\mathrm{H}\rangle$ and $|\mathrm{V}\rangle$. (b) The modal weights are obtained by first separating the $|\mathrm{H}\rangle$ and $|\mathrm{V}\rangle$ modes via a polarizing beam splitter (PBS) followed by two polarization-insensitive detectors. When the field is polarized, $I_{\mathrm{H}}=|E_{\mathrm{H}}|^{2}$ and $I_{\mathrm{V}}=|E_{\mathrm{V}}|^{2}$, and when the field is partially polarized, $I_{\mathrm{H}}=G_{\mathrm{HH}}$ and $I_{\mathrm{V}}=G_{\mathrm{VV}}$. (c) Polarization interference is observed by placing in the field path the unitary $\hat{U}_{\mathrm{p}}$ (Eq.~\ref{eq:PolInterference}), which comprises a quarter-wave plate (QWP) rotated by an angle $\tfrac{\pi}{4}$ with respect to $|\mathrm{H}\rangle$, a half-wave plate (HWP) rotated by an angle $\varphi$, followed by a PBS. Only one modal weight is needed to produce polarization interference whose visibility corresponds to $2|G_{\mathrm{HV}}|$.}
\label{fig:PolarizationBasic}
\end{figure}

Many of the concepts described above are familiar once applied to the polarization DoF. The two modes here are the horizontal and vertical polarized field components, $|\mathrm{H}\rangle$ and $|\mathrm{V}\rangle$, respectively [Fig.~\ref{fig:PolarizationBasic}(a)]. In this modal basis, a \textit{fully polarized} field is written as $|E\rangle=E_{\mathrm{H}}|\mathrm{H}\rangle+E_{\mathrm{V}}|E_{\mathrm{V}}\rangle=\left(\begin{array}{c}E_{\mathrm{H}}\\E_{\mathrm{V}}\end{array}\right)$, where $|\mathrm{H}\rangle=\left(\begin{array}{c}1\\0\end{array}\right)$ and $|\mathrm{V}\rangle=\left(\begin{array}{c}0\\1\end{array}\right)$. The modal weights can be measured in the setup shown in Fig.~\ref{fig:PolarizationBasic}(b), where the polarization modes are separated spatially by a polarizing beam splitter (PBS) followed by polarization-insensitive detectors: $I_{\mathrm{H}}=|E_{\mathrm{H}}|^{2}$ and $I_{\mathrm{V}}=|E_{\mathrm{V}}|^{2}$. Normalizing the field vector $\langle E|E\rangle=1$ entails that $|E_{\mathrm{H}}|^{2}+|E_{\mathrm{V}}|^{2}=1$, which allows us to write $|E\rangle=\left(\begin{array}{c}\cos\tfrac{\theta}{2}\\e^{i\varphi}\sin\tfrac{\theta}{2}\end{array}\right)$. The most general form of polarization is elliptical, which degenerates into linear polarization when $\varphi=0$, and into circular polarization when $\varphi=\tfrac{\pi}{2}$ and $\theta=\pm\tfrac{\pi}{2}$ \cite{Brosseau98Book,SalehBook07}.

In the case of a \textit{partially polarized} field, the polarization coherence matrix is:
\begin{equation}
\mathbf{G}_{\mathrm{p}}=\left(\begin{array}{cc}G_{\mathrm{HH}}&G_{\mathrm{HV}}\\G_{\mathrm{VH}}&G_{\mathrm{VV}}\end{array}\right),
\end{equation}
where $G_{jk}=\langle E_{j}E_{k}^{*}\rangle$, $\langle\cdot\rangle$ represents averaging over a statistical ensemble, and $j,k=\mathrm{H},\mathrm{V}$. Because $\mathbf{G}_{\mathrm{p}}$ is Hermitian, $G_{\mathrm{HH}}$ and $G_{\mathrm{VV}}$ are real, and $G_{\mathrm{HV}}=G_{\mathrm{VH}}^{*}$. The diagonal elements $I_{\mathrm{H}}=G_{\mathrm{HH}}$ and $I_{\mathrm{V}}=G_{\mathrm{VV}}$ are the fractions of power associated with the $|\mathrm{H}\rangle$ and $|\mathrm{V}\rangle$ modes, respectively, and $\mathrm{Tr}\{\mathbf{G}\}=G_{\mathrm{HH}}+G_{\mathrm{VV}}=1$. The off-diagonal element $G_{\mathrm{HV}}$, which represents the correlation between the $|\mathrm{H}\rangle$ and $|\mathrm{V}\rangle$ modes, is not measured directly in the setup in Fig.~\ref{fig:PolarizationBasic}(b). The coherence matrix can be diagonalized by a unitary to take the form $\mathbf{G}_{\mathrm{p}}^{\mathrm{D}}=\left(\begin{array}{cc}\lambda_{\mathrm{H}}&0\\0&\lambda_{\mathrm{V}}\end{array}\right)$, where the real eigenvalues $\lambda_{\mathrm{H}}$ and $\lambda_{\mathrm{V}}$ are arranged so that $\lambda_{\mathrm{H}}\geq\lambda_{\mathrm{V}}\geq0$. A restricted polarization unitary,
\begin{equation}\label{eq:RestrictedPolarizationUnitary}
\hat{U}_{\mathrm{p}}=\left(\begin{array}{cc}\cos\tfrac{\theta}{2}&-e^{-i\varphi}\sin\tfrac{\theta}{2}\\e^{i\varphi}\sin\tfrac{\theta}{2}&\cos\tfrac{\theta}{2}\end{array}\right)    
\end{equation}
casts the diagonal polarization coherence matrix $\mathbf{G}_{\mathrm{p}}^{\mathrm{D}}$ into the general form:
\begin{equation}
\mathbf{G}_{\mathrm{p}}=\hat{U}_{\mathrm{p}}\mathbf{G}^{\mathrm{D}}\hat{U}_{\mathrm{p}}^{\dagger}=\frac{1}{2}\left(\begin{array}{cc}1+(\lambda_{\mathrm{H}}-\lambda_{\mathrm{V}})\cos\theta&(\lambda_{\mathrm{H}}-\lambda_{\mathrm{V}})e^{-i\varphi}\sin\theta\\(\lambda_{\mathrm{H}}-\lambda_{\mathrm{V}})e^{i\varphi}\sin\theta&1-(\lambda_{\mathrm{H}}-\lambda_{\mathrm{V}})\cos\theta\end{array}\right),
\end{equation}
and the polarization unitary $\hat{U}_{\mathrm{p}}^{\dagger}$ diagonalizes $\mathbf{G}_{\mathrm{p}}$: $\hat{U}_{\mathrm{p}}^{\dagger}\mathbf{G}_{\mathrm{p}}\hat{U}_{\mathrm{p}}=\mathbf{G}_{\mathrm{p}}^{\mathrm{D}}$. The degree of polarization $D_{\mathrm{p}}$ is then defined as:
\begin{equation}
D_{\mathrm{p}}=\lambda_{\mathrm{H}}-\lambda_{\mathrm{V}}=\sqrt{1-4\mathrm{det}\{\mathbf{G}_{\mathrm{p}}\}},
\end{equation}
and the polarization entropy is $S_{\mathrm{p}}=-\mathrm{Tr}\{\mathbf{G}_{\mathrm{p}}\log_{2}\mathbf{G}_{\mathrm{p}}\}=-\lambda_{\mathrm{H}}\log_{2}\lambda_{\mathrm{H}}-\lambda_{\mathrm{V}}\log_{2}\lambda_{\mathrm{V}}$. A polarized field corresponds to $\lambda_{\mathrm{H}}=1$ and $\lambda_{\mathrm{V}}=0$, so that $D_{\mathrm{p}}=1$ and $S_{\mathrm{p}}=0$. In this case, $\mathbf{G}_{\mathrm{p}}$ can be expressed as an outer product, $\mathbf{G}_{\mathrm{p}}=|E\rangle\langle E|$ with $|E\rangle=\left(\begin{array}{c}\cos\tfrac{\theta}{2}\\e^{i\varphi}\sin\tfrac{\theta}{2}\end{array}\right)$. In contrast, an unpolarized field corresponds to $\lambda_{\mathrm{H}}=\lambda_{\mathrm{V}}=\tfrac{1}{2}$, so that $D_{\mathrm{p}}=0$ and $S_{\mathrm{p}}=1$~bit, and $\mathbf{G}_{\mathrm{p}}=\tfrac{1}{2}\hat{\mathbb{I}}_{2}$. The polarization DoF (as an example of a binary DoF) can thus carry at most 1~bit of entropy.

From the above it is clear that a \textit{polarized} field, which is in general elliptical, can always be converted into the $|\mathrm{H}\rangle$ polarization mode via a unitary. If $|E\rangle=\left(\begin{array}{c}\cos\tfrac{\theta}{2}\\e^{i\varphi}\sin\tfrac{\theta}{2}\end{array}\right)$, then $\hat{U}_{\mathrm{p}}^{\dagger}|E\rangle=|\mathrm{H}\rangle$, where $\hat{U}_{\mathrm{p}}$ is given in Eq.~\ref{eq:RestrictedPolarizationUnitary}. That is, one can always unitarily eliminate one of the polarization modes from a polarized field. However, when the field is \textit{partially polarized}, the total power initially distributed between the two polarization modes can\textit{not} be concentrated into $|\mathrm{H}\rangle$ via a polarization unitary; some power must always remain in the $|\mathrm{V}\rangle$ polarization mode. At best, a fraction $\lambda_{1}<1$ of the input power can be concentrated into $|\mathrm{H}\rangle$ via a unitary (by diagonalizing $\mathbf{G}_{\mathrm{p}}$).

The question remains as how to construct the general polarization unitary $\hat{U}_{\mathrm{p}}$ in Eq.~\ref{eq:RestrictedPolarizationUnitary}, which requires a `rotator' $\hat{R}(\theta)$ and a phase operator $\hat{S}(\varphi)$ (Eq.~\ref{eq:PhaseAndRotationUnitaries}). Polarization rotators are typically bulky because they rely on the Faraday effect and thus require a magnetic field. Instead, unitary manipulations of polarization utilize `wave plates' that introduce a phase $\varphi$ between two orthogonal polarization modes identified by the fast and slow axes (or principal axes) of the wave plate. Aligning these two axes with $|\mathrm{H}\rangle$ and $|\mathrm{V}\rangle$ yields the unitary $\hat{S}(\varphi)=\left(\begin{array}{cc}e^{i\varphi/2}&0\\0&e^{-i\varphi/2}\end{array}\right)$. However, a general wave plate with tunable $\varphi$ typically requires a liquid-crystal device. Instead, the most common phase plates are the half-wave plate (HWP) and the quarter-wave plate (QWP) with $\varphi=\pi$ and $\tfrac{\pi}{2}$, respectively represented by the unitaries:
\begin{equation}
\hat{U}_{\mathrm{HWP}}=i\left(\begin{array}{cc}1&0\\0&-1\end{array}\right),\;\;\hat{U}_{\mathrm{QWP}}=e^{i\pi/4}\left(\begin{array}{cc}1&0\\0&-i\end{array}\right).
\end{equation}

Polarization unitaries are typically constructed out of combinations of HWPs and QWPs rotated with respect to their principal axes. Rotating the wave plates by a physical angle $\theta$ with respect to $|\mathrm{H}\rangle$ yields:
\begin{eqnarray}\label{eq:RotatedWavePlates}
\hat{U}_{\mathrm{HWP}}(\theta)\!\!&=&\!\!\hat{R}(2\theta)\hat{U}_{\mathrm{HWP}}\hat{R}(-2\theta)=i\left(\begin{array}{cc}\cos2\theta&\sin2\theta\\\sin2\theta&-\cos2\theta\end{array}\right),\nonumber\\
\hat{U}_{\mathrm{QWP}}(\theta)\!\!&=&\!\!\hat{R}(2\theta)\hat{U}_{\mathrm{QWP}}\hat{R}(-2\theta)=\frac{1}{\sqrt{2}}\left(\begin{array}{cc}1+i\cos2\theta&i\sin2\theta\\i\sin2\theta&1-i\cos2\theta\end{array}\right).
\end{eqnarray}
Useful special cases include $\hat{U}_{\mathrm{HWP}}(\tfrac{\pi}{8})=\tfrac{i}{\sqrt{2}}\left(\begin{array}{cc}1&1\\1&-1\end{array}\right)$, $\hat{U}_{\mathrm{QWP}}(\tfrac{\pi}{4})=\tfrac{1}{\sqrt{2}}\left(\begin{array}{cc}1&i\\i&1\end{array}\right)$, and a polarization unitary to exhibit polarization interference fringes comprises a QWP rotated an angle $\tfrac{\pi}{4}$ followed by a HWP rotated by an angle $\varphi$:
\begin{equation}\label{eq:PolInterference}
\hat{U}_{\mathrm{p}}=\hat{U}_{\mathrm{HWP}}(\varphi)\hat{U}_{\mathrm{QWP}}(\tfrac{\pi}{4})=\tfrac{i}{\sqrt{2}}\left(\begin{array}{cc}e^{i2\varphi}&ie^{-i2\varphi}\\-ie^{i2\varphi}&-e^{-i2\varphi}\end{array}\right).
\end{equation}

Such wave plates and their combinations enable us to carry out several important tasks. For example, we can utilize the polarization unitary in Eq.~\ref{eq:PolInterference} to observe polarization interference fringes. The off-diagonal element $G_{\mathrm{HV}}$ determines the visibility of the polarization interference fringes, in analogy with double-slit interference for the spatial DoF. By varying the angle $\varphi$ of the HWP over $[0,\tfrac{\pi}{4}]$ and measuring the $|\mathrm{H}\rangle$ modal weight after a PBS, we obtain the interferogram pattern $I_{\mathrm{H}}=\tfrac{1}{2}[1+2|G_{\mathrm{HV}}|\sin(4\varphi-\mathrm{arg}\{G_{\mathrm{HV}}\})]$, whose visibility is $V=2|G_{\mathrm{HV}}|$; see Fig.~\ref{fig:PolarizationBasic}(c). The most general $2\times2$ polarization unitary $\hat{U}_{\mathrm{p}}$ can be constructed by cascading a sequence of HWPs and QWPs \cite{Sit17JO}; typically, at least 3~such wave plates are needed.

\begin{figure}[t!]
\centering
\includegraphics[width=13.3cm]{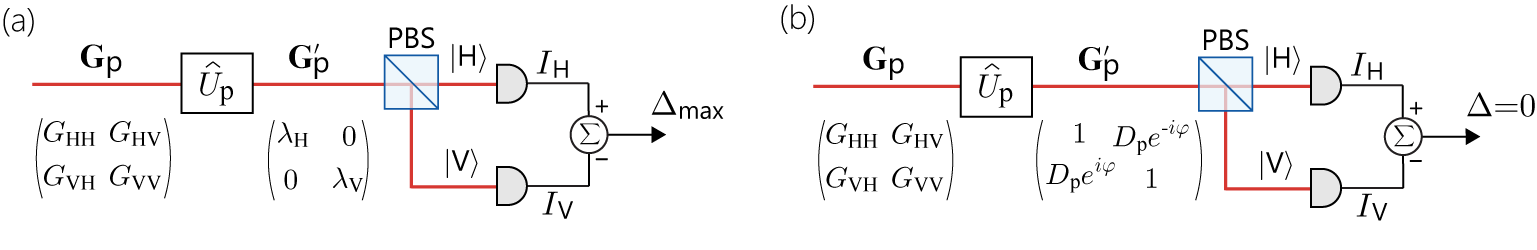}
\caption{(a) Determining the degree of polarization $D_{\mathrm{p}}$ through diagonalization. A polarization unitary $\hat{U}_{\mathrm{p}}$ precedes a PBS, and $\hat{U}_{\mathrm{p}}$ is tuned to maximize $\Delta=I_{\mathrm{H}}-I_{\mathrm{V}}$. Once $\Delta$ is maximized, $\mathbf{G}_{\mathrm{p}}'$ is diagonalized, $I_{\mathrm{H}}$ and $I_{\mathrm{V}}$ correspond to the eigenvalues $\lambda_{\mathrm{H}}$ and $\lambda_{\mathrm{V}}$, respectively, and $D_{\mathrm{p}}=\Delta_{\mathrm{max}}=I_{\mathrm{H}}-I_{\mathrm{V}}$. (b) Determining $D_{\mathrm{p}}$ through equalization. The parameters of $\hat{U}$ are tuned to equalize $I_{\mathrm{H}}$ and $I_{\mathrm{V}}$ and reach $\Delta=I_{\mathrm{H}}-I_{\mathrm{V}}=0$. Once $\Delta=0$, the magnitude of the off-diagonal element is $D_{\mathrm{p}}/2$, which can be revealed via the polarization interference configuration in Fig.~\ref{fig:PolarizationBasic}(c).}
\label{fig:DegreeOfPolarization}
\end{figure}

The degree of polarization $D_{\mathrm{p}}$ may be measured in one of several ways. First, the field traverses a general polarization unitary $\hat{U}_{\mathrm{p}}$ (Eq.~\ref{eq:RestrictedPolarizationUnitary}) whose parameters are scanned to maximize the difference between the modal weights $\Delta=I_{\mathrm{H}}-I_{\mathrm{V}}$. Once $\Delta_{\mathrm{max}}$ is reached, $\mathbf{G}_{\mathrm{p}}'$ is diagonalized, and $\Delta_{\mathrm{max}}=\lambda_{\mathrm{H}}-\lambda_{\mathrm{V}}=D_{\mathrm{p}}$ [Fig.~\ref{fig:DegreeOfPolarization}(a)]. Alternatively, the parameters of the general polarization unitary $\hat{U}_{\mathrm{p}}$ are scanned to reach $\Delta=0$, whereupon the modal weights are equalized, $G_{\mathrm{HH}}=G_{\mathrm{VV}}=\tfrac{1}{2}$, and $\mathbf{G}_{\mathrm{p}}'=\tfrac{1}{2}\left(\begin{array}{cc}1&D_{\mathrm{p}}e^{-i\varphi}\\D_{\mathrm{p}}e^{i\varphi}&1\end{array}\right)$ [Fig.~\ref{fig:DegreeOfPolarization}(b)]. To observe the off-diagonal element $G_{\mathrm{HV}}$, we can use the polarization interference configuration in Fig.~\ref{fig:PolarizationBasic}(c) with $\hat{U}_{\mathrm{p}}$ from Eq.~\ref{eq:PolInterference}, whereupon the visibility of the polarization interference reveals the off-diagonal term, $V=2|G_{\mathrm{HV}}|=D_{\mathrm{p}}$.

\begin{figure}[t!]
\centering
\includegraphics[width=13.3cm]{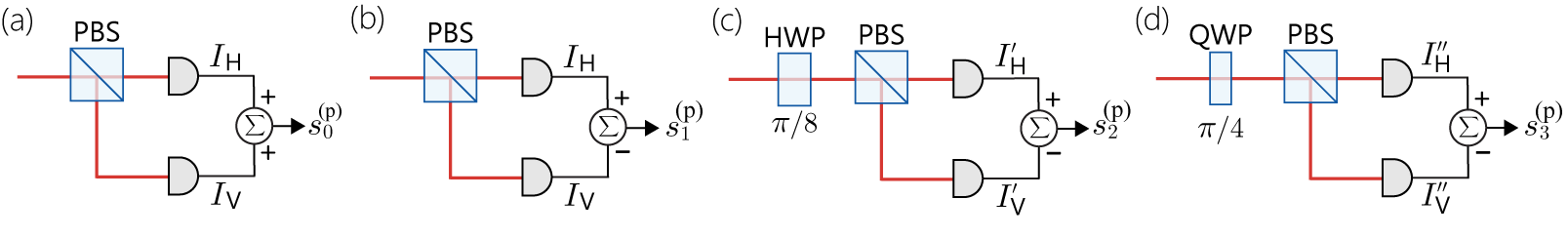}
\caption{Polarization Stokes parameters. (a) Measuring $s_{0}^{(\mathrm{p})}=I_{\mathrm{H}}+I_{\mathrm{V}}$, (b) $s_{1}^{(\mathrm{p})}=I_{\mathrm{H}}-I_{\mathrm{V}}$, (c) $s_{2}^{(\mathrm{p})}=I_{\mathrm{H}}'-I_{\mathrm{V}}'$ after a unitary $\hat{U}_{\mathrm{p}}=\hat{U}_{\mathrm{HWP}}(\tfrac{\pi}{8})$, and (d) $s_{3}^{(\mathrm{p})}=I_{\mathrm{H}}''-I_{\mathrm{V}}''$ after a unitary $\hat{U}_{\mathrm{p}}=\hat{U}_{\mathrm{QWP}}(\tfrac{\pi}{4})$.}
\label{fig:PolarizationStokes}
\end{figure}

A third approach relies on reconstructing $\mathbf{G}_{\mathrm{p}}$ by measuring the Stokes parameters using the configurations illustrated in Fig.~\ref{fig:PolarizationStokes}. The Stokes parameter $s_{0}$ required for normalization is obtained by adding the modal weights $s_{0}^{(\mathrm{p})}=I_{\mathrm{H}}+I_{\mathrm{V}}$ [Fig.~\ref{fig:PolarizationStokes}(a)], and $s_{1}^{(\mathrm{p})}$ is obtained from their difference $s_{1}^{(\mathrm{p})}=I_{\mathrm{H}}-I_{\mathrm{V}}$ after normalization [Fig.~\ref{fig:PolarizationStokes}(b)]. The Stokes parameter $s_{2}^{(\mathrm{p})}$ is obtained after the field traverses a HWP with $\hat{U}_{\mathrm{HWP}}(\tfrac{\pi}{8})=\tfrac{i}{\sqrt{2}}\left(\begin{array}{cc}1&1\\1&-1\end{array}\right)$, whereupon $s_{2}^{(\mathrm{p})}=I_{\mathrm{H}}'-I_{\mathrm{V}}'$ [Fig.~\ref{fig:PolarizationStokes}(c)]. Finally, the Stokes parameter $s_{3}^{(\mathrm{p})}$ is obtained after the field traverses a QWP with $\hat{U}_{\mathrm{QWP}}(-\tfrac{\pi}{4})=\tfrac{1}{\sqrt{2}}\left(\begin{array}{cc}1&-i\\-i&1\end{array}\right)$, whereupon $s_{3}^{(\mathrm{p})}=I_{\mathrm{H}}''-I_{\mathrm{V}}''$ [Fig.~\ref{fig:PolarizationStokes}(d)]. The degree of polarization is then given by $D_{\mathrm{p}}=\sqrt{(s_{1}^{(\mathrm{p})})^{2}+(s_{2}^{(\mathrm{p})})^{2}+(s_{3}^{(\mathrm{p})})^{2}}$.

A polarization unitary does not change $D_{\mathrm{p}}$ nor $S_{\mathrm{p}}$. Instead, `partial polarizers' are needed to modify $D_{\mathrm{p}}$ and $S_{\mathrm{p}}$. A partial polarizer is a non-unitary filtering operation (Section~\ref{sec:Filtering}) that reduces the amplitudes of the two orthogonal polarization modes by different amounts. When these two modes are aligned with $|\mathrm{H}\rangle$ and $|\mathrm{V}\rangle$, the partial polarizer is represented by the non-unitary operator $\hat{F}^{\mathrm{D}}=\left(\begin{array}{cc}\sqrt{d_{\mathrm{H}}}&0\\0&\sqrt{d_{\mathrm{V}}}\end{array}\right)$. One can use a partial polarizer to produce from a diagonalized polarization coherence matrix $\mathbf{G}_{\mathrm{p}}^{\mathrm{D}}$ a polarized field ($d_{\mathrm{V}}=0\rightarrow D_{\mathrm{p}}=1$), a fully unpolarized field ($d_{\mathrm{H}}\lambda_{\mathrm{H}}=d_{\mathrm{V}}\lambda_{\mathrm{V}}\rightarrow D_{\mathrm{p}}=0$), or any other desired value of $D_{\mathrm{p}}$.

Finally, one may implement a simple depolarizer (Section~\ref{sec:Decohering}) using a HWP whose physical rotation angle $\theta$ (Eq.~\ref{eq:RotatedWavePlates}) is varied randomly. The elements of the new polarization coherence matrix is given in terms of the initial Stokes parameters as:
\begin{eqnarray}
G_{\mathrm{HH}}'\!\!&=&\!\!\frac{1}{2}\left\{1+s_{1}\cos4\theta+s_{2}\sin4\theta\right\},\nonumber\\
G_{\mathrm{VV}}'\!\!&=&\!\!\frac{1}{2}\left\{1-s_{1}\cos4\theta-s_{2}\sin4\theta\right\},\nonumber\\
G_{\mathrm{HV}}'\!\!&=&\!\!\frac{1}{2}\left\{s_{1}\sin4\theta-s_{2}\cos4\theta+is_{3}\right\}.
\end{eqnarray}
If $\theta$ is varied over the range $[0,\tfrac{\pi}{2})$ with uniform probability, then $G_{\mathrm{HH}}'=G_{\mathrm{VV}}'=\tfrac{1}{2}$ and $G_{\mathrm{HV}}'=\tfrac{i}{2}s_{3}$. As long as $s_{3}=0$, the field is rendered unpolarized, $\mathbf{G}_{\mathrm{p}}=\tfrac{1}{2}\hat{\mathbb{I}}_{2}$. Alternatively, we can randomly select (with equal probabilities) one of 4~unitaries that can be realized using HWPs:
\begin{eqnarray}
\hat{U}_{1}\!\!&=&\!\!\left(\begin{array}{cc}1&0\\0&1\end{array}\right)=\hat{U}_{\mathrm{HWP}}(0)\hat{U}_{\mathrm{HWP}}(0),\;\;\hat{U}_{2}=\left(\begin{array}{cc}0&-1\\1&0\end{array}\right)=\hat{U}_{\mathrm{HWP}}(0)\hat{U}_{\mathrm{HWP}}(\frac{\pi}{4}),\nonumber\\
\hat{U}_{3}\!\!&=&\!\!i\left(\begin{array}{cc}1&0\\0&-1\end{array}\right)=\hat{U}_{\mathrm{HWP}}(0),\;\;\hat{U}_{4}=i\left(\begin{array}{cc}0&1\\1&0\end{array}\right)=\hat{U}_{\mathrm{HWP}}(\tfrac{\pi}{4}).
\end{eqnarray}
Alternative selections of unitaries can of course be made as depolarizes.

\subsection{Example~2: Two-mode spatial DoF}\label{sec:SpatialDoF}

Utilizing the matrix formalism for partial coherence outlined above for the \textit{spatial} DoF is less familiar than for polarization where it is routinely used. However, it is crucial to adopt this approach for structured spatial coherence, especially as the utilization of partially coherent optical fields moves from conventional free-space settings to on-chip implementations \cite{Hashemi26TwoModes,Hashemi26FourModes}.

\begin{figure}[t!]
\centering
\includegraphics[width=2.4in]{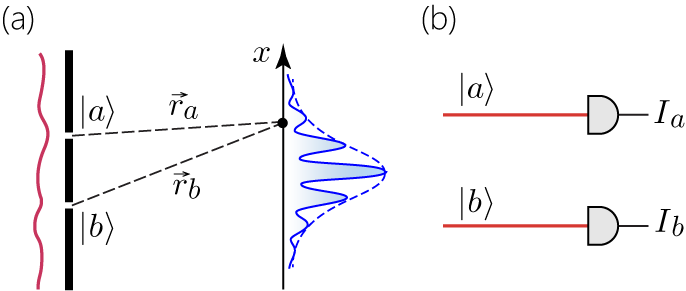}
\caption{(a) Traditional double-slit interference. The fields from $|a\rangle$ and $|b\rangle$ traverse distances $\vec{r}_{a}$ and $\vec{r}_{b}$ to an observation point. (b) A generic field comprising two spatial modes $|a\rangle$ and $|b\rangle$. Detectors record the modal weights. When the field is coherent, we have $I_{a}=|E^{a}|^{2}$ and $I_{b}=|E^{b}|^{2}$; and when the field is partially coherent, we have $I_{a}=G^{aa}$ and $I_{b}=G^{bb}$.}
\label{fig:SpatialBasic}
\end{figure}

Spatial coherence with two spatial modes, denoted $|a\rangle$ and $|b\rangle$, has been considered in multiple settings. Most commonly, when two points in a continuous field identified by a pair of identical slits are taken as the spatial modes \cite{Gori06OL,Abouraddy17OE}, and partial spatial coherence is manifest in the reduced visibility of double-slit interference fringes after equalizing the field amplitudes at $|a\rangle$ and $|b\rangle$. There are technical drawbacks to this approach [Fig.~\ref{fig:SpatialBasic}(a)]; e.g., (1) the fields from $|a\rangle$ and $|b\rangle$ travel different distances ($\vec{r}_{a}$ and $\vec{r}_{b}$, respectively), so they are weighted by slightly different amplitudes, and (2) the overall interference pattern is modulated by the diffraction pattern from a single slit. The impact of these two factors can be reduced by maintaining the slits and the interference fringes close to the optical axis. More fundamental issues are: (1) the dimensional mismatch between the input field (spanned by two modes $|a\rangle$ and $|b\rangle$) and the output field (extending continuously along one spatial dimension); (2) only two points (having the maximum and minimum intensity values $I_{\mathrm{max}}$ and $I_{\mathrm{min}}$, respectively) contribute to the estimation of the visibility; and (3) an imbalance in the values of the intensity at the two slits reduces the estimate of $D$ (i.e., $V\leq D$) \cite{Zernicke}.

Another setting comprising two spatial modes (a binary spatial DoF) is that of a pair of overlapping modes in free space (e.g., parity modes \cite{Abouraddy07PRA,Yarnall07PRL1,Yarnall07PRL2,Yarnall08OE}) or in a two-mode fiber or waveguide. A less-common setting is the field in a pair of single-mode waveguides \cite{Hashemi26TwoModes,Hashemi26FourModes}. Because on-chip implementations for the manipulation of spatial coherence are expected to become a major research endeavor in the next few years, we consider here this latter configuration: two spatial modes confined to a pair of single-mode waveguide [Fig.~\ref{fig:SpatialBasic}(b)]. However, the analysis applies to any other realization of bimodal spatial coherence. 

When the field is \textit{spatially coherent}, it can be written as a field vector:
\begin{equation}
|E\rangle=E^{a}|a\rangle+E^{b}|b\rangle=E^{a}\left(\begin{array}{c}1\\0\end{array}\right)+E^{b}\left(\begin{array}{c}0\\1\end{array}\right)=\left(\begin{array}{c}E^{a}\\E^{b}\end{array}\right).
\end{equation}
We make use here of superscripts to identify the spatial modal amplitudes for future convenience in Section~\ref{sec:2DoFs} where we combine the spatial and polarization DoFs. The modal weights can be measured in the configuration illustrated in Fig.~\ref{fig:SpatialBasic}(b), where $I_{a}=|E^{a}|^{2}$ and $I_{b}=|E^{b}|^{2}$. When the field is \textit{partially coherent}, we make use of a $2\times2$ spatial coherence matrix given by:
\begin{equation}
\mathbf{G}_{\mathrm{s}}=\left(\begin{array}{cc}G^{aa}&G^{ab}\\G^{ba}&G^{bb}\end{array}\right),
\end{equation}
where $G^{jk}=\langle E^{j}(E^{k})^{*}\rangle$, $j,k=a,b$, and $\langle\cdot\rangle$ is a statistical average over an ensemble. This spatial coherence matrix is Hermitian, $\mathbf{G}_{\mathrm{s}}^{\dagger}=\mathbf{G}_{\mathrm{s}}$, so that $G^{aa}$ and $G^{bb}$ are real, and $G^{ba}=(G^{ab})^{*}$. Moreover, the coherence matrix has unity trace, $\mathrm{Tr}\{\mathbf{G}_{\mathrm{s}}\}=G^{aa}+G^{bb}=1$. The diagonal elements are measured using the setup in Fig.~\ref{fig:SpatialBasic}(b), where $I_{a}=G^{aa}$ and $I_{b}=G^{bb}$, which thus correspond to the fractions of power in modes $|a\rangle$ and $|b\rangle$, respectively. The off-diagonal element $G^{ab}$ represents the statistical correlation between the $|a\rangle$ and $|b\rangle$ modes, and thus cannot be measured directly using the configuration in Fig.~\ref{fig:SpatialBasic}(b). Rather, $|G^{ab}|$ is related to the visibility of interference fringes formed from the overlap of the fields associated with the two modes, and the phase of $G^{ab}$ shifts the fringes. We define the degree of spatial coherence $D_{\mathrm{s}}$ as: 
\begin{equation}
D_{\mathrm{s}}=\lambda_{a}-\lambda_{b}=\sqrt{1-4\mathrm{det}\{\mathbf{G}_{\mathrm{s}}\}},
\end{equation}
where $\lambda_{a}$ and $\lambda_{b}$ are the (real, positive) eigenvalues of $\mathbf{G}_{\mathrm{s}}$, and we take $1\geq\lambda_{a}\geq\lambda_{b}\geq0$. The spatial entropy is defined as $S_{\mathrm{s}}=-\mathrm{Tr}\{\mathbf{G}_{\mathrm{s}}\log_{2}\mathbf{G}_{\mathrm{s}}\}=\lambda_{a}\log_{2}\lambda_{a}-\lambda_{b}\log_{2}\lambda_{b}$, with $0\leq S_{\mathrm{s}}\leq1$. The spatial DoF in a two-mode field thus carries at most 1~bit of entropy.

Spatially \textit{coherent} light ($D_{\mathrm{s}}=1$ and $S_{\mathrm{s}}=0$) corresponds to $\lambda_{a}=1$ and $\lambda_{b}=0$, so that $\mathbf{G}_{\mathrm{s}}$ factors into an outer product: $\mathbf{G}_{\mathrm{s}}=|E\rangle\langle E|$. Spatially \textit{incoherent} light ($D_{\mathrm{s}}=0$ and $S_{\mathrm{s}}=1$~bit), corresponds to $\lambda_{a}=\lambda_{b}=\tfrac{1}{2}$ and $\mathbf{G}_{\mathrm{s}}=\tfrac{1}{2}\hat{\mathbb{I}}_{2}$. Partial spatial coherence corresponds to $0<D_{\mathrm{s}}<1$. The spatial coherence matrix $\mathbf{G}_{\mathrm{s}}$ upon traversing a spatial unitary $\hat{U}_{\mathrm{s}}$ is transformed to $\mathbf{G}_{\mathrm{s}}'=\hat{U}_{\mathrm{s}}\mathbf{G}_{\mathrm{s}}\hat{U}_{\mathrm{s}}^{\dagger}$, which of course leaves both $D_{\mathrm{s}}$ and $S_{\mathrm{s}}$ invariant.

\begin{figure}[t!]
\centering
\includegraphics[width=5.25in]{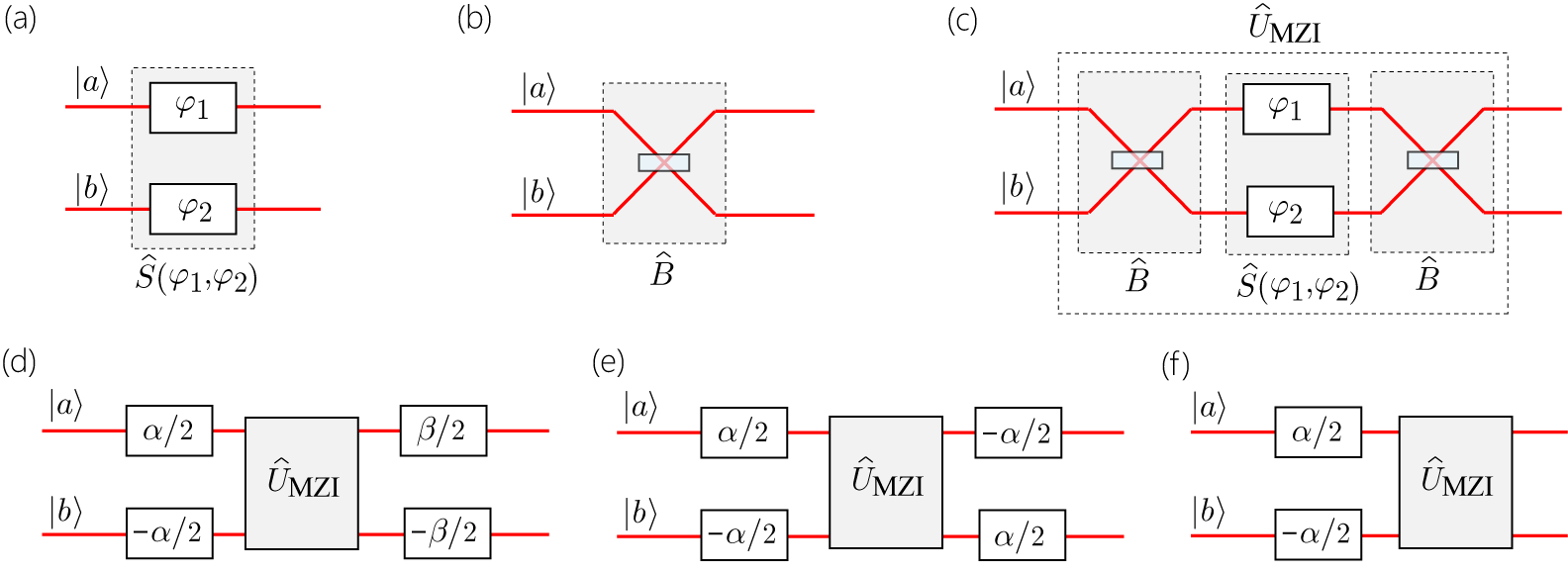}
\caption{(a) A phase operator $\hat{S}(\varphi_{1},\varphi_{2})$ where a phase $\varphi_{1}$ is inserted in mode $|a\rangle$ and a phase $\varphi_{2}$ in the mode $|b\rangle$. (b) A symmetric beam splitter implements the unitary operator $\hat{B}$. (c) A variable coupler represented by the unitary $\hat{U}_{\mathrm{MZI}}(\delta,\varphi)$ is formed of the phase operator $\hat{S}(\varphi_{1},\varphi_{2})$ sandwiched between two symmetric beam splitters; here $\delta=\varphi_{1}-\varphi_{2}$ and $\varphi=\tfrac{1}{2}(\varphi_{1}+\varphi_{2})$. (d) A general spatial unitary $\hat{U}_{\mathrm{s}}$ is formed of a variable coupler $\hat{U}_{\mathrm{MZI}}(\delta,\varphi)$ sandwiched between two phase operators $\hat{S}(\alpha)$ and $\hat{S}(\beta)$. (e) A restricted unitary obtained from the general unitary in (d) after setting $\beta=-\alpha$. (f) A restricted unitary obtained from the general unitary in (d) after setting $\beta=0$.}
\label{fig:SpatialUnitaries}
\end{figure}

As described in Section~\ref{sec:UnitarySingleDoF}, constructing a general unitary requires a variable rotation operator $\hat{R}(\theta)$ [Eq.~\ref{eq:PhaseAndRotationUnitaries} and Fig.~\ref{fig:2x2Unitary}]. However, it is challenging to construct an on-chip coupler $\hat{R}(\theta)$ with tunable coupling strength. Rather, $\hat{R}(\theta)$ is constructed from simpler basic blocks. Specifically, it is straightforward to implement phase shifts (via the thermo-optic or electro-optic effects) in a single waveguide. Two such phase shifts implemented on two waveguides produce the transformation $\hat{S}(\varphi_{1},\varphi_{2})=\left(\begin{array}{cc}e^{i\varphi_{1}}&0\\0&e^{i\varphi_{2}}\end{array}\right)$ [Fig.~\ref{fig:SpatialUnitaries}(a)]. A basic unitary that can also be readily implemented on-chip is a fixed symmetric beam splitter corresponding to the unitary $\hat{B}=\tfrac{1}{\sqrt{2}}\left(\begin{array}{cc}1&i\\i&1\end{array}\right)$ [Fig.~\ref{fig:SpatialUnitaries}(b)]. Forming a Mach-Zehnder interferometer (MZI) by inserting the phase operator $\hat{S}(\varphi_{1},\varphi_{2})$ between two symmetric beam splitters yields the unitary [Fig.~\ref{fig:SpatialUnitaries}(c)]:
\begin{equation}
\hat{U}_{\mathrm{MZI}}(\delta,\varphi)=\hat{B}\hat{S}(\varphi_{1},\varphi_{2})\hat{B}=ie^{i\varphi}\left(\begin{array}{cc}\sin\tfrac{\delta}{2}&\cos\tfrac{\delta}{2}\\\cos\tfrac{\delta}{2}&-\sin\tfrac{\delta}{2}\end{array}\right),
\end{equation}
where $\varphi=\tfrac{1}{2}(\varphi_{1}+\varphi_{2})$ and $\delta=\varphi_{1}-\varphi_{2}$. This MZI therefore corresponds to a variable coupler with an overall phase. 

Sandwiching $\hat{U}_{\mathrm{MZI}}$ between phase transformations $\hat{S}(\alpha)$ and $\hat{S}(\beta)$ [Fig.~\ref{fig:SpatialUnitaries}(d)] yields the general unitary (compare to Eq.~\ref{eq:2x2U}):
\begin{equation}
\hat{U}_{\mathrm{s}}=\hat{S}(\beta)\hat{U}_{\mathrm{MZI}}(\varphi,\delta)\hat{S}(\alpha)=ie^{i\varphi}\left(\begin{array}{cc}e^{i\xi_{1}}\sin\tfrac{\delta}{2}&e^{-i\xi_{2}}\cos\tfrac{\delta}{2}\\e^{i\xi_{2}}\cos\tfrac{\delta}{2}&-e^{-i\xi_{1}}\sin\tfrac{\delta}{2}\end{array}\right),
\end{equation}
where $\xi_{1}=\tfrac{\alpha+\beta}{2}$ and $\xi_{2}=\tfrac{\alpha-\beta}{2}$. A restricted unitary [Fig.~\ref{fig:SpatialUnitaries}(e)] is realized by setting $\beta=-\alpha$: 
\begin{equation}\label{eq:SpatialUnitary}
\hat{U}_{\mathrm{s}}=\hat{S}(-\alpha)\hat{U}_{\mathrm{MZI}}(\varphi,\delta)\hat{S}(\alpha)=ie^{i\varphi}\left(\begin{array}{cc}\sin\tfrac{\delta}{2}&e^{-i\alpha}\cos\tfrac{\delta}{2}\\e^{i\alpha}\cos\tfrac{\delta}{2}&-\sin\tfrac{\delta}{2}\end{array}\right).
\end{equation}
An alternate reduced unitary that requires future elements is obtained by eliminating the phase shift $\hat{S}(\beta)$, yielding [Fig.~\ref{fig:SpatialUnitaries}(f)]:
\begin{equation}
\hat{U}_{\mathrm{s}}=\hat{U}_{\mathrm{MZI}}(\delta,\varphi)\hat{S}(\alpha)=ie^{i(\varphi+\alpha/2)}\left(\begin{array}{cc}\sin\tfrac{\delta}{2}&e^{-i\alpha}\cos\tfrac{\delta}{2}\\\cos\tfrac{\delta}{2}&-e^{-i\alpha}\sin\tfrac{\delta}{2}\end{array}\right).
\end{equation}

\begin{figure}[t!]
\centering
\includegraphics[width=4.9in]{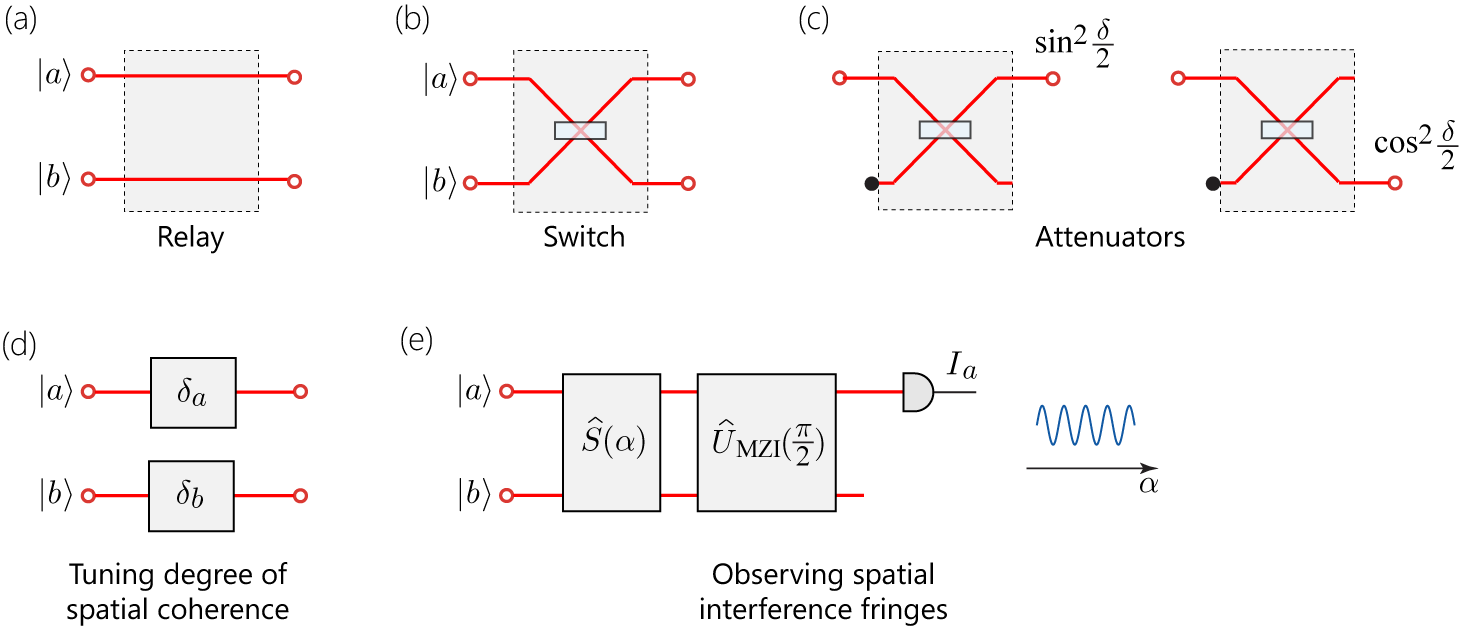}
\caption{Optical devices operating on the spatial DoF obtained from $U_{\mathrm{MZI}}$ in Fig.~\ref{fig:SpatialUnitaries}(c). (a) A relay with $|a\rangle\rightarrow|a\rangle$ and $|b\rangle\rightarrow|b\rangle$; (b) a switch with $|a\rangle\rightarrow|b\rangle$ and $|b\rangle\rightarrow|a\rangle$; and (c) attenuators with one empty input port and one blocked output port. (d) Tuning the degree of spatial coherence utilizing two attenuators from (c), one placed in each mode. (e) Spatial unitary to produce interference whose fringe visibility corresponds to the off-diagonal element of $\mathbf{G}_{\mathrm{s}}$, $V=2|G_{12}|$.}
\label{fig:SpatialDevices}
\end{figure}

The MZI in Fig.~\ref{fig:SpatialUnitaries}(c) may be operated in one of several useful configurations:
\begin{enumerate}
\item \textit{Optical relay.} By setting $\varphi_{1}=\varphi_{2}+\pi$, we have $\delta=\pi$ and $\varphi=\varphi_{1}+\tfrac{\pi}{2}$, so that $\hat{U}_{\mathrm{MZI}}(\pi,\varphi_{1}+\tfrac{\pi}{2})=-e^{i\varphi_{1}}\left(\begin{array}{cc}1&0\\0&-1\end{array}\right)$, and the field is transformed as:
\begin{equation}
|E\rangle\rightarrow|E'\rangle=\left(\begin{array}{c}E^{a}\\-E^{b}\end{array}\right),\;\mathbf{G}\rightarrow\mathbf{G}'=\left(\begin{array}{cc}G^{aa}&-G^{ab}\\-G^{ba}&G^{bb}\end{array}\right).
\end{equation}
The modal weights and magnitude of the correlation term are unchanged [Fig.~\ref{fig:SpatialDevices}(a)].
\item \textit{Optical switch.} By setting $\varphi_{1}=\varphi_{2}$, we have $\delta=0$, $\varphi=\varphi_{1}$, and $\hat{U}_{\mathrm{MZI}}(0,\varphi_{1})=ie^{i\varphi_{1}}\left(\begin{array}{cc}0&1\\1&0\end{array}\right)$, which switches the modes $|a\rangle$ and $|b\rangle$ [Fig.~\ref{fig:SpatialDevices}(b)]:
\begin{equation}
|E\rangle\rightarrow|E'\rangle=\left(\begin{array}{c}E^{b}\\E^{a}\end{array}\right),\;\mathbf{G}\rightarrow\mathbf{G}'=\left(\begin{array}{cc}G^{bb}&G^{ba}\\G^{ab}&G^{aa}\end{array}\right).
\end{equation}

\item \textit{Variable attenuator.} Whereas the relay and switch described above are unitary operations, an attenuator is a non-unitary operation associated with loss. By sending only one input to the MZI at mode $|a\rangle$ for example (no input is provided to mode $|b\rangle$) and discarding the output at modes $|b\rangle$, then the modal weight for $|a\rangle$ acquires an attenuation factor $\sin^{2}\tfrac{\delta}{2}$ [Fig.~\ref{fig:SpatialDevices}(c)]. Alternatively, by discarding the output at mode $|a\rangle$ and taking instead output mode $|b\rangle$, we have an attenuating switch $|a\rangle\rightarrow|b\rangle$ with an attenuation factor $\cos^{2}\tfrac{\delta}{2}$ in the modal weight [Fig.~\ref{fig:SpatialDevices}(c)].
\end{enumerate}

The transformations outlined above can be utilized to carry out several useful tasks:
\begin{enumerate}
\item By placing two variable attenuators in modes $|a\rangle$ and $|b\rangle$, we produce a spatial filter: $\hat{F}^{\mathrm{D}}=\left(\begin{array}{cc}\sin\tfrac{\delta_{a}}{2}&0\\0&\sin\tfrac{\delta_{b}}{2}\end{array}\right)$, where mode $|a\rangle$ is attenuated by a factor $\sin^{2}\tfrac{\delta_{a}}{2}$ and mode $|b\rangle$ by a factor $\sin^{2}\tfrac{\delta_{b}}{2}$. This device allows us to tune the degree of spatial coherence $D_{\mathrm{s}}$ and the spatial entropy $S_{\mathrm{s}}$. Starting with spatially incoherent light $\mathbf{G}_{\mathrm{s}}=\tfrac{1}{2}\hat{\mathbb{I}}_{2}$, $\hat{F}^{\mathrm{D}}$ produces a diagonal coherence matrix $\mathbf{G}_{\mathrm{s}}^{\mathrm{D}}=\left(\begin{array}{cc}\lambda_{a}&0\\0&\lambda_{b}\end{array}\right)$ after renormalization, where $\lambda_{a}=\tfrac{\sin^{2}\tfrac{\delta_{a}}{2}}{\sin^{2}\tfrac{\delta_{a}}{2}+\sin^{2}\tfrac{\delta_{b}}{2}}$ and $\lambda_{b}=\tfrac{\sin^{2}\tfrac{\delta_{b}}{2}}{\sin^{2}\tfrac{\delta_{a}}{2}+\sin^{2}\tfrac{\delta_{b}}{2}}$, thereby yielding a degree of spatial coherence $D_{\mathrm{s}}=\tfrac{\cos\delta_{a}-\cos\delta_{b}}{2-\cos\delta_{a}-\cos\delta_{b}}$ [Fig.~\ref{fig:SpatialDevices}(d)]. 

\item Starting from a diagonal coherence matrix $\mathbf{G}_{\mathrm{s}}^{\mathrm{D}}=\left(\begin{array}{cc}\lambda_{a}&0\\0&\lambda_{b}\end{array}\right)$, $1\geq\lambda_{a}\geq\lambda_{b}\geq0$, the restricted spatial unitary $\hat{U}_{\mathrm{s}}$ in Eq.~\ref{eq:SpatialUnitary} transforms $\mathbf{G}_{\mathrm{s}}^{\mathrm{D}}$ into the general spatial coherence matrix,
\begin{equation}\label{eq:SpatialCoherenceFunction}
\mathbf{G}_{\mathrm{s}}=\hat{U}_{\mathrm{s}}\mathbf{G}_{\mathrm{s}}^{\mathrm{D}}\hat{U}_{\mathrm{s}}^{\dagger}=\frac{1}{2}\left(\begin{array}{cc}1-(\lambda_{a}-\lambda_{b})\cos\delta&(\lambda_{a}-\lambda_{b})e^{-i\alpha}\sin\delta\\(\lambda_{a}-\lambda_{b})e^{i\alpha}\sin\delta&1+(\lambda_{a}-\lambda_{b})\cos\delta\end{array}\right),
\end{equation}
while maintaining the same entropy and the degree of spatial coherence, $D_{\mathrm{s}}=\lambda_{a}-\lambda_{b}$ as in $\mathbf{G}_{\mathrm{s}}^{\mathrm{D}}$. For a coherent field ($\lambda_{a}=1$ and $\lambda_{b}=0$), the spatial coherence matrix is reduced to an outer product $\mathbf{G}_{\mathrm{s}}=|E\rangle\langle E|$ with $|E\rangle=\left(\begin{array}{c}\sin\tfrac{\delta}{2}\\e^{i\alpha}\cos\tfrac{\delta}{2}\end{array}\right)$.

\item \textit{Spatial interference.} Rather than rely on the usual double-slit interferogram [Fig.~\ref{fig:SpatialBasic}(a)], spatial interference can be observed instead using the unitary
\begin{equation}
\hat{U}_{\mathrm{s}}=\frac{i}{\sqrt{2}}e^{i\varphi}\left(\begin{array}{cc}1&e^{-i\alpha}\\e^{i\alpha}&-1\end{array}\right)
\end{equation}
after setting $\delta=\tfrac{\pi}{2}$ in Eq.~\ref{eq:SpatialUnitary}, or equivalently cascading a beam splitter after the phase operator $\hat{S}(\alpha)$. After this unitary, we have $I_{a}=\tfrac{1}{2}[1+2|G^{ab}|\cos(\alpha+\chi)]$ and $I_{b}=\tfrac{1}{2}[1-2|G^{ab}|\cos(\alpha+\chi)]$, where $G^{ab}=|G^{ab}|e^{i\chi}$, thus yielding an interference visibility of $V=2|G^{ab}|$, $V=2|G^{ab}|$; see Fig.~\ref{fig:SpatialDevices}(e).
\end{enumerate}

Note that partial coherence entails that when the input field is initially distributed between the two waveguides, the field cannot be unitarily concentrated into a single waveguide. In this case, the maximum fraction of the total power that can be concentrated in one waveguide is $\lambda_{a}$ (the larger eigenvalue). Only in the case of a coherent field can the input power always be concentrated via a unitary into one waveguide.

\begin{figure}[t!]
\centering
\includegraphics[width=4.4in]{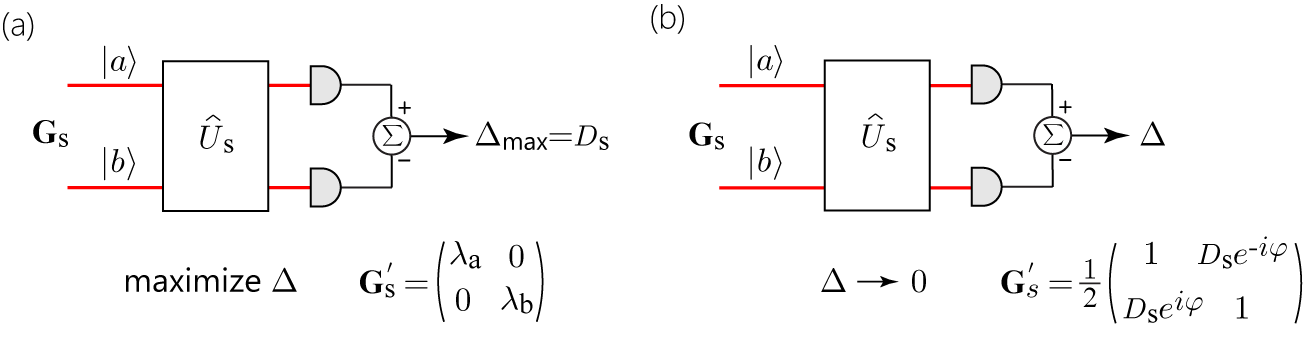}
\caption{(a) Diagonalizing $\mathbf{G}_{\mathrm{s}}$ via a unitary $\hat{U}_{\mathrm{s}}$ by maximizing $\Delta=I_{a}-I_{b}$. (b) Equalization ($\Delta=0$) of the diagonal elements of $\mathbf{G}_{\mathrm{s}}$ via a unitary $\hat{U}_{\mathrm{s}}$.}
\label{fig:DigonalizingGs}
\end{figure}

One may measure the degree of spatial coherence by one of several techniques that follow the schemes outlined earlier and applied above to the polarization DoF. First, one may diagonalize $\mathbf{G}_{\mathrm{s}}$ [Fig.~\ref{fig:DigonalizingGs}(a)] via a unitary $\hat{U}_{\mathrm{s}}=\left(\begin{array}{cc}\sin\tfrac{\delta_{\mathrm{o}}}{2}&e^{-i\alpha_{\mathrm{o}}}\cos\tfrac{\delta_{\mathrm{o}}}{2}\\e^{i\alpha_{\mathrm{o}}}\cos\tfrac{\delta_{\mathrm{o}}}{2}&-\sin\tfrac{\delta_{\mathrm{o}}}{2}\end{array}\right)$. The resulting spatial coherence matrix is $\mathbf{G}_{\mathrm{s}}'=\hat{U}_{\mathrm{s}}\mathbf{G}_{\mathrm{s}}\hat{U}_{\mathrm{s}}^{\dagger}$. By scanning the values of $\delta_{\mathrm{o}}$ and $\alpha_{\mathrm{o}}$ and evaluating $\Delta=I_{a}-I_{b}$, the maximum value $\Delta_{\mathrm{max}}$ is reached once $\delta_{\mathrm{o}}=-\delta$ and $\alpha_{\mathrm{o}}=\alpha$, whereupon $\mathbf{G}_{\mathrm{s}}$ is diagonalized and $\Delta_{\mathrm{max}}=\lambda_{a}-\lambda_{b}=D_{\mathrm{s}}$.

Alternatively, one may obtain $D_{\mathrm{s}}$ by first equalizing the modal weights via the unitary $\hat{U}_{\mathrm{s}}$ [Fig.~\ref{fig:DigonalizingGs}(b)]. Once $\alpha_{\mathrm{o}}=\alpha$ and $\delta_{\mathrm{o}}=-\delta+\pi$, we can have $\Delta=I_{a}-I_{b}\rightarrow0$, and the coherence matrix takes the form $\mathbf{G}_{\mathrm{s}}=\tfrac{1}{2}\left(\begin{array}{cc}1&D_{\mathrm{s}}e^{-i\varphi}\\D_{\mathrm{s}}e^{i\varphi}&1\end{array}\right)$. The off-diagonal term can be observed as the visibility of spatial interference fringes after traversing the unitary in Eq.~\ref{eq:SpatialCoherenceFunction} [Fig.~\ref{fig:SpatialDevices}(e)].

In a third approach, the spatial coherence matrix $\mathbf{G}_{\mathrm{s}}$ is reconstructed by measuring the \textit{spatial} Stokes parameters. Three measurement settings (involving unitaries $\hat{U}_{1}$, $\hat{U}_{2}$, and $\hat{U}_{3}$) are required:
\begin{enumerate}
\item First, setting $\hat{U}_{1}=\hat{\mathbb{I}}_{2}$, the modal weights are measured, $I_{a}=G^{aa}$ and $I_{b}=G^{bb}$, from which we obtain the first two Stokes parameters: $s_{0}^{(\mathrm{s})}=I_{a}+I_{b}$ and $s_{1}^{(\mathrm{s})}=I_{a}-I_{b}$ [Fig.~\ref{fig:SpatialStokes}(a)], and we normalize all the Stokes parameters with respect to $I_{a}+I_{b}$, so that $s_{0}^{(\mathrm{s})}=1$. 

\begin{figure}[t!]
\centering
\includegraphics[width=13.3cm]{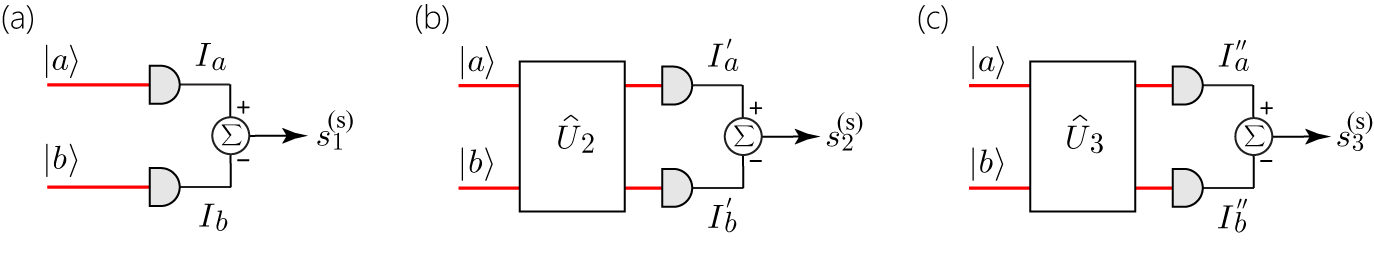}
\caption{Reconstruction of the spatial coherence matrix $\mathbf{G}_{\mathrm{s}}$ via measurements of the spatial Stokes parameters. (a) Measuring $s_{1}^{\mathrm{(s)}}$, (b) $s_{2}^{\mathrm{(s)}}$, and (c) $s_{3}^{\mathrm{(s)}}$.}
\label{fig:SpatialStokes}
\end{figure}

\item The field traverses $\hat{U}_{2}=\hat{U}_{\mathrm{MZI}}(\tfrac{\pi}{2},\varphi)=\tfrac{1}{\sqrt{2}}e^{i\varphi}\left(\begin{array}{cc}1&1\\1&-1\end{array}\right)$ by setting $\alpha=0$ and $\delta=\tfrac{\pi}{2}$ in Eq.~\ref{eq:SpatialUnitary}, from which we obtain $s_{2}^{(\mathrm{s})}=I_{a}'-I_{b}'$ [Fig.~\ref{fig:SpatialStokes}(b)].

\item The field traverses $\hat{U}_{3}=\tfrac{1}{\sqrt{2}}\left(\begin{array}{cc}1&-i\\i&-1\end{array}\right)$ by setting $\alpha=\tfrac{\pi}{2}$ and $\delta=\tfrac{\pi}{2}$ in Eq.~\ref{eq:SpatialUnitary}, from which we obtain $s_{3}^{(\mathrm{s})}=I_{a}''-I_{b}''$ [Fig.~\ref{fig:SpatialStokes}(c)].
\end{enumerate}

Obtaining the spatial Stokes parameters $\{s_{0}^{(\mathrm{s})},s_{1}^{(\mathrm{s})},s_{2}^{(\mathrm{s})},s_{3}^{(\mathrm{s})}\}$ allows us to reconstruct $\mathbf{G}_{\mathrm{s}}=\tfrac{1}{2}\left(\begin{array}{cc}s_{0}^{(\mathrm{s})}+s_{1}^{(\mathrm{s})}&s_{2}^{(\mathrm{s})}-is_{3}^{(\mathrm{s})}\\s_{2}^{(\mathrm{s})}+is_{3}^{(\mathrm{s})}&s_{0}^{(\mathrm{s})}-s_{1}^{(\mathrm{s})}\end{array}\right)$, from which we obtain the degree of spatial coherence $D_{\mathrm{s}}=\sqrt{\left(s_{1}^{(\mathrm{s})}\right)^{2}+\left(s_{2}^{(\mathrm{s})}\right)^{2}+\left(s_{3}^{(\mathrm{s})}\right)^{2}}$.

\begin{figure}[b!]
\centering
\includegraphics[width=13.3cm]{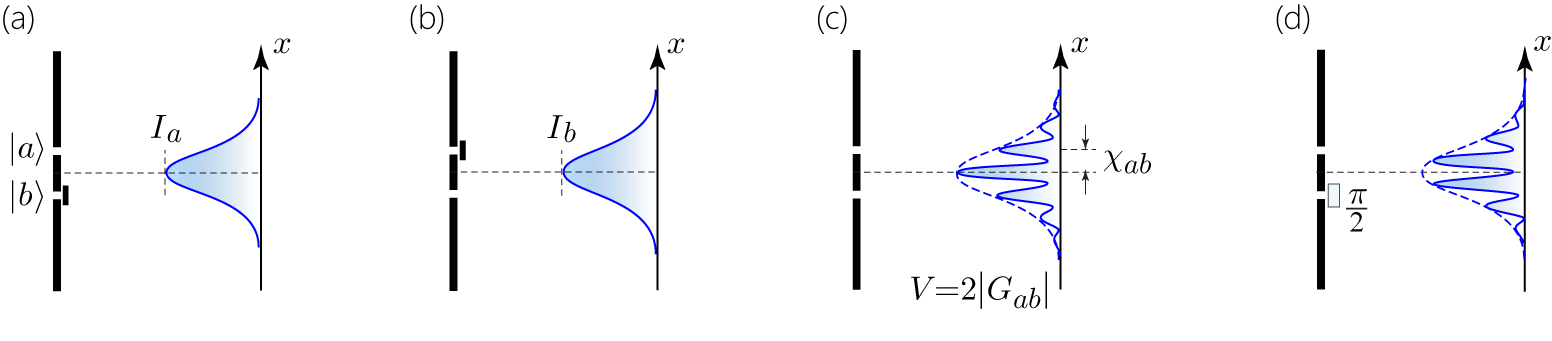}
\caption{Reconstruction of the spatial coherence matrix $\mathbf{G}_{\mathrm{s}}$ in the double-slit configuration. (a) Block $|b\rangle$ to isolate $I_{a}$. (b) Block $|a\rangle$ to isolate $I_{b}$. (c) Record the interference pattern with both $|a\rangle$ and $|b\rangle$. (d) Same as (c) after inserting a relative phase $\tfrac{\pi}{2}$ between $|a\rangle$ and $|b\rangle$.}
\label{fig:DoubleSlitStokes}
\end{figure}

Finally, one may construct a spatial decohering device (to modify the degree of spatial coherence without loss of energy) using an MZI after randomizing $\delta$ in the unitary $\hat{U}_{\mathrm{MZI}}$. This system decoheres any spatial field for which $s_{3}^{\mathrm{(s)}}=0$. The most general spatial decoherence system makes use of a restricted unitary (Eq.~\ref{eq:SpatialUnitary}) after randomizing $\delta$ and $\alpha$.

It is useful to also describe for comparison the process for reconstructing $\mathbf{G}$ in the familiar case of a two-point field defined by double slits as shown in Fig.~\ref{fig:DoubleSlitStokes}. The double-slit interference pattern is $I(\varphi_{ab})=I_{a}+I_{b}+2|G_{ab}|\cos(\varphi_{ab}+\chi_{ab})$, where $G_{ab}=|G_{ab}|e^{i\chi_{ab}}$, and $\varphi_{ab}$ is the relative phase at position $x$ at the output plane incurred along the two paths from $|a\rangle$ and $|b\rangle$. First by blocking $|b\rangle$, we record the peak intensity $I_{a}$ produced by $|a\rangle$ [Fig.~\ref{fig:DoubleSlitStokes}(a)], and similarly by blocking $|a\rangle$ we obtain $I_{b}$. From $I_{a}$ and $I_{b}$ we obtain $s_{0}^{\mathrm{(s)}}=2(I_{a}+I_{b})$ and $s_{1}^{\mathrm{(s)}}=2(I_{a}-I_{b})$ [Fig.~\ref{fig:DoubleSlitStokes}(b)]. By unblocking both $|a\rangle$ and $|b\rangle$, we record the interference pattern $I(\varphi_{ab})$ [Fig.~\ref{fig:DoubleSlitStokes}(c)], whose visibility is $V=\tfrac{I_{\mathrm{max}}-I_{\mathrm{min}}}{I_{\mathrm{max}}+I_{\mathrm{min}}}=\tfrac{2|G_{ab}|}{I_{a}+I_{b}}$, from which we obtain $|G_{ab}|$, and the shift in the peak fringe from the peak intensity when $|a\rangle$ or $|b\rangle$ are blocked is $\chi_{ab}$; $s_{2}=|G_{ab}|\cos\chi_{ab}$ and $s_{3}=-|G_{ab}|\sin\chi_{ab}$. We have thus reconstructed the spatial coherence matrix $\mathbf{G}_{\mathrm{s}}$. There are alternatives to determining the off-diagonal element $G_{ab}$. With $|a\rangle$ and $|b\rangle$ unblocked, one may obtain the intensity at the location corresponding to that of the peak when $|a\rangle$ or $|b\rangle$ is blocked (the geometric center of the system), which yields $I_{a}+I_{b}+2|G_{ab}|\cos\chi_{ab}$ [Fig.~\ref{fig:DoubleSlitStokes}(c)]. Next, we obtain the intensity at the point midway between the peak and the first minimum in the intensity pattern, which corresponds to $\varphi_{ab}=\tfrac{\pi}{2}$ and thus an intensity $I_{a}+I_{b}+2|G_{ab}|\sin\chi_{ab}$. From these two measurements we obtain $|G_{ab}|$ and $\chi_{ab}$. The last measurement can also be captured by measuring the intensity at the geometric center of the system after adding a $\tfrac{\pi}{2}$ phase shift between $|a\rangle$ and $|b\rangle$ [Fig.~\ref{fig:DoubleSlitStokes}(d)].

\subsection{Applications: Communications using optical coherence}\label{sec:Communications1DoF}

Traditionally, partially coherent optical fields have \textit{not} been exploited in optical communications. Rather, coherent fields (laser beams and pulses) have been the mainstay of optical communication links -- whether in optical fibers or in free space. Exceptions include short-range free-space links (Li-Fi) where LEDs are utilized \cite{Tsonev14ProcSPIE}. However, the partial coherence of the field plays no role in the communications protocol. This situation has changed in the past few years where novel protocols for optical communications \cite{Nardi22OL} and cryptography \cite{Peng21P,Liu25LPR} have been proposed and implemented that rely explicitly on partial coherence to overcome a challenge faced by coherent light fields in the same setting. We describe here one communication protocol where partial coherence plays a beneficial role. 

Logical bits are commonly encoded in orthogonal physical states to enhance their distinguishability after traversing a communications channel that may introduce noise into the transmitted signal or feature scattering. Consider utilizing the polarization DoF as an information carrier: by encoding logical bit~0 in the $|\mathrm{H}\rangle$ polarization mode ($0\rightarrow|\mathrm{H}\rangle$) and logical bit~1 in the $|\mathrm{V}\rangle$ polarization mode ($1\rightarrow|\mathrm{V}\rangle$). Polarization scattering in the communication channel can thus corrupt the data by introducing errors in the decoding at the channel output.  

We make the following assumptions about the communications channel:
\begin{enumerate}
\item \textit{Single-DoF channel}. The channel impacts only the polarization DoF.

\item \textit{Unitarity}. The channel can be represented for any bit during data transmission by a $2\times2$ polarization unitary $\hat{U}_{\mathrm{p}}$.

\item \textit{Channel losses}. An overall loss factor can be included, which is assumed to be polarization-independent.

\item \textit{Strong polarization scattering}. $\hat{U}_{\mathrm{p}}$ is selected randomly from the entire family of $2\times2$ unitaries. This indicates that the input and output polarization states differ significantly, and may even be orthogonal. 

\item \textit{Rapidly varying scattering}. $\hat{U}_{\mathrm{p}}$ changes from bit to bit, which precludes utilizing adaptive optics techniques that rely on probing a slowly varying channel and pre-conditioning the transmitted field to compensate for the effect of the channel.

\item \textit{No-memory channel}. The unitaries $\hat{U}_{\mathrm{p}}$ at any two moments in time are uncorrelated; a measurement in one time slot has no bearing on any other time slot. 
\end{enumerate}

We use the moniker `Channel-1' for this collective set of assumptions, which represents of course an extreme case. Most communication channels do not feature all of these assumptions simultaneously. Nevertheless, it is useful to consider this extreme channel to highlight the unique utility of optical coherence as an information carrier.

\begin{figure}[t!]
\centering
\includegraphics[width=13.3cm]{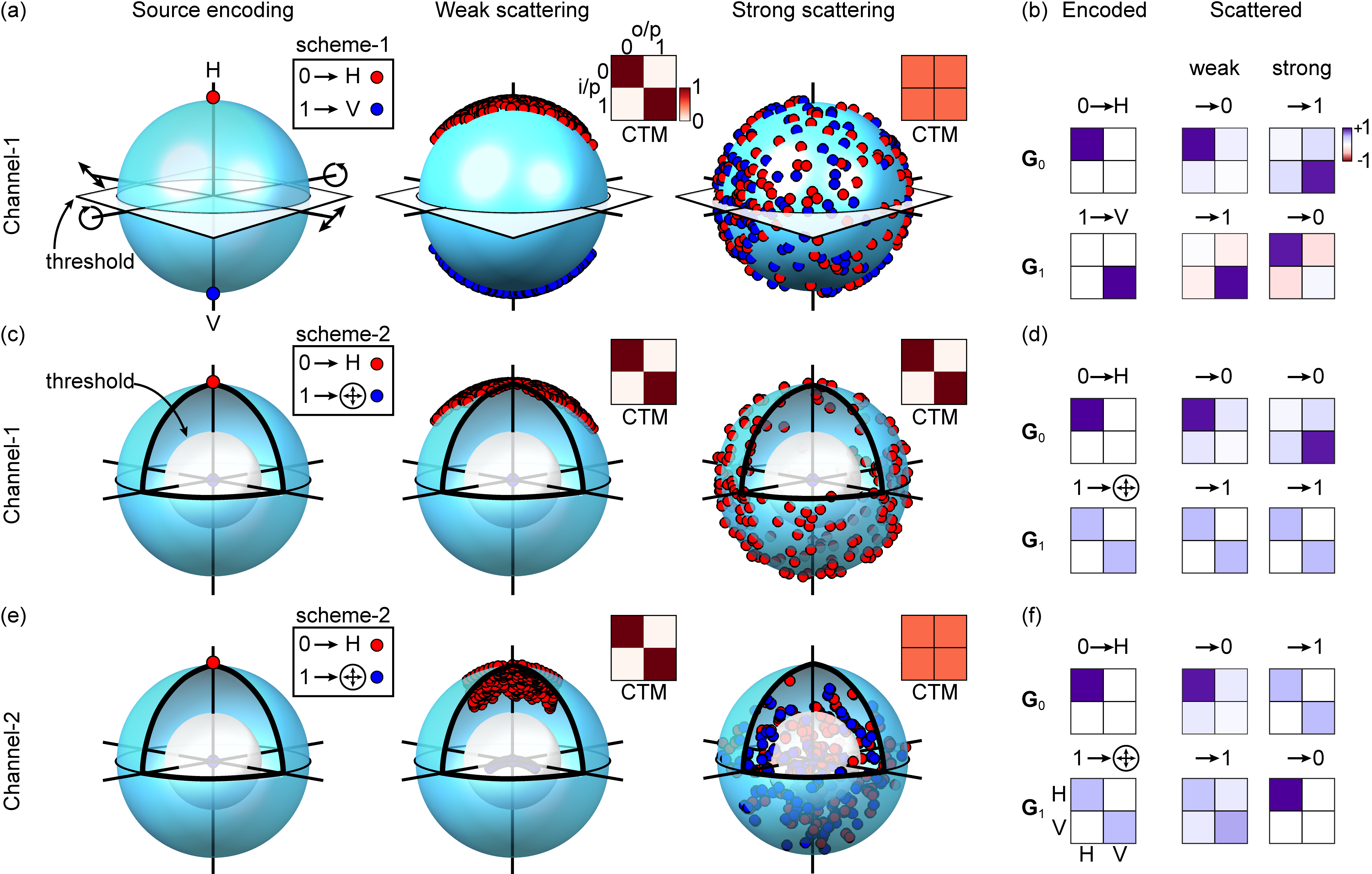}
\caption{Optical communications across a scattering channel (from Ref.~\cite{Harling25APLP}). (a) Encoding scheme-1 ($0\rightarrow\mathrm{H}$ and $1\rightarrow\mathrm{V}$); the decision threshold is the equator plane. The middle and right panels depict the output polarization on the PS after traversing a channel that weakly and strongly scatters, polarization (Channel-1), respectively, along with the CTMs as insets. (b) The polarization coherence matrices $\mathbf{G}_{\mathrm{p}}$ associated with the encoded and scattered (weakly and strongly) bits through Channel-1. The strongly scattered coherence matrices cannot be decoded correctly. (c) Same as (a) after utilizing encoding scheme-2, where the logical bits are encoded in polarized ($D_{\mathrm{p}}=1$) and unpolarized ($D_{\mathrm{p}}=0$) fields. The decision threshold is the spherical surface of radius $\tfrac{1}{2}$. (d) Same as (b); however, encoding scheme-2 succeeds where encoding scheme-1 fails. (e) Same as (c) utilizing encoding scheme-2, but the scattering channel couples the polarization DoF to an unused spatial DoF (Channel-2); see Section~\ref{sec:Communications2DoFs}. This channel can increase or decrease $D_{\mathrm{p}}$. Therefore the point representing the field can move anywhere across the PS volume. (f) Same as (d); however the strongly scattered coherence matrices can no longer be decoded.}
\label{fig:PScommunications}
\end{figure}

We illustrate in Fig.~\ref{fig:PScommunications}(a,b) the impact of Channel-1 on coherent fields exploited for information encoding. Specifically, $0\rightarrow|\mathrm{H}\rangle$ and $1\rightarrow|\mathrm{V}\rangle$ correspond to the north and south poles on the PS surface, with corresponding coherence matrices $\mathbf{G}_{0}=\left(\begin{array}{cc}1&0\\0&0\end{array}\right)$ and $\mathbf{G}_{1}=\left(\begin{array}{cc}0&0\\0&1\end{array}\right)$, respectively. The equator plane is the decision threshold for \textit{decoding}: any output polarization detected above the equator is assigned to bit~0, otherwise it is assigned to bit~1. \textit{Weak scattering} in the communications channel moves the point representing the field vector on the PS surface while remaining close to the ideal starting point (the PS north or south poles). Consequently, the cross-talk matrix (CTM), which depicts the probabilities of detecting bits~0 or~1 at the output for input bits~0 and~1 is diagonal: $0\rightarrow0$ and $1\rightarrow1$. Minimal corruption occurs in the transmitted polarization coherence matrices. However, in presence of \textit{strong scattering}, the bit~0 encoded as $|\mathrm{H}\rangle$ after the channel may move to \textit{any point} on the PS surface. Major corruption may occur to the transmitted polarization coherence matrix. Consequently, the CTM is flat [Fig.~\ref{fig:PScommunications}(b)].

Utilizing coherent polarization states for optical communications across such an extreme channel is not useful. However, optical coherence here provides an alternative by adopting an encoding scheme in which bit~0 is assigned to a pure polarization state (say, $|\mathrm{H}\rangle$), whereas bit~1 is assigned to the maximal unpolarized state, $0\rightarrow\mathbf{G}_{0}=\left(\begin{array}{cc}1&0\\0&0\end{array}\right)$ and $1\rightarrow\mathbf{G}_{1}=\tfrac{1}{2}\hat{\mathbb{I}}_{2}$, so that bits~0 and~1 are associated with different degrees of polarization, $0\rightarrow D_{\mathrm{p}}=1$ and $1\rightarrow D_{\mathrm{p}}=0$, respectively, corresponding to the north pole and center of the PS [Fig.~\ref{fig:PScommunications}(c,d)]. The decision threshold is the spherical surface of radius $\tfrac{1}{2}$ (corresponding to $D_{\mathrm{p}}=0.5$). In presence of weak scattering, the point representing $|\mathrm{H}\rangle$ the field at the north pole remains in its vicinity, whereas the point at the center remains invariant (the CTM is diagonal). Strong scattering moves the point at the PS north pole to \textit{any} other point on the PS, while the point at the center is impervious to scattering. Because the decision threshold distinguishes the degree of polarization coherence $D_{\mathrm{p}}$, and because the point on the PS surface remains on the surface while the point at the center is invariant in both cases of weak and strong scattering, the CTM consequently is still diagonal. Of course, one could use \textit{any} pure polarization state besides $|\mathrm{H}\rangle$ to encode the bit~0 because we detect only $D_{\mathrm{p}}$.

\begin{figure}[t!]
\centering
\includegraphics[width=13.3cm]{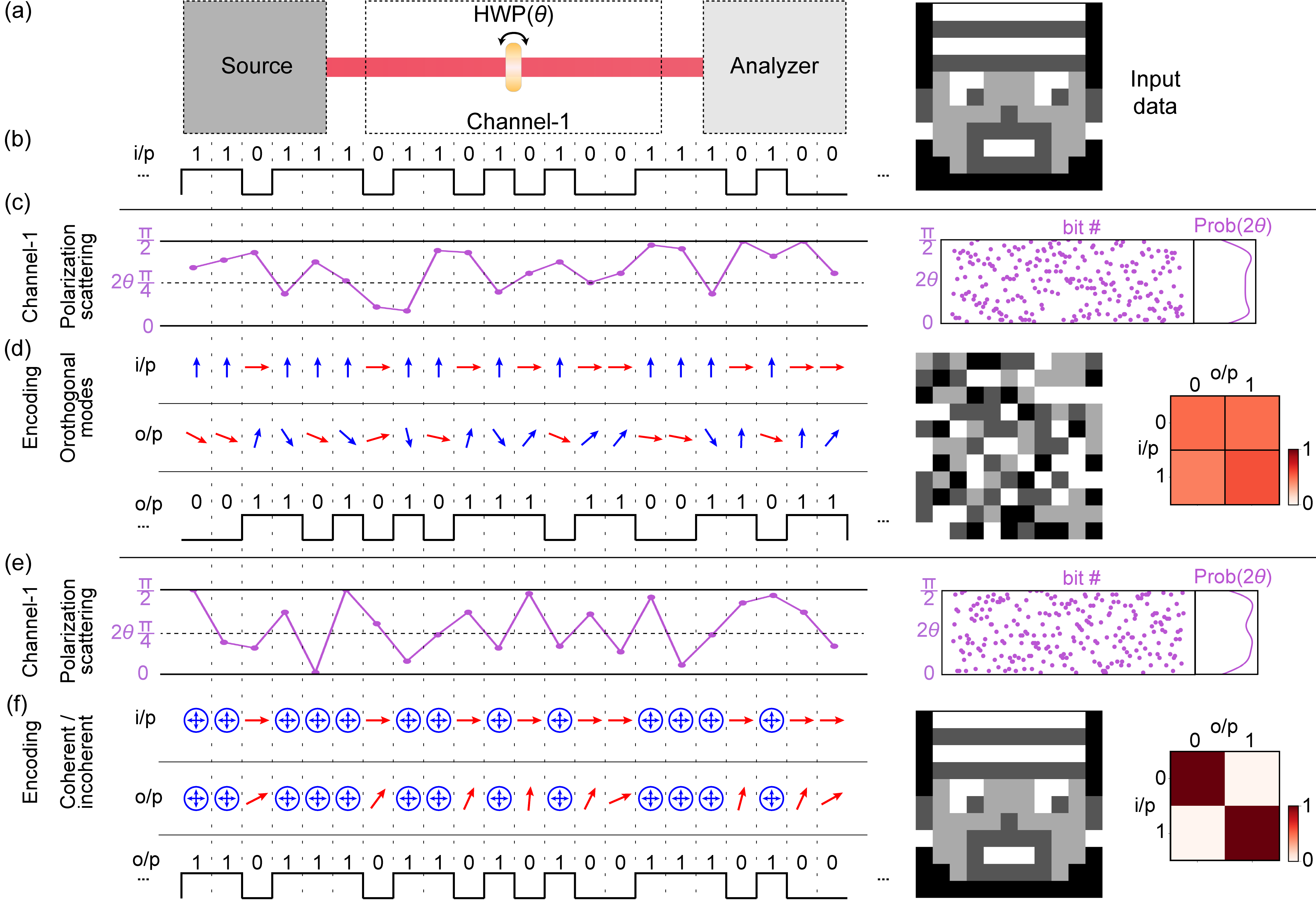}
\caption{Realization of optical communications through a polarization scattering channel (channel-1). (a) Schematic of the setup for channel-1 comprising a randomly rotating HWP (Section~\ref{sec:polarizationDoF}). (b) A portion of the data stream produced by the image on the right. (c,d) Encoding scheme-1, $0\rightarrow$H and $1\rightarrow|\mathrm{V}\rangle$ [Fig.~\ref{fig:PScommunications}(a)]. (c) A portion of the settings for the HWP angle $\theta$ with the probability distribution of $\theta$ plotted on the right. (d) Input and output polarizations corresponding to the data stream in (c). The measured reconstructed image and CTM are plotted on the right. (e,f) Same as (c,d) for the encoding scheme: $0\rightarrow|\mathrm{H}\rangle$ and $1\rightarrow$XXX, respectively.}
\label{fig:PolCommData}
\end{figure}

We now consider another adverse assumption for the communications channel: the channel may also change the degree of polarization $D_{\mathrm{p}}$ via polarization filtering (e.g., polarization-dependent losses), depolarization, or trace-preserving decoherence that can decrease or increase $D_{\mathrm{p}}$. In other words, $D_{\mathrm{p}}$ can also change, and the point representing the field vector can potentially leave the PS surface and move within the PS volume. We refer to this communications channel as `Channel-2'. We use the same encoding scheme, $0\rightarrow\mathbf{G}_{0}=\left(\begin{array}{cc}1&0\\0&0\end{array}\right)$ and $1\rightarrow\mathbf{G}_{1}=\tfrac{1}{2}\hat{\mathbb{I}}_{2}$, and set the same decision threshold at $D_{\mathrm{p}}=\tfrac{1}{2}$ [Fig.~\ref{fig:PScommunications}(e,f)]. Channel-2 can move the point representing logical bit~0 on the PS north pole either on the PS surface (whereupon $D_{\mathrm{p}}=1$ is maintained) or within its volume ($D_{\mathrm{p}}<1$ decreases due to depolarization). Similarly for the point at the PS center representing logical bit~1: it is no longer invariant, and may indeed move within the PS volume or even reach its surface (due to polarization filtering). When the scattering in  this channel is weak, the CTM is diagonal, and the transmitted coherence matrices are not corrupted significantly. However, when the scattering is strong, the CTM is flat and the transmitted coherence matrices are significantly corrupted. In this extreme case (Channel-2), neither polarized nor unpolarized fields can be utilized as information carriers.

Proof-of-principle experimental confirmation for the prediction regarding Channel-1 is presented in Fig.~\ref{fig:PolCommData}. The data stream [Fig.~\ref{fig:PolCommData}(b)] is in the form of bits corresponding to the gray-scale image shown in Fig.~\ref{fig:PolCommData}(a), which has dimensions $11\times11$~pixels, and each pixel has four gray-scale levels (corresponding to 2 bits per pixel). The polarization channel corresponds here to a half-wave plate (HWP) whose fast/slow axes are rotated randomly by an angle $\theta$ measured with respect to $|\mathrm{H}\rangle$, with a uniform probability distribution over the range $[0,\tfrac{\pi}{2}]$ (Section~\ref{sec:polarizationDoF}), corresponding to the strong scattering regime [Fig.~\ref{fig:PolCommData}(b)]. The angle $\theta$ is changed from bit to bit so that the correlation length of the channel is 1~bit (no memory in the channel). 

In the first encoding scheme [Fig.~\ref{fig:PolCommData}(c,d)] we employ $0\rightarrow|\mathrm{H}\rangle$ and $1\rightarrow|\mathrm{V}\rangle$. We implement the settings for $\theta$ shown in Fig.~\ref{fig:PolCommData}(c), and we plot in Fig.~\ref{fig:PolCommData}(d) the polarization states at the channel input, the detected polarization states at the channel output, along with the decoded bits. The bit-error-rate is high, the image is not reconstructed after the channel, and the CMT is flat. In such a channel, encoding scheme~1 (polarized fields) cannot transmit any information.

In the second encoding scheme [Fig.~\ref{fig:PolCommData}(e,f)] we employ $0\rightarrow|\mathrm{H}\rangle$ and $1\rightarrow\mathbf{G}_{1}=\tfrac{1}{2}\hat{\mathbb{I}}_{2}$; that is, bit~0 is encoded in a polarized field and bit~1 in an unpolarized field. We implement the settings for $\theta$ shown in Fig.~\ref{fig:PolCommData}(e), and we plot in Fig.~\ref{fig:PolCommData}(f) the polarization states at the channel input and output. Although the transmitted $|\mathrm{H}\rangle$ polarization changes at the output, it nevertheless remains polarized, and is decoded at bit~0. The bit-error-rate is low, the image is reconstructed after the channel, and the CMT is diagonal. In such a strongly scattering channel, encoding scheme~2 enables high-fidelity information transfer. However, it is expected that even encoding scheme~2 fails when the channel features depolarization (Channel-2). We tackle that challenge in Section~\ref{sec:Communications2DoFs}. 

\subsection{Correspondence between a binary DoF in a classical field and a qubit in quantum mechanics}\label{sec:QuantumSingleDoF}

The correspondence between certain \textit{mathematical} features of optical coherence on the one hand and of quantum mechanics on the other has long been appreciated -- despite the obvious disparity between the \textit{physical interpretation} of this shared mathematical structure \cite{Fano54PR}. Fundamentally, because the propagation of light is governed by Maxwell's equations, which give rise to a linear homogeneous wave equation in linear media, the principle of superposition applies to solutions of this wave equation just as it applied to wave-function solutions of Schr{\"o}dinger's equation in quantum mechanics. Furthermore, in absence of loss and gain, the wave equation can be posed as a Hermitian eigenvalue problem \cite{Joannopoulos08book}, leading to unitary evolution of the optical field. These features are all in correspondence with the standard machinery of wave functions in single-particle quantum mechanics.

This isomorphism between the quantum wave functions and classical optical fields underpins the development of photonics crystals, for example, where the periodic refractive index variation utilized in Maxwell's equations replaces the periodic crystal potential utilized in Schr{\"o}dinger's equation. This correspondence leads to the emergence of \textit{photonic} bandgaps in periodic photonic structures in analogy with \textit{electronic} bandgaps in crystalline solids. This has consequently led to the migration of concepts from condensed matter physics into optics over the past three decades; including the concepts of defect states \cite{Yablonovitch91PRL}, Anderson localization in disordered systems \cite{Schwartz07Nature}, and most recently topological states \cite{Lu14NP,Leykam26NRP} and bound states in the continuum \cite{Hsu16NRM}. Moreover, a mathematical correspondence can be established between classical optics on the one hand and quantum mechanics, quantum field theory, and even non-Hermitian extensions to quantum mechanics \cite{Bender98PRL} on the other hand, which has led to the development of Airy beams \cite{Siviloglou07OL}, supersymmetric optical systems \cite{Hokmabadi19Science}, PT-symmetric systems \cite{Ruter10NP}, and non-Hermitian photonics in general \cite{Elganainy18NP}.

It is thus well-established that the mathematical structure of a classical optical field is isomorphic to the wave functions representing the quantum state associated with a quantum system. The description of an optical field characterized by a binary DoF therefore  maps directly to the quantum state of a two-state quantum system known as a `qubit' \cite{Schumacher95PRA}. Both systems can be expressed in a state spanned by two orthonormal `modes' or `states' $|\psi_{1}\rangle=\left(\begin{array}{c}1\\0\end{array}\right)$ and $|\psi_{2}\rangle=\left(\begin{array}{c}0\\1\end{array}\right)$. When the classical field is \textit{coherent} or the quantum state is \textit{pure}, both system can be described by a vector of the form:$|\psi\rangle=\cos\tfrac{\theta}{2}|\psi_{1}\rangle+e^{i\varphi}\sin\tfrac{\theta}{2}|\psi_{2}\rangle=\left(\begin{array}{cc}\cos\tfrac{\theta}{2}\\e^{i\varphi}\sin\tfrac{\theta}{2}\end{array}\right)$, with $\langle\psi|\psi\rangle=1$. Consequently, the representation of a binary optical DoF on the PS corresponds to the representation of a spin-$\tfrac{1}{2}$ quantum particle or two-level quantum systems on the `Bloch sphere'. Moreover, a \textit{partially coherent} field characterized by this binary DoF is described by a $2\times2$ Hermitian, unity-trace, positive sem-definite coherence matrix $\mathbf{G}$, just asa qubit in a \textit{mixed} state is represented by a $2\times2$ Hermitian, unity-trace, positive semi-definite density matrix $\mathbf{\rho}$. The two descriptions are indeed mathematically isomorphic,
\begin{equation}
\mathbf{G}=\left(\begin{array}{cc}G_{11}&G_{12}\\G_{21}&G_{22}\end{array}\right)\Leftrightarrow\mathbf{\rho}=\left(\begin{array}{cc}\rho_{11}&\rho_{12}\\\rho_{21}&\rho_{22}\end{array}\right)
\end{equation}
This mapping further extends to: (1) unitary evolution of a freely evolving classical field or qubit; (2) the degree of coherence of $\mathbf{G}$ corresponding to the degree of purity of $\mathbf{\rho}$; and (3) reconstructing either matrix via measurements of the Stokes parameters \cite{Stokes51TCPS,Stone63Book,Brosseau98Book}. This correspondence was identified early on by U.~Fano \cite{Fano54PR,Fano57RMP} and others, which allows for the fruitful cross-pollination between these two disparate branches of physics.

\begin{figure}[t!]
\centering
\includegraphics[width=2.8in]{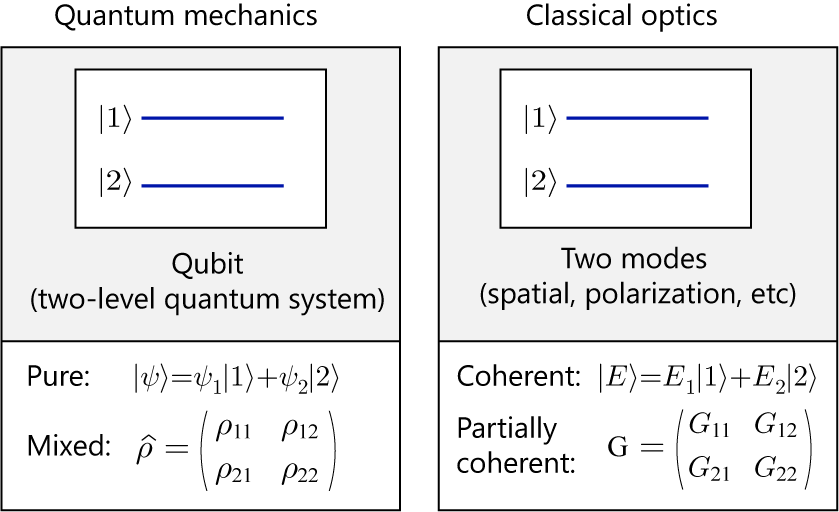}
\caption{Correspondence between a qubit in quantum mechanics and a classical optical field with a binary DoF.}
\label{fig:CQcorrespondence}
\end{figure}

However, there are nevertheless fundamental differences between these two different physical settings that must always be kept in mind. First, the no-cloning theorem in quantum mechanics indicates that a single qubit in an unknown quantum state cannot be cloned (or multiple identical copies produced) \cite{Wootters82Nature}. In the classical setting, the field can be amplified, or the power split into multiple copies on which independent measurements can be implemented. One can thus reconstruct $\mathbf{G}$ in a single-shot by parallelizing the measurements of the Stokes parameters. The density matrix $\mathbf{\rho}$ cannot be reconstructed if only a single qubit is available (rather than an ensemble of identically prepared qubits). The quantum/classical correspondence is thus maintained between the classical field and an \textit{ensemble} of qubits (rather than a single instantiation). Second, it is straightforward to extend the family of transformations of the classical field to non-unitary evolution; examples include: (1) nonlinear evolution in nonlinear optical materials; (2) non-Hermitian evolution in media endowed with optical loss or gain \cite{Elganainy18NP} (which undergirds the realization PT-symmetry and other phenomena associated with non-Hermitian photonics); (3) non-unitary filtering to increase or decrease the degree of coherence; among a host of other possibilities. Such transformations are challenging to realize for quantum systems. For example, although PT-symmetry \cite{Bender98PRL} and the fractional Schr{\"o}dinger equation were initially proposed in the context of quantum field theory, they were realized experimentally first in classical optics \cite{Ruter10NP}.

\section{Two binary DoFs: Basic Definition}\label{sec:2DoFs}

\begin{figure}[t!]
\centering
\includegraphics[width=5.8cm]{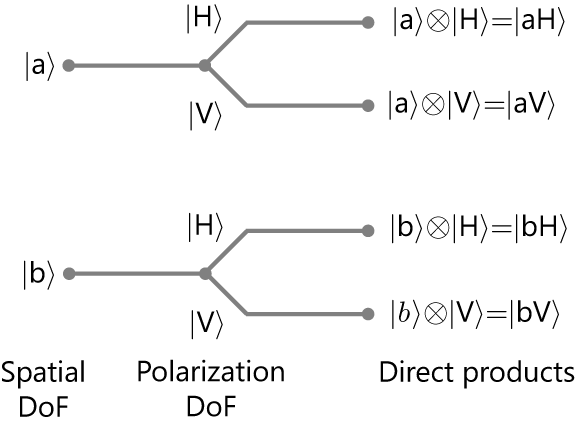}
\caption{The direct-product formulation of the composite or joint modal basis for a field characterized by two binary DoFs: a spatial DoF spanned by the modes $\{|a\rangle,|b\rangle\}$ and a polarization DoF spanned by the modes $\{|\mathrm{H}\rangle,|\mathrm{V}\rangle\}$. The field is spanned by the composite modes $\{|a\mathrm{H}\rangle,|a\mathrm{V}\rangle,|b,\mathrm{H}\rangle,|b\mathrm{V}\rangle\}$, where $|a\mathrm{H}\rangle=|a\rangle\otimes|\mathrm{H}\rangle$, etc.}
\label{fig:DualDoFDefinition}
\end{figure}

\subsection{Coherent fields: Basis definition}

We proceed to examine optical fields characterized by \textit{two} binary DoFs. For concreteness, we consider the polarization DoF spanned by polarization modes $|\mathrm{H}\rangle$ and $|\mathrm{V}\rangle$ (Section~\ref{sec:polarizationDoF}) and a spatial DoF spanned by spatial modes $|a\rangle$ and $|b\rangle$ (e.g., a two-point field or two single-mode waveguides, Section~\ref{sec:SpatialDoF}). However, the concepts elucidated here apply to any pair of binary DoFs.

How do we construct the joint modal basis for the field from the individual modal bases $\{|a\rangle,|b\rangle\}$ and $\{|\mathrm{H}\rangle,|\mathrm{V}\rangle\}$ associated with these two DoFs? The construction follows a route that is analogous to the sample-space concept familiar from probability theory. When a die is thrown, the exhaustive list of all possible outcomes $\{1,2,3,4,5,6\}$ comprises the sample space. When two dice are thrown, the list of all possible outcomes is exhausted by the set of $6\times6=36$ composite outcomes $\{(1,1),(1,2),(1,3),\cdots,(6,6)\}$, which constitutes the new sample space. A similar procedure is followed here for the dual-DoF field.

The modal basis for the field then comprises \textit{composite} modes formed of all the combinations of the basis modes associated with each DoF. We assume that each modal basis is orthonormal: $\langle a|a\rangle=\langle b|b\rangle=1$ and $\langle a|b\rangle=0$ for the spatial DoF; similarly, $\langle\mathrm{H}|\mathrm{H}\rangle=\langle\mathrm{V}|\mathrm{V}\rangle=1$ and $\langle\mathrm{H}|\mathrm{V}\rangle=0$ for the polarization DoF. The dual-DoF field is thus spanned by the composite or joint basis $\{|a,\mathrm{H}\rangle,|a,\mathrm{V}\rangle,|b,\mathrm{H}\rangle,|b,\mathrm{V}\rangle\}$. Each composite mode in the new basis comprises a pair of modes, one for each DoF. Mathematically, this structure is known as a `direct product' of the two bases; for example, the mode $|a,\mathrm{H}\rangle$ is shorthand for the direct product $|a,\mathrm{H}\rangle=|a\rangle\otimes|\mathrm{H}\rangle$. This modal basis exhausts all the mutually exclusive possibilities for the two DoFs. A pictorial representation of this procedure is provided in Fig.~\ref{fig:DualDoFDefinition}. The same procedure is followed in quantum mechanics when constructing quantum states describing a composite system comprising two subsystems when the basis for the Hilbert space each subsystem is known.

Operations using the direct-product modal basis typically separate into products of operations on each basis. For example, the inner product of a basis vector with itself is $\langle a,\mathrm{H}|a,\mathrm{H}\rangle=(\langle a|\otimes\langle\mathrm{H}|)(|a\rangle\otimes|\mathrm{H}\rangle)=\langle a|a\rangle\langle\mathrm{H}|\mathrm{H}\rangle=1$; similarly, $\langle a,\mathrm{V}|a,\mathrm{V}\rangle=\langle b,\mathrm{H}|b,\mathrm{H}\rangle=\langle b,\mathrm{V}|b,\mathrm{V}\rangle=1$. In other words, the normalization of the underlying basis sets $\{|a\rangle,|b\rangle\}$ and $\{|\mathrm{H}\rangle,|\mathrm{V}\rangle\}$ guarantees that the direct-product modes are also normalized. Additionally, $\langle a,\mathrm{H}|b,\mathrm{H}\rangle=(\langle a|\otimes\langle\mathrm{H}|)(|b\rangle\otimes|\mathrm{H}\rangle)=\langle a|b\rangle\langle\mathrm{H}|\mathrm{H}\rangle=0$, $\langle a,\mathrm{H}|a,\mathrm{V}\rangle=(\langle a|\otimes\langle\mathrm{H}|)(|a\rangle\otimes|\mathrm{V}\rangle)=\langle a|a\rangle\langle\mathrm{H}|\mathrm{V}\rangle=0$, and similarly for any pair of distinct composite modes. That is, the orthogonality of the underlying modal basis for each DoF is inherited by the direct-product composite modal basis.

In general, a \textit{coherent} field in this basis can be written as the modal superposition:
\begin{eqnarray}\label{eq:General4x1vector}
|E\rangle\!\!&=&\!\!E_{\mathrm{H}}^{a}|a,\mathrm{H}\rangle+E_{\mathrm{V}}^{a}|a,\mathrm{V}\rangle+E_{\mathrm{H}}^{b}|b,\mathrm{H}\rangle+E_{\mathrm{V}}^{b}|b,\mathrm{V}\rangle\nonumber\\&=&\!\!
E_{\mathrm{H}}^{a}\left(\begin{array}{c}1\\0\\0\\0\end{array}\right)+
E_{\mathrm{V}}^{a}\left(\begin{array}{c}0\\1\\0\\0\end{array}\right)+
E_{\mathrm{H}}^{b}\left(\begin{array}{c}0\\0\\1\\0\end{array}\right)+
E_{\mathrm{V}}^{b}\left(\begin{array}{c}0\\0\\0\\1\end{array}\right)=\left(\begin{array}{c}E_{\mathrm{H}}^{a}\\E_{\mathrm{V}}^{a}\\E_{\mathrm{H}}^{b}\\E_{\mathrm{V}}^{b}\end{array}\right).
\end{eqnarray}
The subscript and superscript identify the polarization and spatial modes, respectively, and we make use of $|a\rangle=\left(\begin{array}{c}1\\0\end{array}\right)$, $|b\rangle=\left(\begin{array}{c}0\\1\end{array}\right)$, $|\mathrm{H}\rangle=\left(\begin{array}{c}1\\0\end{array}\right)$, and $|\mathrm{V}\rangle=\left(\begin{array}{c}0\\1\end{array}\right)$, whereupon:
\begin{equation}
|a,\mathrm{V}\rangle=|a\rangle\otimes|\mathrm{V}\rangle=\left(\begin{array}{c}1\\0\end{array}\right)\otimes\left(\begin{array}{c}0\\1\end{array}\right)=\left(\begin{array}{c}1\times\left(\begin{array}{c}0\\1\end{array}\right)\\0\times\left(\begin{array}{c}0\\1\end{array}\right)\end{array}\right)=\left(\begin{array}{c}0\\1\\0\\0\end{array}\right),
\end{equation}
and similarly for $|a,\mathrm{H}\rangle$, $|b,\mathrm{H}\rangle$, and $|b,\mathrm{V}\rangle$, as given in Eq.~\ref{eq:General4x1vector}. The normalization $\langle E|E\rangle=1$ entails that $|E_{\mathrm{H}}^{a}|^{2}+|E_{\mathrm{V}}^{a}|^{2}+|E_{\mathrm{H}}^{b}|^{2}+|E_{\mathrm{V}}^{b}|^{2}=1$.

\begin{figure}[t!]
\centering
\includegraphics[width=11.3cm]{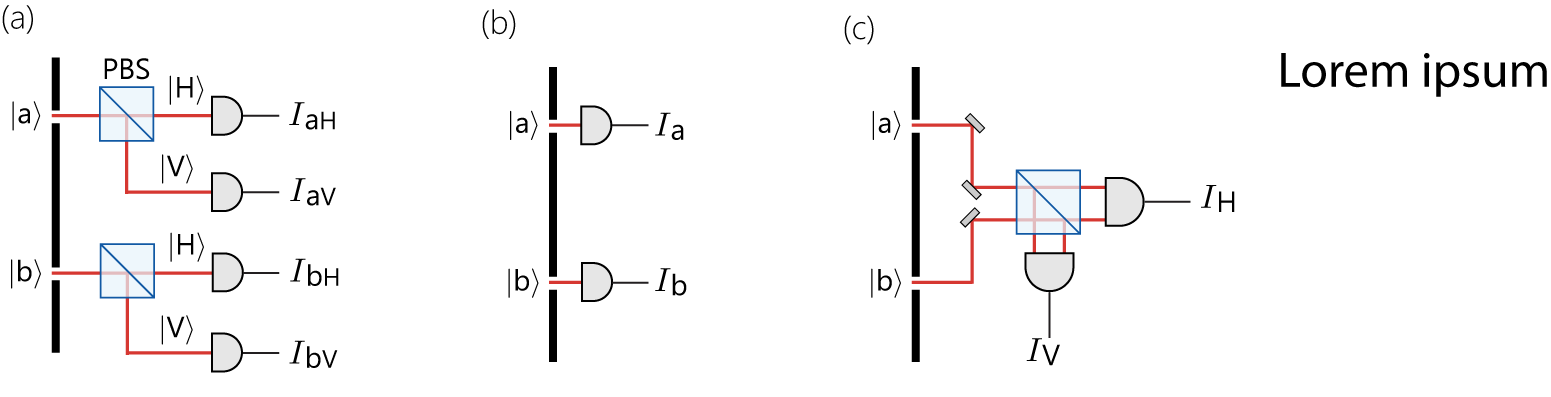}
\caption{(a) Measurements of the modal weights via four detectors. The polarization components at $|a\rangle$ and $|b\rangle$ are separated with a polarizing beam splitter (PBS). When the field is coherent, the modal weights are $I_{a\mathrm{H}}=|E_{\mathrm{H}}^{a}|^{2}$, $I_{a\mathrm{V}}=|E_{\mathrm{V}}^{a}|^{2}$, $I_{b\mathrm{H}}=|E_{\mathrm{H}}^{b}|^{2}$, and $I_{b\mathrm{V}}=|E_{\mathrm{V}}^{b}|^{2}$. When the field is partially coherent, $I_{a\mathrm{H}}=G_{\mathrm{HH}}^{aa}$, $I_{a\mathrm{V}}=G_{\mathrm{VV}}^{aa}$, $I_{b\mathrm{H}}=G_{\mathrm{HH}}^{bb}$, and $I_{b\mathrm{V}}=G_{\mathrm{VV}}^{bb}$. (b) Purely spatial measurements via detectors that are not sensitive to polarization; $I_{a}=|E_{\mathrm{H}}^{a}|^{2}+|E_{\mathrm{V}}^{a}|^{2}$ and $I_{b}=|E_{\mathrm{H}}^{b}|^{2}+|E_{\mathrm{V}}^{b}|^{2}$ when the field is coherent; whereas $I_{a}=G_{\mathrm{HH}}^{aa}+G_{\mathrm{VV}}^{aa}$ and $I_{b}=G_{\mathrm{HH}}^{bb}+G_{\mathrm{VV}}^{bb}$ when the field is partially coherent. (c) Purely polarization measurements via polarization-sensitive detectors (by separating the $|\mathrm{H}\rangle$ and $|\mathrm{V}\rangle$ polarization modes via a PBS) that are not sensitive to spatial position (i.e., bucket detectors). Here $I_{\mathrm{H}}=|E_{\mathrm{H}}^{a}|^{2}+|E_{\mathrm{H}}^{b}|^{2}$ and $I_{\mathrm{V}}=|E_{\mathrm{V}}^{a}|^{2}+|E_{\mathrm{V}}^{b}|^{2}$ when the field is coherent; whereas $I_{\mathrm{H}}=G_{\mathrm{HH}}^{aa}+G_{\mathrm{HH}}^{bb}$ and $I_{\mathrm{V}}=G_{\mathrm{VV}}^{aa}+G_{\mathrm{VV}}^{bb}$ when the field is partially coherent.}
\label{fig:ReducedMeasurements}
\end{figure}

The interpretation of the basis states is straightforward: $|a,\mathrm{H}\rangle$ corresponds to the $|\mathrm{H}\rangle$ polarization mode at $|a\rangle$; $|a,\mathrm{V}\rangle$ corresponds to the $|\mathrm{V}\rangle$ polarization mode at $|a\rangle$; $|b,\mathrm{H}\rangle$ corresponds to the $|\mathrm{H}\rangle$ polarization mode at $|b\rangle$; and $|b,\mathrm{V}\rangle$ corresponds to the $|\mathrm{V}\rangle$ polarization mode at $|b\rangle$. The first vector in Eq.~\ref{eq:General4x1vector} can be interpreted in this new modal basis as follows: the coefficient $E_{\mathrm{H}}^{a}$ is the complex modal amplitude of the basis vector $|a,\mathrm{H}\rangle$, and thus $|E_{\mathrm{H}}^{a}|^{2}$ represents the modal weight or fraction of the power in the $|\mathrm{H}\rangle$ polarization mode for the spatial mode $|a\rangle$. Figure~\ref{fig:ReducedMeasurements}(a) presents the measurement configuration to determine the modal weights. Placing a PBS at $|a\rangle$ to separate out the $|\mathrm{H}\rangle$ and $|\mathrm{V}\rangle$ polarization modes, the measurements are thus $I_{a\mathrm{H}}=|E_{a}^{\mathrm{H}}|^{2}$ and $I_{a\mathrm{V}}=|E_{a}^{\mathrm{V}}|^{2}$. A similar procedure at $|b\rangle$ reveals $I_{b\mathrm{H}}=|E_{b}^{\mathrm{H}}|^{2}$ and $I_{b\mathrm{V}}=|E_{b}^{\mathrm{V}}|^{2}$.

Placing a polarization-independent detector at $|a\rangle$ and another at $|b\rangle$, the measurements are $I_{a}=I_{a\mathrm{H}}+I_{a\mathrm{V}}=|E_{\mathrm{H}}^{a}|^{2}+|E_{\mathrm{V}}^{a}|^{2}$ and $I_{b}=I_{b\mathrm{H}}+I_{b\mathrm{V}}=|E_{\mathrm{H}}^{b}|^{2}+|E_{\mathrm{V}}^{b}|^{2}$; see Fig.~\ref{fig:ReducedMeasurements}(b). Alternatively, separating the $|\mathrm{H}\rangle$ and $|\mathrm{V}\rangle$ polarization modes at $|a\rangle$ and $|b\rangle$ jointly with a PBS, and then using `bucket detectors' (detectors with no spatial resolution) at the PBS output ports that integrate over space, $I_{\mathrm{H}}=I_{a\mathrm{H}}+I_{b\mathrm{H}}=|E_{\mathrm{H}}^{a}|^{2}+|E_{\mathrm{H}}^{b}|^{2}$ and $I_{\mathrm{V}}=I_{a\mathrm{V}}+I_{b\mathrm{V}}=|E_{\mathrm{V}}^{a}|^{2}+|E_{\mathrm{V}}^{b}|^{2}$; see Fig.~\ref{fig:ReducedMeasurements}(c). 

\subsection{The coherence matrix}

\subsubsection{Definition of the coherence matrix}

When this dual-DoF field is partially coherent, we follow the same procedure outlined in Section~\ref{sec:CoherenceMatrix1DoF} to encode the statistical field correlations in a \textit{coherence matrix}. Rather than the field vector $|E\rangle$ for a dual-DoF coherent field (Eq.~\ref{eq:General4x1vector}), we take an ensemble of field vectors $\{|E(\xi)\rangle\}$, where $\xi$ is a set of random variables characterizing the statistical fluctuations undergirding the optical field. The stochasticity of the field may be confined to one DoF or to the other; it may comprise both DoFs with the random variables underpinning the spatial and polarization DoFs statistically independent of each other; or, most generally, it may comprise both DoFs, which are -- moreover -- partially correlated. The coherence matrix is then written as an average over this ensemble,
\begin{equation}
\mathbf{G}=\int\!d\xi\;P(\xi)|E(\xi)\rangle\langle E(\xi)|,
\end{equation}
where $\xi$ may run over either DoF, or over both DoFs, $P(\xi)$ is the probability density function over $\xi$, and $\int\!P(\xi)d\xi=1$, in which case $\mathbf{G}$ is a $4\times4$ matrix cast in the general form:
\begin{equation}\label{sec:General4X4CoherenceMatrix}
\mathbf{G}=\left(\begin{array}{cccc}
G_{\mathrm{HH}}^{aa}&G_{\mathrm{HV}}^{aa}&
G_{\mathrm{HH}}^{ab}&G_{\mathrm{HV}}^{ab}\\
G_{\mathrm{VH}}^{aa}&G_{\mathrm{VV}}^{aa}&
G_{\mathrm{VH}}^{ab}&G_{\mathrm{VV}}^{ab}\\
G_{\mathrm{HH}}^{ba}&G_{\mathrm{HV}}^{ba}&
G_{\mathrm{HH}}^{bb}&G_{\mathrm{HV}}^{bb}\\
G_{\mathrm{VH}}^{ba}&G_{\mathrm{VV}}^{ba}&
G_{\mathrm{VH}}^{bb}&G_{\mathrm{VV}}^{bb}
\end{array}\right),
\end{equation}
where $G_{k\ell}^{ij}=\langle E_{k}^{i}(E_{\ell}^{j})^{*}\rangle$, with spatial-mode indices $i,j=a,b$, and polarization-mode indices $k,\ell=\mathrm{H},\mathrm{V}$, and $\langle\cdot\rangle$ indicates an ensemble average.

\subsubsection{Properties of the coherence matrix}

The $4\times4$ coherence matrix $\mathbf{G}$ has the following relevant mathematical properties:
\begin{enumerate}
\item $\mathbf{G}^{\dagger}=\mathbf{G}$ is Hermitian, so that:
    \begin{enumerate}
    \item The diagonal elements are real $G_{jj}=G_{jj}^{*}$, and the off-diagonal elements are complex conjugate pairs $G_{ij}=G_{ji}^{*}$ ($i\neq j$).
    \item The eigenvalues of $\mathbf{G}$, $\{\lambda_{j}\}_{j=1}^{4}$, are real.
    \item The associated eigenvectors corresponding to different eigenvalues are orthogonal.
\end{enumerate}
\item The normalization of the field vectors $\{|E(\xi)\rangle\}$ entails that $\mathbf{G}$ is a unity-trace matrix: $\mathrm{Tr}\{\mathbf{G}\}=1$, and thus $\sum_{j=1}^{4}\lambda_{j}=1$.
\item $\mathbf{G}$ is positive semi-definite, so that its eigenvalues $\lambda_{j}\geq0$. Moreover, the diagonal elements are always positive semi-definite, $G_{jj}\geq0$.
\end{enumerate}

\subsubsection{Physical significance of the diagonal elements of $\mathbf{G}$}

The diagonal elements of $\mathbf{G}$ are revealed by measurements of the modal weights and correspond to the fraction of power in each of the four modes; see Fig.~\ref{fig:ReducedMeasurements}(a). The diagonal element $G_{\mathrm{HH}}^{aa}$ corresponds to the power of the $|\mathrm{H}\rangle$ polarization mode at $|a\rangle$, $I_{a\mathrm{H}}=G_{aa}^{\mathrm{HH}}$, which is obtained by placing a PBS at $|a\rangle$ and recording the power at the $|\mathrm{H}\rangle$ output port; $I_{a\mathrm{V}}=G_{\mathrm{VV}}^{aa}$ corresponds to the power of the $|\mathrm{V}\rangle$ polarization mode at $|a\rangle$; and similarly for $I_{b\mathrm{H}}=G_{\mathrm{HH}}^{bb}$ and $I_{b\mathrm{V}}=G_{\mathrm{VV}}^{bb}$ at $|b\rangle$. Polarization-independent detectors placed at $|a\rangle$ and $|b\rangle$ record $I_{a}=G^{aa}_{\mathrm{HH}}+G^{aa}_{\mathrm{VV}}$ and $I_{b}=G^{bb}_{\mathrm{HH}}+G^{bb}_{\mathrm{VV}}$ [Fig.~\ref{fig:ReducedMeasurements}(b)], whereas a `bucket detector' with no spatial resolution placed at the output ports of a PBS record $I_{H}=G^{aa}_{\mathrm{HH}}+G^{bb}_{\mathrm{HH}}$ and $I_{V}=G^{aa}_{\mathrm{VV}}+G^{bb}_{\mathrm{VV}}$ [Fig.~\ref{fig:ReducedMeasurements}(c)].

\subsubsection{Physical significance of the off-diagonal elements of $\mathbf{G}$}

\begin{figure}[t!]
\centering
\includegraphics[width=11cm]{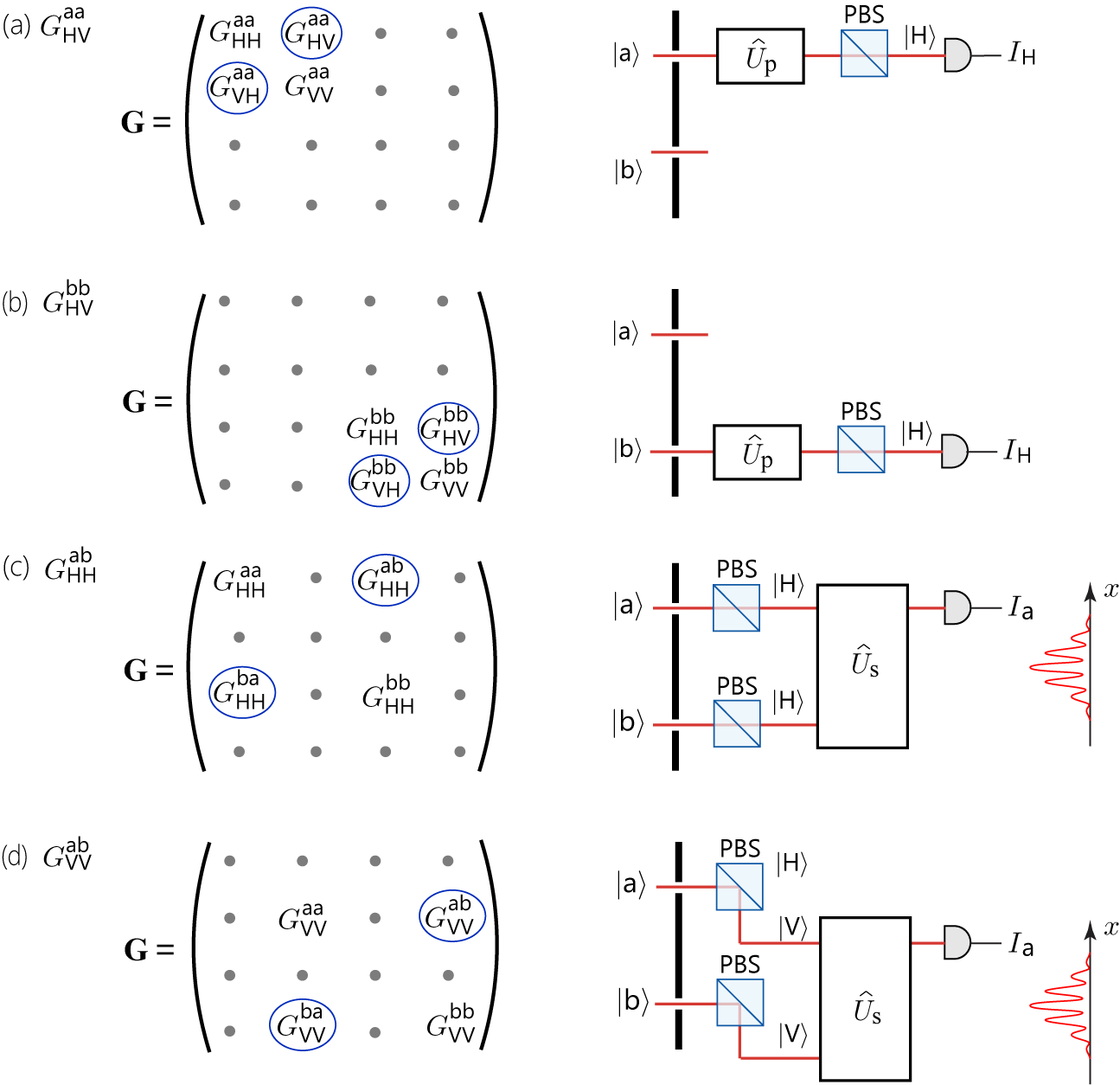}
\caption{Interpretation of the off-diagonal elements of the $4\times4$ coherence matrix $\mathbf{G}$ in terms of appropriately designed interference experiments: (a) $G_{\mathrm{HV}}^{aa}$ is determined by polarization interference at $|a\rangle$; (b) $G_{\mathrm{HV}}^{bb}$ is determined by polarization interference at $|b\rangle$; (c) $G_{\mathrm{HH}}^{ab}$ is determined by spatial interference for the $|\mathrm{H}\rangle$ component; and (d) $G_{\mathrm{VV}}^{ab}$ by spatial interference for the $|\mathrm{V}\rangle$ component. On the left we present the relevant elements in $\mathbf{G}$, and we encircle the off-diagonal element to be measured. On the right we illustrate schematically the interference setup. PBS: Polarizing beam splitter; $\hat{U}_{\mathrm{p}}$: polarization unitary (Eq.~\ref{eq:RestrictedPolarizationUnitary}); $\hat{U}_{\mathrm{s}}$ spatial unitary (Eq.~\ref{eq:SpatialUnitary}).}
\label{fig:OffDiagonalElementsI}
\end{figure}

\begin{figure}[t!]
\centering
\includegraphics[width=13.3cm]{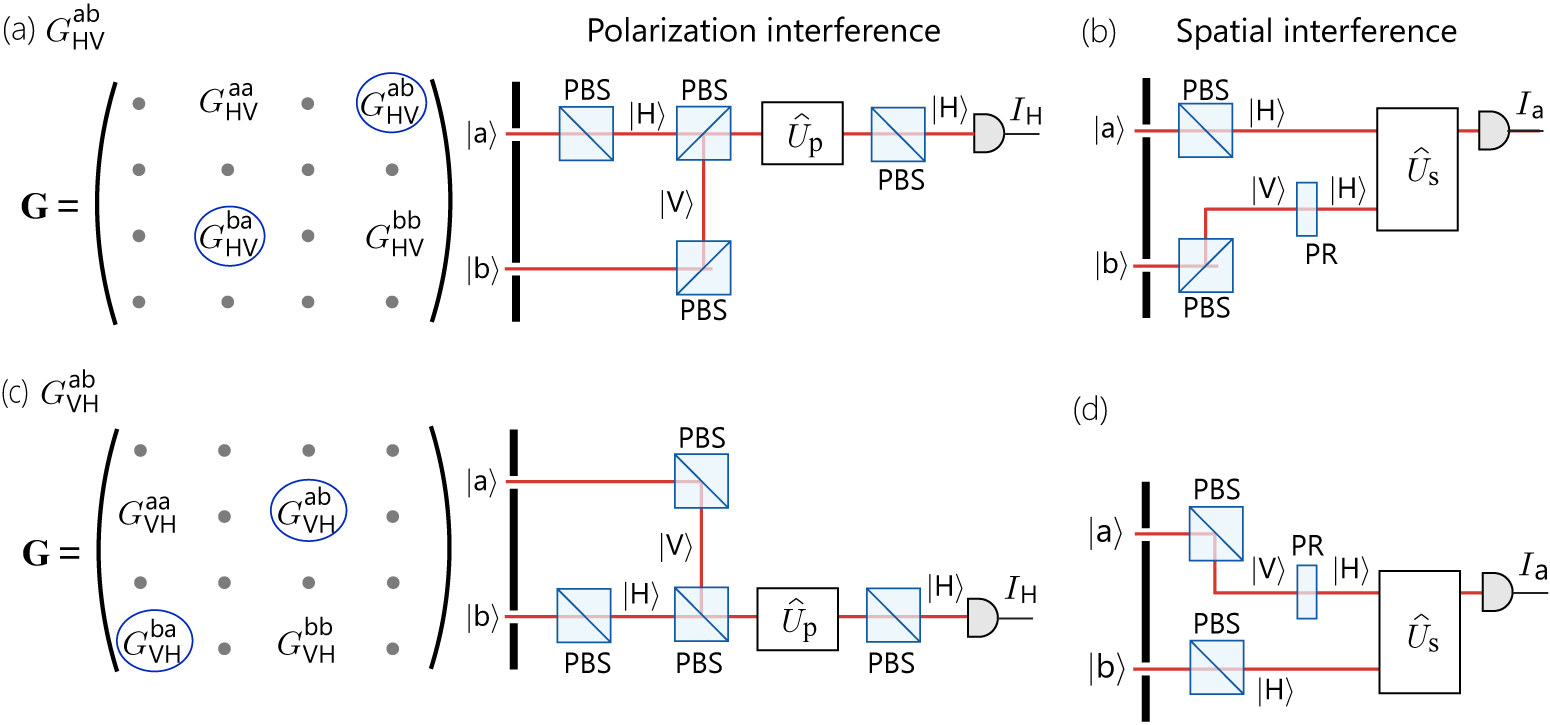}
\caption{Interpretation of the off-diagonal terms of the coherence matrix in terms of appropriately designed interference experiments. (a,b) $G_{\mathrm{HV}}^{ab}$; and (c,d) $G_{\mathrm{VH}}^{ab}$. In (a) and (c) we present on the left the relevant elements of the $4\times4$ coherence matrix $\mathbf{G}$ and encircle the off-diagonal element to be measured. On the right we illustrate schematically the interference setup. PBS: Polarizing beam splitter; PR: polarization rotator; $\hat{U}_{\mathrm{p}}$: polarization unitary; $\hat{U}_{\mathrm{s}}$ spatial unitary.}
\label{fig:OffDiagonalElementsII}
\end{figure}

We have shown that the off-axis elements of a $2\times2$ coherence matrix (Eq.~\ref{eq:General2x2CoherenceMatrix} and Eq.~\ref{eq:GeneralGwithD}) determine the interference visibility observed when the fields associated with the two modes are overlapped (Section~\ref{sec:CoherenceMatrix1DoF}). Similarly, the 6~off-diagonal elements of a $4\times4$ coherence matrix (Eq.~\ref{sec:General4X4CoherenceMatrix}) correspond to the 6~distinct interference experiments illustrated in Fig.~\ref{fig:OffDiagonalElementsI} and Fig.~\ref{fig:OffDiagonalElementsII}. The magnitude of the off-diagonal element is related to the visibility of the interference fringes, and its phase to the shift in the fringes.

\begin{enumerate}
\item $G_{\mathrm{HV}}^{aa}$: The interpretation of the off-diagonal element $G_{\mathrm{HV}}^{aa}$ can be appreciated by isolating the sub-matrix highlighted in Fig.~\ref{fig:OffDiagonalElementsI}(a), which is the \textit{polarization} coherence matrix at $|a\rangle$ and requires appropriate normalization. From this we see that $\tfrac{2|G_{\mathrm{HV}}^{aa}|}{G_{\mathrm{HH}}^{aa}+G_{\mathrm{VV}}^{aa}}$ is the visibility of polarization interference between the $|\mathrm{H}\rangle$ and $|\mathrm{V}\rangle$ components at $|a\rangle$ [Fig.~\ref{fig:OffDiagonalElementsI}(a)].
\item $G_{\mathrm{HV}}^{bb}$: By isolating the sub-matrix highlighted in Fig.~\ref{fig:OffDiagonalElementsI}(b), which is the \textit{polarization} coherence matrix at $|b\rangle$ (without normalization guaranteed), we see that $\tfrac{2|G_{\mathrm{HV}}^{bb}|}{G_{\mathrm{HH}}^{bb}+G_{\mathrm{VV}}^{bb}}$ is the visibility of polarization interference between the $|\mathrm{H}\rangle$ and $|\mathrm{V}\rangle$ components at point $|b\rangle$ [Fig.~\ref{fig:OffDiagonalElementsI}(b)]. 
\item $G_{\mathrm{HH}}^{ab}$: By isolating the sub-matrix highlighted in Fig.~\ref{fig:OffDiagonalElementsI}(c), which is the \textit{spatial} coherence matrix associated with the $|\mathrm{H}\rangle$ polarization component (without normalization guaranteed), we see that $\tfrac{2|G_{\mathrm{HH}}^{ab}|}{G_{\mathrm{HH}}^{aa}+G_{\mathrm{HH}}^{bb}}$ is the visibility of spatial interference formed by the $|\mathrm{H}\rangle$ polarization from $|a\rangle$ and $|b\rangle$. This can be obtained by a spatial unitary $\hat{U}_{\mathrm{s}}$ (Eq.~\ref{eq:SpatialUnitary}), or by overlapping the $|\mathrm{H}\rangle$ polarized field from $|a\rangle$ and $|b\rangle$ in a double-slit interference configuration [Fig.~\ref{fig:OffDiagonalElementsI}(c)].  
\item  $G_{\mathrm{VV}}^{ab}$: By isolating the sub-matrix highlighted in Fig.~\ref{fig:OffDiagonalElementsI}(d), which is the \textit{spatial} coherence matrix associated with the $|\mathrm{V}\rangle$ polarization component (without normalization guaranteed), we can see that $\tfrac{2|G_{\mathrm{VV}}^{ab}|}{G_{\mathrm{VV}}^{aa}+G_{\mathrm{VV}}^{bb}}$ is the visibility of spatial interference formed by the $|\mathrm{V}\rangle$ polarization from $|a\rangle$ and $|b\rangle$. This can be obtained by a spatial unitary $\hat{U}_{\mathrm{s}}$ or overlapping the $|\mathrm{V}\rangle$ polarization from $|a\rangle$ and $|b\rangle$ in a double-slit interference configuration [Fig.~\ref{fig:OffDiagonalElementsI}(d)]. 
\item $G_{\mathrm{HV}}^{ab}$: The interpretation of $G_{\mathrm{HV}}^{ab}$ can be appreciated by isolating the sub-matrix highlighted in Fig.~\ref{fig:OffDiagonalElementsII}(a), which is a hybrid spatial-polarization coherence matrix. This element is not a visibility of spatial interference (because the two spatially interfering fields have orthogonal polarization), nor is it a visibility of polarization interference (because the two polarization components are associated with different spatial modes). Rather, the two fields associated with the modes $|a\mathrm{H}\rangle$ and $|b\mathrm{V}\rangle$ must be combined after first introducing an appropriate modification to enable them to interfere. One first isolates the $|\mathrm{H}\rangle$ polarization component at $|a\rangle$ and the $|\mathrm{V}\rangle$ polarization component at $|b\rangle$. Two options then present themselves. In one approach, we combine the $|\mathrm{H}\rangle$ polarization component at $|a\rangle$ and the $|\mathrm{V}\rangle$ polarization component at $|b\rangle$ into a single spatial mode via a PBS, after which polarization interference reveals a visibility $\tfrac{2|G_{\mathrm{HV}}^{ab}|}{G_{\mathrm{HH}}^{aa}+G_{\mathrm{VV}}^{bb}}$ [Fig.~\ref{fig:OffDiagonalElementsII}(a)]. Alternatively, we rotate the polarization at $|b\rangle$ from $|\mathrm{V}\rangle$ to $|\mathrm{H}\rangle$, at which point we can form the spatial interference pattern from $|a\rangle$ and $|b\rangle$, whose visibility is $\tfrac{2|G_{\mathrm{HV}}^{ab}|}{G_{\mathrm{HH}}^{aa}+G_{\mathrm{VV}}^{bb}}$; one may have rotated the polarization at $|a\rangle$ from $|\mathrm{H}\rangle$ to $|\mathrm{V}\rangle$ with the same outcome [Fig.~\ref{fig:OffDiagonalElementsII}(b)]. 
\item $G_{\mathrm{VH}}^{ab}$: By isolating the sub-matrix highlighted in Fig.~\ref{fig:OffDiagonalElementsII}(c), which is a hybrid spatial-polarization coherence matrix, $G_{\mathrm{VH}}^{ab}$ is related to the visibility of interfering the two fields associated with the $|a\mathrm{V}\rangle$ and $|b\mathrm{H}\rangle$ modes -- after first introducing an appropriate modification. In one approach, we combine the $|\mathrm{V}\rangle$ polarization at $|a\rangle$ and the $|\mathrm{H}\rangle$ polarization at $|b\rangle$ into a single spatial mode via a PBS, after which polarization interference reveals a visibility $\tfrac{2|G_{\mathrm{VH}}^{ab}|}{G_{\mathrm{VV}}^{aa}+G_{\mathrm{HH}}^{bb}}$ [Fig.~\ref{fig:OffDiagonalElementsII}(c)]. Alternatively, one may rotate the polarization at $|b\rangle$ from $|\mathrm{H}\rangle$ to $|\mathrm{V}\rangle$, at which point we can form the spatial interference pattern from $|a\rangle$ and $|b\rangle$, which share the $|\mathrm{V}\rangle$ polarization, whose visibility is $\tfrac{2|G_{\mathrm{VH}}^{ab}|}{G_{\mathrm{VV}}^{aa}+G_{\mathrm{HH}}^{bb}}$; one may have rotated the polarization at $|a\rangle$ from $|\mathrm{V}\rangle$ to $|\mathrm{H}\rangle$ with the same outcome [Fig.~\ref{fig:OffDiagonalElementsII}(d)].
\end{enumerate}

Care must be taken with the conjugate pairs of elements in $\mathbf{G}$. For example, although the element $G_{\mathrm{VH}}^{ab}$ may seem related to the element $G_{\mathrm{HV}}^{ab}$, these two off-diagonal elements are in fact independent of each other. Only when the indices of \textit{both} spatial and polarization modes are switched do we obtain the complex conjugate; e.g., $(G_{\mathrm{VH}}^{ba})^{*}=G_{\mathrm{HV}}^{ab}$. Each of the six interference experiments outlined above yields two real parameters: the visibility that provides the magnitude of the corresponding complex off-diagonal element, and a shift in the fringes that provides its phase. These 12~real parameters [Fig.~\ref{fig:OffDiagonalElementsI} and Fig.~\ref{fig:OffDiagonalElementsII}], along with the 4~real parameters associated with the modal-weight measurements that correspond to the real diagonal elements of $\mathbf{G}$ [Fig.~\ref{fig:ReducedMeasurements}], are sufficient to reconstruct any coherence matrix $\mathbf{G}$. To the best of our knowledge, the six independent interference measurements outlined above and required to reconstruct $\mathbf{G}$ have not been performed to date. We will see in Section~\ref{sec:Stokes2DoFs} below an alternative approach to reconstructing $\mathbf{G}$ that generalizes to a dual-DoF field the Stokes-parameters approach outlined in Section~\ref{sec:StokesSingleDoF} above for a single-DoF field. 

\subsubsection{Block-matrix form of the coherence matrix}

Further understanding of the interpretation of the elements in $\mathbf{G}$ can be gleaned after first expressing $\mathbf{G}$ in block-matrix form as follows:
\begin{equation}
\mathbf{G}=\left(\begin{array}{cc}|\alpha|^{2}\mathbf{G}_{a}&\mathbf{G}_{ab}\\\mathbf{G}_{ba}&|\beta|^{2}\mathbf{G}_{b}\end{array}\right),
\end{equation}
where $\mathbf{G}_{a}=\tfrac{1}{G_{\mathrm{HH}}^{aa}+G_{\mathrm{VV}}^{aa}}\left(\begin{array}{cc}G_{\mathrm{HH}}^{aa}&G_{\mathrm{HV}}^{aa}\\G_{\mathrm{VH}}^{aa}&G_{\mathrm{VV}}^{aa}\end{array}\right)$ and $\mathbf{G}_{b}=\tfrac{1}{G_{\mathrm{HH}}^{bb}+G_{\mathrm{VV}}^{bb}}\left(\begin{array}{cc}G_{\mathrm{HH}}^{bb}&G_{\mathrm{HV}}^{bb}\\G_{\mathrm{VH}}^{bb}&G_{\mathrm{VV}}^{bb}\end{array}\right)$ are unity-trace, Hermitian, polarization coherence matrices associated with $|a\rangle$ and $|b\rangle$, $|\alpha|^{2}=G_{\mathrm{HH}}^{aa}+G_{\mathrm{VV}}^{aa}$ and $|\beta|^{2}=G_{\mathrm{HH}}^{bb}+G_{\mathrm{VV}}^{bb}$ are the modal weights (or fractions of power) at $|a\rangle$ and $|b\rangle$ from the total power, $|\alpha|^{2}+|\beta|^{2}=1$. The matrices $\mathbf{G}_{ab}=\left(\begin{array}{cc}G_{\mathrm{HH}}^{ab}&G_{\mathrm{HV}}^{ab}\\G_{\mathrm{VH}}^{ab}&G_{\mathrm{VV}}^{ab}\end{array}\right)$ and $\mathbf{G}_{ba}=\left(\begin{array}{cc}G_{\mathrm{HH}}^{ba}&G_{\mathrm{HV}}^{ba}\\G_{\mathrm{VH}}^{ba}&G_{\mathrm{VV}}^{ba}\end{array}\right)$ capture the correlations between the $|\mathrm{H}\rangle$ and $|\mathrm{V}\rangle$ polarization modes at $|a\rangle$ and $|b\rangle$ as described above in terms of various interference experiments [Fig.~\ref{fig:OffDiagonalElementsI} and Fig.~\ref{fig:OffDiagonalElementsII}]. Neither of these two correlation matrices, $\mathbf{G}_{ab}$ nor $\mathbf{G}_{ba}$, is necessarily Hermitian; e.g., the off-diagonal element $G_{\mathrm{VH}}^{ab}$ is not necessarily equal to $(G_{\mathrm{HV}}^{ab})^{*}$, where only the polarization indices are flipped, and the diagonal elements $G_{\mathrm{HH}}^{ab}$ and $G_{\mathrm{VV}}^{ab}$ are not necessarily real. Indeed, as shown above, $G_{\mathrm{HV}}^{ab}$ and $G_{\mathrm{VH}}^{ab}$ are related to independent interference experiments. However, the two off-diagonal sub-matrices form a Hermitian pair, $\mathbf{G}_{ab}=\mathbf{G}_{ba}^{\dagger}$, so that $G_{\mathrm{HV}}^{ab}=(G_{\mathrm{VH}}^{ba})^{*}$, $G_{\mathrm{VH}}^{ab}=(G_{\mathrm{HV}}^{ba})^{*}$, $G_{\mathrm{HH}}^{ab}=(G_{\mathrm{HH}}^{ba})^{*}$, and $G_{\mathrm{VV}}^{ab}=(G_{\mathrm{VV}}^{ba})^{*}$; that is both the spatial \textit{and} the polarization indices are flipped.

\subsubsection{Diagonal coherence matrices}

In general, the Hermitian coherence matrix $\mathbf{G}$ can be diagonalized via a $4\times4$ unitary,
\begin{equation}\label{eq:4x4Diagonal}
\mathbf{G}^{\mathrm{D}}=\left(\begin{array}{cccc}
\lambda_{1}&0&0&0\\
0&\lambda_{2}&0&0\\
0&0&\lambda_{3}&0\\
0&0&0&\lambda_{4}
\end{array}\right)=\mathrm{diag}\{\lambda_{1},\lambda_{2},\lambda_{3},\lambda_{4}\},
\end{equation}
and we assume throughout (unless stated otherwise) that the eigenvalues are arranged in descending order: $\lambda_{1}\geq\lambda_{2}\geq\lambda_{3}\geq\lambda_{4}\geq0$. We use here the shorthand $\mathrm{diag}\{\cdots\}$ to indicate a diagonal matrix with the given entries representing the diagonal elements of the matrix. Any coherence matrix $\mathbf{G}$ can thus be expressed as a unitary transformation of a diagonal matrix, $\mathbf{G}=\hat{U}\mathbf{G}^{\mathrm{D}}\hat{U}^{\dagger}$. The diagonal matrix $\mathbf{G}^{\mathrm{D}}$ in which all the off-diagonal elements have been eliminated corresponds to adopting a modal basis in which all correlations between the modes have been suppressed; so that the visibilities of the 6~interference experiments illustrated in Fig.~\ref{fig:OffDiagonalElementsI} and Fig.~\ref{fig:OffDiagonalElementsII} all vanish. For example, the $|\mathrm{H}\rangle$ mode at $|a\rangle$ is now uncorrelated with the $|\mathrm{V}\rangle$ mode at $|a\rangle$, and is also uncorrelated to the $|\mathrm{H}\rangle$ and $|\mathrm{V}\rangle$ modes at $|b\rangle$. We can interpret the eigenvalues as follows: $\lambda_{1}$ is the power in the $|\mathrm{H}\rangle$ component at $|a\rangle$; $\lambda_{2}$ is the power of the $|\mathrm{V}\rangle$ component at $|b\rangle$; and $\lambda_{3}$ and $\lambda_{4}$ correspond to the power of the $|\mathrm{H}\rangle$ and $|\mathrm{V}\rangle$ components at $|b\rangle$, respectively, \textit{when the correlations between all modes have been eliminated}. In this setting, the total power at $|a\rangle$ is $\lambda_{1}+\lambda_{2}$, and that at $|b\rangle$ is $\lambda_{3}+\lambda_{4}$. Alternatively, the total power of the $|\mathrm{H}\rangle$ component at both $|a\rangle$ and $|b\rangle$ is $\lambda_{1}+\lambda_{3}$, while the total power of the $|V\rangle$ component is $\lambda_{2}+\lambda_{4}$. It is clear that the diagonal representation for the coherence matrix is the discrete counterpart to the coherent-mode representation in traditional coherence theory.


\subsubsection{Coherence entropy}

We define the entropy $S$ for the $4\times4$ coherence matrix $\mathbf{G}$ as we did earlier for the $2\times2$ coherence matrix:
\begin{equation}
S=-\mathrm{Tr}\{\mathbf{G}\log_{2}\mathbf{G}\}=-\sum_{j=1}^{4}\lambda_{j}\log_{2}\lambda_{j},
\end{equation}
where $0\leq S\leq2$. The limit $S=0$ occurs when $\lambda_{1}=1$ and $\lambda_{2}=\lambda_{3}=\lambda_{3}=0$, corresponding to a \textit{fully coherent} field. The opposite limit $S=2$~bits occurs when $\lambda_{1}=\lambda_{2}=\lambda_{3}=\lambda_{4}=\tfrac{1}{4}$ corresponding to a \textit{maximally incoherent} field. This is to be expected for two binary DoFs since each binary DoF alone can carry 1~bit of entropy. In contrast to the single-DoF scenario, it is \textit{not} straightforward to devise here a single scalar parameter to characterize the `degree of coherence' for the field represented by $\mathbf{G}$.

\subsection{Unitary transformations}\label{sec:Unitary2DoFs}

The unitaries $\hat{U}$ in the space of two binary DoFs are represented by $4\times4$ unitary matrices, and are characterized once again by $\hat{U}\hat{U}^{\dagger}=\hat{\mathbb{I}}_{4}=\hat{U}^{\dagger}\hat{U}$, or $\hat{U}^{\dagger}=\hat{U}^{-1}$, where $\hat{\mathbb{I}}_{4}$ is the $4\times4$ identity matrix. A unitary transforms the field vector as $|E\rangle\rightarrow|E'\rangle=\hat{U}|E\rangle$, and transforms the coherence matrix as $\mathbf{G}\rightarrow\mathbf{G}'=\hat{U}\mathbf{G}\hat{U}^{\dagger}$.  

\begin{figure}[t!]
\centering
\includegraphics[width=10.4cm]{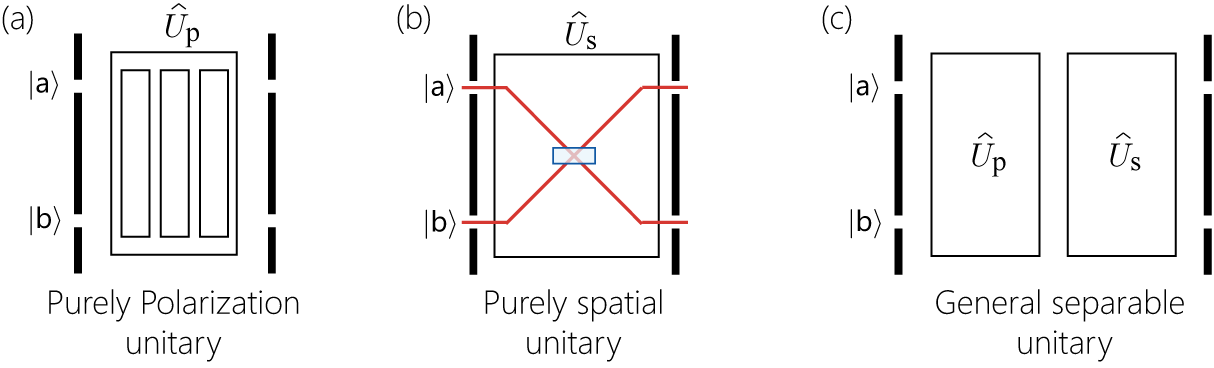}
\caption{(a) A purely polarization unitary $\hat{U}=\hat{\mathbb{I}}_{2}\otimes\hat{U}_{\mathrm{p}}$ formed of a cascade of wave plates impacting both $|a\rangle$ and $|b\rangle$. (b) A purely spatial unitary $\hat{U}=\hat{U}_{\mathrm{s}}\otimes\hat{\mathbb{I}}_{2}$ is independent of polarization. Both polarization modes $|\mathrm{H}\rangle$ and $|\mathrm{V}\rangle$ undergo the \textit{same} spatial unitary. The particular example here is of a polarization-independent beam splitter (BS) corresponding to Eq.~\ref{eq:BeamSplitter}. (c) Cascades of purely polarization and purely spatial unitaries constitute a general separable unitary.}
\label{fig:4x4SeparableUnitaries}
\end{figure}

We classify such unitaries broadly as \textit{separable} and \textit{non-separable}:
\begin{enumerate}
\item \textit{Separable unitaries}. A separable unitary can be expressed as a direct product of two unitaries, each impacting only one DoF: $\hat{U}=\hat{U}_{\mathrm{s}}\otimes\hat{U}_{\mathrm{p}}$, where $\hat{U}_{\mathrm{s}}$ is a $2\times2$ spatial unitary that does \textit{not} impact the polarization DoF (Section~\ref{sec:SpatialDoF}), and $\hat{U}_{\mathrm{p}}$ is a $2\times2$ polarization unitary that does \textit{not} impact the spatial DoF (Section~\ref{sec:polarizationDoF}).
\begin{enumerate}
\item \textit{Polarization unitaries}. A purely polarization unitary is a separable unitary transformation that impacts only the polarization DoF and not the spatial DoF, in which case $\hat{U}=\hat{\mathbb{I}}_{2}\otimes\hat{U}_{\mathrm{p}}$. An example is depicted in Fig.~\ref{fig:4x4SeparableUnitaries}(a) of a polarization unitary $\hat{U}_{\mathrm{p}}$ formed of wave plates extending over $|a\rangle$ and $|b\rangle$, so that the same polarization transformation is implemented for both. For $\hat{U}_{\mathrm{p}}=\left(\begin{array}{cc}u_{1}&-u_{2}\\u_{2}^{*}&u_{1}^{*}\end{array}\right)$, we have:
\begin{equation}
\hat{U}=\hat{\mathbb{I}}\otimes\hat{U}_{\mathrm{p}}=\left(\begin{array}{cc}\hat{U}_{\mathrm{p}}&\hat{\mathbf{0}}_{2}\\\hat{\mathbf{0}}_{2}&\hat{U}_{\mathrm{p}}\end{array}\right)=\left(\begin{array}{cccc}
u_{1}&-u_{2}&0&0\\
u_{2}^{*}&u_{1}^{*}&0&0\\
0&0&u_{1}&-u_{2}\\
0&0&u_{2}^{*}&u_{1}^{*}
\end{array}\right),
\end{equation}
where $\hat{\mathbf{0}}_{2}$ is a $2\times2$ matrix with all zero elements. Utilizing such a polarization unitary, the block matrix-form of the coherence matrix is transformed as follows:
\begin{equation}
\mathbf{G}=\left(\begin{array}{cc}|\alpha|^{2}\mathbf{G}_{a}&\mathbf{G}_{ab}\\\mathbf{G}_{ba}&|\beta|^{2}\mathbf{G}_{b}\end{array}\right)\rightarrow\hat{U}\mathbf{G}\hat{U}^{\dagger}=\left(\begin{array}{cc}|\alpha|^{2}\hat{U}_{\mathrm{p}}\mathbf{G}_{a}\hat{U}_{\mathrm{p}}^{\dagger}&\hat{U}_{\mathrm{p}}\mathbf{G}_{ab}\hat{U}_{\mathrm{p}}^{\dagger}\\\hat{U}_{\mathrm{p}}\mathbf{G}_{ba}\hat{U}_{\mathrm{p}}^{\dagger}&|\beta|^{2}\hat{U}_{\mathrm{p}}\mathbf{G}_{b}\hat{U}_{\mathrm{p}}^{\dagger}\end{array}\right).
\end{equation}
This separable unitary does \textit{not} mix the block matrices in $\mathbf{G}$, and all the blocks undergo the same transformation.

\item \textit{Spatial unitaries}. A purely spatial unitary is a separable unitary that impacts only the spatial DoF and not the polarization DoF, in which case $\hat{U}=\hat{U}_{\mathrm{s}}\otimes\hat{\mathbb{I}}_{2}$, and $\hat{U}_{\mathrm{s}}$ combines the fields from $|a\rangle$ and $|b\rangle$ independently of polarization. An example of a polarization-independent spatial unitary $\hat{U}_{\mathrm{s}}$ is shown in Fig.~\ref{fig:4x4SeparableUnitaries}(b), corresponding to a beam splitter that superposes the fields from $|a\rangle$ and $|b\rangle$ independently of polarization. For $\hat{U}_{\mathrm{s}}=\left(\begin{array}{cc}v_{1}&-v_{2}\\v_{2}^{*}&v_{1}^{*}\end{array}\right)$, we have:
\begin{equation}
\hat{U}=\hat{U}_{\mathrm{s}}\otimes\hat{\mathbb{I}}_{2}=\left(\begin{array}{cc}v_{1}\hat{\mathbb{I}}_{2}&-v_{2}\hat{\mathbb{I}}_{2}\\v_{2}^{*}\hat{\mathbb{I}}_{2}&v_{1}^{*}\hat{\mathbb{I}}_{2}\end{array}\right)=\left(\begin{array}{cccc}
v_{1}&0&-v_{2}&0\\
0&v_{1}&0&-v_{2}\\
v_{2}^{*}&0&u_{1}^{*}&0\\
0&v_{2}^{*}&0&u_{1}^{*}
\end{array}\right).
\end{equation}
For example, the $4\times4$ unitary $\hat{U}_{\mathrm{BS}}=\hat{B}\otimes\mathbb{I}$ for a symmetric (balanced) non-polarizing BS, where $\hat{B}=\tfrac{1}{\sqrt{2}}\left(\begin{array}{cc}1&i\\i&1\end{array}\right)$, is given by:
\begin{equation}\label{eq:BeamSplitter}
\hat{U}_{\mathrm{BS}}=\frac{1}{\sqrt{2}}\left(\begin{array}{cccc}1&0&i&0\\0&1&0&i\\i&0&1&0\\0&i&0&1\end{array}\right).
\end{equation}

\item \textit{General separable unitaries}. In general, a separable unitary $\hat{U}=\hat{U}_{\mathrm{s}}\otimes\hat{U}_{\mathrm{p}}$ can be produced by cascading purely spatial and purely polarization unitaries: $\hat{U}=\hat{U}_{\mathrm{s}}\otimes\hat{U}_{\mathrm{p}}$, where $\hat{U}_{\mathrm{s}}=\hat{U}_{\mathrm{s}1}\hat{U}_{\mathrm{s}2}\cdots$ is composed of all the purely spatial unitaries and $\hat{U}_{\mathrm{p}}=\hat{U}_{\mathrm{p}1}\hat{U}_{\mathrm{p}2}\cdots$ is composed of all the purely polarization unitaries, as depicted in Fig.~\ref{fig:4x4SeparableUnitaries}(c). In the former we take the sequence of spatial unitaries while ignoring the interspersed polarization unitaries, and in the latter we ignore the spatial unitaries interspersed between the polarization unitaries. This is made possible by the fact that the two operators $\hat{U}_{\mathrm{s}}\otimes\hat{\mathbb{I}}_{2}$ and $\hat{\mathbb{I}}_{2}\otimes\hat{U}_{\mathrm{p}}$ `commute'; that is, their order can be changed without impacting the implemented transformation:
\begin{equation}
(\hat{U}_{\mathrm{s}}\otimes\hat{\mathbb{I}}_{2})(\hat{\mathbb{I}}_{2}\otimes\hat{U}_{\mathrm{p}})=\hat{U}_{\mathrm{s}}\otimes\hat{U}_{\mathrm{p}}=(\hat{\mathbb{I}}_{2}\otimes\hat{U}_{\mathrm{p}})(\hat{U}_{\mathrm{s}}\otimes\hat{\mathbb{I}}_{2}).
\end{equation}

\end{enumerate}

\item \textit{Non-separable unitaries}. A non-separable unitary cannot be expressed as a direct product of purely spatial and polarization unitaries: $\hat{U}\neq\hat{U}_{\mathrm{s}}\otimes\hat{U}_{\mathrm{p}}$. Three particular classes of non-separable unitaries are of interest here.

\begin{figure}[t!]
\centering
\includegraphics[width=9cm]{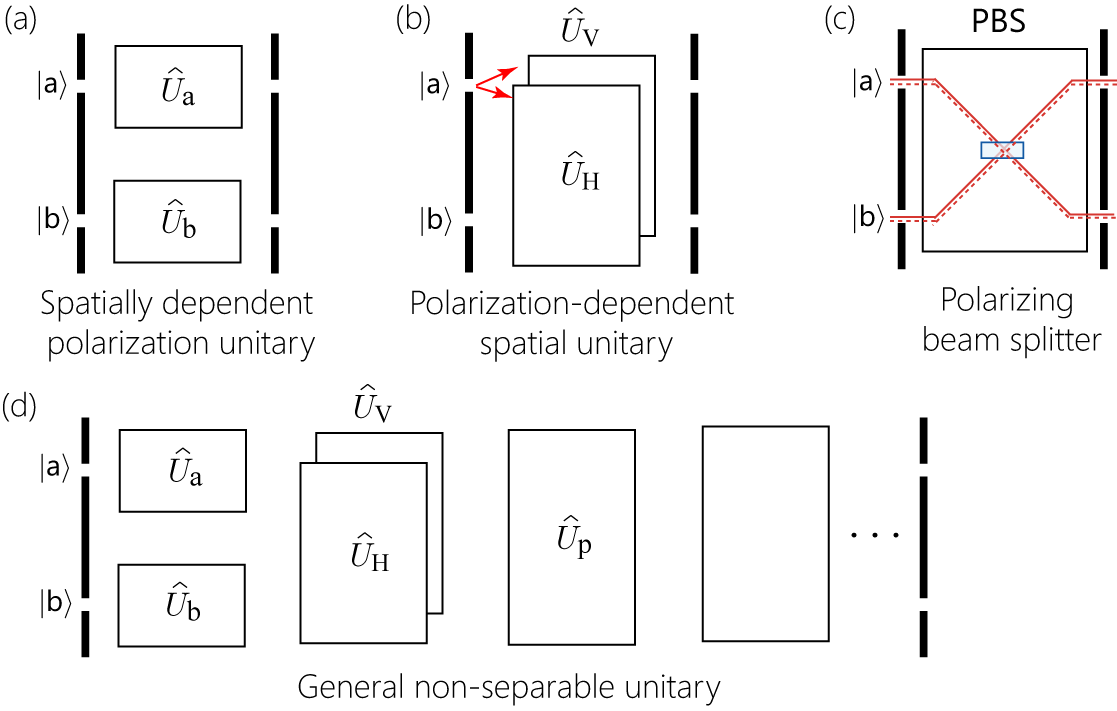}
\caption{(a) A spatially dependent polarization unitary. Polarization unitary $\hat{U}_{a}$ transforms the polarization of the field at $|a\rangle$ and polarization unitary $\hat{U}_{b}$ transforms the polarizatuion of the field at $|b\rangle$. (b) A polarization-dependent spatial unitary. Spatial unitary $\hat{U}_{\mathrm{H}}$ transforms the $|\mathrm{H}\rangle$ polarization component, and spatial unitary $\hat{U}_{\mathrm{V}}$ transforms the $|\mathrm{V}\rangle$ polarization component. (c) A polarizing beam splitter (PBS) as an example of a polarization-dependent spatial unitary corresponding to Eq.~\ref{eq:Polarizing BS}. (d) A general non-separable unitary $\hat{U}$ formed of a cascade of other unitaries.}
\label{fig:4x4NonseparableUnitaries}
\end{figure}

\begin{enumerate}
\item \textit{Spatially dependent polarization unitaries}. One may place a polarization unitary $\hat{U}_{a}$ at $|a\rangle$ and a different polarization unitary $\hat{U}_{b}$ at $|b\rangle$, as depicted at Fig.~\ref{fig:4x4NonseparableUnitaries}(a). This non-separable $4\times4$ unitary $\hat{U}$ can be expressed in block matrix form as:
\begin{equation}
\hat{U}=\left(\begin{array}{cc}1&0\\0&0\end{array}\right)\otimes\hat{U}_{a}+\left(\begin{array}{cc}0&0\\0&1\end{array}\right)\otimes\hat{U}_{b}=\left(\begin{array}{cc}\hat{U}_{a}&\hat{\mathbf{0}}_{2}\\\hat{\mathbf{0}}_{2}&\hat{U}_{b}\end{array}\right).
\end{equation}
Therefore, $\hat{U}$ transforms the block-matrix form of $\mathbf{G}$ as follows:
\begin{equation}
\mathbf{G}=\left(\begin{array}{cc}|\alpha|^{2}\mathbf{G}_{a}&\mathbf{G}_{ab}\\\mathbf{G}_{ba}&|\beta|^{2}\mathbf{G}_{b}\end{array}\right)\rightarrow\hat{U}\mathbf{G}\hat{U}^{\dagger}=\left(\begin{array}{cc}|\alpha|^{2}\hat{U}_{a}\mathbf{G}_{a}\hat{U}_{a}^{\dagger}&\hat{U}_{a}\mathbf{G}_{ab}\hat{U}_{b}^{\dagger}\\\hat{U}_{b}\mathbf{G}_{ba}\hat{U}_{a}^{\dagger}&|\beta|^{2}\hat{U}_{b}\mathbf{G}_{b}\hat{U}_{b}^{\dagger}\end{array}\right).
\end{equation}
Although $\hat{U}$ does not mix the block matrices together, $\hat{U}$ nevertheless implements a \textit{different} transformation for each block.

\item\textit{Polarization-dependent spatial unitaries}. Such a unitary presents different \textit{spatial} unitaries $\hat{U}_{\mathrm{H}}$ and $\hat{U}_{\mathrm{V}}$ for the $|\mathrm{H}\rangle$ and $|\mathrm{V}\rangle$ modes, respectively [Fig.~\ref{fig:4x4NonseparableUnitaries}(b)]. The $4\times4$ unitary $\hat{U}$ takes the form:
\begin{equation}
\hat{U}=\hat{U}_{\mathrm{H}}\otimes\left(\begin{array}{cc}1&0\\0&0\end{array}\right)+\hat{U}_{\mathrm{V}}\otimes\left(\begin{array}{cc}0&0\\0&1\end{array}\right).
\end{equation}
An example of such a unitary is a polarizing beam splitter (PBS), in which the $|\mathrm{H}\rangle$ mode is transmitted and the $|\mathrm{V}\rangle$ mode is reflected Fig.~\ref{fig:4x4NonseparableUnitaries}(c), whereupon $\hat{U}_{\mathrm{H}}=\hat{\mathbb{I}}_{2}$ and $\hat{U}_{\mathrm{V}}=i\left(\begin{array}{cc}0&1\\1&0\end{array}\right)$, in which case the unitary is given by:
\begin{equation}\label{eq:Polarizing BS}
\hat{U}_{\mathrm{PBS}}=\left(\begin{array}{cccc}1&0&0&0\\0&0&0&i\\0&0&1&0\\0&i&0&0\end{array}\right),
\end{equation}

\item \textit{General non-separable unitary}. More general non-separable unitaries $\hat{U}$ can be constructed by cascading the above three classes of two-separable unitaries, along with separable unitaries, as shown in Fig.~\ref{fig:4x4NonseparableUnitaries}(d).

\end{enumerate}

\end{enumerate}

\begin{figure}[t!]
\centering
\includegraphics[width=13.3cm]{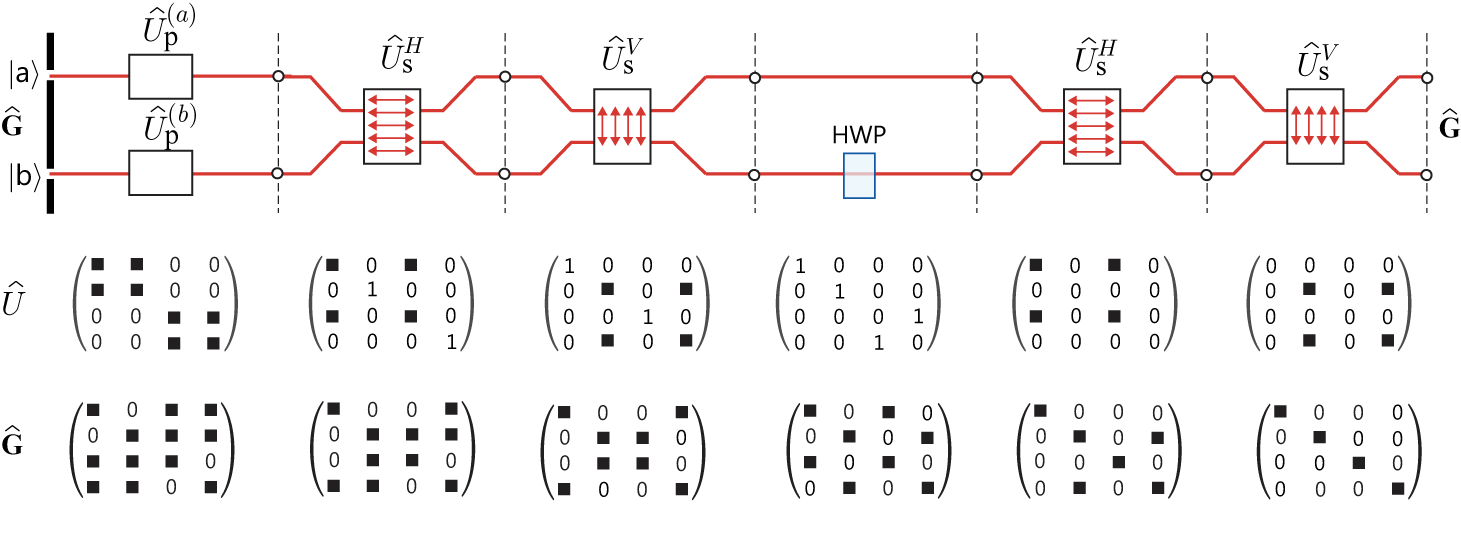}
\caption{A unitary $\hat{U}$ capable of diagonalizing any $4\times4$ coherence matrix is decomposed into a sequence of basic unitaries. From left to right: spatially dependent polarization unitary [Fig.~\ref{fig:4x4NonseparableUnitaries}(a)], two polarization-dependent spatial unitaries [Fig.~\ref{fig:4x4NonseparableUnitaries}(b)], a HWP to flip $|\mathrm{H}\rangle$ and $|\mathrm{V}\rangle$ at $|b\rangle$, and then two polarization-dependent unitaries. In the first row under the figure, we display the structure of the unitaries; the solid black boxes are potentially non-zero elements of the unitary. In the second row under the figure, we display the structure of the coherence matrix after each unitary (highlighting which elements become 0 through the action of the unitary).}
\label{fig:diagonalization}
\end{figure}

The question arises as to the efficient construction of the most general $4\times4$ unitary $\hat{U}$. Can such a unitary \textit{always} be decomposed in terms of a sequence of polarization-dependent spatial unitaries and spatially dependent polarization unitaries? If so, what is the minimum number of such unitaries required to construct a general $4\times4$ unitary $\hat{U}$? This question can be cast in a mathematical form: can a $4\times4$ unitary be decomposed into a sequence of $2\times2$ unitaries operating on pairs of modes at a time? This question has been tackled extensively in mathematics and is especially relevant in the area of quantum computing and quantum information processing \cite{Reck94PRL}, where a general non-separable unitary is to be implemented on a multi-qubit system, but only single-qubit operators can be realized. We plot in Fig.~\ref{fig:diagonalization} the most general form of this construction, where we have made use of the decomposition of $4\times4$ unitaries to be explained in Section~\ref{sec:ExtenstioToN}. It is sufficient here to state that a sequence of 6~$2\times 2$ unitaries (operating on a single DoF at a time) are sufficient to accomplish this task.

Finally, it can be challenging to construct the matrix for a unitary for a given setup. The simplest approach is to evaluate the output field vector emerging from the device for each of the four basis field vectors provided at the input:
\begin{equation}
|u_{a\mathrm{H}}\rangle=\hat{U}\left(\begin{array}{c}1\\0\\0\\0\end{array}\right),|u_{a\mathrm{V}}\rangle=\hat{U}\left(\begin{array}{c}0\\1\\0\\0\end{array}\right),|u_{b\mathrm{H}}\rangle=\hat{U}\left(\begin{array}{c}0\\0\\1\\0\end{array}\right),|u_{b\mathrm{V}}\rangle=\hat{U}\left(\begin{array}{c}0\\0\\0\\1\end{array}\right).
\end{equation}
The unitary matrix $\hat{U}$ representing the system is thus formed by assembling the vectors $|u_{a\mathrm{H}}\rangle$, $|u_{a\mathrm{V}}\rangle$, $|u_{b\mathrm{H}}\rangle$, and $|u_{b\mathrm{V}}\rangle$ as columns of $\hat{U}$.

Consider for example the PBS in Fig.~\ref{fig:4x4NonseparableUnitaries}(c). We assume that the $|\mathrm{H}\rangle$ polarization mode is transmitted by the PBS whereas the $|\mathrm{V}\rangle$ polarization mode is reflected. When the input field is $|a\mathrm{H}\rangle$, the $|\mathrm{H}\rangle$ mode is transmitted through the PBS yielding $|u_{a\mathrm{H}}\rangle=|a\mathrm{H}\rangle$ unchanged from the input. For the input field $|a\mathrm{V}\rangle$, $|\mathrm{V}\rangle$ is reflected by the PBS to yield $|u_{a\mathrm{V}}\rangle=i|b\mathrm{V}\rangle$. Similarly, $|b\mathrm{H}\rangle\rightarrow|u_{b\mathrm{H}}\rangle=|b\mathrm{H}\rangle$, whereas $|b\mathrm{V}\rangle\rightarrow|u_{b\mathrm{V}}\rangle=i|a\mathrm{V}\rangle$, so that:
\begin{equation}
|u_{a\mathrm{H}}\rangle=\left(\begin{array}{c}1\\0\\0\\0\end{array}\right),
|u_{a\mathrm{V}}\rangle=\left(\begin{array}{c}0\\0\\0\\i\end{array}\right),
|u_{b\mathrm{H}}\rangle=\left(\begin{array}{c}0\\0\\1\\0\end{array}\right),
|u_{b\mathrm{V}}\rangle=\left(\begin{array}{c}0\\i\\0\\0\end{array}\right).
\end{equation}
Assembling the vector $|u_{a\mathrm{H}}\rangle$, $|u_{a\mathrm{V}}\rangle$, $|u_{b\mathrm{H}}\rangle$, and $|u_{b\mathrm{V}}\rangle$ as columns of a $4\times4$ matrix, we obtain the PBS unitary in Eq.~\ref{eq:Polarizing BS}.

\subsection{Coherence matrices for one DoF}\label{sec:ReducedRestricted}

\subsubsection{Reduced coherence matrices: The partial trace}\label{sec:ReducedCoherenceMatrices}

\begin{figure}[t!]
\centering
\includegraphics[width=12cm]{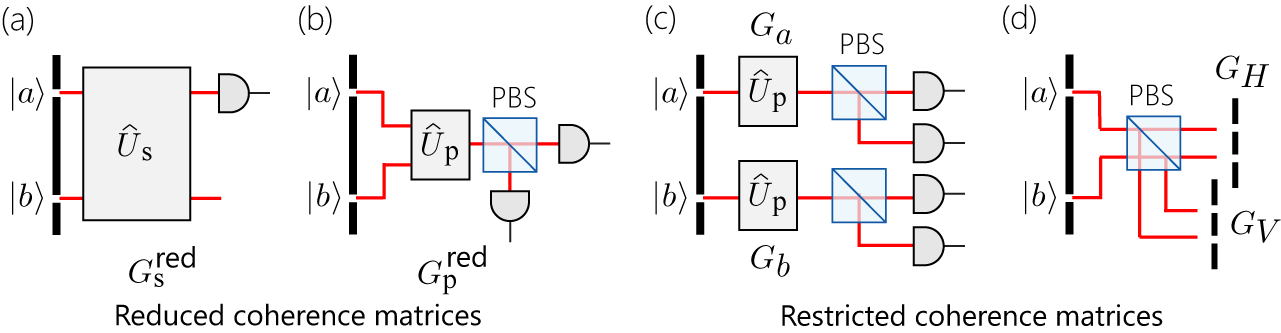}
\caption{(a,b) Reduced and (c,d) restricted coherence matrices. (a) Reduced \textit{spatial} coherence matrix $\mathbf{G}_{\mathrm{s}}^{\mathrm{red.}}$, obtained when all transformations and detectors are insensitive to polarization, and (b) reduced \textit{polarization} coherence matrix $\mathbf{G}_{\mathrm{p}}^{\mathrm{red.}}$, obtained when all transformations and (bucket) detectors are insensitive to the spatial DoF. (c) Restricted \textit{polarization} coherence matrices $\mathbf{G}_{a}$ and $\mathbf{G}_{b}$ associated separately with the spatial modes $|a\rangle$ and $|b\rangle$, respectively. (d) Reduced \textit{spatial} coherence matrices $\mathbf{G}_{\mathrm{H}}$ and $\mathbf{G}_{\mathrm{V}}$ associated separately with the polarization modes $|\mathrm{H}\rangle$ and $|\mathrm{V}\rangle$, respectively.}
\label{fig:ReducedRestricted}
\end{figure}

Several $2\times2$ coherence matrices pertaining to a single DoF can be extracted from the $4\times4$ coherence matrix $\mathbf{G}$ pertaining to two DoFs. One can obtain a \textit{reduced spatial coherence matrix} $\mathbf{G}_{\mathrm{s}}^{\mathrm{red.}}$ that describes the spatial coherence of the two-point field when all unitaries and detectors are independent of polarization [Fig.~\ref{fig:ReducedRestricted}(a)]. We obtain $\mathbf{G}_{\mathrm{s}}^{\mathrm{red.}}$ via a procedure known in quantum mechanics as the `partial trace' \cite{Peres93Book}:
\begin{equation}\label{eq:ReducedSpatialG}
\mathbf{G}_{\mathrm{s}}^{\mathrm{red.}}=\mathrm{Tr}_{\mathrm{p}}\{\mathbf{G}\}=\left(\begin{array}{cc}G_{\mathrm{HH}}^{aa}+G_{\mathrm{VV}}^{aa}&
G_{\mathrm{HH}}^{ab}+G_{\mathrm{VV}}^{ab}\\
G_{\mathrm{HH}}^{ba}+G_{\mathrm{VV}}^{ba}&
G_{\mathrm{HH}}^{bb}+G_{\mathrm{VV}}^{bb}\end{array}\right),
\end{equation}
We can evaluate the degree of spatial coherence $D_{\mathrm{s}}=\sqrt{1-4\mathrm{det}\{\mathbf{G}_{\mathrm{s}}^{\mathrm{red.}}\}}$ and spatial entropy $S_{\mathrm{s}}=-\mathbf{G}_{\mathrm{s}}^{\mathrm{red.}}\log_{2}\mathbf{G}_{\mathrm{s}}^{\mathrm{red.}}=-\lambda_{a}\log_{2}\lambda_{a}-\lambda_{b}\log_{2}\lambda_{b}$ from $\mathbf{G}_{\mathrm{s}}^{\mathrm{red.}}$, where $\lambda_{a}$ and $\lambda_{b}$ are its eigenvalues.

Some of the relevant properties of the reduced spatial coherence matrix are as follows:
\begin{enumerate}
\item $\mathbf{G}_{\mathrm{s}}^{\mathrm{red.}}$ is a Hermitian $2\times2$ spatial coherence matrix.
\item The partial trace operation guarantees that $\mathrm{Tr}\{\mathbf{G}_{\mathrm{s}}^{\mathrm{red.}}\}=1$ if $\mathrm{Tr}\{\mathbf{G}\}=1$; i.e., the partial trace operation is trace-preserving.
\item Only half the elements in $\mathbf{G}$ appear in $\mathbf{G}_{\mathrm{s}}^{\mathrm{red.}}$, so that the process of partial trace is accompanied by loss of information regarding the state of the field.
\item Upon traversing a purely spatial unitary $\Hat{U}=\hat{U}_{\mathrm{s}}\otimes\hat{\mathbb{I}}_{2}$, the transformed coherence matrices are $\mathbf{G}'=(\hat{U}_{\mathrm{s}}\otimes\hat{\mathbb{I}}_{2})\mathbf{G}(\hat{U}_{\mathrm{s}}^{\dagger}\otimes\hat{\mathbb{I}}_{2})$ and $\mathbf{G}_{\mathrm{s}}^{\mathrm{red.}'}=\hat{U}_{\mathrm{s}}\mathbf{G}_{\mathrm{s}}^{\mathrm{red.}}\hat{U}_{\mathrm{s}}^{\dagger}$. That is, $\mathbf{G}_{\mathrm{s}}^{\mathrm{red.}}$ undergoes a unitary transformation via $\hat{U}_{\mathrm{s}}$, which conserves the entropy $S_{\mathrm{s}}'=S_{\mathrm{s}}$.
\item Upon traversing a purely polarization unitary $\hat{U}=\hat{\mathbb{I}}_{2}\otimes\hat{U}_{\mathrm{p}}$, the transformed coherence matrices are $\mathbf{G}'=(\hat{\mathbb{I}}_{2}\otimes\hat{U}_{\mathrm{p}})\mathbf{G}(\hat{\mathbb{I}}_{2}\otimes\hat{U}_{\mathrm{p}}^{\dagger})$ and $\mathbf{G}_{\mathrm{s}}^{\mathrm{red.}'}=\mathbf{G}_{\mathrm{s}}^{\mathrm{red.}}$. That is, $\mathbf{G}_{\mathrm{s}}^{\mathrm{red.}}$ is invariant with respect to purely polarization unitaries, and once again the entropy is conserved $S_{\mathrm{s}}'=S_{\mathrm{s}}$.
\item Upon traversing a \textit{non-separable} unitary $\hat{U}$, the coherence matrix for the field is transformed as $\mathbf{G}'=\hat{U}\mathbf{G}\hat{U}^{\dagger}$, but there is no straightforward transformation of $\mathbf{G}_{\mathrm{s}}^{\mathrm{red.}}$. Although the trace of $\mathbf{G}_{\mathrm{s}}^{\mathrm{red.}}$ is preserved, it does \textit{not} necessarily undergo a unitary transformation itself. Indeed, the spatial entropy $S_{\mathrm{s}}'$ of the transformed coherence matrix may increase or decrease with respect to its initial value $S_{\mathrm{s}}$.
\end{enumerate}

One can similarly obtain a \textit{reduced polarization coherence matrix} $\mathbf{G}_{\mathrm{p}}^{\mathrm{red.}}$ by performing a partial trace over the spatial DoF:
\begin{equation}\label{eq:ReducedPolG}
\mathbf{G}_{\mathrm{p}}^{\mathrm{red.}}=\mathrm{Tr}_{\mathrm{s}}\{\mathbf{G}\}=\left(\begin{array}{cc}G_{\mathrm{HH}}^{aa}+G_{\mathrm{HH}}^{bb}&
G_{\mathrm{HV}}^{aa}+G_{\mathrm{HV}}^{bb}\\
G_{\mathrm{VH}}^{aa}+G_{\mathrm{VH}}^{bb}&
G_{\mathrm{VV}}^{aa}+G_{\mathrm{VV}}^{bb}\end{array}\right),
\end{equation}
This reduced polarization coherence matrix describes the polarization of the field when the unitaries and detectors have no spatial resolution and thus cannot discriminate between the fields at $|a\rangle$ and $|b\rangle$ [Fig.~\ref{fig:ReducedRestricted}(b)]. The degree of polarization coherence is $D_{\mathrm{p}}=\sqrt{1-4\mathrm{det}\{\mathbf{G}_{\mathrm{p}}^{\mathrm{red.}}\}}$, and the polarization entropy $S_{\mathrm{p}}=-\mathbf{G}_{\mathrm{p}}^{\mathrm{red.}}\log_{2}\mathbf{G}_{\mathrm{p}}^{\mathrm{red.}}=-\lambda_{\mathrm{H}}\log_{2}\lambda_{\mathrm{H}}-\lambda_{\mathrm{V}}\log_{2}\lambda_{\mathrm{V}}$ can be obtained from $\mathbf{G}_{\mathrm{p}}^{\mathrm{red.}}$, where $\lambda_{\mathrm{H}}$ and $\lambda_{\mathrm{V}}$ are its eigenvalues. The properties of $\mathbf{G}_{\mathrm{p}}^{\mathrm{red.}}$ are similar to those listed above for $\mathbf{G}_{\mathrm{s}}^{\mathrm{red.}}$ after switching the spatial and polarization DoFs.

Because only a subset of the elements of $\mathbf{G}$ appear in $\mathbf{G}_{\mathrm{s}}^{\mathrm{red.}}$ and $\mathbf{G}_{\mathrm{p}}^{\mathrm{red.}}$, they are \textit{not} sufficient to reconstruct $\mathbf{G}$, except when $\mathbf{G}$ is separable, $\mathbf{G}=\mathbf{G}_{\mathrm{s}}\otimes\mathbf{G}_{\mathrm{p}}$, in which case $\mathbf{G}_{\mathrm{s}}=\mathbf{G}_{\mathrm{s}}^{\mathrm{red.}}$ and $\mathbf{G}_{\mathrm{p}}=\mathbf{G}_{\mathrm{p}}^{\mathrm{red.}}$. In this special case, reconstructing $\mathbf{G}_{\mathrm{s}}^{\mathrm{red.}}$ and $\mathbf{G}_{\mathrm{p}}^{\mathrm{red.}}$ separately as outlined above in Section~\ref{sec:polarizationDoF} and Section~\ref{sec:SpatialDoF}, respectively, is sufficient to reconstruct $\mathbf{G}$. Indeed, defining $\mathbf{G}_{\mathrm{s}}^{\mathrm{red.}}=\left(\begin{array}{cc}G^{aa}&G^{ab}\\G^{ba}&G^{bb}\end{array}\right)$ and $\mathbf{G}_{\mathrm{p}}^{\mathrm{red.}}=\left(\begin{array}{cc}G_{\mathrm{HH}}&G_{\mathrm{HV}}\\G_{\mathrm{VH}}&G_{\mathrm{VV}}\end{array}\right)$, for a separable field we have $G_{\mathrm{HH}}^{aa}=G_{\mathrm{HH}}G^{aa}$, $G_{\mathrm{VV}}^{aa}=G_{\mathrm{VV}}G^{aa}$, etc.

There are two distinct ways to diagonalize the reduced coherence matrix $\mathbf{G}_{\mathrm{s}}^{\mathrm{red.}}$.
\begin{enumerate}
\item One may simply implement a purely spatial unitary $\hat{U}=\hat{U}_{\mathrm{s}}\otimes\hat{\mathbb{I}}_{2}$, were $\hat{U}_{\mathrm{s}}$ diagonalizes $\mathbf{G}_{\mathrm{s}}^{\mathrm{red.}}\rightarrow(\mathbf{G}_{\mathrm{s}}^{\mathrm{red.}})^{\mathrm{D}}=\hat{U}_{\mathrm{s}}\mathbf{G}_{\mathrm{s}}^{\mathrm{red.}}\hat{U}_{\mathrm{s}}^{\dagger}$. This operation does not necessarily diagonalize $\mathbf{G}$ simultaneously with $\mathbf{G}_{\mathrm{s}}^{\mathrm{red.}}$. In this case, $\mathbf{G}_{\mathrm{s}}^{\mathrm{red.}}$ is diagonalized without change in its entropy $S_{\mathrm{s}}$, which is invariant under $\hat{U}_{\mathrm{s}}$.
\item Alternatively, one may implement the unitary $\hat{U}$ (in general a non-separable unitary) that diagonalizes $\mathbf{G}$, $\mathbf{G}^{\mathrm{D}}=\hat{U}\mathbf{G}\hat{U}^{\dagger}=\mathrm{diag}\{\lambda_{1},\lambda_{2},\lambda_{3},\lambda_{4}\}$, which guarantees that $\mathbf{G}_{\mathrm{s}}^{\mathrm{red.}}$ is diagonalized simultaneously with $\mathbf{G}$, $\mathbf{G}_{\mathrm{s}}^{\mathrm{red.}}\rightarrow(\mathbf{G}_{\mathrm{s}}^{\mathrm{red.}})^{\mathrm{D}}=\mathrm{diag}\{\lambda_{1}+\lambda_{2},\lambda_{3}+\lambda_{4}\}$. 
\end{enumerate}

The two diagonalized reduced coherence matrices $\mathbf{G}_{\mathrm{s}}^{\mathrm{red.}}$ produced by these two approaches need not be equal to each other. Crucially, although $\hat{U}$ does not change the entropy $S$ of $\mathbf{G}$ upon diagonalization, it can nevertheless change the entropy of $\mathbf{G}_{\mathrm{s}}^{\mathrm{red.}}$ upon its diagonalization when $\hat{U}$ is not separable. Consequently, the entropy of $(\mathbf{G}_{\mathrm{s}}^{\mathrm{red.}})^{\mathrm{D}}$ can be lower than the entropy of the initial $\mathbf{G}_{\mathrm{s}}^{\mathrm{red.}}$, so that higher visibility of spatial interference fringes are recorded after diagonalizing $\mathbf{G}_{\mathrm{s}}^{\mathrm{red.}}$ in this approach jointly with $\mathbf{G}$. We discuss this phenomenon in more detail below (Section~\ref{sec:MaximizingDoubleSlits}).

\subsubsection{Restricted coherence matrices}

Other coherence matrices that characterize a single DoF can also be extracted from $\mathbf{G}$. For example, we have \textit{restricted polarization coherence matrices} associated with $|a\rangle$ and $|b\rangle$,
\begin{equation}
\mathbf{G}_{a}=\frac{1}{G_{\mathrm{HH}}^{aa}+G_{\mathrm{VV}}^{aa}}\left(\begin{array}{cc}G_{\mathrm{HH}}^{aa}&G_{\mathrm{HV}}^{aa}\\G_{\mathrm{VH}}^{aa}&G_{\mathrm{VV}}^{aa}\end{array}\right),\;\;\;\mathbf{G}_{b}=\frac{1}{G_{\mathrm{HH}}^{bb}+G_{\mathrm{VV}}^{bb}}\left(\begin{array}{cc}G_{\mathrm{HH}}^{bb}&G_{\mathrm{HV}}^{bb}\\G_{\mathrm{VH}}^{bb}&G_{\mathrm{VV}}^{bb}\end{array}\right),
\end{equation}
respectively. Unlike the \textit{reduced} coherence matrices resulting from the partial trace that have $\mathrm{Tr}\{\mathbf{G}_{\mathrm{s}}^{\mathrm{red.}}\}=\mathrm{Tr}\{\mathbf{G}_{\mathrm{p}}^{\mathrm{red.}}\}=1$, the coherence matrices $\mathbf{G}_{a}$ and $\mathbf{G}_{b}$ need to be first re-normalized. Once normalized, these two reduced coherence matrices describe the polarization coherence at $|a\rangle$ and $|b\rangle$ separately, while ignoring their mutual correlations. Alternatively, the restricted polarization coherence matrix $\mathbf{G}_{a}$ results from implementing a spatial filter $\hat{F}_{\mathrm{s}}^{\mathrm{D}}=\left(\begin{array}{cc}1&0\\0&0\end{array}\right)$ that blocks the field at $|b\rangle$. Similarly, $\mathbf{G}_{b}$ is the restricted polarization coherence matrix at $|b\rangle$ when the field at $|a\rangle$ is blocked via a spatial filter $\hat{F}_{\mathrm{s}}^{\mathrm{D}}=\left(\begin{array}{cc}0&0\\0&1\end{array}\right)$; see Fig.~\ref{fig:ReducedRestricted}(c).

Alternatively, one can obtain \textit{restricted spatial coherence matrices} for the $|\mathrm{H}\rangle$ and $|\mathrm{V}\rangle$ polarization modes,
\begin{equation}
\mathbf{G}_{\mathrm{H}}=\frac{1}{G_{\mathrm{HH}}^{aa}+G_{\mathrm{HH}}^{bb}}\left(\begin{array}{cc}G_{\mathrm{HH}}^{aa}&G_{\mathrm{HH}}^{ab}\\G_{\mathrm{HH}}^{ba}&G_{\mathrm{HH}}^{bb}\end{array}\right),\;\;\;\mathbf{G}_{\mathrm{V}}=\frac{1}{G_{\mathrm{VV}}^{aa}+G_{\mathrm{VV}}^{bb}}\left(\begin{array}{cc}G_{\mathrm{VV}}^{aa}&G_{\mathrm{VV}}^{ab}\\G_{\mathrm{VV}}^{ba}&G_{\mathrm{VV}}^{bb}\end{array}\right),
\end{equation}
respectively. These reduced coherence matrices describe the \textit{spatial coherence} if only the $|\mathrm{H}\rangle$ or $|\mathrm{V}\rangle$ components are detected separately [Fig.~\ref{fig:ReducedRestricted}(d)]. That is, $\mathbf{G}_{\mathrm{H}}$ and $\mathbf{G}_{\mathrm{V}}$ are the spatial coherence matrices for the $|\mathrm{H}\rangle$ and $|\mathrm{V}\rangle$ modes, respectively, at the output ports of a PBS. Note that some elements of $\mathbf{G}$ are missing from the restricted coherence matrices $\mathbf{G}_{a}$, $\mathbf{G}_{b}$, $\mathbf{G}_{\mathrm{H}}$, and $\mathbf{G}_{\mathrm{V}}$, so that knowledge of all~4 of these $2\times2$ coherence matrices does not suffice to reconstruct $\mathbf{G}$.

\subsection{Relationship between the entropy of $\mathbf{G}$ and the entropy for one DoF}

Consider a general $4\times4$ coherence matrix $\mathbf{G}$ with entropy $S$. This entropy is invariant under both separable and non-separable unitaries: if $\mathbf{G}'=\hat{U}\mathbf{G}\hat{U}^{\dagger}$, then $S(\mathbf{G}')=S(\mathbf{G})$. However, it does \textit{not} follow that the entropies of the reduced and restricted coherence matrices ($\mathbf{G}_{\mathrm{s}}^{\mathrm{red.}}$, $\mathbf{G}_{\mathrm{p}}^{\mathrm{red.}}$, $\mathbf{G}_{a}$, $\mathbf{G}_{b}$, $\mathbf{G}_{\mathrm{H}}$, and $\mathbf{G}_{\mathrm{V}}$) are invariant under unitaries $\hat{U}$ on $\mathbf{G}$.

\subsubsection{Entropy of reduced coherence matrices}

When the field is separable with respect to the two DoFs, $\mathbf{G}=\mathbf{G}_{\mathrm{s}}\otimes\mathbf{G}_{\mathrm{p}}=\mathbf{G}_{\mathrm{s}}^{\mathrm{red.}}\otimes\mathbf{G}_{\mathrm{p}}^{\mathrm{red.}}$, then the entropy of $\mathbf{G}$ is the sum of the spatial and polarization entropies:
\begin{equation}
S(\mathbf{G})=S(\mathbf{G}_{\mathrm{s}}^{\mathrm{red.}}\otimes\mathbf{G}_{\mathrm{p}}^{\mathrm{red.}})=S(\mathbf{G}_{\mathrm{s}}^{\mathrm{red.}})+S(\mathbf{G}_{\mathrm{p}}^{\mathrm{red.}})=S_{\mathrm{s}}+S_{\mathrm{p}}.
\end{equation}
However, when the field is \textit{not} separable, then $S_{\mathrm{s}}+S_{\mathrm{p}}>S$. One can intuitively understand this result by noting that the procedure of partial trace that yields the reduced coherence matrices can be viewed as `ignoring' information about the field. This loss of information manifests itself in an apparent increase in entropy, which is clear from the absence of some elements of $\mathbf{G}$ from both $\mathbf{G}_{\mathrm{s}}^{\mathrm{red.}}$ and $\mathbf{G}_{\mathrm{p}}^{\mathrm{red.}}$; consequently, they are \textit{not} sufficient to reconstruct $\mathbf{G}$. When $\mathbf{G}$ is separable (the two DoFs are independent of each other), there is no loss of information; consequently, $S=S_{\mathrm{s}}+S_{\mathrm{p}}$. 

As an example, consider the coherent field $|E\rangle=\cos\frac{\theta}{2}|a,\mathrm{H}\rangle+\sin\tfrac{\theta}{2}|b,\mathrm{V}\rangle$ ($S=0$), whereupon:
\begin{equation}
\mathbf{G}=|E\rangle\langle E|=\left(\begin{array}{cccc}
\cos^{2}\tfrac{\theta}{2}&0&0&\sin\tfrac{\theta}{2}\cos\tfrac{\theta}{2}\\
0&0&0&0\\
0&0&0&0\\
\sin\tfrac{\theta}{2}\cos\tfrac{\theta}{2}&0&0&\cos^{2}\tfrac{\theta}{2}
\end{array}\right), \, \mathbf{G}_{\mathrm{s}}^{\mathrm{red.}}=\mathbf{G}_{\mathrm{p}}^{\mathrm{red.}}=\left(\begin{array}{cc}\cos^{2}\tfrac{\theta}{2}&0\\0&\sin^{2}\tfrac{\theta}{2}\end{array}\right);
\end{equation}
with degrees of coherence $D_{\mathrm{s}}=D_{\mathrm{p}}=|\cos\theta|$ and entropies $S_{\mathrm{s}}=S_{\mathrm{p}}=-2\{\cos^{2}\tfrac{\theta}{2}\log_{2}|\cos\tfrac{\theta}{2}|+\sin^{2}\tfrac{\theta}{2}\log_{2}|\sin\tfrac{\theta}{2}|\}$. Tuning $\theta$ varies the outcome dramatically. When $\theta=0$, $|E\rangle=|a,\mathrm{H}\rangle$, the field is separable, both reduced coherence matrices correspond to a fully coherent DoF, $\mathbf{G}_{\mathrm{s}}^{\mathrm{red.}}=\mathbf{G}_{\mathrm{p}}^{\mathrm{red.}}=\mathrm{diag}\{1,0\}$, $S_{\mathrm{s}}=S_{\mathrm{p}}=0$ and $D_{\mathrm{s}}=D_{\mathrm{p}}=1$, so that $S_{\mathrm{s}}+S_{\mathrm{p}}=S=0$ as expected for a separable field. At the other extreme when $\theta=\tfrac{\pi}{2}$, $\mathbf{G}$ is no longer separable, $\mathbf{G}_{\mathrm{s}}^{\mathrm{red.}}=\mathbf{G}_{\mathrm{p}}^{\mathrm{red.}}=\tfrac{1}{2}\hat{\mathbb{I}}_{2}$ corresponding to fully incoherent DoFs, $D_{\mathrm{s}}=D_{\mathrm{p}}=0$ and $S_{\mathrm{s}}=S_{\mathrm{p}}=1$, with $S_{\mathrm{s}}+S_{\mathrm{p}}=2>S=0$. Although the field is coherent ($S=0$), the correlation between the two DoFs when $\theta\neq0$ leads to each DoF appearing incoherent when ignoring the other DoF. In general, when $\theta\neq0$, $\mathbf{G}_{\mathrm{s}}^{\mathrm{red.}}$ and $\mathbf{G}_{\mathrm{p}}^{\mathrm{red.}}$ correspond in general to partially coherent DoFs, although the field vector $|E\rangle$ is associated with a coherent field. This phenomenon has been dubbed `classical entanglement' \cite{Spreeuw98FP,Kagalwala13NP}. 

Of course the field vector $|E\rangle=\cos\tfrac{\theta}{2}|a,\mathrm{H}\rangle+\sin\tfrac{\theta}{2}|b,\mathrm{V}\rangle$ can be viewed as the result of the transformation of the separable field vector $|E\rangle=|a,\mathrm{H}\rangle$ by the \textit{non-separable} unitary:
\begin{equation}
\hat{U}=\left(\begin{array}{cccc}
\cos\tfrac{\theta}{2}&0&0&-\sin\tfrac{\theta}{2}\\
0&1&0&0\\
0&0&1&0\\
\sin\tfrac{\theta}{2}&0&0&\cos\tfrac{\theta}{2}
\end{array}\right),
\end{equation}
which implements a rotation on the composite modes $|a,\mathrm{H}\rangle$ and $|b,\mathrm{V}\rangle$. Varying $\theta$ in this unitary leaves the entropy of the field vector invariant. Moreover, this unitary preserves the traces of $\mathbf{G}_{\mathrm{s}}^{\mathrm{red.}}$ and $\mathbf{G}_{\mathrm{p}}^{\mathrm{red.}}$. However, $\hat{U}$ does \textit{not} preserve the degree of coherence nor the entropy for $\mathbf{G}_{\mathrm{s}}^{\mathrm{red.}}$ or $\mathbf{G}_{\mathrm{p}}^{\mathrm{red.}}$. Starting with $|E\rangle=|a,\mathrm{H}\rangle$, this non-separable unitary implements a trace-preserving non-unitary transformation on $\mathbf{G}_{\mathrm{s}}^{\mathrm{red.}}$:
\begin{equation}
\mathbf{G}_{\mathrm{s}}^{\mathrm{red.}}=\left(\begin{array}{cc}1&0\\0&0\end{array}\right)\xrightarrow{\hat{U}}\mathbf{G}_{\mathrm{s}}^{\mathrm{red.}}=\left(\begin{array}{cc}\cos^{2}\tfrac{\theta}{2}&0\\0&\sin^{2}\tfrac{\theta}{2}\end{array}\right),
\end{equation}
and similarly for $\mathbf{G}_{\mathrm{p}}^{\mathrm{red.}}$. In this scenario, a coherent reduced spatial coherence matrix $\mathbf{G}_{\mathrm{s}}^{\mathrm{red.}}$ with $S_{\mathrm{s}}=0$ undergoes a decohering process ($S_{\mathrm{s}}$ increases to $S_{\mathrm{s}}=1$ when $\theta=\tfrac{\theta}{2}$) by implementing a trace-preserving non-separable unitary $\hat{U}$ on the field. If we start with a partially coherent field, this non-separable unitary can increase or decrease the entropies of $\mathbf{G}_{\mathrm{s}}^{\mathrm{red.}}$ and $\mathbf{G}_{\mathrm{p}}^{\mathrm{red.}}$.

Starting with a separable coherence matrix $\mathbf{G}=\mathbf{G}_{\mathrm{s}}^{\mathrm{red.}}\otimes\mathbf{G}_{\mathrm{p}}^{\mathrm{red.}}$ and $S=S_{\mathrm{s}}+S_{\mathrm{p}}$, one can always implement a non-separable unitary $\hat{U}$ that couples the two DoFs, thus rendering the coherence matrix in turn non-separable $\mathbf{G}\neq\mathbf{G}_{\mathrm{s}}^{\mathrm{red.}}\otimes\mathbf{G}_{\mathrm{P}}^{\mathrm{red.}}$ and consequently $S_{\mathrm{s}}+S_{\mathrm{p}}>S$. The reverse question can now be posed: starting with a non-separable coherence matrix ($S_{\mathrm{s}}+S_{\mathrm{p}}>S$), can we always find a unitary $\hat{U}$ that renders the coherence matrix separable $\mathbf{G}=\mathbf{G}_{\mathrm{s}}^{\mathrm{red.}}\otimes\mathbf{G}_{\mathrm{p}}^{\mathrm{red.}}$ ($S=S_{\mathrm{s}}+S_{\mathrm{p}}$)? We show below (Section~\ref{sec:CoherenceRank}) that there are entire classes of coherence matrices that cannot be rendered separable via unitaries. Such fields are intrinsically non-separable. We will explore the consequences of this feature for the distribution of the entropy between the DoFs.

We note in passing that full coherence places a strict constraint on the reduced coherence matrices. If $\mathbf{G}$ corresponds to a coherent field ($S=0$), whether separable ($S_{\mathrm{s}}+S_{\mathrm{p}}=S$) or non-separable ($S_{\mathrm{s}}+S_{\mathrm{p}}>S$), then we must have $S_{\mathrm{s}}=S_{\mathrm{p}}$ (and $D_{\mathrm{s}}=D_{\mathrm{p}}$). Although the reduced coherence matrices $\mathbf{G}_{\mathrm{s}}^{\mathrm{red.}}$ and $\mathbf{G}_{\mathrm{p}}^{\mathrm{red.}}$ can of course be different, the degree of coherence of both must be equal if $\mathbf{G}$ is coherent. This conclusion does \textit{not} hold for a partially coherent field, where $S_{\mathrm{s}}$ and $S_{\mathrm{p}}$ can vary widely from each other. Indeed, if $\mathbf{G}=\mathrm{diag}\{\tfrac{1}{2},\tfrac{1}{2},0,0\}$, then $\mathbf{G}_{\mathrm{s}}^{\mathrm{red.}}=\mathrm{diag}\{1,0\}$ with $D_{\mathrm{s}}=1$ and $S_{\mathrm{s}}=0$, and $\mathbf{G}_{\mathrm{s}}^{\mathrm{red.}}=\tfrac{1}{2}\hat{\mathbb{I}}_{2}$ with $D_{\mathrm{p}}=0$ and $S_{\mathrm{p}}=1$. Such a result cannot be produced for a coherent field.

\subsubsection{Entropy of restricted coherence matrices}

The behavior of the entropies for the restricted coherence matrices $\mathbf{G}_{a}$ and $\mathbf{G}_{b}$ (or $\mathbf{G}_{\mathrm{H}}$ and $\mathbf{G}_{\mathrm{V}}$) has not been previously studied, so we make here only a few comments in anticipation of future research on this topic. For the entropy of a $2\times2$ coherence matrix to be meaningful, the matrix must have unity trace, so that its entropy varies in the range $[0,1]$. Because the restricted coherence matrices are not automatically normalized to unity trace, we form the weighted entropy $P_{a}S_{a}+P_{b}S_{b}$, where $P_{a}=\mathrm{Tr}\{\mathbf{G}_{a}\}$, $P_{b}=\mathrm{Tr}\{\mathbf{G}_{b}\}$, with $P_{a}+P_{b}=\mathrm{Tr}\{\mathbf{G}\}=1$, and $S_{a}$ and $S_{b}$ are the entropies of the \textit{normalized} restricted coherence matrices.

With this definition, one immediately finds a stark contrast between the entropy of restricted and reduced coherence matrices. Whereas $S_{\mathrm{s}}+S_{\mathrm{p}}\geq S$ in the case of reduced coherence matrices, with equality holding only for a separable $\mathbf{G}$, this constraint does \textit{not} apply to the entropies of the restricted coherence matrices. Instead, the weighted entropy satisfies the constraint $P_{a}S_{a}+P_{b}S_{b}\leq S$. 

If the field is coherent ($S=0$), then all restricted coherence matrices correspond to coherent fields, whether $\mathbf{G}$ is separable or non-separable. Therefore, for a coherent field $S_{a}=S_{b}=S_{\mathrm{H}}=S_{\mathrm{V}}=0$ and thus $P_{a}S_{a}+P_{b}S_{b}=P_{\mathrm{H}}S_{\mathrm{H}}+P_{\mathrm{V}}S_{\mathrm{V}}=S=0$. This is in contradistinction to the reduced coherence matrices, where $\mathbf{G}_{\mathrm{s}}^{\mathrm{red.}}$ and $\mathbf{G}_{\mathrm{p}}^{\mathrm{red.}}$ correspond to partially coherent fields even when $\mathbf{G}=|E\rangle\langle E|$ is coherent -- as long as $\mathbf{G}$ is non-separable.

For a partially coherent field, consider the following three coherence matrices as examples:
\begin{equation}
\mathbf{G}_{1}=\left(\begin{array}{cccc}\tfrac{1}{2}&0&0&0\\0&\tfrac{1}{2}&0&0\\0&0&0&0\\0&0&0&0\end{array}\right),\;
\mathbf{G}_{2}=\left(\begin{array}{cccc}\tfrac{1}{2}&0&0&0\\0&0&0&0\\0&0&\tfrac{1}{2}&0\\0&0&0&0\end{array}\right),\;
\mathbf{G}_{3}=\frac{1}{4}\left(\begin{array}{cccc}2&0&0&0\\0&1&1&0\\0&1&1&0\\0&0&0&0\end{array}\right).
\end{equation}
The coherence matrix $\mathbf{G}_{1}$ corresponds to an unpolarized field at $|a\rangle$ and zero field amplitude at $|b\rangle$, so that $S_{a}=1$, $S_{b}=0$, and $S=1$. The weights of the restricted coherence matrices are $P_{a}=1$ and $P_{b}=0$, so that $P_{a}S_{a}+P_{b}S_{b}=S=1$. The coherence matrix $\mathbf{G}_{2}$ is obtained by splitting off the $|\mathrm{V}\rangle$ mode from $|a\rangle$, transferring it to $|b\rangle$, and converting it to $|\mathrm{H}\rangle$. Now we have $\mathbf{G}_{a}=\mathbf{G}_{b}=\mathrm{diag}\{1,0\}$, $S_{a}=S_{b}=0$, $P_{a}=P_{b}=\tfrac{1}{2}$, and $P_{a}S_{a}+P_{b}S_{b}=0<S=1$. Finally, $\mathbf{G}_{3}$ is obtained from $\mathbf{G}_{1}$ in a manor similar to that followed to obtain $\mathbf{G}_{2}$, except that only half the power from $|\mathrm{V}\rangle$ in $|a\rangle$ is transferred to $|\mathrm{H}\rangle$ at $|b\rangle$. We now have $P_{a}=\tfrac{3}{4}$, $P_{b}=\tfrac{1}{4}$, $S_{a}\approx0.918$, and $S_{b}=0$, so that $P_{a}S_{a}+P_{b}S_{b}\approx0.69<1$~bit.

\subsection{Reconstruction of the coherence matrix: Stokes tomography}\label{sec:Stokes2DoFs}

We described in Section~\ref{sec:StokesSingleDoF} a methodology to reconstruct the $2\times2$ coherence matrix for a binary DoF that relies on measuring the modal Stokes parameters, whether for polarization modes (Section~\ref{sec:polarizationDoF}), or spatial modes (Section~\ref{sec:SpatialDoF}). To accomplish this, three different configurations are required, each involves implementing a unitary that is then followed by a measurement of the modal weights (in addition to one measurement to ensure normalization). These 4~measurements provide the modal Stokes parameters that uniquely identify the $2\times2$ coherence matrix.

In reconstructing a $4\times4$ coherence matrix for two binary DoFs, two questions arise. First, how many measurements are required? A $4\times4$ coherence matrix is uniquely identified by 16~real parameters (including the normalization). Therefore, one needs at least 16~measurements to reconstruct $\mathbf{G}$. Second, what are the measurements needed? It has been shown that reconstructing $\mathbf{G}$ for two binary DoFs is possible through the concatenation or cascade of the measurements necessary for characterizing each binary DoF separately \cite{Wootters90article}. This corresponds to 4~measurement configurations for the spatial DoF and 4~measurement configurations for the polarization DoF. When concatenated, this generates $4\times4=16$ different measurements, from which 16~modal Stokes parameters are obtained that comprise both binary DoFs, are obtained as an intermediary step. The coherence matrix is then reconstructed from these modal Stokes parameters with the help of Kronecker-Pauli matrices. 

\subsubsection{Definition of the modal Stokes parameters}

\begin{figure}[t!]
\centering
\includegraphics[width=10 cm]{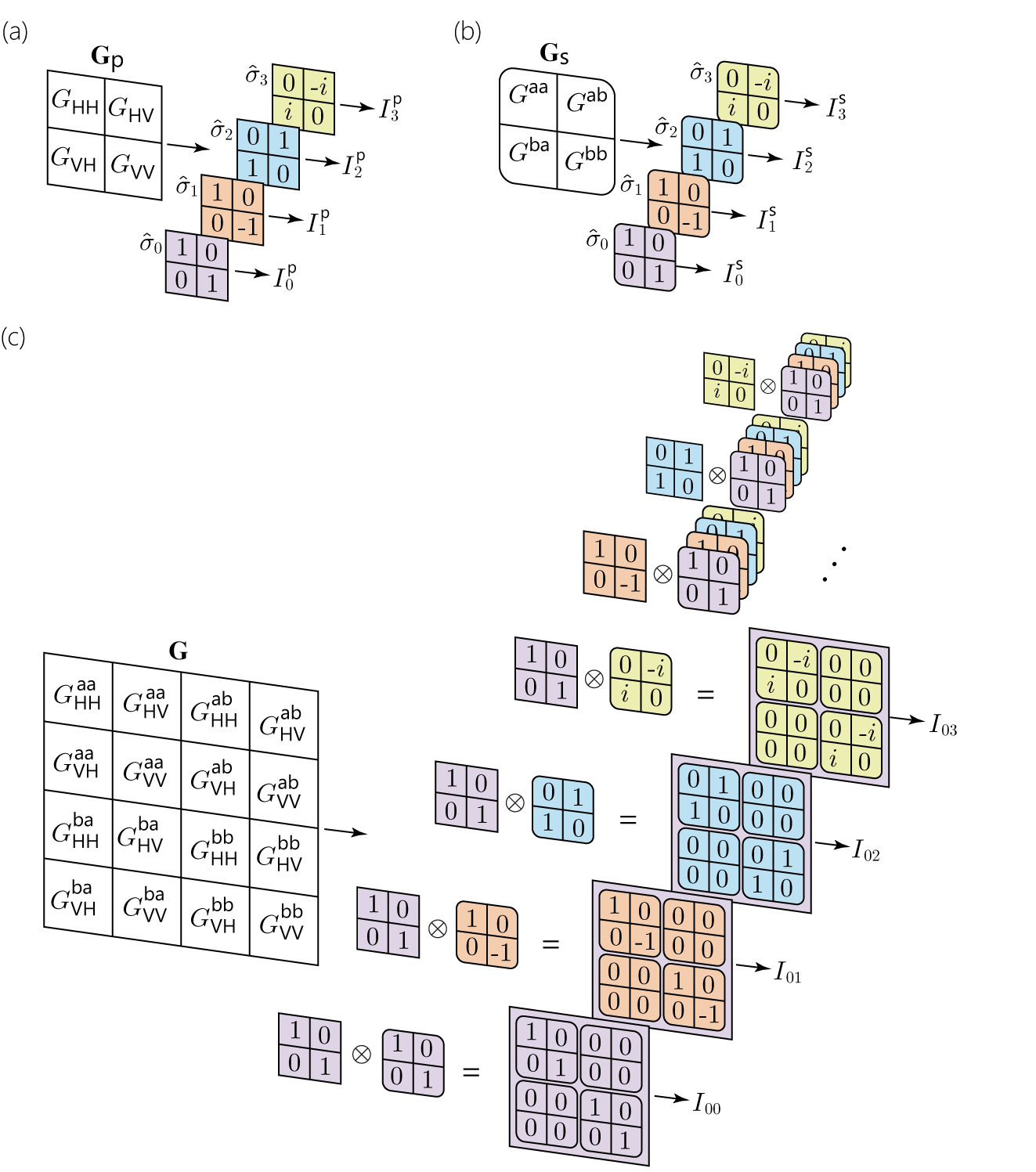}
\caption{Definition of Kronecker-Pauli matrices. (a) The $2\times2$ Pauli matrices $\hat{\sigma}_{j}^{(\mathrm{s})}$ for the spatial DoF. Using each such Pauli matrix, we define the projections $\mathrm{Tr}\{\hat{\sigma}_{j}^{\mathrm{(s)}}\mathbf{G}_{\mathrm{s}}\}$ on the $2\times2$ spatial coherence matrix $\mathbf{G}_{\mathrm{s}}$. These projections are the spatial Stokes parameters $s_{j}^{(\mathrm{s})}$. (b) Same as (a) for the polarization DoF. The four $2\times2$ Pauli matrices $\hat{\sigma}_{k}^{(\mathrm{p})}$ for the polarization DoF are used to defined projections $\mathrm{Tr}\{\hat{\sigma}_{k}^{\mathrm{(p)}}\mathbf{G}_{\mathrm{p}}\}$ on the $2\times2$ polarization coherence matrix $\mathbf{G}_{\mathrm{p}}$. These projections are the polarization Stokes parameters $s_{k}^{(\mathrm{p})}$. (c) The $4\times4$ Kronecker-Pauli matrices $\hat{\sigma}_{jk}=\hat{\sigma}_{j}^{(\mathrm{s})}\otimes\hat{\sigma}_{k}^{(\mathrm{p})}$ are direct products of the $2\times2$ spatial and polarization Pauli matrices from (a) and (b), respectively. Projections $\mathrm{Tr}\{\hat{\sigma}_{jk}\mathbf{G}\}$ of the $4\times4$ coherence matrix $\mathbf{G}$ are defined, which correspond to the modal Stokes parameters $s_{jk}$.}
\label{fig:OCmT1}
\end{figure}

We first introduce a generalization of the modal Stokes parameters to two binary DoFs by expressing the coherence matrix as follows:
\begin{equation}
\mathbf{G}=\frac{1}{4}\sum_{j,k=0}^{3}s_{jk}\left(\hat{\sigma}_{j}^{(\mathrm{s)}}\otimes\hat{\sigma}_{k}^{\mathrm{(p)}}\right)=\frac{1}{4}\sum_{j,k=0}^{3}s_{jk}\hat{\sigma}_{jk},
\end{equation}
where $\{\hat{\sigma}_{j}^{\mathrm{(s)}}\}_{j=0}^{3}$ and $\{\hat{\sigma}_{k}^{\mathrm{(p)}}\}_{k=0}^{3}$ are the Pauli matrices for the spatial and polarization DoFs, respectively, and the Kronecker-Pauli matrices $\{\hat{\sigma}_{jk}\}_{j,k=0}^{3}$ are formed of separable direct products of the Pauli matrices spanning both binary DoFs, $\hat{\sigma}_{jk}=\hat{\sigma}_{j}^{\mathrm{(s)}}\otimes\hat{\sigma}_{k}^{\mathrm{(p)}}$ [Fig.~\ref{fig:OCmT1}]. The Kronecker-Pauli matrices have the following properties:
\begin{enumerate}
\item $\left(\hat{\sigma}_{jk}\right)^{\dagger}=\hat{\sigma}_{jk}$.
\item The eigenvalues of any Kronecker-Pauli matrix are $\{1,1,-1,-1\}$.
\item $\mathrm{Tr}\left(\hat{\sigma}_{jk}\right)=\mathrm{Tr}\left\{\hat{\sigma}_{j}^{\mathrm{(s)}}\right\}\cdot\mathrm{Tr}\left\{\hat{\sigma}_{k}^{\mathrm{(p)}}\right\}=0$, except if $j=k=0$.
\item $\hat{\sigma}_{jk}^{2}=\left(\hat{\sigma}_{j}^{(\mathrm{s})}\otimes\hat{\sigma}_{k}^{(\mathrm{p})}\right)\left(\hat{\sigma}_{j}^{(\mathrm{s})}\otimes\hat{\sigma}_{k}^{(\mathrm{p})}\right)=\left(\hat{\sigma}_{j}^{(\mathrm{s})}\right)^{2}\otimes\left(\hat{\sigma}_{k}^{(\mathrm{p})}\right)^{2}=\hat{\mathbb{I}}_{2}\otimes\hat{\mathbb{I}}_{2}=\hat{\mathbb{I}}_{4}$.
\item $\mathrm{det}\left(\hat{\sigma}_{jk}\right)=\mathrm{det}\left(\hat{\sigma}_{j}^{(\mathrm{s})}\otimes\hat{\sigma}_{k}^{(\mathrm{p})}\right)=\mathrm{det}\left(\hat{\sigma}_{j}^{(\mathrm{s})}\right)\mathrm{det}\left(\hat{\sigma}_{k}^{(\mathrm{p})}\right)=(-1)^{2}=1$.
\end{enumerate}

The 16~modal Stokes parameters $\{s_{jk}\}$ for the two DoFs are the expansion coefficients of $\mathbf{G}$ in terms of the Kronecker-Pauli matrices, which can be extracted through a mathematical projection:
\begin{equation}\label{eq:ModalStokesParameters}
s_{jk}=\mathrm{Tr}\left\{\left(\hat{\sigma}_{j}^{(\mathrm{s})}\otimes\hat{\sigma}_{k}^{(\mathrm{p})}\right)\mathbf{G}\right\}=\mathrm{Tr}\left\{\hat{\sigma}_{jk}\mathbf{G}\right\}.
\end{equation}
The Kronecker-Pauli matrices are given explicitly as follows [Fig.~\ref{fig:OCmT1}]:
\begin{equation}
\hat{\sigma}_{00}=\left(\begin{array}{cccc}1&0&0&0\\0&1&0&0\\0&0&1&0\\0&0&0&1\end{array}\right),
\hat{\sigma}_{01}=\left(\begin{array}{cccc}1&0&0&0\\0&-1&0&0\\0&0&1&0\\0&0&0&-1\end{array}\right),
\hat{\sigma}_{02}=\left(\begin{array}{cccc}0&1&0&0\\1&0&0&0\\0&0&0&1\\0&0&1&0\end{array}\right),
\hat{\sigma}_{03}=\left(\begin{array}{cccc}0&-i&0&0\\i&0&0&0\\0&0&0&-i\\0&0&i&0\end{array}\right);
\end{equation}

\begin{equation}
\hat{\sigma}_{10}=\left(\begin{array}{cccc}1&0&0&0\\0&1&0&0\\0&0&-1&0\\0&0&0&-1\end{array}\right),
\hat{\sigma}_{11}=\left(\begin{array}{cccc}1&0&0&0\\0&-1&0&0\\0&0&-1&0\\0&0&0&1\end{array}\right),
\hat{\sigma}_{12}=\left(\begin{array}{cccc}0&1&0&0\\1&0&0&0\\0&0&0&-1\\0&0&-1&0\end{array}\right),
\hat{\sigma}_{13}=\left(\begin{array}{cccc}0&-i&0&0\\i&0&0&0\\0&0&0&i\\0&0&-i&0\end{array}\right);
\end{equation}

\begin{equation}
\hat{\sigma}_{20}=\left(\begin{array}{cccc}0&0&1&0\\0&0&0&1\\1&0&0&0\\0&1&0&0\end{array}\right),
\hat{\sigma}_{21}=\left(\begin{array}{cccc}0&0&1&0\\0&0&0&-1\\1&0&0&0\\0&-1&0&0\end{array}\right),
\hat{\sigma}_{22}=\left(\begin{array}{cccc}0&0&0&1\\0&0&1&0\\0&1&0&0\\1&0&0&0\end{array}\right),
\hat{\sigma}_{23}=\left(\begin{array}{cccc}0&0&0&-i\\0&0&i&0\\0&-i&0&0\\i&0&0&0\end{array}\right);
\end{equation}

\begin{equation}
\hat{\sigma}_{30}=\left(\begin{array}{cccc}0&0&-i&0\\0&0&0&-i\\i&0&0&0\\0&i&0&0\end{array}\right),
\hat{\sigma}_{31}=\left(\begin{array}{cccc}0&0&-i&0\\0&0&0&i\\i&0&0&0\\0&-i&0&0\end{array}\right),
\hat{\sigma}_{32}=\left(\begin{array}{cccc}0&0&0&-i\\0&0&-i&0\\0&i&0&0\\i&0&0&0\end{array}\right),
\hat{\sigma}_{33}=\left(\begin{array}{cccc}0&0&0&-1\\0&0&1&0\\0&1&0&0\\-1&0&0&0\end{array}\right);
\end{equation}

The coherence matrix $\mathbf{G}$ can thus be expressed in terms of the modal Stokes parameters by direct substitution in Eq.~\ref{eq:ModalStokesParameters}:
\begin{equation}
\mathbf{G}=\tfrac{1}{4}\left(\begin{array}{cccc}
s_{00}+s_{01}+s_{10}+s_{11}&
s_{02}+s_{12}-i(s_{03}+s_{13})&
s_{20}+s_{21}-i(s_{30}+s_{31})&
s_{22}-s_{33}-i(s_{23}+s_{32})\\
s_{02}+s_{12}+i(s_{03}+s_{13})&
s_{00}-s_{01}+s_{10}-s_{11}&
s_{22}+s_{33}+i(s_{23}-s_{32})&
s_{20}-s_{21}-i(s_{30}-s_{31})\\
s_{20}+s_{21}+i(s_{30}+s_{31})&
s_{22}+s_{33}-i(s_{23}-s_{32})&
s_{00}+s_{01}-s_{01}-s_{11}&
s_{02}-s_{12}-i(s_{03}-s_{13})\\
s_{22}-s_{33}+i(s_{23}+s_{32})&
s_{20}-s_{21}+i(s_{30}-s_{31})&
s_{02}-s_{12}+i(s_{03}-s_{13})&
s_{00}-s_{01}-s_{10}+s_{11}
\end{array}\right).
\end{equation}
Alternatively, the modal Stokes parameters can be obtained in terms of the elements of $\mathbf{G}$:
\begin{align}
s_{00}&=
G_{\mathrm{HH}}^{aa}+G_{\mathrm{VV}}^{aa}
+G_{\mathrm{HH}}^{bb}+G_{\mathrm{VV}}^{bb}, &s_{01}&=
G_{\mathrm{HH}}^{aa}-G_{\mathrm{VV}}^{aa}
+G_{\mathrm{HH}}^{bb}-G_{\mathrm{VV}}^{bb},\nonumber\\
s_{02}&=
G_{\mathrm{HV}}^{aa}+G_{\mathrm{VH}}^{aa}
+G_{\mathrm{HV}}^{bb}+G_{\mathrm{VH}}^{bb}, &s_{03}&=
i(G_{\mathrm{HV}}^{aa}-G_{\mathrm{VH}}^{aa}
+G_{\mathrm{HV}}^{bb}-G_{\mathrm{VH}}^{bb}),\nonumber\\
s_{10}&=
G_{\mathrm{HH}}^{aa}+G_{\mathrm{VV}}^{aa}
-G_{\mathrm{HH}}^{bb}-G_{\mathrm{VV}}^{bb}, &s_{11}&=
G_{\mathrm{HH}}^{aa}-G_{\mathrm{VV}}^{aa}
-G_{\mathrm{HH}}^{bb}+G_{\mathrm{VV}}^{bb},\nonumber\\
s_{12}&=
G_{\mathrm{HV}}^{aa}+G_{\mathrm{VH}}^{aa}
-G_{\mathrm{HV}}^{bb}-G_{\mathrm{VH}}^{bb}, &s_{13}&=
i(G_{\mathrm{HV}}^{aa}-G_{\mathrm{VH}}^{aa}
-G_{\mathrm{HV}}^{bb}+G_{\mathrm{VH}}^{bb}),\nonumber\\
s_{20}&=
G_{\mathrm{HH}}^{ab}+G_{\mathrm{VV}}^{ab}
+G_{\mathrm{HH}}^{ba}+G_{\mathrm{VV}}^{ba}, &s_{21}&=
G_{\mathrm{HH}}^{ab}-G_{\mathrm{VV}}^{ab}
+G_{\mathrm{HH}}^{ba}-G_{\mathrm{VV}}^{ba},\nonumber\\
s_{22}&=
G_{\mathrm{HV}}^{ab}+G_{\mathrm{VH}}^{ab}
+G_{\mathrm{HV}}^{ba}+G_{\mathrm{VH}}^{ba}, &s_{23}&=
i(G_{\mathrm{HV}}^{ab}-G_{\mathrm{VH}}^{ab}
+G_{\mathrm{HV}}^{ba}-G_{\mathrm{VH}}^{ba}),\nonumber\\
s_{30}&=
i(G_{\mathrm{HH}}^{ab}+G_{\mathrm{VV}}^{ab}
-G_{\mathrm{HH}}^{ba}-G_{\mathrm{VV}}^{ba}), &s_{31}&=
i(G_{\mathrm{HH}}^{ab}-G_{\mathrm{VV}}^{ab}
-G_{\mathrm{HH}}^{ba}+G_{\mathrm{VV}}^{ba}),\nonumber\\
s_{32}&=
i(G_{\mathrm{HV}}^{ab}+G_{\mathrm{VH}}^{ab}
-G_{\mathrm{HV}}^{ba}-G_{\mathrm{VH}}^{ba}), &s_{33}&=
-G_{\mathrm{HV}}^{ab}+G_{\mathrm{VH}}^{ab}
+G_{\mathrm{HV}}^{ba}-G_{\mathrm{VH}}^{ba}.
\end{align}

The (real) modal Stokes parameters for two binary DoFs have the following properties:
\begin{enumerate}
\item A unity-trace coherence matrix matrix enforces the normalization $s_{00}=1$.
\item The modal Stokes parameters are real.
\item $0\leq|s_{jk}|\leq1$.
\end{enumerate}

\begin{figure}[t!]
\centering
\includegraphics[width=3.5in]{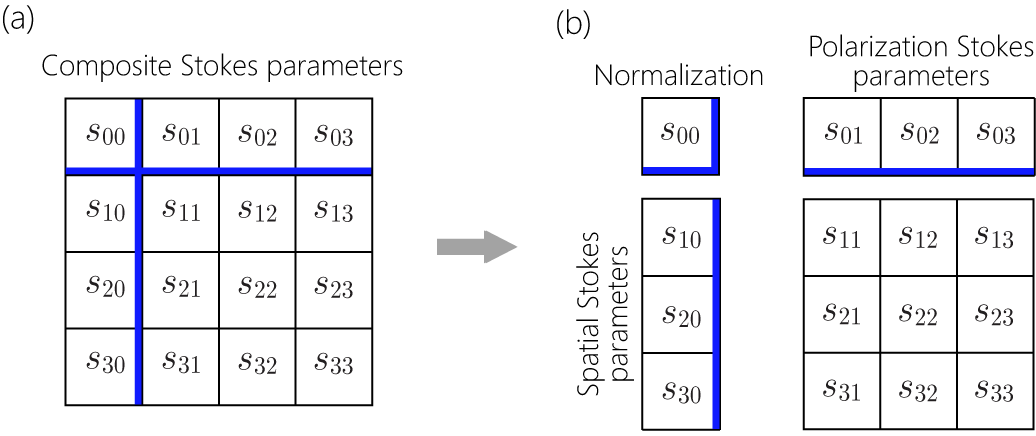}
\caption{(a) The modal Stokes parameters $\{s_{jk}\}$ arranged in a $4\times4$ matrix form. (b) Exploded form of the matrix in (a). The corner element $s_{00}=1$ represents the normalization; the top row $\{s_{0k}\}$ correspond to the purely polarization Stokes parameters $\{s_{k}^{(\mathrm{p})}\}$ obtained from the reduced polarization coherence matrix $\mathbf{G}_{\mathrm{p}}^{\mathrm{red.}}$ (Eq.~\ref{eq:ReducedStokesPol}); the leftmost column $\{s_{j0}\}$ correspond to the purely spatial Stokes parameters $\{s_{j}^{(\mathrm{s})}\}$ obtained from the reduced spatial coherence matrix $\mathbf{G}_{\mathrm{s}}^{\mathrm{red.}}$ (Eq.~\ref{eq:ReducedStokesSpatial}); and the remaining $3\times3$ matrix comprises the modal Stokes parameters $\{s_{jk}\}_{j,k\neq0}$ describing the correlations between the spatial and polarization DoFs \cite{Abouraddy02OC,Abouraddy14OL,Kagalwala15SR}. When the coherence matrix $\mathbf{G}$ is separable with respect to the spatial and temporal DoFs, we have $s_{jk}=s_{j0}s_{0k}$, and there is no extra information in the $3\times3$ matrix of modal Stokes parameters $\{s_{jk}\}_{j,k\neq0}$ that does not already exist in the purely spatial and polarization Stokes parameters.}
\label{fig:CompositeStokesMatrix}
\end{figure}

A useful feature of this formulation is that the purely spatial Stokes parameters $\{s_{j}^{(\mathrm{s})}\}$ associated with the reduced spatial coherence matrix $\mathbf{G}_{\mathrm{s}}^{\mathrm{red.}}$ and the purely polarization Stokes parameters $\{s_{k}^{(\mathrm{p})}\}$ associated with the reduced polarization coherence matrix $\mathbf{G}_{\mathrm{p}}^{\mathrm{red}}$ are included within the modal Stokes parameters $\{s_{jk}\}$ [Fig.~\ref{fig:CompositeStokesMatrix}]. The reduced spatial coherence matrix $\mathbf{G}_{\mathrm{s}}^{\mathrm{red.}}$ (Eq.~\ref{eq:ReducedSpatialG}) and the associated spatial Stokes parameters are given by [Fig.~\ref{fig:OCmT1}(a)]:
\begin{equation}\label{eq:ReducedStokesSpatial}
\mathbf{G}_{\mathrm{s}}^{\mathrm{red}}=\left(\begin{array}{cc}G_{\mathrm{HH}}^{aa}+G_{\mathrm{VV}}^{aa}&
G_{\mathrm{HH}}^{ab}+G_{\mathrm{VV}}^{ab}\\
G_{\mathrm{HH}}^{ba}+G_{\mathrm{VV}}^{ba}&
G_{\mathrm{HH}}^{bb}+G_{\mathrm{VV}}^{bb}\end{array}\right)\Rightarrow\mathbf{S}^{(\mathrm{s})}=\left(\begin{array}{c}s_{0}^{(\mathrm{s})}\\s_{1}^{(\mathrm{s})}\\s_{2}^{(\mathrm{s})}\\s_{3}^{(\mathrm{s})}\end{array}\right)=\left(\begin{array}{c}
G_{\mathrm{HH}}^{aa}+G_{\mathrm{VV}}^{aa}+
G_{\mathrm{HH}}^{bb}+G_{\mathrm{VV}}^{bb}\\
G_{\mathrm{HH}}^{aa}+G_{\mathrm{VV}}^{aa}-
G_{\mathrm{HH}}^{bb}-G_{\mathrm{VV}}^{bb}\\
G_{\mathrm{HH}}^{ab}+G_{\mathrm{VV}}^{ab}+
G_{\mathrm{HH}}^{ba}+G_{\mathrm{VV}}^{ba}\\
i(G_{\mathrm{HH}}^{ab}+G_{\mathrm{VV}}^{ab}-
G_{\mathrm{HH}}^{ba}-G_{\mathrm{VV}}^{ba})
\end{array}\right)=\left(\begin{array}{c}s_{00}\\s_{10}\\s_{20}\\s_{30}\end{array}\right).
\end{equation}
In other words, after extracting the spatial Stokes parameters from the reduced spatial coherence matrix, we find that they are in fact a subset of the modal Stokes parameters: $\{s_{0}^{(\mathrm{s})},s_{1}^{(\mathrm{s})},s_{2}^{(\mathrm{s})},s_{3}^{(\mathrm{s})}\}=\{s_{00},s_{10},s_{20},s_{30}\}$.

Similarly, the reduced polarization coherence matrix (Eq.~\ref{eq:ReducedPolG}) and the associated polarization Stokes parameters are given by:
\begin{equation}\label{eq:ReducedStokesPol}
\mathbf{G}_{\mathrm{p}}^{\mathrm{red}}=\left(\begin{array}{cc}
G_{\mathrm{HH}}^{aa}+G_{\mathrm{HH}}^{bb}&
G_{\mathrm{HV}}^{aa}+G_{\mathrm{HV}}^{bb}\\
G_{\mathrm{VH}}^{aa}+G_{\mathrm{VH}}^{bb}&
G_{\mathrm{VV}}^{aa}+G_{\mathrm{VV}}^{bb}\end{array}\right)\Rightarrow\mathbf{S}^{(\mathrm{p})}=\left(\begin{array}{c}s_{0}^{(\mathrm{p})}\\s_{1}^{(\mathrm{p})}\\s_{2}^{(\mathrm{p})}\\s_{3}^{(\mathrm{p})}\end{array}\right)=\left(\begin{array}{c}
G_{\mathrm{HH}}^{aa}+G_{\mathrm{VV}}^{aa}+
G_{\mathrm{HH}}^{bb}+G_{\mathrm{VV}}^{bb}\\
G_{\mathrm{HH}}^{aa}+G_{\mathrm{HH}}^{bb}-
G_{\mathrm{VV}}^{aa}-G_{\mathrm{VV}}^{bb}\\
G_{\mathrm{HV}}^{aa}+G_{\mathrm{HV}}^{bb}+
G_{\mathrm{VH}}^{aa}+G_{\mathrm{VH}}^{bb}\\
i(G_{\mathrm{HV}}^{aa}+G_{\mathrm{HV}}^{bb}-
G_{\mathrm{VH}}^{aa}-G_{\mathrm{VH}}^{bb})
\end{array}\right)=\left(\begin{array}{c}s_{00}\\s_{01}\\s_{02}\\s_{03}\end{array}\right).
\end{equation}
In other words, after extracting the polarization Stokes parameters from the reduced polarization coherence matrix, we find that they are in fact a subset of the modal Stokes parameters: $\{s_{0}^{(\mathrm{p})},s_{1}^{(\mathrm{p})},s_{2}^{(\mathrm{p})},s_{3}^{(\mathrm{p})}\}=\{s_{00},s_{01},s_{02},s_{03}\}$ [Fig.~\ref{fig:OCmT1}(b)].

Note that for coherence matrices that are separable with respect to the spatial and polarization DoFs, $\mathbf{G}=\mathbf{G}_{\mathrm{s}}\otimes\mathbf{G}_{\mathrm{p}}$, where $\mathbf{G}_{\mathrm{s}}=\mathbf{G}_{\mathrm{s}}^{\mathrm{red.}}$ and $\mathbf{G}_{\mathrm{p}}=\mathbf{G}_{\mathrm{p}}^{\mathrm{red.}}$, only measurements for the spatial and polarization Stokes parameters $\{s_{00},s_{10},s_{20},s_{30}\}$ and $\{s_{00},s_{01},s_{02},s_{03}\}$, respectively, are sufficient. Indeed, in this case we have $s_{jk}=s_{j0}\cdot s_{0k}=s_{j}^{(\mathrm{s})}\cdot s_{k}^{(\mathrm{p})}$. More generally for non-separable coherence matrices, all 16~modal Stokes parameters must be measured [Fig.~\ref{fig:CompositeStokesMatrix}].

\begin{figure}[t!]
\centering
\includegraphics[width=11cm]{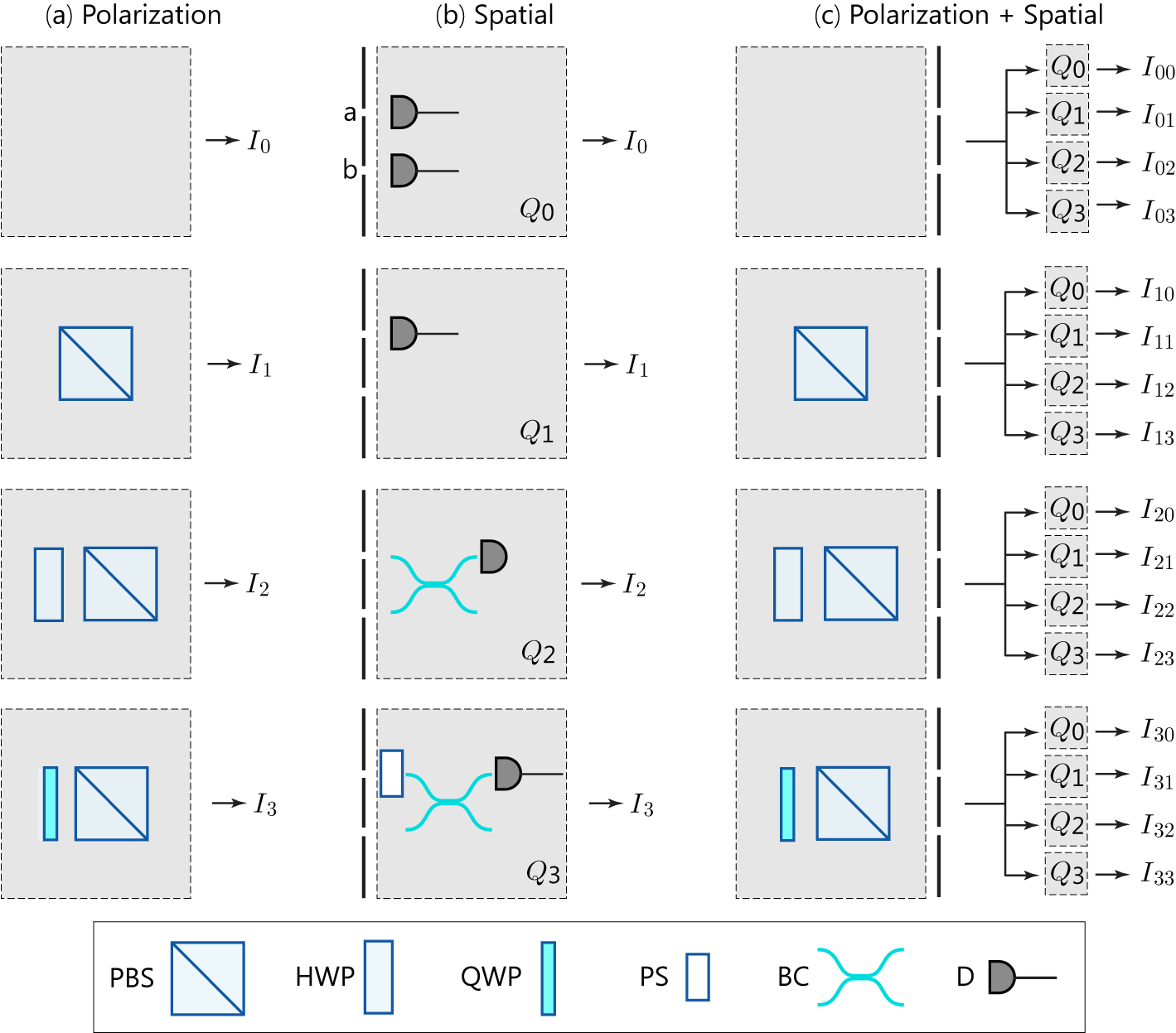}
\caption{Measurements for acquiring the modal Stokes parameters. (a) Measuring the 4 spatial Stokes parameters $\{s_{j}^{(\mathrm{s})}\}$. Here $s_{0}^{(\mathrm{s})}$ corresponds to measuring $I_{a}+I_{b}$ and $s_{1}^{(\mathrm{s})}$ to $I_{a}-I_{b}$, where $I_{a}$ and $I_{b}$ are the powers at $|a\rangle$ and $|b\rangle$. ... (b) Measuring the 4 polarization Stokes parameters $\{s_{k}^{(\mathrm{p})}\}$. (c) Measuring the 16~modal Stokes parameters $\{s_{jk}\}$. Each modal Stokes parameter $s_{jk}$ is measured using a concatenation of two measurement setups: one associated with the spatial Stokes parameter $s_{j}^{(\mathrm{s})}$ and that associated with the polarization Stokes parameter $s_{k}^{(\mathrm{s})}$.}
\label{fig:OCmT2}
\end{figure}

The modal Stokes parameter $s_{jk}$ is measured by concatenating the setup for measuring $s_{j}^{(\mathrm{s})}$ for the spatial DoF and that for measuring $s_{k}^{(\mathrm{p})}$ for the polarization DoF. We depict in Fig.~\ref{fig:OCmT2}(a) the measurement configurations to measure the polarization Stokes parameters: (1) Measuring the total power $I=I_{\mathrm{H}}+I_{\mathrm{V}}=s_{0}^{(\mathrm{p})}$ to be used for normalization; (2) a PBS splits the $|\mathrm{H}\rangle$ and $|\mathrm{V}\rangle$ modes in two spatial paths, from which we obtain $s_{1}^{(\mathrm{p})}=I_{\mathrm{H}}-I_{\mathrm{V}}$; (3) a HWP rotated by $\tfrac{\pi}{8}$ with respect to $|\mathrm{H}\rangle$, $\hat{U}_{\mathrm{HWP}}(\tfrac{\pi}{8})=\tfrac{1}{\sqrt{2}}\left(\begin{array}{cc}1&1\\1&-1\end{array}\right)$, followed by a PBS yields $s_{2}^{(\mathrm{p})}=I_{\mathrm{H}}'-I_{\mathrm{V}}'$; and (4) a QWP rotated by $-\tfrac{\pi}{4}$ with respect to $|\mathrm{H}\rangle$, $\hat{U}_{\mathrm{QWP}}(-\tfrac{\pi}{4})=\tfrac{1}{\sqrt{2}}\left(\begin{array}{cc}1&-i\\-i&1\end{array}\right)$, followed by a PBS yields $s_{3}^{(\mathrm{p})}=I_{\mathrm{H}}''-I_{\mathrm{V}}''$.

We depict in Fig.~\ref{fig:OCmT2}(b) the measurement configurations to measure the spatial Stokes parameters: (1) a bucket detector provides the total power $I=I_{a}+I_{b}=s_{0}^{(\mathrm{s})}$ for normalization; (2) two detectors at $|a\rangle$ and $|b\rangle$ yield $s_{1}^{(\mathrm{s})}=I_{a}-I_{b}$; (3) a spatial rotator (an MZI), $\hat{U}_{2}^{(\mathrm{s})}=\tfrac{1}{2}\left(\begin{array}{cc}1&1\\1&-1\end{array}\right)$, yields $s_{2}^{(\mathrm{s})}=I_{a}'-I_{b}'$; and (4) a unitary formed of phase operators and a spatial rotator, $\hat{U}_{3}^{(\mathrm{s})}=\tfrac{1}{2}\left(\begin{array}{cc}1&-i\\i&-1\end{array}\right)$, yields $s_{3}^{(\mathrm{s})}=I_{a}''-I_{b}''$. These are the requisite measurement configurations for an on-chip setting involving a pair of single-mode waveguides. One could also have selected the configurations illustrated in Fig.~\ref{fig:DoubleSlitStokes} if the double-slit setting is the relevant one.

To measure the 16~modal Stokes parameters $\{s_{jk}\}$, we form in Fig.~\ref{fig:OCmT2}(c) the $4\times4=16$ combinations of the polarization configurations in Fig.~\ref{fig:OCmT2}(a) and the spatial configurations in Fig.~\ref{fig:OCmT2}(b). In each configuration we use 4~detectors to obtain the modal weights $I_{a\mathrm{H}}$, $I_{a\mathrm{V}}$, $I_{b\mathrm{H}}$, and $I_{b\mathrm{V}}$. The normalization is obtained from $s_{00}$, where we can place one polarization-independent bucket detector that measures the total power, or add the measurements for all modal weights, $s_{00}=I_{a\mathrm{H}}+I_{a\mathrm{V}}+I_{b\mathrm{H}}+I_{b\mathrm{V}}$. For modal Stokes parameters of the form $\{s_{j0}\}$, we perform a spatial measurement [Fig.~\ref{fig:OCmT2}(b)] and make use of polarization-insensitive detectors at $|a\rangle$ and $|b\rangle$, or add the relevant measurements $I_{a}=I_{a\mathrm{H}}+I_{a\mathrm{V}}$ and $I_{b}=I_{b\mathrm{H}}+I_{b\mathrm{V}}$. For modal Stokes parameters of the form $\{s_{0k}\}$, we perform a polarization measurement [Fig.~\ref{fig:OCmT2}(a)] and make use of bucket detectors for $|\mathrm{H}\rangle$ and $|\mathrm{V}\rangle$, or add the relevant measurements $I_{\mathrm{H}}=I_{a\mathrm{H}}+I_{b\mathrm{H}}$ and $I_{\mathrm{V}}=I_{a\mathrm{V}}+I_{b\mathrm{V}}$.

For modal Stokes parameters of the form $\{s_{jk}\}$, with $j\neq0$ and $k\neq0$, we implement a cascade of polarization and spatial measurement configurations, which have 4~measured modal weights. From these modal weights we obtain the modal Stokes parameters, from which in turn we reconstruct the coherence marix $\mathbf{G}$.

\subsection{What is the maximum visibility in Young's double-slit interference for vector fields}\label{sec:MaximizingDoubleSlits}

The visibility $V$ in double-slit interference with a scalar field is bound by the degree of spatial coherence $V\leq D_{\mathrm{s}}$, with equality holding only when the field amplitudes are equal at the slits, whereupon $V_{\mathrm{max}}=D_{\mathrm{s}}$ (Section~\ref{sec:SpatialDoF}). The situation is more complicated in a vector field. A trivial example is when the field is spatially coherent and the amplitudes at $|a\rangle$ and $|b\rangle$ are equal, \textit{but} the polarizations modes are orthogonal, whereupon $V=0$. Changing the polarization at $|b\rangle$ to coincide with the polarization at $|a\rangle$ yields $V=1$ as expected. However, it is difficult to determine in general how to obtain the maximum visibility $V_{\mathrm{max}}$ when the field is partially coherent at $|a\rangle$ and $|b\rangle$ with respect to both the spatial and polarization DoFs. This problem has been tackled multiple times, with different estimates for $V_{\mathrm{max}}$ reported under distinct constraints on the permitted transformations to be implemented on the field. The matrix formulation facilitates answering the following question: what is the maximum visibility $V_{\mathrm{max}}$ obtainable in double-slit interference in a vector field when \textit{arbitrary unitaries} -- spatial (Section~\ref{sec:SpatialDoF}), polarization (Section~\ref{sec:SpatialDoF}), or joint spatial-polarization (Section~\ref{sec:Unitary2DoFs}) -- can be implemented?

\subsubsection{Definition of the problem}

In a scalar field described by a spatial coherence matrix $\mathbf{G}_{\mathrm{s}}$, we define $V_{\mathrm{max}}$ as the maximum visibility observed after implementing an arbitrary spatial unitary $\hat{U}_{\mathrm{s}}$ on $\mathbf{G}_{\mathrm{s}}$. As shown in Section~\ref{sec:SpatialDoF}, $V_{\mathrm{max}}=D_{\mathrm{s}}=\lambda_{a}-\lambda_{b}$ can be extracted by implementing the unitary $\hat{U}_{\mathrm{s}}$ that diagonalizes $\mathbf{G}_{\mathrm{s}}$. Because the off-diagonal elements of $\mathbf{G}_{\mathrm{s}}$ has been eliminated, interference cannot be observed. Nevertheless, implementing the unitary $\hat{U}_{\mathrm{s}}=\tfrac{1}{\sqrt{2}}\left(\begin{array}{cc}1&-e^{-i\varphi}\\e^{i\varphi}&1\end{array}\right)$ after diagonalization yields $\mathbf{G}_{\mathrm{s}}=\tfrac{1}{2}\left(\begin{array}{cc}1&D_{\mathrm{s}}e^{-i\varphi}\\D_{\mathrm{s}}e^{i\varphi}&1\end{array}\right)$, thus displaying an interferogram of visibility $V_{\mathrm{max}}=\lambda_{a}-\lambda_{b}=D_{\mathrm{s}}$.

In the case of a vector field at $|a\rangle$ and $|b\rangle$ described by a $4\times4$ coherence matrix $\mathbf{G}$, what is the \textit{maximum visibility} attainable in the far-field intensity when the intensity is recorded by a polarization-independent detector (i.e., purely spatial intensity distribution) and the field is subject to an arbitrary $4\times4$ \textit{unitary} transformation? 

\subsubsection{Maximum visibility in terms of the eigenvalues of $\mathbf{G}$}

Because we record a purely spatial intensity distribution, the visibility can be obtained from $\mathbf{G}_{\mathrm{s}}^{\mathrm{red.}}$ via diagonalization. However, as described in Section~\ref{sec:ReducedCoherenceMatrices}, there are two approaches to diagonalizing $\mathbf{G}_{\mathrm{s}}^{\mathrm{red.}}$: (1) utilizing a purely spatial unitary $\hat{U}_{\mathrm{s}}$ (which does not guarantee that $\mathbf{G}$ is diagonalized); or (2) implementing a unitary $\hat{U}$ to diagonalize $\mathbf{G}$ (which guarantees that $\mathbf{G}_{\mathrm{s}}^{\mathrm{red.}}$ is also diagonalized). The diagonalized reduced spatial coherence matrices are \textit{not} the same after following these two routes. 

We have shown in Ref.~\cite{Abouraddy17OE} that diagonalizing $\mathbf{G}$ via a $4\times4$ unitary -- rather than diagonalizing only $\mathbf{G}_{\mathrm{s}}^{\mathrm{red.}}$ via a $2\times2$ spatial unitary -- produces the maximum visibility: $\mathbf{G}\rightarrow\mathbf{G}^{\mathrm{D}}=\mathrm{diag}\{\lambda_{1},\lambda_{2},\lambda_{3},\lambda_{4}\}$, $1\geq\lambda_{1}\geq\lambda_{2}\geq\lambda_{3}\geq\lambda_{4}\geq0$, $\mathbf{G}_{\mathrm{s}}^{\mathrm{red.}}\rightarrow(\mathbf{G}_{\mathrm{s}}^{\mathrm{red.}})^{\mathrm{D}}=\mathrm{diag}\{\lambda_{1}+\lambda_{2},\lambda_{3}+\lambda_{4}\}$, and the maximum visibility is:
\begin{equation}\label{eq:Vmax}
V_{\mathrm{max}}=(\lambda_{1}+\lambda_{2})-(\lambda_{3}+\lambda_{4}),
\end{equation}
In this form, $V_{\mathrm{max}}$ generalizes to a $4\times4$ coherence matrix the corresponding relationship for a $2\times2$ spatial coherence matrix where $V_{\mathrm{max}}=D_{\mathrm{s}}=\lambda_{a}-\lambda_{b}$. Of course, when the field is fully coherent $\lambda_{1}=1$ and $\lambda_{2}=\lambda_{3}=\lambda_{4}=0$ (in which case $S=0$), then $V_{\mathrm{max}}=1$ as expected. However, the formula in Eq.~\ref{eq:Vmax} makes a counter-intuitive claim: full coherence of the field ($S=0$) is \textit{not} necessary to observe high-visibility fringes; even if the field is partially coherent ($S\neq0$), then $V_{\mathrm{max}}=1$ as long as $\lambda_{3}=\lambda_{4}=0$. Rather, the only condition is $\lambda_{3}=\lambda_{4}=0$ (independently of the values of $\lambda_{1}$ and $\lambda_{2}$) to yield full-visibility fringes.

This observation leads to the following surprising result.  Consider a \textit{scalar} field (e.g., linearly polarized along $|\mathrm{H}\rangle$) with equal modal weights at $|a\rangle$ and $|b\rangle$ that is spatially incoherent, whereupon $\mathbf{G}=\mathrm{diag}\{\tfrac{1}{2},0,\tfrac{1}{2},0\}$ and $\mathbf{G}_{\mathrm{s}}^{\mathrm{red.}}=\tfrac{1}{2}\hat{\mathbb{I}}_{2}$. Such a field displays no interference fringes. Nevertheless, the formula in Eq.~\ref{eq:Vmax} predicts that $V_{\mathrm{max}}=1$ because $\lambda_{3}=\lambda_{4}=0$ after arranging the eigenvalues in descending order; i.e., this scalar, spatially incoherent field can in fact display full interference visibility. The formula for $V_{\mathrm{max}}$ thus indicates that although there does not exist a $2\times2$ spatial unitary $\hat{U}_{\mathrm{s}}$ on $\mathbf{G}_{\mathrm{s}}^{\mathrm{red.}}$ that increases the visibility (because $\mathbf{G}_{\mathrm{s}}^{\mathrm{red.}}=\tfrac{1}{2}\hat{\mathbb{I}}_{2}$), there nevertheless exists a $4\times4$ unitary on both the spatial and polarization DoFs that transforms this spatially incoherent field -- without filtering or loss of energy -- into one that is \textit{fully spatially coherent} and displaying full-visibility fringes. 

\subsubsection{Maximizing the visibility via coherence or entropy conversion}\label{eq:VisEntropyConversion}

We first examine the example above mathematically. Starting with the coherence matrix $\mathbf{G}_{1}=\mathrm{diag}\{\tfrac{1}{2},0,\tfrac{1}{2},0\}$, we rotate the polarization at $|b\rangle$ from $|\mathrm{H}\rangle$ to $|\mathrm{V}\rangle$ via unitary $\hat{U}_{12}=\left(\begin{array}{cc}\hat{\mathbb{I}}_{2}&\hat{\mathbf{0}}_{2}\\\hat{\mathbf{0}}_{2}&\hat{U}_{\mathrm{HWP}}\end{array}\right)$, which is a (non-separable) spatially dependent polarization unitary, with $\hat{U}_{\mathrm{HWP}}$ represents a HWP that rotates the polarization $|\mathrm{H}\rangle\rightarrow|\mathrm{V}\rangle$, thereby yielding the coherence matrix $\mathbf{G}_{2}$. We then bring the fields from $|a\rangle$ and $|b\rangle$ together to a PBS (unitary $\hat{U}_{\mathrm{PBS}}$ in Eq.~\ref{eq:Polarizing BS}), in which case the fields are combined into a single spatial mode corresponding to the coherence matrix $\mathbf{G}_{3}$, which is subsequently split into two spatial modes via a non-polarizing beam splitter (unitary $\hat{U}_{\mathrm{BS}}$ in Eq.~\ref{eq:BeamSplitter})corresponding to the coherence matrix $\mathbf{G}_{4}$. These transformations are given explicitly as follows:

\begin{figure}[t!]
\centering
\includegraphics[width=10.3cm]{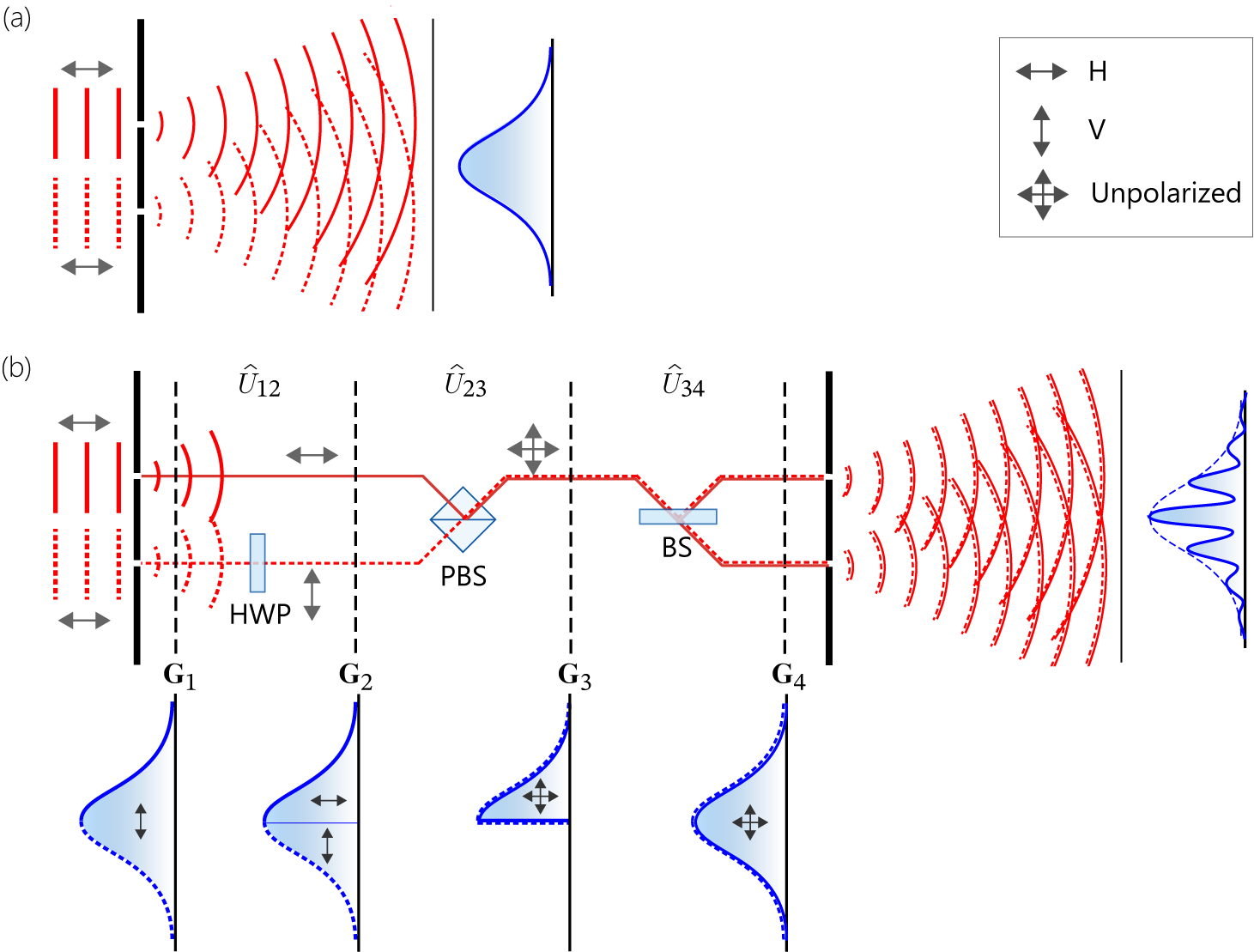}
\caption{(a) A spatially incoherent field at $|a\rangle$ and $|b\rangle$ that is linearly polarized along $|\mathrm{H}\rangle$ displays no fringes in a double-slit configuration. (b) Rotating the polarization at $|b\rangle$ from $|\mathrm{H}\rangle$ to $|\mathrm{V}\rangle$, combining the fields from $|a\rangle$ and $|b\rangle$ at a PBS into mode $|a\rangle$, and then splitting the resulting field between spatial modes $|a\rangle$ and $|b\rangle$ via a BS produces a field that is unpolarized but spatially coherent. Full-visibility fringes are now produced after a lossless, reversible procedure. (c) Graphical depiction of the field structure at the different stages of the procedure in (b).}
\label{fig:DoubleSlitEntropyConversion}
\end{figure}

\begin{equation}
\mathbf{G}_{1}\overset{\hat{U}_{12}}{\longrightarrow}
\mathbf{G}_{2}=\frac{1}{2}\mathrm{diag}\{1,0,0,1\}\overset{\hat{U}_{\mathrm{PBS}}}{\longrightarrow}
\mathbf{G}_{3}=\frac{1}{2}\mathrm{diag}\{1,1,0,0\}\overset{\hat{U}_{\mathrm{BS}}}{\longrightarrow}
\mathbf{G}_{4}=\frac{1}{4}\left(\begin{array}{cccc}
1&0&1&0\\0&1&0&1\\1&0&1&0\\0&1&0&1\end{array}\right),
\end{equation}
which is associated with the following transformations of the reduced spatial coherence matrix:
\begin{equation}
\mathbf{G}_{\mathrm{s}}^{\mathrm{red.}}=\underbrace{\frac{1}{2}\left(\begin{array}{cc}1&0\\0&1\end{array}\right)}_{V=0}\overset{\hat{U}_{12}}{\longrightarrow}
\underbrace{\frac{1}{2}\left(\begin{array}{cc}1&0\\0&1\end{array}\right)}_{V=0}\overset{\hat{U}_{\mathrm{PBS}}}{\longrightarrow}\underbrace{\left(\begin{array}{cc}1&0\\0&0\end{array}\right)}_{V=0}\overset{\hat{U}_{\mathrm{BS}}}{\longrightarrow}\underbrace{\frac{1}{2}\left(\begin{array}{cc}1&1\\1&1\end{array}\right)}_{V=1},
\end{equation}

The coherence matrix $\mathbf{G}_{1}$ describes a spatially incoherent field that is fully polarized along $|\mathrm{H}\rangle$, with reduced coherence matrices $\mathbf{G}_{\mathrm{s}1}^{\mathrm{red.}}=\tfrac{1}{2}\hat{\mathbb{I}}_{2}$ and $\mathbf{G}_{\mathrm{p}}^{\mathrm{red.}}=\left(\begin{array}{cc}1&0\\0&0\end{array}\right)$. Note that $\mathbf{G}_{1}$ is separable with respect to the spatial and polarization DoFs, $\mathbf{G}_{1}=\frac{1}{2}{\left(\begin{array}{cc}1&0\\0&1\end{array}\right)}_{\mathrm{s}}\otimes{\left(\begin{array}{cc}1&0\\0&0\end{array}\right)}_{\mathrm{p}}$ (the subscripts `s' and `p' refer to the spatial and polarization DoF's, respectively). The field is linearly polarized but is spatially incoherent. No interference fringes can be observed with this field.

After the non-separable unitary $\hat{U}_{12}$, the new coherence matrix $\mathbf{G}_{2}$ describes a partially coherent field where polarization is now correlated with position and is no longer factorizable. The polarizing beam splitter $\hat{U}_{\mathrm{PBS}}$ combines the fields from $|a\rangle$ and $|b\rangle$ to produce an unpolarized field at $|a\rangle$. Here $\mathbf{G}_{3}=\left(\begin{array}{cc}1&0\\0&0\end{array}\right)_{\!\mathrm{s}}\otimes\frac{1}{2}\left(\begin{array}{cc}1&0\\0&1\end{array}\right)_{\mathrm{p}}$ is again separable with respect to its DoFs; the field is now spatially coherent but unpolarized. The procedure is fully reversible and there has been no optical energy lost.

The beam splitter $\hat{U}_{\mathrm{BS}}$ splits the field at $|a\rangle$ into equal-amplitude spatially coherent fields at $|a\rangle$ and $|b\rangle$, $\mathbf{G}_{4}$. Therefore, a polarized but spatially incoherent field $\mathbf{G}_{1}$ that displays zero interference visibility has thus been transformed to a spatially coherent but unpolarized field $\mathbf{G}_{4}$ that displays full visibility. We refer to this process as `coherence conversion' [Fig.~\ref{fig:DoubleSlitEntropyConversion}(b)].

The general procedure to observe the maximum visibility $V_{\mathrm{max}}$ is therefore as follows. First, diagonalize $\mathbf{G}$ via a unitary $\hat{U}$, $\mathbf{G}^{\mathrm{D}}=\hat{U}\mathbf{G}\hat{U}^{\dagger}=\mathrm{diag}\{\lambda_{1},\lambda_{2},\lambda_{3},\lambda_{4}\}$. Second, implement a symmetric beam splitter that combines the fields from $|a\rangle$ and $|b\rangle$, so that $\mathbf{G}'=\hat{U}_{\mathrm{BS}}\mathbf{G}\hat{U}_{\mathrm{BS}}^{\dagger}$, where: 
\begin{equation}
\mathbf{G}'=\frac{1}{2}\left(\begin{array}{cccc}
\lambda_{1}+\lambda_{3}&0&-i\lambda_{1}+i\lambda_{3}&0\\
0&\lambda_{2}+\lambda_{4}&0&-i\lambda_{2}+i\lambda_{4}\\
i\lambda_{1}-i\lambda_{3}&0&\lambda_{1}+\lambda_{3}&0\\
0&i\lambda_{2}-i\lambda_{4}&0&\lambda_{2}+\lambda_{4}
\end{array}\right)\rightarrow \mathbf{G}_{\mathrm{s}}^{\mathrm{red.}}=\frac{1}{2}\left(\begin{array}{cc}
1&-iV_{\mathrm{max}}\\iV_{\mathrm{max}}&1
\end{array}\right).
\end{equation}

\subsubsection{Comparison to previous measures}

Several efforts preceding Ref.~\cite{Abouraddy17OE} attempted to obtain expressions for $V_{\mathrm{max}}$ that fall short of the optimal value. These attempts did not consider the role of the eigenvalues of $\mathbf{G}$ nor exploited the full family of $4\times4$ unitaries $\hat{U}$.

The basic expression for the visibility starting from $\mathbf{G}_{\mathrm{s}}^{\mathrm{red.}}$ in Eq.~\ref{eq:ReducedSpatialG} is:
\begin{equation}\label{eq:VisibiltyGori}
V=2|G_{ab}^{\mathrm{HH}}+G_{ab}^{\mathrm{VV}}|=2|\mathrm{Tr}\{\mathbf{G}_{ab}\}|,
\end{equation}
where $\mathbf{G}_{ab}$ is the off-diagonal block-matrix of $\mathbf{G}$, which is related to the spectral degree of coherence as defined by E.~Wolf \cite{Wolf03PLA} and Karczewski \cite{Karczewski63INC}. Rather than diagonalizing $\mathbf{G}$, previous attempts maximized the expression in Eq.~\ref{eq:VisibiltyGori} subject to various families of field transformations.

\begin{figure}[t!]
\centering
\includegraphics[width=10.3cm]{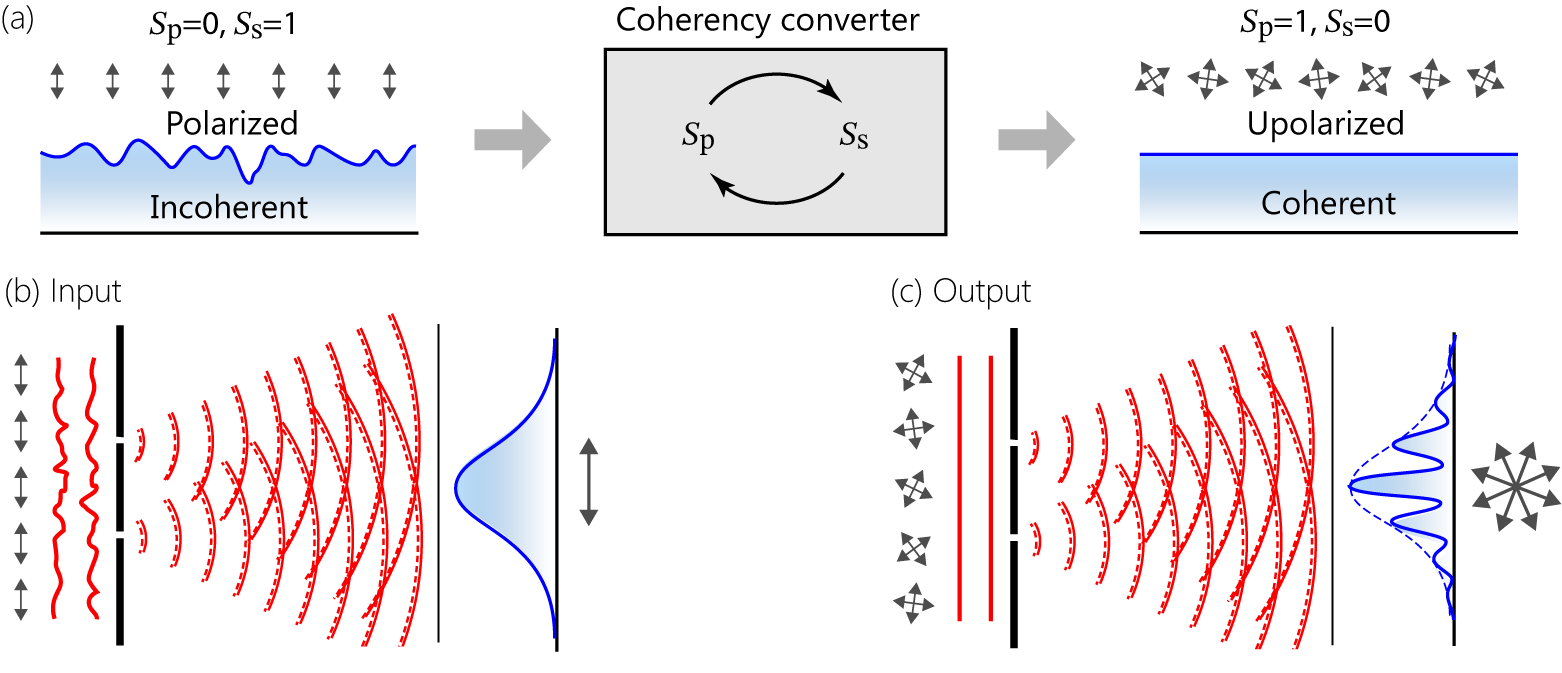}
\caption{Concept of coherence conversion. (a) Starting with a polarized (scalar) field that is spatially incoherent (the field is separable with respect to the two DoFs), with $D_{\mathrm{p}}=1$ ($S_{\mathrm{p}}=0$) and $D_{\mathrm{s}}=0$ ($S_{\mathrm{s}}=1$). (b) Such a field displays no interference fringes in a Young double-slit configuration. (c) The coherence converter unitarily transfers coherence from the polarization to the spatial DoFs (or transfers the entropy from the spatial to the polarization DoFs) without loss of energy \cite{Okoro17Optica,Harling22OE,Harling23JO} The field is again separable, but now with $D_{\mathrm{p}}=0$ ($S_{\mathrm{p}}=1$) and $D_{\mathrm{s}}=1$ ($S_{\mathrm{s}}=0$); that is, the field is spatially coherent but unpolarized. (d) Such a field produces fringes with full visibility in a Young double-slit configuration.}
\label{fig:}
\end{figure}

\subsubsection{Unitary measures}

Although a global polarization unitary $\hat{U}=\mathbb{I}\otimes\hat{U}_{\mathrm{p}}$ does not change $V$, because $\mathbf{G}_{ab}'=\hat{U}_{\mathrm{p}}\mathbf{G}_{ab}\hat{U}_{\mathrm{p}}^{\dagger}$, $\hat{U}$ does not change $V$: 
\begin{equation}
V'=2|\mathrm{Tr}\{\mathbf{G}_{ab}'\}|=2|\mathrm{Tr}\{\hat{U}_{\mathrm{p}}\mathbf{G}_{ab}\hat{U}_{\mathrm{p}}^{\dagger}\}|=2|\mathrm{Tr}\{\mathbf{G}_{ab}\}|=V,
\end{equation}
but applying different polarization unitaries at $|a\rangle$ and $|b\rangle$ ($\hat{U}_{a}$ and $\hat{U}_{b}$, respectively), $\hat{U}=\left(\begin{array}{cc}\hat{U}_{a}&\mathbf{0}\\\mathbf{0}&\hat{U}_{b}\end{array}\right)$, can indeed change $V$: $\mathbf{G}_{ab}'=\hat{U}_{a}\mathbf{G}_{ab}\hat{U}_{b}^{\dagger}$ and $V'=2|\mathrm{Tr}\{\hat{U}_{a}\mathbf{G}_{ab}\hat{U}_{b}^{\dagger}\}|\neq V$. In other words, the traditional visibility is \textit{not} a unitary invariant under spatially dependent (or local) polarization unitaries. This has prompted introducing a different measure for spatial coherence in a vector field \cite{Tervo03OE,Setala04OL} called `the electromagnetic degree of coherence' $\gamma$, where:
\begin{equation}
\gamma^{2}=\frac{\mathrm{Tr}\{\mathbf{G}_{ab}\mathbf{G}_{ab}^{\dagger}\}}{\mathrm{Tr}\{\mathbf{G}_{a}\}\mathrm{Tr}\{\mathbf{G}_{b}\}},
\end{equation}
which is invariant under local polarization unitaries. Nevertheless, $\gamma$ is not related directly to the double-slit interference visibility. Crucially, if $\mathbf{G}_{ab}=\hat{\mathbf{0}}_{2}$, then $\gamma^{2}=0$ and $V=0$, although $V$ may still reach a value $V\rightarrow V_{\mathrm{max}}=1$ (when $\lambda_{3}=\lambda_{4}=0$).

Gori \textit{et al.} maximized $V=2|\mathrm{Tr}\{\hat{U}_{a}\mathbf{G}_{ab}\hat{U}_{b}^{\dagger}\}$ over all polarization unitaries $\hat{U}_{a}$ and $\hat{U}_{b}$ in Ref.~\cite{Gori07OL}. It was found that the maximum value of $V$ over local polarization unitaries corresponds to the so-called Ky-Fan 1-norm \cite{Horn90Book} of $\mathbf{G}$:
\begin{equation}
\mathrm{max}\{V_{\mathrm{o}}\}\xrightarrow[\hat{U}_{a},\hat{U}_{b}]{}V_{\mathrm{LPU}}=2(\mu_{1}+\mu_{2})=2\sqrt{\mathrm{Tr}\{\mathbf{G}_{ab}\mathbf{G}_{ab}^{\dagger}\}+2|\mathrm{det}\{\mathbf{G}_{ab}\}|},
\end{equation}
where $\mu_{1}$ and $\mu_{2}$ are the singular values of $\mathbf{G}_{ab}$ \cite{Gori07OL}. Once again, if $\mathbf{G}_{ab}=\hat{\mathbf{0}}_{2}$ (diagonalized $\mathbf{G}$), then $V_{\mathrm{LPU}}=0$ even when $V_{\mathrm{max}}=1$. Both of these approaches suffer from not considering the full family of unitaries $\hat{U}$, including transformations of the spatial DoF.

\subsubsection{Non-unitary measures}

The visibility may of course be increased via \textit{non}-unitary filtering, which reduces the energy. The use of such transformations involves an element of arbitrariness, in contrast to reliance on unitary transformations that conserve energy. Nevertheless, some interesting studies have been reported along this vein.

(1) The work by R\'efr\'egier and Goudail on so-called `intrinsic degrees of coherence' \cite{Refregier07OPN}  provides an \textit{algorithm} for extracting two identified unitary invariants $0\!\leq\!\mu_{\mathrm{S}},\mu_{\mathrm{I}}\!\leq\!1$ ($\mu_{\mathrm{S}}\!\geq\!\mu_{\mathrm{I}}$) \cite{Refregier05OE}: (1) spatially dependent polarization unitaries $\hat{U}=\left(\begin{array}{cc}\hat{U}_{a}&\hat{\mathbf{0}}_{2}\\\hat{\mathbf{0}}_{2}&\hat{U}_{b}\end{array}\right)$ diagonalize $\mathbf{G}_{a}$ and $\mathbf{G}_{b}$; (2) the eigenvalues of $\mathbf{G}_{a}$ and $\mathbf{G}_{b}$ are `equalized' by implementing non-unitary partial polarizers at $|a\rangle$ and $|b\rangle$, $\hat{F}^{\mathrm{D}}=\left(\begin{array}{cc}\hat{F}_{a}^{\mathrm{D}}&\hat{\mathbf{0}}_{2}\\\hat{\mathbf{0}}_{2}&\hat{F}_{b}^{\mathrm{D}}\end{array}\right)$; and (3) implementing a second spatially dependent polarization unitary $\hat{U}=\left(\begin{array}{cc}\hat{U}_{a}'&\hat{\mathbf{0}}_{2}\\\hat{\mathbf{0}}_{2}&\hat{U}_{b}'\end{array}\right)$ to diagonalize $\mathbf{G}_{\mathrm{ab}}$. The resulting coherency matrix has the form
\begin{equation}\label{RG}
\mathbf{G}=\frac{1}{4}\left(\begin{array}{cccc}
1&0&\mu_{\mathrm{S}}&0\\
0&1&0&\mu_{\mathrm{I}}\\
\mu_{\mathrm{S}}&0&1&0\\
0&\mu_{\mathrm{I}}&0&1\end{array}\right),
\end{equation}
in which case $V=V_{\mathrm{LPU}}=(\mu_{\mathrm{S}}+\mu_{\mathrm{I}})/2$. The eigenvalues of $\mathbf{G}$ in this form $\{\lambda\}=\frac{1}{4}\{1+\mu_{\mathrm{S}},1+\mu_{\mathrm{I}},1-\mu_{\mathrm{I}},1-\mu_{\mathrm{S}}\}$, from which we have $V_{\mathrm{max}}=V$. An implicit assumption in this approach is that the power at $|a\rangle$ is equal to that at $|b\rangle$.

(2) A different analysis by Luis \cite{Luis07OL} puts forth the definition $V_{\mathrm{L}}\!=\!\frac{\lambda_{1}-\lambda_{4}}{\lambda_{1}+\lambda_{4}}$ for maximum visibility. This expression is reached by first diagonalizing $\mathbf{G}$ and then filtering out the modes associated with the eigenvalues $\lambda_{2}$ and $\lambda_{3}$ (eliminating a fraction $\lambda_{2}+\lambda_{3}$ of the total power), but yields $V_{\mathrm{L}}\leq V_{\mathrm{max}}$. In contrast, the analysis presented here suggests an alternative \textit{optimal} filtering methodology to maximize $V$: filter out the modes associated with $\lambda_{3}$ and $\lambda_{4}$ (instead of $\lambda_{2}$ and $\lambda_{3}$). This procedure eliminates a smaller fraction of energy since $\lambda_{3}+\lambda_{4}\leq\lambda_{2}+\lambda_{3}$, and the resulting visibility is \textit{always} $V_{\mathrm{max}}\!=\!1$.

(3) Another approach to determining the double-slit visibility involves a generalized form of the Fresnel-Arago interference laws \cite{Mujat04JOSAA}, but this requires \textit{first} placing linear polarizers at $|a\rangle$ and $|b\rangle$. Within our approach, placing linear polarizers at $|a\rangle$ and $|b\rangle$ always produces $V_{\mathrm{max}}\!=\!1$ independently of the state of coherence.

\subsection{Classification of coherence matrices according to their rank}\label{sec:CoherenceRank}

A new perspective on the properties of $4\times4$ coherence matrices $\mathbf{G}$ can be attained by considering a fourfold classification scheme based on the `coherence rank' \cite{Harling24PRA,Harling24PRA2}. We define the coherence rank of $\mathbf{G}$ as the number of its non-zero eigenvalues, so that the rank can take on the values 1, 2, 3, or~4.

\subsubsection{Rank-1 fields}

Rank-1 fields (denoted by $\mathbf{G}_{1}$) are those whose coherence matrix takes the following form after diagonalization: 
\begin{equation}
\mathbf{G}_{1}^{\mathrm{D}}=\mathrm{diag}\{1,0,0,0\}=\left(\begin{array}{cc}1&0\\0&0\end{array}\right)_{\mathrm{s}}\otimes\left(\begin{array}{cc}1&0\\0&0\end{array}\right)_{\mathrm{p}};
\end{equation}
that is, only one eigenvalue is non-zero: $\lambda_{1}=1$ and $\lambda_{2}=\lambda_{3}=\lambda_{3}=0$. The entropy of a rank-1 field is thus $S=0$, so that rank-1 fields are coherent and free of statistical fluctuations. 

Although the field entropy is $S=0$, the entropy for the reduced coherence matrices may \textit{not} vanish. Here, the separability of $\mathbf{G}_{1}$ plays a key role. Starting from the unique diagonalized form $\mathbf{G}_{1}^{\mathrm{D}}$, any rank-1 field can be produced via a unitary: $\mathbf{G}_{1}=\hat{U}\mathbf{G}_{1}^{\mathrm{D}}\hat{U}^{\dagger}$. If $\hat{U}=\hat{U}_{\mathrm{s}}\otimes\hat{U}_{\mathrm{p}}$ is separable, then the initially separable diagonalized coherence matrix $\mathbf{G}_{1}^{\mathrm{D}}$ remains separable $\mathbf{G}_{1}=(\hat{U}_{\mathrm{s}}\otimes\hat{U}_{\mathrm{p}})\mathbf{G}_{1}^{\mathrm{D}}(\hat{U}_{\mathrm{s}}^{\dagger}\otimes\hat{U}_{\mathrm{p}}^{\dagger})=\mathbf{G}_{\mathrm{s}}\otimes\mathbf{G}_{\mathrm{p}}=\mathbf{G}_{\mathrm{s}}^{\mathrm{red.}}\otimes\mathbf{G}_{\mathrm{p}}^{\mathrm{red.}}$. In this case, $\mathbf{G}_{\mathrm{s}}=\hat{U}_{\mathrm{s}}\left(\begin{array}{cc}1&0\\0&0\end{array}\right)\hat{U}_{\mathrm{s}}^{\dagger}$ and $\mathbf{G}_{\mathrm{p}}=\hat{U}_{\mathrm{p}}\left(\begin{array}{cc}1&0\\0&0\end{array}\right)\hat{U}_{\mathrm{p}}^{\dagger}$, so that both $\mathbf{G}_{\mathrm{s}}$ and $\mathbf{G}_{\mathrm{p}}$ correspond to coherent spatial and polarization coherence matrices, respectively, with $S(\mathbf{G}_{\mathrm{s}})=S(\mathbf{G}_{\mathrm{p}})=0$. The two DoFs are independent of each other and both are free of random fluctuations.

If instead $\hat{U}$ is \textit{not} separable, then $\mathbf{G}_{1}=\hat{U}\mathbf{G}_{1}^{\mathrm{D}}\hat{U}^{\dagger}$ in turn becomes non-separable, $\mathbf{G}_{1}\neq\mathbf{G}_{\mathrm{s}}\otimes\mathbf{G}_{\mathrm{p}}$. However, in this scenario we can write $\mathbf{G}_{1}=|u_{1}\rangle\langle u_{1}|$, where $|u_{1}\rangle$ is the first column in $\hat{U}$, $\langle u_{1}|u_{1}\rangle=1$, so that $\mathbf{G}_{1}$ is a projection operator (outer product) as expected for a fully coherent field \cite{Gamo64PO}. In this case, $\mathbf{G}_{\mathrm{s}}^{\mathrm{red.}}$ and $\mathbf{G}_{\mathrm{p}}^{\mathrm{red.}}$ no longer correspond to coherent DoFs. That is, $\mathbf{G}_{\mathrm{s}}^{\mathrm{red.}}$ and $\mathbf{G}_{\mathrm{p}}^{\mathrm{red.}}$ represent partially coherent DoFs; indeed $S(\mathbf{G}_{\mathrm{s}}^{\mathrm{red.}})\neq0$ and $S(\mathbf{G}_{\mathrm{p}}^{\mathrm{red.}})\neq0$ but $S(\mathbf{G}_{\mathrm{s}}^{\mathrm{red.}})=S(\mathbf{G}_{\mathrm{p}}^{\mathrm{red.}})$ as noted earlier. The non-separability of $\mathbf{G}_{1}$ induces \textit{effective partial coherence} in each DoF when the other DoF is traced out. This phenomenon has been coined `classical entanglement', whereby a coherent optical field that is not separable with respect to two DoFs displays partial coherence in each DoF when considered separately \cite{Spreeuw98FP,Kagalwala13NP,Forbes19PO}, and is quantified by a `degree of classical entanglement' $C$. The \textit{restricted} coherence matrices, on the other hand, remain coherent, whether $\mathbf{G}_{a}$ and $\mathbf{G}_{b}$ for the polarization DoF, or $\mathbf{G}_{\mathrm{H}}$ and $\mathbf{G}_{\mathrm{V}}$ for the spatial DoF.

Because the overall field is fully coherent ($S=0$), and yet each DoF considered separately may not be coherent, there is a complementarity between the degree of coherence of either DoF and the degree of classical entanglement. For coherent fields, this complementarity takes the form of an equality \cite{Eberly17Optica,Abouraddy19Optica}:
\begin{equation}
D^{2}+C^{2}=1,
\end{equation}
which can apply equally to the spatial and polarization DoFs (an earlier version of this equality was developed in quantum optics \cite{Wootters79PRD,Horne89PRL,Jaeger93,Jaeger95PRA,Horne97InBook,Saleh00PRA,Abouraddy01PRA2}, where $C$ corresponds to the concurrence \cite{Hill97PRL}). When $\mathbf{G}_{1}$ is separable, $D_{\mathrm{s}}=D_{\mathrm{p}}=1$ and $C=0$. When $\mathbf{G}_{1}$ has maximal classical entanglement, $D_{\mathrm{s}}=D_{\mathrm{p}}=0$ and $C=1$.

The condition $S=0$ is a unique identifier of rank-1 fields: any field for which $S=0$ is rank-1 and vice versa. Furthermore, all rank-1 fields can be inter-converted into each other via unitaries.

\subsubsection{Rank-2 fields}

Rank-2 fields have two non-zero eigenvalues $\lambda_{1},\lambda_{2}\neq0$, $\lambda_{1}+\lambda_{2}=1$, and $\lambda_{3}=\lambda_{4}=0$. In diagonalized form, a rank-2 field $\mathbf{G}_{2}^{\mathrm{D}}$ is given by: 
\begin{equation}\label{eq:Rank2}
\mathbf{G}_{2}^{\mathrm{D}}=\mathrm{diag}\{\lambda_{1},\lambda_{2},0,0\}=\left(\begin{array}{cc}1&0\\0&0\end{array}\right)_{\mathrm{s}}\otimes\left(\begin{array}{cc}\lambda_{1}&0\\0&\lambda_{2}\end{array}\right)_{\mathrm{p}},
\end{equation}
and the field entropy is $S=-\lambda_{1}\log_{2}\lambda_{1}-\lambda_{2}\log_{2}\lambda_{2}$, with $0\leq S\leq1$, so that a rank-2 field can carry at most 1-bit of entropy, with the maximum-entropy condition associated with equal eigenvalues, $\lambda_{1}=\lambda_{2}=\tfrac{1}{2}$. In general, $S_{\mathrm{s}}+S_{\mathrm{p}}\geq S$, where $S_{\mathrm{s}}$ and $S_{\mathrm{p}}$ are the spatial and polarization entropies, respectively, obtained from $\mathbf{G}_{\mathrm{s}}^{\mathrm{red.}}$ and $\mathbf{G}_{\mathrm{p}}^{\mathrm{red.}}$, and equality is achieved when the field is separable with respect to its two DoFs.

\begin{figure}[t!]
\centering
\includegraphics[width=3.8in]{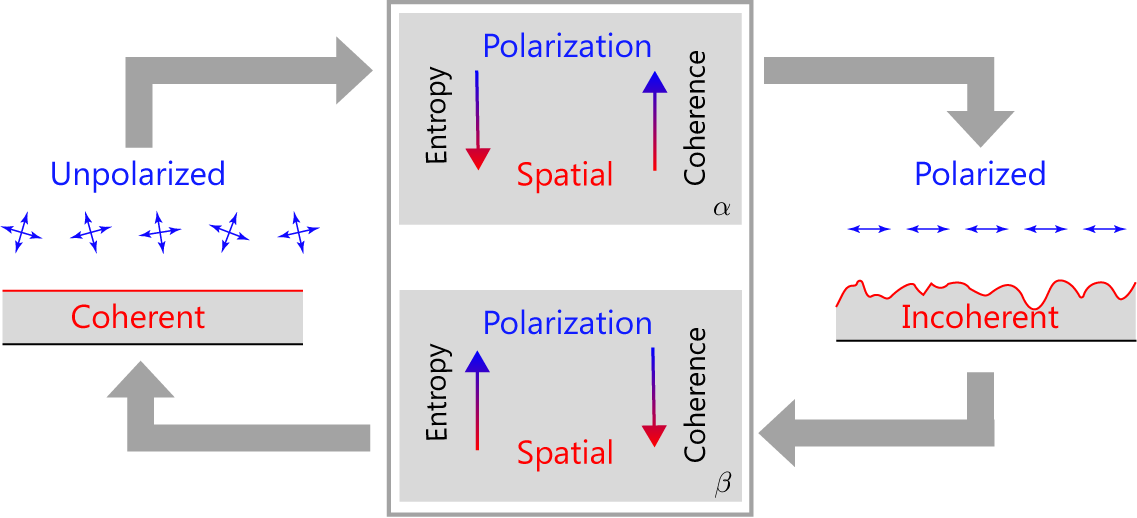}
\caption{Reversible coherence conversion or entropy swapping between two DoFs of a rank-2 field. The maximum-entropy rank-2 field ($S=1$~bit) that is initially separable (on the left) with coherent spatial DoF and incoherent polarization DoF; that is, the field is unpolarized but displays full-visibility spatial interference fringes (e.g., in a Young double-slit interferometer). The entropy can be swapped between the two DoFs, resulting in a separable field (on the right) that is polarized but spatially incoherent, in which case no spatial interference fringes are observed. Moreover, the procedure is reversible \cite{Harling22OE,Harling23JO}.}
\label{fig:RecCoherenceConv}
\end{figure}

Rank-2 fields have several fascinating properties:
\begin{enumerate}
\item The diagonalized rank-2 coherence matrix $\mathbf{G}_{2}^{\mathrm{D}}$ is separable with respect to the two DoFs, $\mathbf{G}_{2}^{\mathrm{D}}=\mathbf{G}_{\mathrm{s}}\otimes\mathbf{G}_{\mathrm{p}}$, with $\mathbf{G}_{\mathrm{s}}^{\mathrm{red.}}=\mathbf{G}_{\mathrm{s}}=\left(\begin{array}{cc}1&0\\0&0\end{array}\right)$ and $\mathbf{G}_{\mathrm{p}}^{\mathrm{red.}}=\mathbf{G}_{\mathrm{p}}=\left(\begin{array}{cc}\lambda_{1}&0\\0&\lambda_{2}\end{array}\right)$. In this specific factorization, the spatial DoF is completely coherent ($D_{\mathrm{s}}=1$ and $S_{\mathrm{s}}=0$), whereas the polarization DoF is partially coherent ($D_{\mathrm{p}}=\lambda_{1}-\lambda_{2}$ and entropy $S_{\mathrm{p}}=-\lambda_{1}\log_{2}\lambda_{1}-\lambda_{2}\log_{2}\lambda_{2}=S$). Because $\mathbf{G}_{2}^{\mathrm{D}}$ is separable, then $S=S_{\mathrm{s}}+S_{\mathrm{p}}=S_{\mathrm{p}}$. The roles of the spatial and polarization DoFs can of course be switched.

\item
Because any rank-2 coherence matrix $\mathbf{G}_{2}$ can be diagonalized via a unitary, $\mathbf{G}_{2}=\hat{U}\mathbf{G}_{2}^{\mathrm{D}}\hat{U}^{\dagger}$, a rank-2 field \textit{can always be rendered separable with respect to its DoFs.}

\item
When $\hat{U}$ renders $\mathbf{G}_{2}$ separable, the entropy is \textit{not} divided between the spatial and polarization DoFs. Rather, once $\mathbf{G}_{2}$ is rendered separable (which is \textit{always} possible for rank-2 fields), one DoF takes on \textit{all} the entropy (becoming partially coherent or incoherent), leaving the other DoF fully coherent. All the entropy associated with $\mathbf{G}_{2}$ is then assigned to only one DoF. The process of diagonalization of a rank-2 field `concentrates' the entropy into one DoF.

\item
When $\mathbf{G}_{2}$ is \textit{not} separable, both DoFs can be partially coherent or indeed fully incoherent, with $S_{\mathrm{s}}+S_{\mathrm{p}}\geq S$.

\item
Rank-2 fields have a remarkable property: the entropy can be fully `swapped' between the two DoFs. Consider a rank-2 field with maximum entropy $S=1$, $\mathbf{G}_{2}^{\mathrm{D}}=\mathrm{diag}\{\tfrac{1}{2},\tfrac{1}{2},0,0\}$. Because the spatial DoF is fully coherent, such a field displays full double-slit interference visibility $V=1$. However, by simply re-arranging the diagonal elements, $\mathbf{G}_{2}^{\mathrm{red}}=\mathrm{diag}\{\tfrac{1}{2},0,\tfrac{1}{2},0\}$, we have an H-polarized (scalar) field that is spatially incoherent, $\mathbf{G}_{2}^{\mathrm{D}}=\tfrac{1}{2}\left(\begin{array}{cc}1&0\\0&1\end{array}\right)_{\mathrm{s}}\otimes\left(\begin{array}{cc}1&0\\0&0\end{array}\right)_{\mathrm{p}}$, and no double-slit interference fringes are observed. In other words, the coherence has been converted (or the entropy swapped) between the two DoFs; see Fig.~\ref{fig:RecCoherenceConv}. This process is \textit{reversible} \cite{Harling22OE,Harling23JO}, and is discussed in more detail below (see also Sec.~\ref{eq:VisEntropyConversion}).

\item
Unlike rank-1 fields where the reduced coherence matrices always yield $D_{\mathrm{s}}=D_{\mathrm{p}}$, this does not necessarily hold for rank-2 fields. As a counter-example, consider $\mathbf{G}_{2}^{\mathrm{D}}$ in Eq.~\ref{eq:Rank2} where $D_{\mathrm{p}}=\lambda_{1}-\lambda_{2}\neq D_{\mathrm{s}}=1$, so that $D_{\mathrm{s}}\neq D_{\mathrm{p}}$.

\item The total entropy determines $\mathbf{G}_{2}$ to within a unitary, so that the entropy is a unique identifier of rank-2 fields. In other words, any two iso-entropy rank-2 fields can always be inter-converted into each other unitarily. Conversely, any rank-2 fields that can be inter-converted into each other unitarily have the same value of entropy. For a rank-2 field, $S(\lambda_{1})=-\lambda_{1}\log_{2}\lambda_{1}-(1-\lambda_{1})\log_{2}(1-\lambda_{1})$ has the same form as in Fig.~\ref{fig:EntropyOneDoF}(a) for a single binary DoF.

\end{enumerate}

\begin{figure}[t!]
\centering
\includegraphics[width=4in]{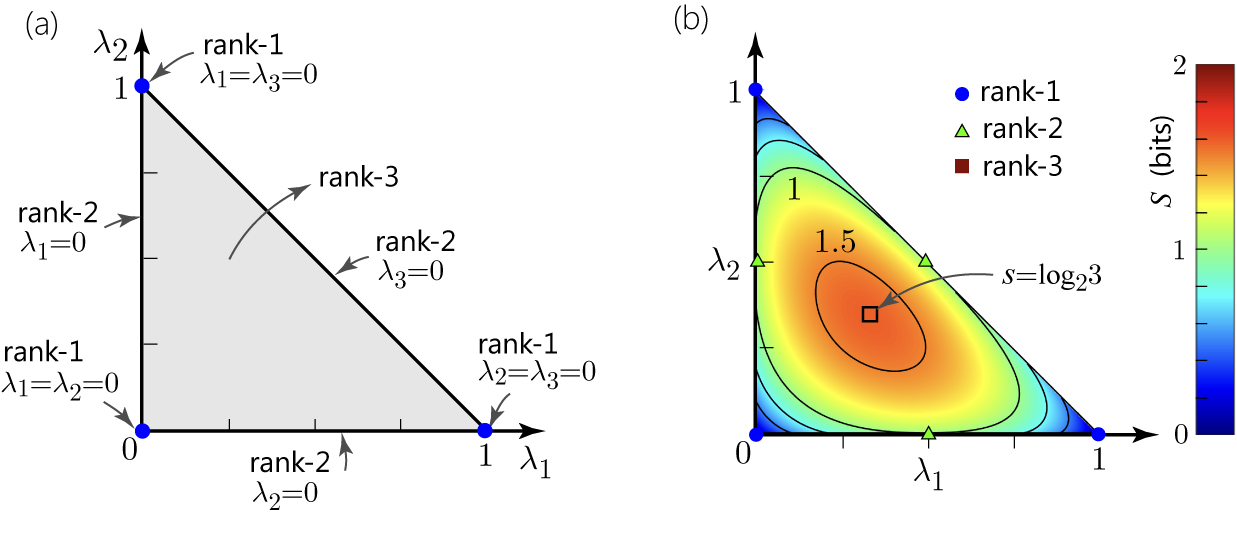}
\caption{Geometric representation of the coherence rank of a $4\times4$ coherence matrix $\mathbf{G}$ limited to rank-3. (a) Each point in the right-angled triangle represents the iso-entropy rank-3 fields that can be inter-converted into each other unitarily. The vertices correspond to the rank-1 limit, the edges to the rank-2 limit, and the face to rank-3 fields. (b) Plot of the entropy $S$ for rank-3 fields on the triangle from (a). We also plot iso-entropy contours for $S=0.5,1$, and 1.5.}
\label{fig:Triangle}
\end{figure}

\subsubsection{Rank-3 fields}

Rank-3 fields are those whose coherence matrix takes the following form after diagonalization: $\mathbf{G}_{3}^{\mathrm{D}}=\mathrm{diag}\{\lambda_{1},\lambda_{2},\lambda_{3},0\}$, with $\lambda_{1}+\lambda_{2}+\lambda_{3}=1$ and $\lambda_{4}=0$. We define the entropy of rank-3 fields as $S=-\mathrm{Tr}\{\mathbf{G}_{3}\log_{2}\mathbf{G}_{3}\}=-\sum_{j=1}^{3}\lambda_{j}\log_{2}\lambda_{j}$, which extends over the span $0<S\leq\log_{2}3\approx1.585$. The maximum-entropy condition is reached when the eigenvalues are equal, $\lambda_{1}=\lambda_{2}=\lambda_{3}=\tfrac{1}{3}$. 

Rank-3 fields behave fundamentally differently from their rank-2 counterparts. This stems from the fact that rank-3 fields are intrinsically \textit{non-separable}: it is impossible to render a rank-3 field in the separable form $\mathbf{G}=\mathbf{G}_{\mathrm{s}}^{\mathrm{red.}}\otimes\mathbf{G}_{\mathrm{p}}^{\mathrm{red.}}$. This has profound implications for entropy concentration and swapping.

Any binary DoF can carry at most 1~bit of entropy. Rank-3 fields can have $S>1$~bit, in which case we would expect that each DoF would be partially coherent. However, we may also expect that in the case of a rank-3 field with $S<1$~bit, one may concentrate all the entropy into one DoF, thus leaving the other DoF coherent (similarly to the case for rank-2 fields). Nevertheless, because rank-3 fields are intrinsically non-separable, each DoF remains partially incoherent no matter how low $S$ is for the field. In other words, for rank-2 fields, the total entropy of the field can always be fully concentrated into one DoF, leaving the other DoF coherent -- no matter how high $S$ is for the field. In contrast, for rank-3 fields, the total entropy can\textit{not} be concentrated into one DoF, thus leaving the other DoF partially coherent -- no matter how low $S$ is for the field. Some entropy must remain associated with each DoF, which we have termed `locked entropy' \cite{Harling24PRA}.

The entropy for a rank-3 field is spanned by two independent eigenvalues $0\leq\lambda_{1},\lambda_{2}\leq1$ once we enforce the normalization $\lambda_{1}+\lambda_{2}+\lambda_{3}=1$, and the entropy can thus be plotted in a plane within a domain defined by a right-angled isosceles triangle [Fig.~\ref{fig:Triangle}(a)]. The vertices correspond to rank-1 fields: $\mathbf{G}_{2}=\mathrm{diag}\{1,0,0,0\}$, $\{0,1,0,0\}$, and $\{0,0,1,0\}$; the sides of the triangle correspond to rank-2 fields, $\mathbf{G}_{2}=\mathrm{diag}\{\lambda_{1},\lambda_{2},0,0\}$, $\{\lambda_{1},0,\lambda_{3},0\}$, and $\{0,\lambda_{2},\lambda_{3},0\}$; and the face of the triangle corresponds to rank-3 fields (with $\lambda_{4}=0$). Every point on this surface represents the family of coherence matrices that can be inter-converted into each other unitarily and thus have the same entropy. However, iso-entropy rank-3 fields correspond to a one-parameter set represented geometrically by a curve in the $(\lambda_{1},\lambda_{2})$-plane; see Fig.~\ref{fig:Triangle}(b). Each point on this curve represents the family of iso-entropy fields that \textit{can} be inter-converted into each other unitarily, whereas different points on this curve represents families of iso-entropy fields that can\textit{not} be inter-converted into each other unitarily. Consequently, moving from one point to another along an iso-entropy curve for rank-3 fields typically requires non-unitary filtering and decohering transformations.

When $S>1$, which precludes rank-2 fields for which $S\leq1$, the iso-entropy curve is closed and fully contained within the triangle [Fig.~\ref{fig:Triangle}(b)]. When $S=1$, the curve is tangential to the triangle sides, touching them at the points corresponding to the rank-2 fields with maximum entropy: $\mathbf{G}_{3}=\mathrm{diag}\{\tfrac{1}{2},\tfrac{1}{2},0,0\}$, $\{\tfrac{1}{2},0,\tfrac{1}{2},0\}$, and $\{0,\tfrac{1}{2},\tfrac{1}{2},0\}$. When $S<1$, the iso-entropy curve becomes disjoint, with ending points of the curved segments on the triangle sides corresponding to the rank-2 field limits that have the same entropy.

\subsubsection{Rank-4 fields}

The most general case is that of rank-4 fields in which all the eigenvalues are non-zero. Rank-4 fields combine features of both rank-2 and rank-3 fields. Crucially, a rank-4 field may or may not be separable, depending on the particular values of the eigenvalues $\{\lambda_{j}\}$. Specifically, if $\lambda_{1}\lambda_{4}=\lambda_{2}\lambda_{3}$, then the rank-4 field is separable \cite{Abouraddy01PRA,Harling24PRA,Harling24PRA2},
\begin{equation}
\mathbf{G}_{4}^{\mathrm{D}}=\mathbf{G}_{\mathrm{s}}\otimes\mathbf{G}_{\mathrm{p}}=
\left(\begin{array}{cc}\gamma_{1}&0\\0&\gamma_{2}\end{array}\right)
\otimes
\left(\begin{array}{cc}\gamma_{3}&0\\0&\gamma_{4}\end{array}\right),  
\end{equation}
where $\lambda_{1}=\gamma_{1}\gamma_{3}$, $\lambda_{2}=\gamma_{1}\gamma_{4}$, $\lambda_{3}=\gamma_{2}\gamma_{3}$, $\lambda_{4}=\gamma_{2}\gamma_{4}$, $\gamma_{1}+\gamma_{2}=1$, and $\gamma_{3}+\gamma_{4}=1$. In this case, we can obtain the factors through $\gamma_{1,2}=\tfrac{1}{2}\{1\pm\sqrt{1-4(\lambda_{1}+\lambda_{2})(\lambda_{3}+\lambda_{4})}\}$, $\gamma_{3}=\lambda_{1}/\gamma_{1}$, and $\gamma_{4}=1-\gamma_{3}$. In contrast, when $\lambda_{1}\lambda_{4}\neq\lambda_{2}\lambda_{3}$, then $\mathbf{G}_{4}^{\mathrm{D}}$ is not separable. We consider below the visual representation of iso-frequency rank-4 fields.

\subsection{Radial versus angular parameters (coordinates)}

When considering a binary DoF, the $2\times2$ coherence matrix could be visualized on or within the PS (Section~\ref{sec:PSandSP}). This is because a unity-trace $2\times2$ Hermitian matrix is identified by~3 real parameters, which can be thought of as the coordinates of a point in a 3D space (e.g., the Stokes parameters). In this $2\times2$ case, only one radial parameter exists, which is the distance of the point on the PS surface or within its volume from the origin, while the two remaining real parameters are angular parameters that vary with unitaries.

\begin{figure}[t!]
\centering
\includegraphics[width=4in]{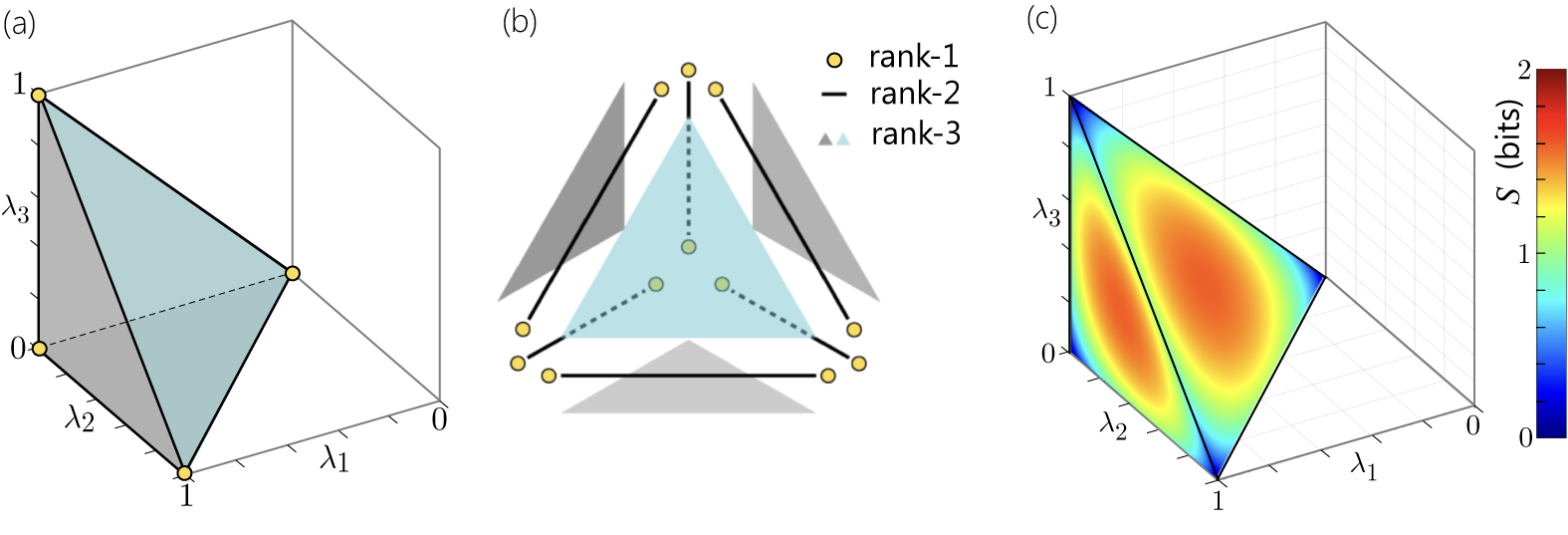}
\caption{(a) Geometric representation of the coherence rank of a $4\times4$ coherence matrix $\mathbf{G}$. Each point in the volume defined by the right-angled pyramid represents all the iso-entropy fields that can be inter-converted into each other unitarily. (b) An exploded view of the volume in (a). The vertices correspond to rank-1 fields, the edges to rank-2, the faces to rank-3, and the volume to rank-4. (c) Plot of the entropy $S$ on the faces of the volume from (a), corresponding to rank-1, rank-2, and rank-3 fields. The entropy of the rank-4 fields within the volume are depicted in Fig.~\ref{fig:IsoEntropySurface}.}
\label{fig:Pyramid}
\end{figure}

This visualizable representation is not possible with the $4\times4$ unity-trace Hermitian coherence matrix for two binary DoFs, which is characterized by 15~real parameters. However, a visualization is made possible by focusing only on the three unitary-invariant radial parameters identified with the eigenvalues of $\mathbf{G}$ (with the fourth fixed by the trace normalization). The remaining 12~real parameters that identify $\mathbf{G}$ are angular parameters. We thus focus on a representation of the coherence matrix $\mathbf{G}$ in a space spanned by the eigenvalues $\{\lambda_{1},\lambda_{2},\lambda_{3},\lambda_{4}\}$, where each eigenvalue is defined over the range $[0,1]$. Each point in this space represents an entire family of coherence matrices that can be interconverted into each other via unitaries. A point in this space is identified with a diagonal coherence matrix $\mathbf{G}^{\mathrm{D}}=\mathrm{diag}\{\lambda_{1},\lambda_{2},\lambda_{3},\lambda_{4}\}$ and coherence matrices of the form $\mathbf{G}=\hat{U}\mathbf{G}^{\mathrm{D}}\hat{U}^{\dagger}$ for all possible $4\times4$ unitaries $\hat{U}$. This is still a 4D space that cannot be visualized directly. However, the subspace of viable coherence matrices is defined by the constraint $\lambda_{1}+\lambda_{2}+\lambda_{3}+\lambda_{4}=1$, $0\leq\lambda_{j}\leq1$, $j=1,\cdots,4$. This is a hyperplane in the 4D space $\{\lambda_{1},\lambda_{2},\lambda_{3},\lambda_{4}\}$. When projected onto the restricted 3D space spanned by $\{\lambda_{1},\lambda_{2},\lambda_{3}\}$, this hyperplane becomes the volume of the right-angled pyramid depicted in Fig.~\ref{fig:Pyramid}(a), with 3 right-angled isosceles triangles and an equilateral triangle as faces, and vertices at $(0,0,0)$, $(0,0,1)$, $(0,1,0)$, and $(1,0,0)$. Each point $(\lambda_{1},\lambda_{2},\lambda_{3})$ in this volume represents the family of coherence matrices $\mathbf{G}=\hat{U}\mathbf{G}^{\mathrm{D}}\hat{U}^{\dagger}$ that can be unitarily inter-converted into each other, with $\mathbf{G}^{\mathrm{D}}=\mathrm{diag}\{\lambda_{1},\lambda_{2},\lambda_{3},1-\lambda_{1}-\lambda_{2}-\lambda_{3})$.

A useful aspect of this geometrical representation is that the different-rank coherence matrices correspond to different features of the pyramid structure, as highlighted in the exploded view depicted in Fig.~\ref{fig:Pyramid}(b). 

(1) Rank-1 coherence matrices correspond to the vertices of the pyramid. Each vertex corresponds to one of the eigenvalues having unity value and all others equal to zero. The vertex at the origin is associated with $\lambda_{4}=1$ and $\lambda_{1}=\lambda_{2}=\lambda_{3}=0.$

(2) Rank-2 coherence matrices correspond to the 6~edges of the pyramid. The edge along the $\lambda_{1}$-axis, where $\lambda_{2}=\lambda_{3}=0$ corresponds to the coherence matrix $\mathbf{G}^{\mathrm{D}}=\mathrm{diag}\{\lambda_{1},0,0,\lambda_{4}\}$ with $\lambda_{1}+\lambda_{4}=1$, and so forth for the other 5~edges: $\{\lambda_{1},\lambda_{2},0,0\}$ along the edge in the $(\lambda_{1},\lambda_{2})$-plane (orthogonal to the $\lambda_{3}$-axis) with $\lambda_{1}+\lambda_{2}=1$; $\{\lambda_{1},0,\lambda_{3},0\}$ along the edge in the $(\lambda_{1},\lambda_{3})$-plane (orthogonal to the $\lambda_{2}$-axis) with $\lambda_{1}+\lambda_{3}=1$; $\{0,\lambda_{2},\lambda_{3},0\}$ along the edge in the $(\lambda_{2},\lambda_{3})$-plane (orthogonal to the $\lambda_{1}$-axis) with $\lambda_{2}+\lambda_{3}=1$; $\{0,\lambda_{2},0,\lambda_{4}\}$ along the edge coinciding with the $\lambda_{2}$-axis, with $\lambda_{2}+\lambda_{4}=1$; and $\{0,0,\lambda_{3},\lambda_{4}\}$ along the edge coinciding with the $\lambda_{3}$-axis, with $\lambda_{3}+\lambda_{4}=1$.

(3) Rank-3 coherence matrices correspond to the 4~faces of the pyramid. The 3 right-angled triangular faces correspond to the coherence matrices $\mathbf{G}^{\mathrm{D}}=\mathrm{diag}\{\lambda_{1},\lambda_{2},0,\lambda_{4}\}$, $\{\lambda_{1},0,\lambda_{3},\lambda_{4}\}$, and $\{0,\lambda_{2},\lambda_{3},\lambda_{4}\}$. The equilateral triangular face corresponds to $\mathbf{G}^{\mathrm{D}}=\mathrm{diag}\{\lambda_{1},\lambda_{2},\lambda_{3},0\}$; see Fig.~\ref{fig:Triangle}(a).

(4) Rank-4 fields with $\lambda_{4}\neq0$ and $\lambda_{1}+\lambda_{2}+\lambda_{3}<1$ correspond to the volume of the pyramid.

\subsection{Non-uniqueness of the entropy}\label{sec:EntropyNonUniqueness}

For $4\times4$ coherence matrices associated with the two binary DoFs, the entropy $S=-\sum_{j=1}^{4}\lambda_{j}\log_{2}\lambda_{j}$ does \textit{not} identify the eigenvalues of $\mathbf{G}$. Indeed, the entropy usually cannot necessarily identify even the rank of $\mathbf{G}$.

Rank-1 fields, associated with the vertices of the pyramid, uniquely correspond to $S=0$. Rank-2 fields $\mathbf{G}^{\mathrm{D}}=\mathrm{diag}\{\lambda_{1},\lambda_{2},0,0\}$ with $\lambda_{1}+\lambda_{2}=1$ have $0<S\leq1$, with the maximum-entropy rank-2 field $S=1$~bit corresponding to $\lambda_{1}=\lambda_{2}=\tfrac{1}{2}$. Therefore, the entropy along the pyramid edges (the locus of rank-2 fields) vary from $S=0$ at the vertices to $S=1$ at the midpoints -- similarly to the entropy for a binary DoF [Fig.~\ref{fig:Triangle}(b) and Fig.~\ref{fig:Pyramid}(c)]. Rank-3 fields $\mathbf{G}^{\mathrm{D}}=\mathrm{diag}\{\lambda_{1},\lambda_{2},\lambda_{3},0\}$ have $0<S\leq\log_{2}3\approx1.585$~bits, with the maximum-entropy rank-3 fields corresponding to $\lambda_{1}=\lambda_{2}=\lambda_{3}=\tfrac{1}{3}$. The entropy on each face is minimum in the vicinity of the vertices (reaching $S=0$ at the vertices themselves) and reach a minimum at the center of the face. Iso-entropy fields correspond to a one-parameter \textit{curve} in the face. Rank-4 fields have $0<S\leq2$, where the maximum-entropy rank-4 field with maximum entropy $S=2$~bits corresponding to $\lambda_{1}=\lambda_{2}=\lambda_{3}=\lambda_{4}=\tfrac{1}{4}$. The entropy in the volume is a 3-parameter function reaching the maximum at the center point. Iso-entropy fields correspond to a 2-parameter \textit{surface} in the volume.

Therefore, the value of the entropy $S$ only partially identifies the coherence rank. The entropy uniquely identifies the coherence rank in two cases: when $1.585<S\leq2$, then the field is rank-4; and when $S=0$, then the field is rank-1. However, when $1<S\leq1.585$, then the field is either rank-3 or rank-4; and when $0<S\leq1$, then the field is either rank-2, rank-3, or rank-4.

\begin{figure}[t!]
\centering
\includegraphics[width=4.80in]{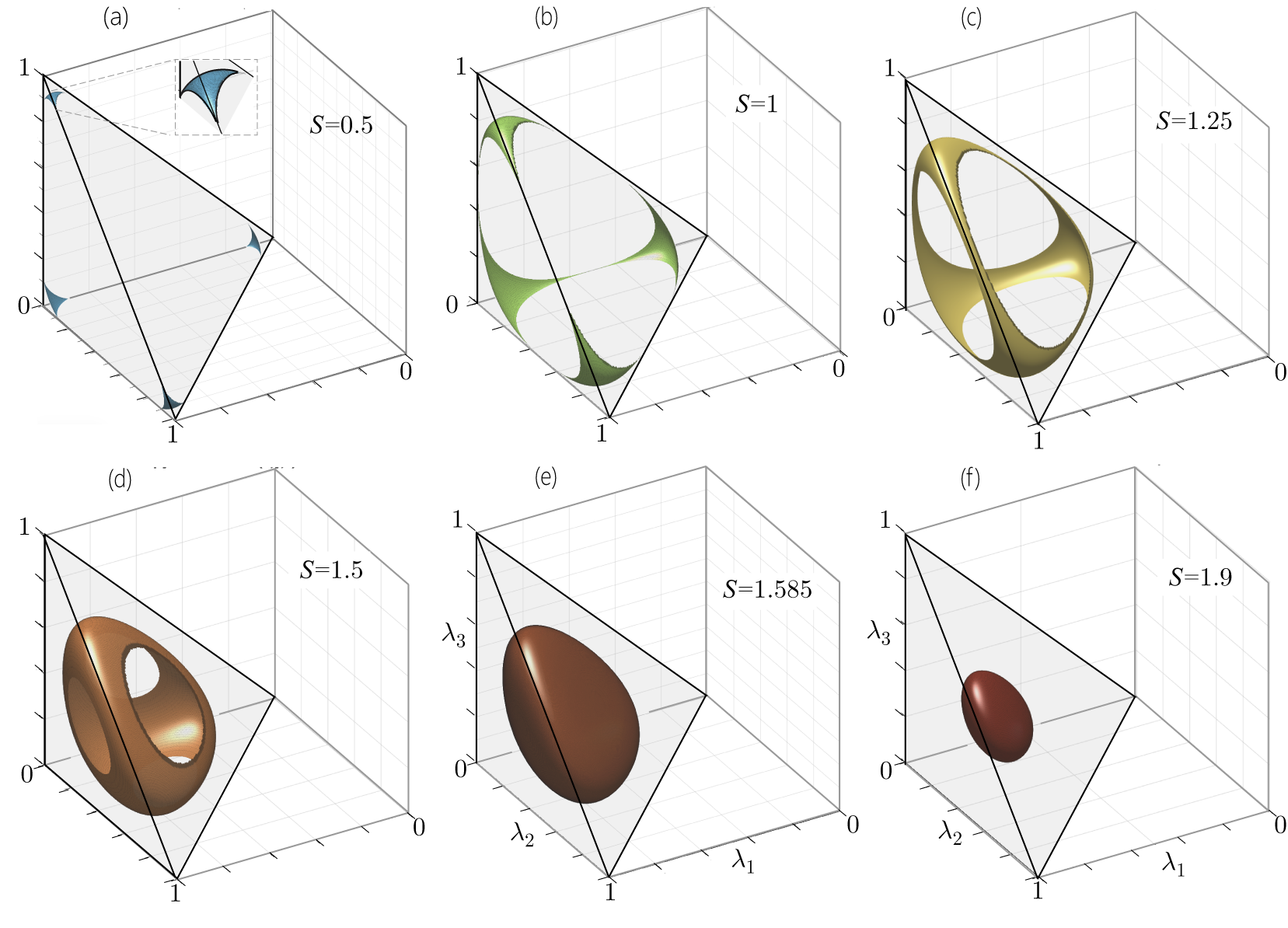}
\caption{Iso-entropy surfaces. (a) $S=0.5$; the surface intersects with the edges. (b) $S=1$; the surface is tangential to the edges. (c) $S=1.25$; the surface no longer reaches the edges but still intersects with the faces. (d) $S=1.5$. (e) $S\approx1.585$; the surface is tangential to the faces of the volume. (f) $S=1.9$~bits; the surface is entirely interior to the volume.}
\label{fig:IsoEntropySurface}
\end{figure}

Each point in the pyramid in the restricted space $(\lambda_{1},\lambda_{2},\lambda_{3})$ represents a family of iso-entropy coherence matrices that can be unitarily inter-converted into each other $\mathbf{G}=\hat{U}\mathbf{G}^{\mathrm{D}}\hat{U}^{\dagger}$. Additionally, permutations of the same eigenvalues yield the same entropy. However, different sets of the eigenvalues, and even different-rank coherence matrices, can have the same entropy, as illustrated in Fig.~\ref{fig:IsoEntropySurface}.

We plot in Fig.~\ref{fig:IsoEntropySurface}(a-f) iso-entropy surfaces in the restricted space $\{\lambda_{1},\lambda_{2},\lambda_{3}\}$. The surface $S=0.5$ [Fig.~\ref{fig:IsoEntropySurface}(a)] is concentrated in the vicinity of the vertices. The surface intersects with each edge at two \textit{points} corresponding to the rank-2 fields with $S=0.5$. This iso-entropy surface intersects with the faces in \textit{curves} corresponding to the rank-3 fields with $S=0.5$. The portion of the iso-entropy \textit{surface} inside the pyramid corresponds to the rank-4 fields with $S=0.5$. When $S=1$ [Fig.~\ref{fig:IsoEntropySurface}(b)], the iso-entropy surface is tangential to the edges at their midpoints, corresponding to the maximum-entropy rank-2 fields, the curves at the intersection with the faces correspond to rank-3 fields, and the surfaces within the volume to rank-4 fields. This reflects the fact that for $0\leq S\leq1$, the field may be rank-2, rank-3, or rank-4.

When $S>1$ [$S=1.25$ in Fig.~\ref{fig:IsoEntropySurface}(c)], the iso-entropy surface no longer intersects with the edges because $S>1$ excludes rank-2 fields. The iso-entropy surface still intersects with the pyramid faces in curves corresponding to rank-3 fields with $S>1$. The size of this curve shrinks as $S$ increases [$S=1.5$~bits in Fig.~\ref{fig:IsoEntropySurface}(d)]. When $S=\log_{2}3$, the surface is tangential to the facets of the pyramid at the points corresponding to the maximum-entropy rank-3 fields $\lambda_{1}=\lambda_{2}=\lambda_{3}=\tfrac{1}{3}$ on the front isosceles facet, $\lambda_{1}=\lambda_{2}=\lambda_{4}=\tfrac{1}{3}$, $\lambda_{1}=\lambda_{3}=\lambda_{4}=\tfrac{1}{3}$, and $\lambda_{2}=\lambda_{3}=\lambda_{4}=\tfrac{1}{3}$ on the right-angled facets [Fig.~\ref{fig:IsoEntropySurface}(e)]. When $S>\log_{2}3$, the iso-entropy surface continues to shrink and is completely enclosed within the pyramid [Fig.~\ref{fig:IsoEntropySurface}(f)]. This reflects the fact that $S>1.585$~bits corresponds to only rank-4 fields. Finally, when $S=2$, the surface shrinks to a points at $\lambda_{1}=\lambda_{2}=\lambda_{3}=\lambda_{4}=\tfrac{1}{4}$.

\subsection{Entropy conversion}\label{sec:EntropySwapping}

A binary DoF can support 1~bit of entropy, and two binary DoFs can support 2~bits. It two DoFs carry entropy~$S$, what is the minimum entropy that can be associated with one DoF after operating with a unitary? This question becomes particularly interesting when $S<1$, because it can be imagined that all the entropy may be converted to one DoF, leaving the other DoF coherent. Can one always concentrate the entropy when $S<1$ into one DoF via a unitary?

Recent results show that answering this question depends on the coherence rank. For rank-1 fields $S=0$, but $S_{\mathrm{s}}$ and $S_{\mathrm{p}}$ can be non-zero in presence of classical entanglement. Nevertheless, there always exist a unitary that renders $\mathbf{G}=\mathbf{G}_{\mathrm{s}}\otimes\mathbf{G}_{\mathrm{p}}$ separable, in which case each DoF is coherent, and $S_{\mathrm{s}}=S_{\mathrm{p}}=0$. Therefore, for rank-1 fields, both DoFs can always be rendered coherent unitarily.

\begin{figure}[t!]
\centering
\includegraphics[width=13.3cm]{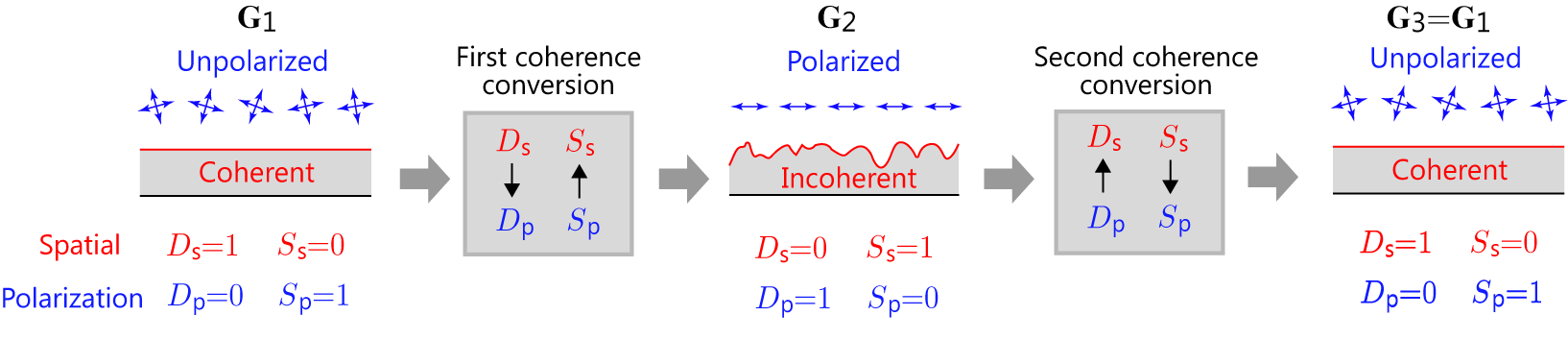}
\caption{Conceptual scheme for entropy swapping between the spatial and polarization DoFs.}
\label{fig:ReversibleConversion}
\end{figure}

Rank-2 fields, where $0\leq S\leq1$, offer interesting prospects. When $\mathbf{G}$ is \textit{not} separable with respect to the two DoFs, then $S_{\mathrm{s}}+S_{\mathrm{p}}>S$. Rank-2 fields are guaranteed to be rendered separable via a unitary transformation, whereupon $S_{\mathrm{s}}+S_{\mathrm{p}}=S$ and one of the DoFs is coherent; that is, either $S_{\mathrm{s}}=0$ and $S_{\mathrm{p}}=S$, or $S_{\mathrm{s}}=S$ and $S_{\mathrm{p}}=0$. In other words, the entropy for rank-2 field -- no matter how high -- can always be concentrated into a single DoF. Moreover, the process of converting the entropy is reversible: the entropy can be shuttled back and forth between the two DoFs via unitary transformations, as illustrated in Fig.~\ref{fig:ReversibleConversion} \cite{Harling22OE,Harling23JO}. Starting with a maximum-entropy ($S=1$) rank-2 field that is spatially coherent but unpolarized ($S_{\mathrm{s}}=0$ and $S_{\mathrm{p}}=1$), a unitary $\hat{U}$ can convert the field into one that is spatially incoherent but polarized ($S_{\mathrm{s}}=1$ and $S_{\mathrm{p}}=0$). The conversion procedure involves first coupling the two DoFs to each other before uncoupling them again after swapping the entropy between the two DoFs. The Hermitian conjugate $\hat{U}^{\dagger}$ of this unitary then converts the spatially incoherent polarized field back to its initial spatially coherent unpolarized field, thereby completing the cycle of entropy swapping.

The properties of rank-3 fields with regards to entropy concentration are in stark contrast with those of rank-2 fields. For a rank-3 field, the entropy lies within the range $0\leq S\leq\log_{2}2\approx1.585$. Of course, when $S>1$~bit, the entropy cannot be all concentrated in one DoF, leaving the other coherent, because a binary DoF can support at most 1~bit. Consider first the case of rank-3 fields with $0\leq S\leq1$. In this special case, can one DoF be rendered coherent with all the entropy associated with one DoF? We have recently shown that for rank-3 fields, no matter how low the entropy~$S$ of the field is, it is impossible to unitarily concentrate the entropy in one DoF, leaving the other DoF coherent \cite{Harling24PRA2,Harling24PRA}. In other words, any DoF must always retain some entropy that cannot be eliminated unitarily (in contradistinction to rank-2 fields), which we refer to as `locked entropy' \cite{Harling24PRA}. This can be understood by noting that rank-3 fields cannot be rendered separable unitarily. In this case, entropy swapping cannot be achieved completely as done for rank-2 fields. The maximum entropy associated with one DoF, say the spatial DoF, in  rank-3 field is $S_{\mathrm{s}}=-\lambda_{1}\log_{2}\lambda_{1}-\lambda_{2}\log_{2}\lambda_{2}<1$, leaving a minimum locked entropy of $S_{\mathrm{p}}=-\lambda_{3}\log_{2}\lambda_{3}$ in the other (polarization) DoF. To achieve entropy swapping, one couples the two DoFs before uncoupling them and exchanging the entropy, so that $S_{\mathrm{s}}=-\lambda_{3}\log_{2}\lambda_{3}$ and $S_{\mathrm{p}}=-\lambda_{1}\log_{2}\lambda_{1}-\lambda_{2}\log_{2}\lambda_{2}$. The maximum entropy that can be associated with one DoF in a rank-3 field only approach $S=1$ asymptotically, which occurs when $\lambda_{3}\rightarrow0$, $\lambda_{1}\rightarrow\tfrac{1}{2}$ and $\lambda_{2}\rightarrow\tfrac{1}{2}$, which is the maximum-entropy rank-2 limit. In the maximum-entropy rank-3 limit $\mathbf{G}_{3}=\mathrm{diag}\{\tfrac{1}{3},\tfrac{1}{3},\tfrac{1}{3},0\}$, the minimum entropies associated with each DoF (the reduced coherence matrices $\mathbf{G}_{\mathrm{s}}^{\mathrm{red.}}$ and $\mathbf{G}_{\mathrm{p}}^{\mathrm{red.}}$), $S_{\mathrm{s}}=S_{\mathrm{p}}=\log_{2}3-\tfrac{2}{3}\approx0.92$~bits, so that $S_{\mathrm{s}}+S_{\mathrm{p}}>S$ (an equality cannot be achieved).  

There are still many questions to be tackled in future research. For example, what is the limit on the change in entropy for either DoF with unitaries? Of course, the entropy of any binary DoF can be increased to the maximum value of 1~bit by coupling the two DoFs via a non-separable unitary. Another important question is the following: what is the minimum entropy that can be associated with a DoF for a given coherence matrix $\mathbf{G}$?

\subsection{Optical cross-purity}\label{sec:CrossPurity}

\subsubsection{Basic definition}

\begin{figure}[t!]
\centering
\includegraphics[width=9cm]{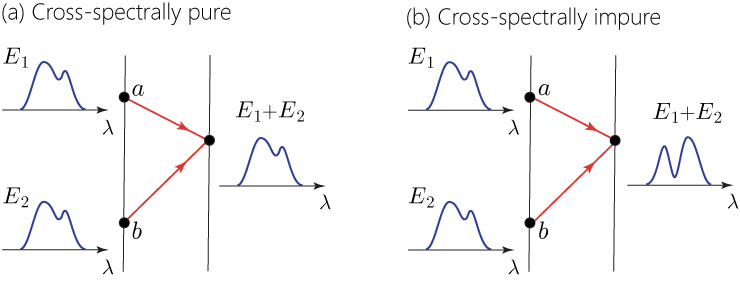}
\caption{The fields $E_{a}$ and $E_{b}$ at $|a\rangle$ and $|b\rangle$ have identical spectra. If the superposition of these two fields yields the same spectrum, we say that the field at these two points is \textit{cross-spectrally pure}. (b) If superposing the fields $E_{a}$ and $E_{b}$ from the points $a$ and $b$ yields a new spectrum, the field is said to be \textit{cross-spectrally impure} at these two points.}
\label{fig:CrossSpectralPurity}
\end{figure}

Spectral cross-purity is a phenomenon first described by L.~Mandel in 1961 using the conventional description of optical coherence in terms of continuous functions in space and time. Consider superposing the spectra from points $|a\rangle$ and $|b\rangle$, $S_{a}(\omega)$ and $S_{b}(\omega)$, respectively, in a scalar, partially coherent field. If the spectra $S_{a}(\omega)$ and $S_{b}(\omega)$ are different, it is expected that their superposition will differ from either $S_{a}(\omega)$ or $S_{b}(\omega)$. Consequently, Mandel restricted himself to cases where the spectra at $|a\rangle$ and $|b\rangle$ are identical, $S_{a}(\omega)=S_{b}(\omega)$ [Fig.~\ref{fig:CrossSpectralPurity}], with equality holding at the level of \textit{normalized} spectra (the field amplitudes at $|a\rangle$ and $|b\rangle$ need not be equal). Will the spectrum resulting from their superposition have the same structure?

Mandel found that under certain conditions, the superposition of identical (normalized) spectra from two points in the field will yield the same spectrum, a scenario he termed `cross-spectral purity' [Fig.~\ref{fig:CrossSpectralPurity}(a)]. If these conditions are not satisfied, the superposition of identical spectra produces a new spectrum, in which case the field is \textit{not} cross-spectrally pure, or is cross-spectrally \textit{impure} [Fig.~\ref{fig:CrossSpectralPurity}(b)]. We consider here the field at only two points. Much effort has been devoted to studying various extensions to \textit{all} pairs of points in the field, to corresponding conditions in space and time \cite{Koivurova25OL,Joshi25JOSAA}, to non-stationary fields \cite{Koivurova19PRA,Joshi24OC}, and adding polarization \cite{Hassinen09OL,Hassinen11APB,Chen14JMO,Peng14Optik} (see also \cite{Lahiri13OL,Hassinen13OL} and Ref.~\cite{Joshi25JOSAA} for a recent comprehensive review).

When restricting our attention to only two points in the field, a cross-spectrally pure field satisfies two conditions:
\begin{enumerate}
\item \textit{Symmetry}: the normalized spectra at $|a\rangle$ and $|b\rangle$ are identical.
\item \textit{Separability}: The coherence function at these two points is independent of the spatial coordinate; i.e., the coherence function is separable with respect to the spatial and spectral DoFs at these two points.
\end{enumerate}
The fundamental concept of cross-spectral purity has been recently generalized in two aspects: (1) it can be applied to any pair of DoFs; and (2) it is equally applicable to discretized DoFs. We thus call this general phenomenon simply `optical cross-purity'.

\subsubsection{Polarization cross-purity}

We apply generalized optical cross-purity to a scenario comprising the spatial and polarization DoFs. We thus consider the following question: in a partially coherent vector field described by a $4\times4$ coherence matrix $\mathbf{G}$, in which the polarizations at $|a\rangle$ and $|b\rangle$ are identical, would superposing the fields from $|a\rangle$ and $|b\rangle$ yield the same or different polarization? If their superposition does indeed yield the same polarization, we call the field polarization cross-pure; if their superposition yields a new polarization, we call the field polarization cross-impure.

We generalize this question to arbitrary $4\times4$ unitaries on the field, yielding two points after the unitary rather than one as typically assumed in the context of cross-spectral purity. In other words, when the field is polarization cross-impure, two new states of polarization are produced that may both differ from that at $|a\rangle$ and $|b\rangle$. When the field is polarization cross-pure, the two new states coincide with the polarization state at $|a\rangle$ and $|b\rangle$. 

\begin{figure}[t!]
\centering
\includegraphics[width=3in]{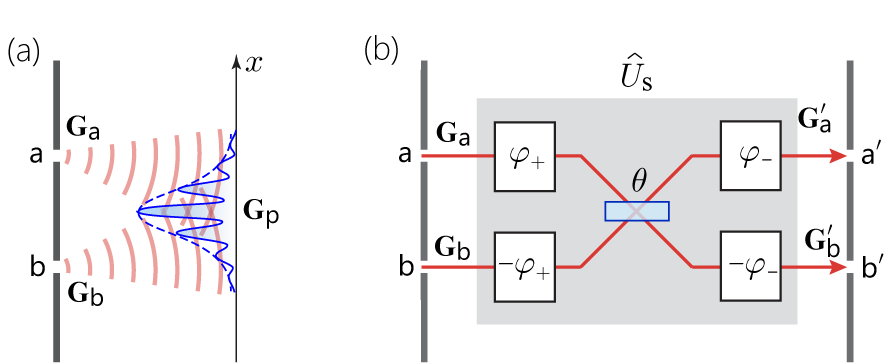} 
\caption{(a) A vector field at $|a\rangle$ and $|b\rangle$, with restricted polarization coherence matrices $\mathbf{G}_{a}$ and $\mathbf{G}_{b}$, respectively, superposed to yield a polarization coherence matrix $\mathbf{G}_{\mathrm{p}}$. (b) A spatial unitary $\hat{U}_{\mathrm{s}}$ maps $\mathbf{G}_{a}$ and $\mathbf{G}_{b}$ to $\mathbf{G}_{a}'$ and $\mathbf{G}_{b}'$; where $\varphi_{+}=(\varphi_{1}+\varphi_{2})/2$ and $\varphi_{-}=(\varphi_{1}-\varphi_{2})/2$, and $\theta$ `rotates' the space of the two spatial modes.}
\label{fig:PolarizationCrossPurity}
\end{figure}

\subsubsection{Conditions for polarization cross-purity}

Rather than superposing the fields from $|a\rangle$ and $|b\rangle$ in the double-slit experiment [Fig.~\ref{fig:PolarizationCrossPurity}(a)], we instead implement a unitary operator $\hat{U}_{\mathrm{s}}$ that impacts the spatial DoF alone and is independent of polarization DoF [Fig.~\ref{fig:PolarizationCrossPurity}(b)]. We start with the block-matrix form for the coherence matrix $\mathbf{G}=\left(\begin{array}{cc}|\alpha|^{2}\mathbf{G}_{a}&\mathbf{G}_{ab}\\\mathbf{G}_{ba}&|\beta|^{2}\mathbf{G}_{a}\end{array}\right)$, in which the polarization is the same at $|a\rangle$ and $|b\rangle$, except for the overall factors $|\alpha|^{2}$ and $|\beta|^{2}$, respectively, where $|\alpha|^{2}+|\beta|^{2}=1$, and we parameterize $\alpha$ and $\beta$ via $|\alpha|^{2}=\cos^{2}\delta$ and $|\beta|^{2}=\sin^{2}\delta$. The trace of $\mathbf{G}$ is unity: $\mathrm{Tr}\{\mathbf{G}\}=|\alpha|^{2}\mathrm{Tr}\{\mathbf{G}_{a}\}+|\beta|^{2}\mathrm{Tr}\{\mathbf{G}_{a}\}=1$.

The field traverses a separable unitary $\hat{U}=\hat{U}_{\mathrm{s}}\otimes\hat{\mathbb{I}}_{2}$, where $\hat{U}_{\mathrm{s}}=\left(\begin{array}{cc}e^{i\varphi_{1}}\cos\theta&-e^{-i\varphi_{2}}\sin\theta\\e^{i\varphi_{2}}\sin\theta&e^{-i\varphi_{1}}\cos\theta\end{array}\right)$ is a $2\times2$ spatial unitary that does not impact the polarization DoF. The unitary $\hat{U}$ produces the transformed coherence matrix:
\begin{equation}
\mathbf{G}'=\hat{U}\mathbf{G}\hat{U}^{\dagger}=(\hat{U}_{\mathrm{s}}\otimes\mathbb{I}_{\mathrm{p}})\mathbf{G}(\hat{U}_{\mathrm{s}}^{\dagger}\otimes\mathbb{I}_{\mathrm{p}})=\left(\begin{array}{cc}|\alpha'|^{2}\mathbf{G}_{a}'&\mathbf{G}_{ab}'\\\mathbf{G}_{ba}'&|\beta'|^{2}\mathbf{G}_{b}'\end{array}\right),
\end{equation}
where the new primed matrices are given by: \begin{eqnarray}
|\alpha'|^{2}\mathbf{G}_{a}'&=&\gamma_{a}\mathbf{G}_{a}-\sin2\theta\mathbf{R}_{ab},\label{eq:NewGa}\\
|\beta'|^{2}\mathbf{G}_{b}'&=&\gamma_{b}\mathbf{G}_{a}+\sin2\theta\mathbf{R}_{ab},\label{eq:NewGb}\\
\mathbf{G}_{ab}'&=&e^{i(\varphi_{1}+\varphi_{2})}\left\{\gamma_{ab}\mathbf{G}_{a}+\mathbf{I}_{ab}+\cos2\theta\mathbf{R}_{ab}\right\};
\end{eqnarray}
the new parameters are $\gamma_{a}=\tfrac{1}{2}\{1-\cos2\delta\cos2\theta\}$, $\gamma_{a}+\gamma_{b}=1$, $\gamma_{ab}=\tfrac{1}{2}\cos2\delta\sin2\theta$, and the new matrices are $\mathbf{R}_{ab}=\tfrac{1}{2}\{\mathbf{N}+\mathbf{N}^{\dagger}\}$ and $\mathbf{I}_{ab}=\tfrac{1}{2}\{\mathbf{N}-\mathbf{N}^{\dagger}\}$, defined in terms of an intermediary matrix $\mathbf{N}=e^{i(\varphi_{1}-\varphi_{2})}\mathbf{G}_{a,b}$, so that $\mathbf{R}_{ab}^{\dagger}=\mathbf{R}_{ab}$ and $\mathbf{I}_{ab}^{\dagger}=-\mathbf{I}_{ab}$. Because $\mathrm{Tr}\{\mathbf{G}\}=1$, after the unitary $\hat{U}$ we have $\mathrm{Tr}\{\mathbf{G}'\}=1$. By requiring that $\mathrm{Tr}\{\mathbf{G}_{a}'\}=\mathrm{Tr}\{\mathbf{G}_{b}'\}=1$, we obtain the new normalization weights $|\alpha'|^{2}=\gamma_{a}-\sin2\theta\mathrm{Re}\{e^{i(\varphi_{1}-\varphi_{2})}\mathrm{Tr}(\mathbf{G}_{ab})\}$ and $|\beta'|^{2}=\gamma_{a}+\sin2\theta\mathrm{Re}\{e^{i(\varphi_{1}-\varphi_{2})}\mathrm{Tr}(\mathbf{G}_{ab})\}$, with $|\alpha'|^{2}+|\beta'|^{2}=1$.

What are the conditions for polarization cross-purity, $\mathbf{G}_{a}'=\mathbf{G}_{a}$ (and thus, concomitantly $\mathbf{G}_{a}'=\mathbf{G}_{b}'$)? We have shown that the necessary condition is $\mathbf{G}_{ba}=\eta\mathbf{G}_{a}$ and $\mathbf{G}_{ab}=\eta^{*}\mathbf{G}_{a}$ (where $\eta$ is a complex constant), for nontrivial settings of $\theta$ ($\sin2\theta\neq0$) \cite{Abouraddy26OL}. In other words, polarization cross-purity requires that the initial coherence matrix $\mathbf{G}$ be separable with respect to the spatial and polarization DoFs: $\mathbf{G}=\left(\begin{array}{cc}|\alpha|^{2}&\eta^{*}\\\eta&\!\!|\beta|^{2}\end{array}\right)_{\mathrm{s}}\otimes\mathbf{G}_{a}$. 

\subsubsection{Optical cross-purity and the coherence rank}

Evaluating cross-purity requires first symmetrizing the field so that $\mathbf{G}_{a}=\mathbf{G}_{b}$. Once the field is symmetrized, is it guaranteed to be separable? Does symmetry ($\mathbf{G}_{a}=\mathbf{G}_{b}$) imply separability (and thus polarization cross-purity)? We have recently shown that the coherence rank \cite{Harling24PRA,Harling24PRA2,Harling25APLP} is crucial in this regard. Indeed, for some coherence ranks, \textit{symmetry} does indeed imply \textit{separability}, so that symmetry $\mathbf{G}_{a}=\mathbf{G}_{b}$ is \textit{sufficient} for polarization cross-purity. We consider here each coherence rank separately.

\textit{Rank-1 fields}. The diagonal form of a rank-1 coherence matrix is $\mathbf{G}^{\mathrm{D}}=\mathrm{diag}\{1,0,0,0\}$, and its general form is $\mathbf{G}=\hat{U}\mathbf{G}^{\mathrm{D}}\hat{U}^{\dagger}=|u_{1}\rangle\langle u_{1}|$, where $|u_{1}\rangle$ is a $4\times1$ vector corresponding to the first column of the $4\times4$ unitary $\hat{U}$. We first write $|u_{1}\rangle$ as a direct sum: $|u_{1}\rangle=|u_{1}'\rangle\oplus|u_{1}''\rangle$, where $|u_{1}'\rangle$ is a $2\times1$ vector formed of the first two elements of $|u_{1}\rangle$, and $|u_{1}''\rangle$ is a $2\times1$ vector formed of the remaining two elements of $|u_{1}\rangle$. Although $\langle u_{1}|u_{1}\rangle=1$, we do \textit{not} have $\langle u_{1}'|u_{1}'\rangle=1$ \textit{and} $\langle u_{1}''|u_{1}''\rangle=1$; rather, we have $\langle u_{1}'|u_{1}'\rangle+\langle u_{1}''|u_{1}''\rangle=\langle u_{1}|u_{1}\rangle=1$. The block-matrix form of $\mathbf{G}$ can thus be expressed as follows:
\begin{equation}
\mathbf{G}=\left(\begin{array}{cc}
|u_{1}'\rangle\langle u_{1}'|&
|u_{1}'\rangle\langle u_{1}''|\\
|u_{1}''\rangle\langle u_{1}'|&
|u_{1}''\rangle\langle u_{1}''|\end{array}\right).
\end{equation}
When compared to the block-matrix form, we have: $|\alpha|^{2}\mathbf{G}_{a}=|u_{1}'\rangle\langle u_{1}'|$ with $|\alpha|^{2}=\langle u_{1}'|u_{1}'\rangle$, $|\beta|^{2}\mathbf{G}_{b}=|u_{1}''\rangle\langle u_{1}''|$ with $|\beta|^{2}=\langle u_{1}''|u_{1}''\rangle$, and $\mathbf{G}_{ab}=|u_{1}'\rangle\langle u_{1}''|$. 

If $\mathbf{G}$ is symmetric ($\mathbf{G}_{a}=\mathbf{G}_{b}$), will a rank-1 $\mathbf{G}$ also be guaranteed to be separable ($\mathbf{G}_{ab}\propto\mathbf{G}_{a}$)? In other words, does symmetry, $\mathbf{G}_{a}=\tfrac{|u_{1}'\rangle\langle u_{1}'|}{\langle u_{1}'|u_{1}'\rangle}=\tfrac{|u_{1}''\rangle\langle u_{1}''|}{\langle u_{1}''|u_{1}''\rangle}=\mathbf{G}_{b}$, entail separability? Evaluating $\langle u_{1}'|\mathbf{G}_{a}|u_{1}'\rangle$ and $\langle u_{1}'|\mathbf{G}_{b}|u_{1}'\rangle$, we find that $|\langle u_{1}'|u_{1}''\rangle|^{2}=\langle u_{1}'|u_{1}'\rangle\langle u_{1}''|u_{1}''\rangle$, which requires that $|u_{1}''\rangle=\eta|u_{1}'\rangle$, where $\eta$ is a complex constant. In this case, $\mathbf{G}_{ab}=|u_{1}'\rangle\langle u_{1}''|\propto\mathbf{G}_{a}=\eta^{*}|u_{1}'\rangle\langle u_{1}'|\propto\mathbf{G}_{a}$, so that $\mathbf{G}$ is separable. In other words, all \textit{symmetric} rank-1 fields are separable, and are thus also polarization cross-pure. This result leads to an interesting observation. Classically entangled fields (non-separable rank-1) cannot be symmetric and are thus \textit{not} cross-pure; i.e., we always have $\mathbf{G}_{a}\neq\mathbf{G}_{b}$ for classically entangled rank-1 fields.

\textit{Rank-2 fields}. The diagonal form of a rank-2 coherence matrix is $\mathbf{G}^{\mathrm{D}}=\mathrm{diag}\{\lambda_{1},\lambda_{2},0,0\}$, with $\lambda_{1}+\lambda_{2}=1$, and its general form if $\mathbf{G}=\hat{U}\mathbf{G}^{\mathrm{D}}\hat{U}^{\dagger}=\lambda_{1}|u_{1}\rangle\langle u_{1}|+\lambda_{2}|u_{2}\rangle\langle u_{2}|$, where $|u_{1}\rangle$ and $|u_{2}\rangle$ are $4\times1$ vectors corresponding to the first and second columns of the $4\times4$ unitary $\hat{U}$; $\langle u_{1}|u_{1}\rangle=\langle u_{2}|u_{2}\rangle=1$ and $\langle u_{1}|u_{2}\rangle=0$. Following the same approach followed above for rank-1 fields, we write $|u_{1}\rangle$ and $|u_{2}\rangle$ as direct sums: $|u_{1}\rangle=|u_{1}'\rangle\oplus|u_{1}''\rangle$ and $|u_{2}\rangle=|u_{2}'\rangle\oplus|u_{2}''\rangle$, with $\langle u_{1}'|u_{1}'\rangle+\langle u_{1}''|u_{1}''\rangle=\langle u_{1}|u_{1}\rangle=1$ and $\langle u_{2}'|u_{2}'\rangle+\langle u_{2}''|u_{2}''\rangle=\langle u_{2}|u_{2}\rangle=1$. We thus have the block submatrices $|\alpha|^{2}\mathbf{G}_{a}=\lambda_{1}|u_{1}'\rangle\langle u_{1}'|+\lambda_{2}|u_{2}'\rangle\langle u_{2}'|$, $|\beta|^{2}\mathbf{G}_{b}=\lambda_{1}|u_{1}''\rangle\langle u_{1}''|+\lambda_{2}|u_{2}''\rangle\langle u_{2}''|$. Ensuring symmetry $\mathbf{G}_{a}$ and $\mathbf{G}_{b}$ requires that $|u_{1}''\rangle=\eta|u_{1}'\rangle$ and $|u_{2}''\rangle=\eta|u_{2}'\rangle$. This implies that $|\alpha|^{2}=|\eta|^{2}|\beta|^{2}$ and $\mathbf{G}_{\mathrm{ba}}=\eta\mathbf{G}_{\mathrm{a}}$; in other words, symmetry here once again implies separability. Therefore, all rank-2 fields are polarization cross-pure.

\textit{Rank-3 fields}. The diagonalized coherence matrix for a rank-3 field is $\mathbf{G}_{3}^{\mathrm{D}}=\mathrm{diag}\{\lambda_{1},\lambda_{2},\lambda_{3},0\}$, with $\lambda_{1}+\lambda_{2}+\lambda_{3}=1$. Rank-3 fields are intrinsically non-separable \cite{Harling24PRA}; no unitary transformation can undo this non-separability. Consequently, even if $\mathbf{G}_{3}=\hat{U}\mathbf{G}_{3}^{\mathrm{D}}\hat{U}^{\dagger}$ is symmetrized ($\mathbf{G}_{a}=\mathbf{G}_{b}$), $\mathbf{G}_{3}$ remains non-separable. That is, \textit{all rank-3 fields are polarization cross-impure}; in this case, symmetry does \textit{not} imply separability.

\begin{figure}[t!]
\centering
\includegraphics[width=13.3cm]{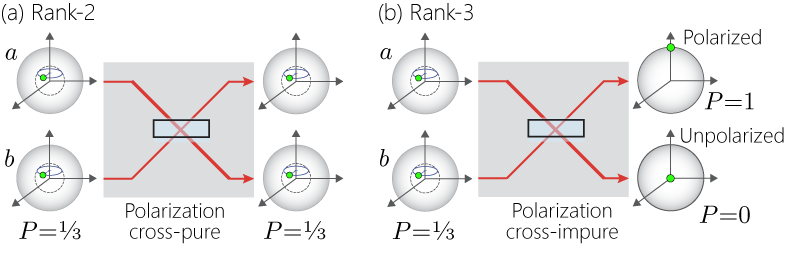}
\caption{(a) The rank-2 coherence matrix $\mathbf{G}_{2}$ (Eq.~\ref{eq:Example}) is polarization cross-\textit{pure} Superposing the fields from $|a\rangle$ and $|b\rangle$ having identical polarizations yields the same polarization (represented on the PS). (b) The rank-3 coherence matrix $\mathbf{G}_{3}$ (Eq.~\ref{eq:Example}) is polarization cross-\textit{impure}. Superposing the fields from $|a\rangle$ and $|b\rangle$ having identical polarizations yields different polarizations:rank-1 (polarized) at $|a\rangle$ and rank-4 (unpolarized) at $|b\rangle$. The unitary $\hat{U}_{\mathrm{s}}$ here corresponds to Fig.~\ref{fig:PolarizationCrossPurity}(b) after setting $\varphi_{1}=\varphi_{2}=0$ and $\theta=\tfrac{\pi}{4}$.}
\label{fig:PCPexample}
\end{figure}

We consider an example to clarify the distinction between rank-2 and rank-3 fields:
\begin{equation}\label{eq:Example}
\mathbf{G}_{2}=\tfrac{1}{2}\left(\begin{array}{cccc}
\tfrac{2}{3}&0&-\tfrac{i2}{3}&0\\
0&\tfrac{1}{3}&0&-\tfrac{i}{3}\\
\tfrac{i2}{3}&0&\tfrac{2}{3}&0\\
0&\tfrac{i}{3}&0&\tfrac{1}{3}
\end{array}\right),
\mathbf{G}_{3}=\tfrac{1}{2}\left(\begin{array}{cccc}
\tfrac{2}{3}&0&0&0\\
0&\tfrac{1}{3}&0&-\tfrac{i}{3}\\
0&0&\tfrac{2}{3}&0\\
0&\tfrac{i}{3}&0&\tfrac{1}{3}
\end{array}\right),
\end{equation}
where $\mathbf{G}_{2}$ is rank-2 with eigenvalues $\{\tfrac{2}{3},\tfrac{1}{3},0,0\}$, and $\mathbf{G}_{3}$ is rank-3 with eigenvalues $\{\tfrac{1}{3},\tfrac{1}{3},\tfrac{1}{3},0\}$. Each coherence matrix is symmetric with $\mathbf{G}_{a}=\mathbf{G}_{b}=\mathrm{diag}\{\tfrac{2}{3},\tfrac{1}{3}\}$ corresponding to a degree of polarization $D_{\mathrm{p}}=\tfrac{1}{3}$ \cite{Wolf07Book}. Whereas $\mathbf{G}_{2}=\tfrac{1}{2}\left(\begin{array}{cc}1&-i\\i&1\end{array}\right)_{\mathrm{s}}\otimes\mathbf{G}_{a}$ is separable, $\mathbf{G}_{3}$ is not (rank-3 fields are intrinsically non-separable). Now, consider superimposing the fields from $|a\rangle$ and $|b\rangle$ with equal weights via a beam splitter that is not sensitive to polarization [Fig.~\ref{fig:PCPexample}]. The resulting polarization for $\mathbf{G}_{2}$ remains unchanged $\mathbf{G}_{a}'=\mathbf{G}_{b}'=\mathbf{G}_{a}$; i.e., the rank-2 field is polarization cross-pure [Fig.~\ref{fig:PCPexample}(a)]. In contrast, the polarization for $\mathbf{G}_{3}$ undergoes a dramatic change: $\mathbf{G}_{a}'=\mathrm{diag}\{1,0\}$ and $\mathbf{G}_{b}'=\tfrac{1}{2}\hat{\mathbb{I}}_{2}$; that is, we have a purely $|\mathrm{H}\rangle$-polarized field at $|a'\rangle$, and an unpolarized field at $|b'\rangle$ [Fig.~\ref{fig:PCPexample}(b)]; i.e., the rank-3 field is polarization cross-impure. Utilizing a general spatial unitary $\hat{U}_{\mathrm{s}}$ to superpose the fields from $|a\rangle$ and $|b\rangle$, and tuning $\theta$ in $\hat{U}_{\mathrm{s}}$ would provide a source of partially polarized light with tunable degree of polarization.

\textit{Rank-4 fields}. For rank-4 fields, all the eigenvalues are non-zero $\mathbf{G}_{4}^{\mathrm{D}}=\mathrm{diag}\{\lambda_{1},\lambda_{2},\lambda_{3},\lambda_{4}\}$, which is separable if and only if $\lambda_{1}\lambda_{4}=\lambda_{2}\lambda_{3}$ \cite{Abouraddy01PRA,Abouraddy17OE}, whereupon $\mathbf{G}_{4}^{\mathrm{D}}=\mathrm{diag}\{\gamma_{1},\gamma_{2}\}\otimes\mathrm{diag}\{\gamma_{3},\gamma_{4}\}$, $\lambda_{2}=\gamma_{1}\gamma_{4}$, $\lambda_{3}=\gamma_{2}\gamma_{3}$, and $\lambda_{4}=\gamma_{2}\gamma_{4}$, where the factors $\gamma_{1}$, $\gamma_{2}$, $\gamma_{3}$, and $\gamma_{4}$ are real and positive. Implementing a separable unitary $\hat{U}=\hat{U}_{\mathrm{s}}\otimes\hat{U}_{\mathrm{p}}$ maintains the separability of a separable $\mathbf{G}_{4}$, and thus its cross-purity. It remains an open question whether a non-separable unitary $\hat{U}$ can yield a non-separable $\mathbf{G}_{4}$ that nevertheless maintains $\mathbf{G}_{a}=\mathbf{G}_{b}$. Non-separable $\mathbf{G}_{4}^{\mathrm{D}}$ are all polarization cross-impure, and no unitary can eliminate this intrinsic non-separability. Therefore, only a subset of rank-4 fields are polarization cross-pure; in the case of rank-4 fields, symmetry does not necessarily imply separability.

\subsection{Applications to communications across a scattering channel with two DoFs}\label{sec:Communications2DoFs}

We showed in Section~\ref{sec:Communications1DoF} that encoding information in polarized and unpolarized light is immune to polarization scattering over a communications channel. This scheme is an example of `coherence-rank communications', where bit~0 is encoded in a rank-1 coherence matrix $0\rightarrow\mathbf{G}_{0}=\mathrm{diag}\{1,0\}$ (degree of polarization $D_{\mathrm{p}}=1$), and bit~1 in a rank-2 coherence matrix $1\rightarrow\mathbf{G}_{1}=\tfrac{1}{2}\hat{\mathbb{I}}_{2}$ ($D_{\mathrm{p}}=0$). The decision threshold in this configuration is set at $D_{\mathrm{p}}=\tfrac{1}{2}$. However, this scheme is not useful for communicating over a \textit{decohering} channel that changes $D_{\mathrm{p}}$. Another setting that corrupts this scheme is a channel in which multiple DoFs are relevant. For example, in a multimode fiber both polarization and spatial modes are relevant, and intermodal scattering occurs, especially for large core diameters. Moreover, both of these settings (decohering channels and multi-DoF channels) are intimately related. A driver of decoherence in optical propagation is coupling between different DoFs followed by tracing over unutilized DoFs. This process can either increase or decrease the degree of coherence of the DoF utilized.

This discussion motivates us to exploit all the DoFs for information transfer. We consider here the polarization DoF and two spatial modes, both of which are binary DoFs, so that the associated coherence matrix is $4\times4$. We make use of the following assumptions about the communications channel:
\begin{enumerate}
\item The channel impacts both the polarization and spatial DoFs but not any further DoFs.
\item The channel can be represented for any bit during data transmission by a $4\times4$ unitary $\hat{U}$ that encompasses both the polarization and spatial DoFs.
\item Rapidly varying channel: $\hat{U}$ changes from bit to bit.
\item Strong scattering:family of $4\times4$ unitaries over both polarization and spatial DoFs.
\item No-memory channel: $\hat{U}$ at any two moments in time are uncorrelated.
\item An overall loss factor can be included, which is assumed to be independent of polarization and spatial modes.
\end{enumerate}
This channel scatters polarization strongly, scatters the spatial modes strongly, and moreover couples the polarization and spatial DoFs. These features vary rapidly from bit to bit with no memory in the channel. Once again, this is an extreme channel, but helps illustrate the advantages of coherence-rank communications.

\begin{figure}[t!]
\centering
\includegraphics[width=13.3cm]{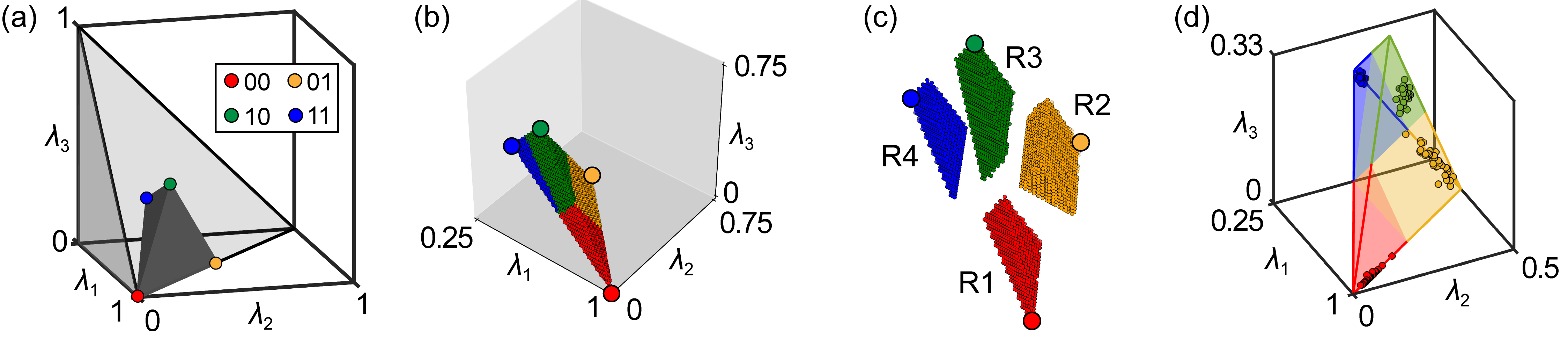}
\caption{(a) Representation of $4\times4$ coherence matrices in a restricted 3D space spanned by $\{\lambda_{1},\lambda_{2},\lambda_{3}\}$. The smaller identified volume is the subspace with descending eigenvalues $\lambda_{1}\geq\lambda_{2}\geq\lambda_{3}\geq\lambda_{4}$. The legend identifies the two-bit encoding scheme, associated with the vertices of the highlighted sub-volume, which correspond to the maximum-entropy fields for each rank. (b) Magnified view of the irreducible volume identified in (a). (c) Exploded view of the volume in (b). The surfaces separating the sub-volumes are the decision thresholds for the logical alphabet code. (d) Same as (b) but with the experimental outcomes for the reconstructed $\mathbf{G}$ (encoding scheme-3 in Fig.~\ref{fig:DualDoFData}) plotted, which cluster around the vertices.}
\label{fig:PyramidDivisiom}
\end{figure}

The coherence matrix $\mathbf{G}$ encompassing both DoFs carries two bits, so we can encode two bits of information per field state. We choose to encode bit pairs $00,01,10$, and 11 in coherence matrices of different rank:
\begin{eqnarray}
00&\rightarrow&\mathrm{rank-1},\mathbf{G}_{00}=\mathrm{diag}\{1,0,0,0\},\nonumber\\
01&\rightarrow&\mathrm{rank}-2,\mathbf{G}_{01}=\mathrm{diag}\{\lambda_{1},\lambda_{2},0,0\},\nonumber\\
10&\rightarrow&\mathrm{rank}-3,\mathbf{G}_{10}=\mathrm{diag}\{\lambda_{1},\lambda_{2},\lambda_{3},0\},\nonumber\\
11&\rightarrow&\mathrm{rank}-4,\mathbf{G}_{11}=\mathrm{diag}\{\lambda_{1},\lambda_{2},\lambda_{3},\lambda_{4}\}.
\end{eqnarray}
This assignment still leaves considerable freedom in selecting the coherence matrix. For example, $\mathbf{G}_{01}=\mathrm{diag}\{\lambda_{1},\lambda_{2},0,0\}$ can span a wide family of fields with entropy $0<S\leq1$, and similarly for $\mathbf{G}_{10}$ with entropy $0<S\leq\log_{2}3$, and $\mathbf{G}_{11}$ with entropy $0<S\leq2$. We select in each case the coherence matrix with the maximum entropy $S$ associated with each rank. Consequently, we make the assignments:
\begin{equation}
\mathbf{G}_{00}=\mathrm{daig}\{1,0,0,0\},
\mathbf{G}_{01}=\mathrm{daig}\{\tfrac{1}{2},\tfrac{1}{2},0,0\},
\mathbf{G}_{10}=\mathrm{daig}\{\tfrac{1}{3},\tfrac{1}{3},\tfrac{1}{3},0\},
\mathbf{G}_{11}=\frac{1}{4}\hat{\mathbb{I}}_{4},
\end{equation}
with associated entropies 0, 1, $\log_{2}3$, and 2~bits, respectively.

The motivation behind this selection can be understood by reference to Fig.~\ref{fig:PyramidDivisiom}. The representation of diagonal coherence matrices in the restricted space $\{\lambda_{1},\lambda_{2},\lambda_{3}\}$, when enforcing a descending order on the eigenvalues correspond to the vertices of the sub-volume highlighted in Fig.~\ref{fig:PyramidDivisiom}(a). The remainder of the volume corresponds to all other permutation of the eigenvalues associated with the sub-volume. The 4~vertices at the points $(\lambda_{1},\lambda_{2},\lambda_{3})=(1,0,0),(\tfrac{1}{2},\tfrac{1}{2},0),(\tfrac{1}{3},\tfrac{1}{3},\tfrac{1}{3}),(\tfrac{1}{4},\tfrac{1}{4},\tfrac{1}{4})$ correspond to $\mathbf{G}_{00}$, $\mathbf{G}_{01}$, $\mathbf{G}_{10}$, and $\mathbf{G}_{11}$. 

\begin{figure}[t!]
\centering
\includegraphics[width=13.3cm]{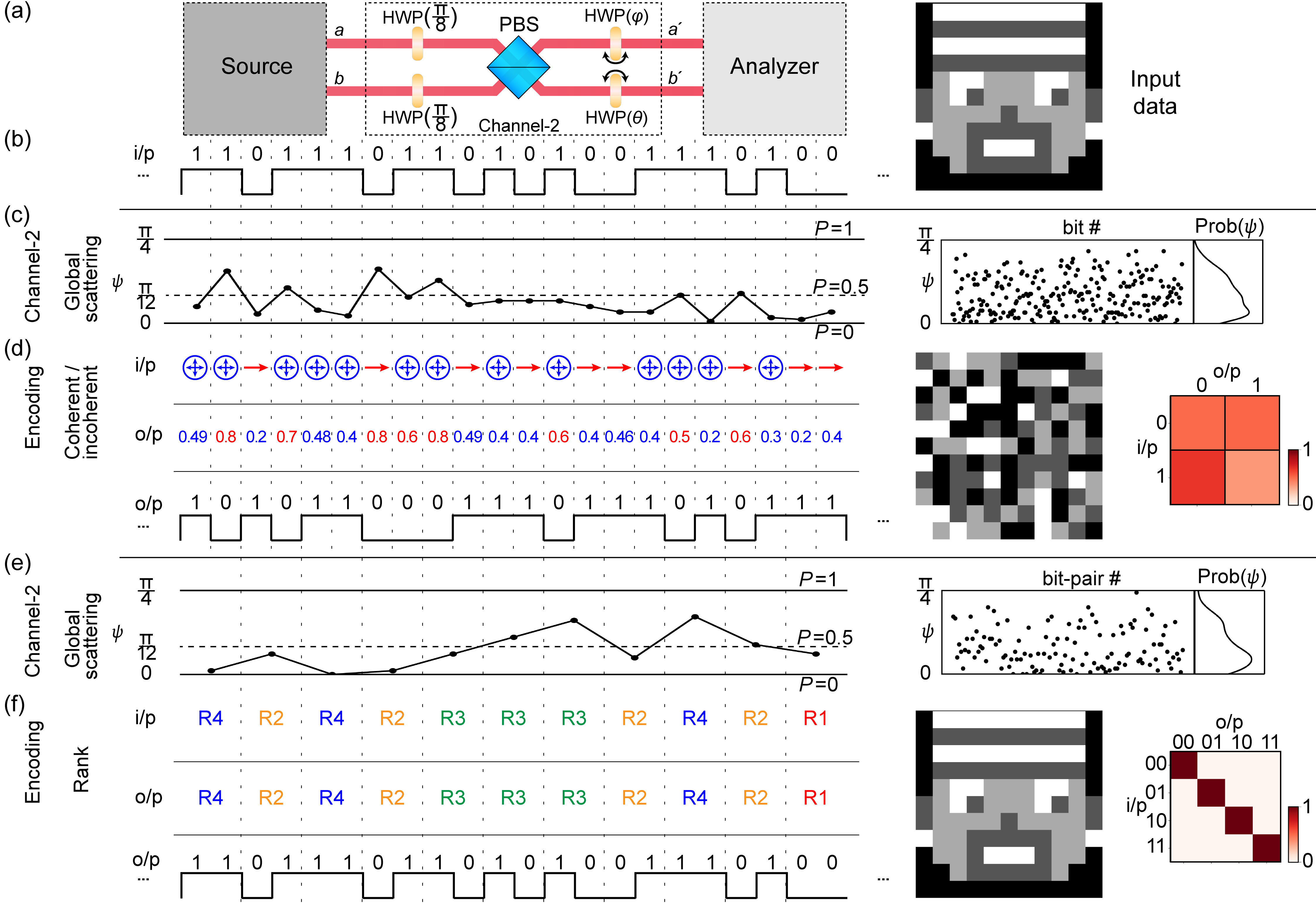}
\caption{(a) Schematic of the setup for channel-2. (b) Portion of the data stream corresponding to the image on the right. (c,d) Polarization encoding scheme-2 from Fig.~\ref{fig:PolCommData}(e,f): $0\rightarrow$H and $1\rightarrow$ $\protect\unpolarizedSymbol$. 
(c) Settings for $\psi\!=\!|\theta-\varphi|$, where $\theta$ and $\varphi$ are the HWP angles in the channel, and the probability distribution of $\psi$ is plotted on the right. (d) Input and output polarization states, the measured CTM is plotted on the right, along with the reconstructed image. (e,f) Same as (c,d) for encoding scheme-3, coherence-rank communications. Pairs of logical bits are encoded in the coherence rank: $00\rightarrow$rank-1, $01\rightarrow$rank-2, $10\rightarrow$rank-3, and $11\rightarrow$rank-4. (f) The measured input and output coherence ranks reconstructed tomographically via Stokes tomography. The measured CTM along with the reconstructed image are given on the right.}
\label{fig:DualDoFData}
\end{figure}

The communications scheme is as follows: 
\begin{enumerate}
\item The bit pairs $00,01,10$, and 11 are encoded in coherence matrices $\mathbf{G}_{00}$, $\mathbf{G}_{01}$, $\mathbf{G}_{10}$, and $\mathbf{G}_{11}$, respectively.
\item The physical field is transmitted across the communications channel.
\item At the channel output, $\mathbf{G}$ is tomographically reconstructed (by measuring the modal Stokes parameters) and its eigenvalues are estimated.
\item The rank of the coherence matrix is estimated.
\item Bit pairs can then be decoded from the estimated rank: rank-1$\rightarrow00$, rank-2$\rightarrow01$, rank-3$\rightarrow10$, and rank$-4\rightarrow11$. 
\end{enumerate}

The difficulty is that imperfections in the synthesis and detection stages, in addition to noise and scattering in the channel, may displace the values of the eigenvalues, so that in general all the eigenvalues will be non-zero. We thus define a `distance' between the reconstructed $\mathbf{G}_{\mathrm{out}}$ and the encoded-coherence matrix $\mathbf{G}_{\mathrm{in}}$ using the Euclidean metric: $d=\sqrt{\sum_{j=1}^{4}(\lambda_{j}-\lambda_{j}')^{2}}$, where $\{\lambda_{j}\}$ and $\{\lambda_{j}'\}$ are the eigenvalues of $\mathbf{G}_{\mathrm{in}}$ and $\mathbf{G}_{\mathrm{out}}$, respectively. This metric thus subdivides the volume depicted in Fig.~\ref{fig:PyramidDivisiom}(a) into 4~sub-volumes. Each sub-volume represents the set of coherence matrices that are `closest' to one of the vertices of the volume. Therefore, finding the detected coherence matrix in one of the 4~sub-volumes indicates the assignment to be followed in the decoding process. In other words, the surfaces separating the sub-volumes represent the decision thresholds for detection. Errors occur when the point representing the detected coherence matrix is located away from the assigned sub-volume. 

We depict the optical setup that emulates the target optical channel in Fig.~\ref{fig:DualDoFData}(a). The system is designed to (1) be rank-preserving for the $4\times4$ coherence matrix, (2) scatter the polarization DoF strongly, (3) scatter the spatial DoF strongly, and (4) couples the spatial and polarization DoFs strongly. Consequently, after tracing out the spatial DoF, the entropy of the polarization coherence matrix may increase or decrease from its initial value. Two encoding schemes are employed. Encoding scheme-2 is the same from Fig.~\ref{fig:PolCommData}(e,f) where bits~0 and~1 are encoded in $|\mathrm{H}\rangle$ and unpolarized light, and we assume the spatial DoF is coherent and separable from polarization. Because the the degree of polarization coherence $D_{\mathrm{p}}$ is no longer maintained across this channel, transmission of data is no longer possible, and the CTM is flat [Fig.~\ref{fig:DualDoFData}(b-d)]. However, when \textit{both} DoFs are exploited, and the data is encoded in the rank of the $4\times4$ coherence matrix, then even this worst-case-scenario channel does not impact the data transmission. The new CTM is diagonal, and the transmitted image is reconstructed at the receiver [Fig.~\ref{fig:DualDoFData}(e,f)].

Coherence-rank optical communications offers a hots of unique features:
\begin{enumerate}
\item Establishing scattering-immune communications over a strongly scattering channel.
\item Solving the problem of frame-sharing: the sender and receiver do not need to have the same shared reference system for polarization or spatial modes (e.g., they may not agree on what constitutes $|\mathrm{H}\rangle$ and $|\mathrm{V}\rangle$.
\item The communication scheme is impervious to any phases introduced between the spatial or polarization modes.
\item Overall losses do not affect the communications scheme.
\end{enumerate}
Potential limitations of coherence-rank communications are:
\begin{enumerate}
\item Modal-dependent losses can introduce errors if they are severe enough to change the coherence rank.
\item It fails if the rate of change in the channel is faster than the data rate.
\item The detection process required reconstruction of the coherence matrix $\mathbf{G}$ rather than detecting the power directly.
\end{enumerate}
The last limitation is currently being alleviated by gradually transitioning to photonic integrated circuits to reconstruct the coherence matrix rather than relying on free-space settings as done in the first demonstrations \cite{Harling25APLP}. 

\subsection{Correspondence with multipartite states in quantum mechanics}\label{sec:Quantum2DoFs}

\subsubsection{Similarities}

\begin{figure}[t!]
\centering
\includegraphics[width=2.8in]{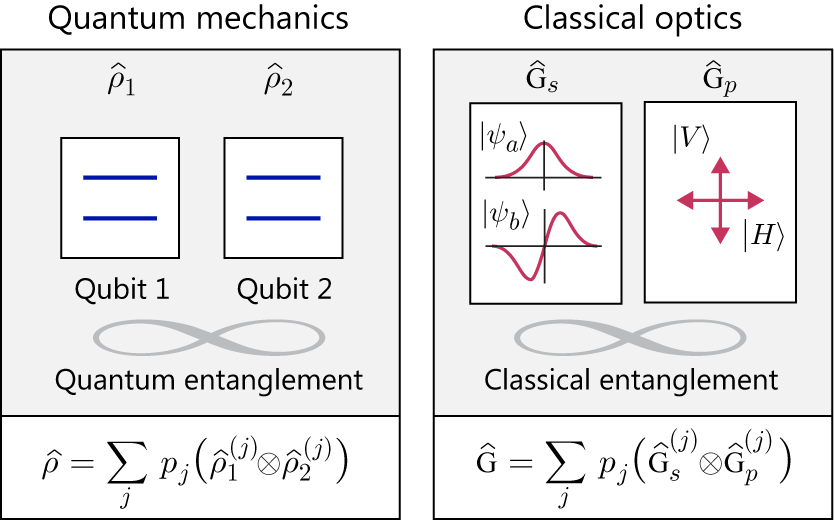}
\caption{Correspondence between quantum mechanics and classical optics with respect to two-qubit systems in the former and two binary DoFs in the latter. The quantum state representing two qubits takes the form of a $4\times4$ density matrix $\hat{\rho}$, which may not separate into a direct product of $2\times2$ density matrices $\hat{\rho}_{1}$ and $\hat{\rho}_{2}$ of the qubits (the state is endowed with quantum entanglement). In analogy, the coherence matrix $\mathbf{G}$ representing two binary DoFs (spatial and polarization) may not be factorizable into coherence matrices $\mathbf{G}_{\mathrm{s}}$ and $\mathbf{G}_{\mathrm{p}}$ for each DoF (classical entanglement).}
\label{fig:CQ2DoFs}
\end{figure}

We explored in Section~\ref{sec:QuantumSingleDoF} the mathematical analogy between a classical optical field characterized by a binary DoF and a qubit (a two-level quantum-mechanical system). This mathematical analogy can be extended to the two binary-DoF scenario, which is analogous to a two-qubit quantum system. Each DoF (which can be described separately with a $2\times2$ coherence matrix) can be mapped to one qubit (described by a $2\times2$ density matrix). Together the two DoFs are described by a $4\times4$ coherence matrix that is in direct correspondence with the $4\times4$ density matrix for two qubits \cite{Kagalwala13NP}; see Fig.~\ref{fig:CQ2DoFs}.

This correspondence entails that several aspects of two-qubit states in quantum mechanics can be carried over to the binary-DoF field. One of these aspects is `quantum entanglement,' which refers to the non-separability of the pure two-qubit state. The corresponding feature has been called `classical entanglement,' which refers to the non-separability of the coherent optical field with respect to the DoFs. This analogy was pointed out early on by Spreeuw \cite{Spreeuw98FP}. Classical entanglement has proven to be a useful concept leading to a variety of insights with respect to optical coherence. For example, Bell's measure that is routinely used in quantum mechanics to demarcate local realism has been shown to be a quantifier of resources needed to construct a given partially coherent two-binary-DoF field \cite{Kagalwala13NP}. Moreover, applications of classical entanglement have been reported in particle tracking \cite{Berg15Optica} and the characterization of optical channels \cite{Ndagano17NP}; see the reviews in Ref.~\cite{Aiello15NJP,Forbes19PO}.

Another analogy that can be drawn between the classical and quantum settings is with regards to state reconstruction. Whereas the measurements needed to reconstruct the density matrix for a qubit or a single binary DoF are well-known (Section~\ref{sec:StokesSingleDoF}), determining the measurements needed to reconstruct a two-qubit density matrix was challenging. The difficulty in the case of two-photon states is that non-separable unitaries implemented on the two photons require photon-photon interactions (mediated by not-yet-available single-photon nonlinearities). In contrast, separable unitaries (where two $2\times2$ unitaries are implemented separately on each photon) are straightforward to construct. Wootters showed that such separable unitaries are sufficient to reconstruct the two-photon state \cite{Wootters90article}. Specifically, the 4~measurements needed to reconstruct the quantum state for one qubit can also reconstruct the two-qubit state when performed in coincidence between the two qubits, thus yielding $4\times4=16$ measurements. Because the coherence matrix $\mathbf{G}$ has the same mathematical structure as the density matrix in quantum mechanics (specifically, the direct product of the basis sets for the two DoFs), the same procedure can be adapted for $\mathbf{G}$, which is the basis for the modal-Stokes-parameters approach described in Section~\ref{sec:Stokes2DoFs} \cite{Abouraddy14OL,Kagalwala15SR}. The reduced coherence matrices $\mathbf{G}_{\mathrm{s}}^{\mathrm{red.}}$ and $\mathbf{G}_{\mathrm{p}}^{\mathrm{red.}}$ discussed in Section~\ref{sec:ReducedRestricted} obtained by a partial trace over one DoF correspond to the reduced density matrices $\hat{\rho}_{1}$ and $\hat{\rho}_{2}$ obtained by partially tracing over one qubit: $\hat{\rho}_{1}=\mathrm{Tr}_{2}\{\hat{\rho}\}$ and $\hat{\rho}_{2}=\mathrm{Tr}_{1}\{\hat{\rho}\}$. The restricted coherence matrices, on the other hand, correspond to heralded quantum states: the state of one photon conditioned on the detection of a particular state of the second photon.

Another classical-quantum analogy can be exploited in the doamin of `system-environment' interaction. A common model of a quantum system decohering is that of a unitary evolution jointly with an environment in thermal equilibrium governed by a Hermitian interaction. Although the evolution of the joint system-environment is unitary, so that the total entropy is constant, tracing over the environment nevertheless reveals that the quantum system evolves gradually to a mixed state (non-unitary evolution). A similar scenario can be set up for a classical optical field. Consider for example a separable, maximum-entropy rank-2 field $S=S_{\mathrm{s}}+S_{\mathrm{p}}=1$ with the spatial DoF coherent ($S_{\mathrm{s}}=0$) and the polarization DoF incoherent ($S_{\mathrm{p}}=1$). A unitary coupling between the two DoFs can gradually increase the spatial entropy.

\subsubsection{Distinctions}

Whereas there are many formal similarities between the mathematical structures of quantum and classical entanglement that have been investigated over the past few decades, there nevertheless exist several critical distinctions.

(a) Quantum entanglement is a valuable resource in quantum information processing. This is because it is difficult to introduce entanglement between two separable qubits, a process that requires a joint unitary. In contrast, disentangling two qubits that are initially entangled via a projective operation or filtering is straightforward. In the classical context, the situation is very different. Indeed, classical entanglement can be readily introduced between two DoFs using a variety of optical systems that couple two DoFs (e.g., Section~\ref{sec:Unitary2DoFs}), and it can also be unitarily eliminated without need for filtering. In other words, it is straightforward to entangle and disentangle two DoFs of a classical field. This stems from the availability of arbitrary $4\times4$ unitaries on two DoFs.

(b) We have shown that the coherence-rank emerges as a useful classifier of the $4\times4$ coherence matrix for two binary DoFs (and of course for larger-dimensional field configurations; Section~\ref{sec:ExtenstioToN}). Crucially, the rank remains invariant after entangling or disentangling the two DoFs, which is a useful feature of the scattering-free communications scheme describe din Section~\ref{sec:Communications2DoFs}. It is perhaps for this reason that the equivalent concept for the density matrix has not found use in the quantum context. The lack of global unitaries on two optical qubits makes extracting the eigenvalues of the density matrix challenging.

(c) No-cloning theorem. The no-cloning theorem \cite{Wootters82Nature} remains a sharp distinction between the classical and quantum domains. Whereas an unknown quantum state cannot be reliably cloned or copied, a classical field can be split by beam splitters into multiple copies and also amplified. The no-cloning theorem is the basis for many applications in quantum communications that have no classical equivalent. However, because classical optics is \textit{not} constrained by the no-cloning theorem, some operations can be carried out that are not available to quantum systems; e.g., single-shot reconstruction of the coherence matrix via Stokes tomography.

(d) With regards to the model for system-environment coupling, the quantum system generally undergoes non-reversible dynamics with the system entropy gradually increasing over time. The environment is assumed to span a large dimensional space (a larger number of degrees of freedom) and to be in thermal equilibrium, so that the interaction is not reversible. Because the dynamics of coupling between the two DoFs in the classical field is reversible, one can explore the dynamics of entropy exchange between the `system' (one DoF of the field) and the `environment' (the second DoF) in both directions. Moreover, selecting the second DoF to be spatial and thus have a large dimension enables studying precisely the impact of the `environment' dimensionality. 

\section{Discussion}\label{sec:Discussion}

\subsection{Extension to an $N$-dimensional modal basis}\label{sec:ExtenstioToN}

We have focused so far on binary DoFs (a modal basis of dimension $N=2$) such as the polarization DoF or the spatial DoF when spanned by a pair of modes. Whereas the dimensionality of the polarization DoF cannot be expanded (except for an extension to polarization in three dimensions in the non-paraxial regime \cite{Wolf59PRSA,Richards59PRSA,Youngworth00OE,Dorn03JMO,Abouraddy06PRL,Alonso23AOP}), the spatial DoF can take on -- in principle -- a large dimensionality $N>2$ by increasing the number of relevant modes. For example, increasing the number of utilized waveguides in on-chip implementations can substantially increase the dimensionality of the modal basis. In the context of spatial modes with \textit{coherent} fields, this regime has indeed been extensively investigated in several contexts: (1) `structured light' with freely propagating fields \cite{Forbes21NP}; (2) structuring fields for multimode fibers \cite{CruzDelgado22NP}; and (3) on-chip `programmable photonics' \cite{Bogaerts20Nature}. Of course, perfect coherence is an idealization that is only approximated in reality. Investigations of the conceptual and technological developments that are made possible by utilizing structured coherence are currently underway. We briefly consider here the scenario of structured coherence with $N>2$ and highlight some of the distinctions to be encountered with respect to binary DoFs.

\subsubsection{Coherence matrix for $N$-mode fields}

\begin{figure}[t!]
\centering
\includegraphics[width=3in]{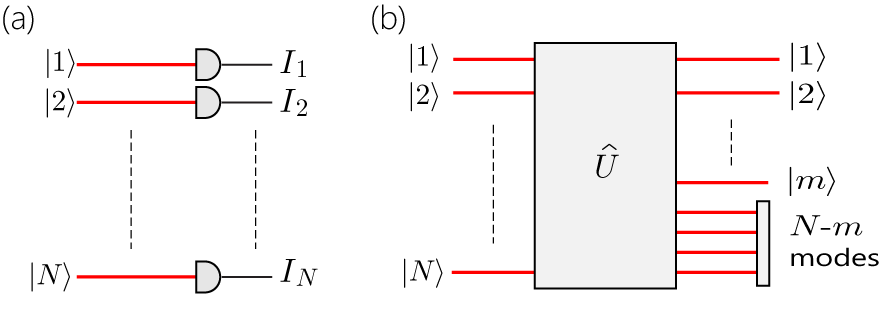}
\caption{(a) Measurement of the modal weights for an optical field comprising an $N$-dimensional modal basis $\{|j\rangle\}_{j=1}^{N}$. When the field is coherent $I_{j}=|E_{j}|^{2}$, and when the field is partially coherent $I_{j}=G_{jj}$, $j=1,\cdots N$. (b) The $N$-dimensional field is directed to an $N\times N$ unitary $\hat{U}$, and we wish to avoid a subset of its output ports. Here we attempt to concentrate the input power unitarily into $m$ modes, $|1\rangle$ through $|m\rangle$, and avoid the remaining $N-m$ modes. For a partially coherent field of coherence matrix $\mathbf{G}$, the \textit{minimum} number of modes $m$ that the power can be concentrated into is the coherence rank, and $\hat{U}$ is then the unitary that diagonalizes $\mathbf{G}$.}
\label{fig:NdimensionalBlock}
\end{figure}

Consider an optical DoF described by a modal basis comprising $N>2$ orthonormal modes, $\{|j\rangle\}_{j=1}^{N}$, so that $\langle j|k\rangle=\delta_{jk}$. A coherent field is associated with an $N\times1$ field vector $|E\rangle$:
\begin{equation}
|E\rangle=\left(\begin{array}{c}E_{1}\\E_{2}\\\vdots\\\vdots\\E_{N}\end{array}\right)=E_{1}\left(\begin{array}{c}1\\0\\\vdots\\\vdots\\0\end{array}\right)+E_{2}\left(\begin{array}{c}0\\1\\\vdots\\\vdots\\0\end{array}\right)+\cdots+E_{N}\left(\begin{array}{c}0\\0\\\vdots\\\vdots\\1\end{array}\right)=E_{1}|1\rangle+E_{2}|2\rangle+\cdots E_{N}|N\rangle.
\end{equation}
Normalizing the field $|E\rangle=\sum_{j=1}^{N}E_{j}|j\rangle$ to $\langle E|E\rangle=1$ entails that the modal coefficients satisfy the constraint $\sum_{j=1}^{N}|E_{j}|^{2}=1$. The modal weights are determined by the detectors as shown in Fig.~\ref{fig:NdimensionalBlock}(a) with $I_{j}=|E_{j}|^{2}$.

When the field is partially coherent, it is represented by an $N\times N$ coherence matrix,
\begin{equation}
\mathbf{G}=\left(\begin{array}{ccccc}
G_{11}&G_{12}&\cdots&\cdots&G_{1N}\\
G_{21}&G_{22}&\cdots&\cdots&G_{2N}\\
\vdots&\vdots&\ddots&&\vdots\\
\vdots&\vdots&&\ddots&\vdots\\
G_{N1}&G_{N2}&\cdots&\cdots&G_{NN}\\
\end{array}\right),
\end{equation}
where $G_{jk}=\langle E_{j}E_{k}^{*}\rangle$, $j,k=1,\cdots,N$, and $\langle\cdot\rangle$ denotes an ensemble average. The coherence matrix is Hermitian, $\mathbf{G}^{\dagger}=\mathbf{G}$, so that:
\begin{enumerate}
\item the diagonal elements are real $G_{jj}=G_{jj}^{*}$;
\item the off-diagonal elements form conjugate pairs $G_{jk}=G_{kj}^{*}$;
\item the eigenvalues $\{\lambda_{1},\lambda_{2},\cdots,\lambda_{N}\}$ of $\mathbf{G}$ are real; 
\item the eigenvectors of $\mathbf{G}$ are orthogonal when their associated eigenvalues are different; 
\item the coherence matrix can be diagonalized via an $N\times N$ unitary: $\mathbf{G}^{\mathrm{D}}=\hat{U}\mathbf{G}\hat{U}^{\dagger}=\mathrm{diag}\{\lambda_{1},\lambda_{2},\cdots,\lambda_{N}\}$; and
\item the coherence matrix is positive semi-definite, so that $\lambda_{j}\geq0$ and $G_{jj}\geq0$.
\end{enumerate}

We normalize the coherence matrix to unity trace, $\mathrm{Tr}\{\mathbf{G}\}=\sum_{j=1}^{N}G_{jj}=\sum_{j=1}^{N}\lambda_{j}=1$. The diagonal elements of $\mathbf{G}$ correspond to the fractions of power associated with each mode. The detectors in Fig.~\ref{fig:NdimensionalBlock}(a) therefore determine the modal weights corresponding to the diagonal elements, $I_{j}=G_{jj}$. The off-diagonal elements represent the correlations between pairs of modes. For an $N\times N$ coherence matrix, $N^{2}-1$ real parameters are required to uniquely identify it (in addition to normalization): $N$~measurements reveal the diagonal elements $G_{jj}$, and $\tfrac{1}{2}N(N-1)$ interference experiments are required on all pairs of modes to obtain 2~real parameters in each experiment, the fringe visibility when superposing the $j^{\mathrm{th}}$ and $k^{\mathrm{th}}$ modes is $2|G_{jk}|,$ and the fringe shift yields the phase of $G_{jk}$, $j,k=1,\cdots N$, $j\neq k$. 

\subsubsection{Coherent, partially coherent, and incoherent fields}

The number of real parameters needed to uniquely identify a Hermitian unity-trace $N\times N$ coherence matrix is $N^{2}-1$. The $N$ eigenvalues are unitary invariants, while the remaining $N(N-1)$ `angular' parameters vary with unitaries. Consequently, an $N$-dimensional DoF cannot be uniquely characterized by a single `degree of coherence'. Rather, $N-1$ real parameters are required to identify the family of all coherence matrices that can be inter-converted into each other unitarily.

We define the field entropy as $S=-\mathrm{Tr}\{\mathbf{G}\log_{2}\mathbf{G}\}=-\sum_{j=1}^{N}\lambda_{j}\log_{2}\lambda_{j}$, $0\leq S(\mathbf{G})\leq\log_{2}N$. By identifying fully coherent fields with absence of random fluctuations $S=0$, full field coherence corresponds to the condition $\lambda_{1}=1$ and $\lambda_{j}=0$ for $2\leq j\leq N$; that is, a rank-1 coherence matrix. The coherence rank varies from~1 to~$N$, with the maximum entropy for rank-$m$ being $S=\log_{2}m$, $1\leq m\leq N$, which is reached when the non-zero eigenvalues are equal, $\lambda_{j}=\tfrac{1}{m}$, $j=1,\cdots,m$. The maximally incoherent field corresponds to a maximum-entropy rank-$N$ field associated with the coherence matrix $\mathbf{G}=\tfrac{1}{N}\mathrm{diag}\{1,1,\cdots,1\}=\tfrac{1}{N}\hat{\mathbb{I}}_{N}$ and an entropy of $\log_{2}N$~bits; here $\hat{\mathbb{I}}_{N}$ is the $N\times N$ identity matrix.

The concept of coherence rank can be given a physical interpretation as depicted in Fig.~\ref{fig:NdimensionalBlock}(b). Consider the following question: can all the input power initially distributed among the $N$ modes be concentrated into a single mode via an $N\times N$ unitary $\hat{U}$? Similarly to the case of a two-mode field [Fig.~\ref{fig:ConcentratingTheField} and Fig.~\ref{fig:PartiallyCoherentBlocked}], the input power can be concentrated into a single mode only if the field is coherent (rank-1 coherence matrix), whereupon the unitary $\hat{U}$ that diagonalizes $\mathbf{G}$ also concentrates all the optical power into one mode. When the field is partially coherent, the field can\textit{not} be concentrated into a single mode. Rather, the input power can be concentrated into a minimum of $m$ modes, where $m$ is the rank of $\mathbf{G}$. This is done once again using the unitary that diagonalizes $\mathbf{G}$ [Fig.~\ref{fig:NdimensionalBlock}(b)]. Consequently, for a rank-$N$ field, it is impossible to eliminate the power from \textit{any} particular mode via a unitary.

\subsubsection{Example: A three-mode field}

For concreteness, we consider explicitly the case of a three-mode field ($N=3$) spanned by a modal basis $\{|1\rangle,|2\rangle,|3\rangle\}$ in which the coherence matrix is expressed as:
\begin{equation}
\mathbf{G}=\left(\begin{array}{ccc}
G_{11}&G_{12}&G_{13}\\
G_{21}&G_{22}&G_{23}\\
G_{31}&G_{32}&G_{33}
\end{array}\right).
\end{equation}
Such a coherence matrix can be diagonalized by a $3\times3$ unitary $\hat{U}$, $\mathbf{G}^{(\mathrm{D})}=\hat{U}\mathbf{G}_{3}\hat{U}^{\dagger}$. It is always simpler from an experimental perspective to construct $2\times2$ unitaries implemented on two modes only [Fig.~\ref{fig:2x2Unitary}]. Can a general $3\times3$ unitary be decomposed into a sequence of $2\times2$ unitaries operating on a pair of modes at a time? If so, what is the minimum number of such $2\times2$ unitaries that is sufficient to construct an arbitrary $3\times3$ unitary? These questions have been tackled extensively in quantum information processing \cite{Reck94PRL,Saleh25book}: it is indeed possible to construct an arbitrary $N\times N$ unitary out of $\tfrac{1}{2}N(N-1)$ restricted $2\times2$ unitaries (Eq.~\ref{eq:UnitaryRestricted}) in addition to $N$ phase shifts implemented on each mode, giving a total of $2\times\tfrac{1}{2}N(N-1)+N=N^{2}$ real parameters that identify an arbitrary $N\times N$ unitary.

\begin{figure}[t!]
\centering
\includegraphics[width=5.25in]{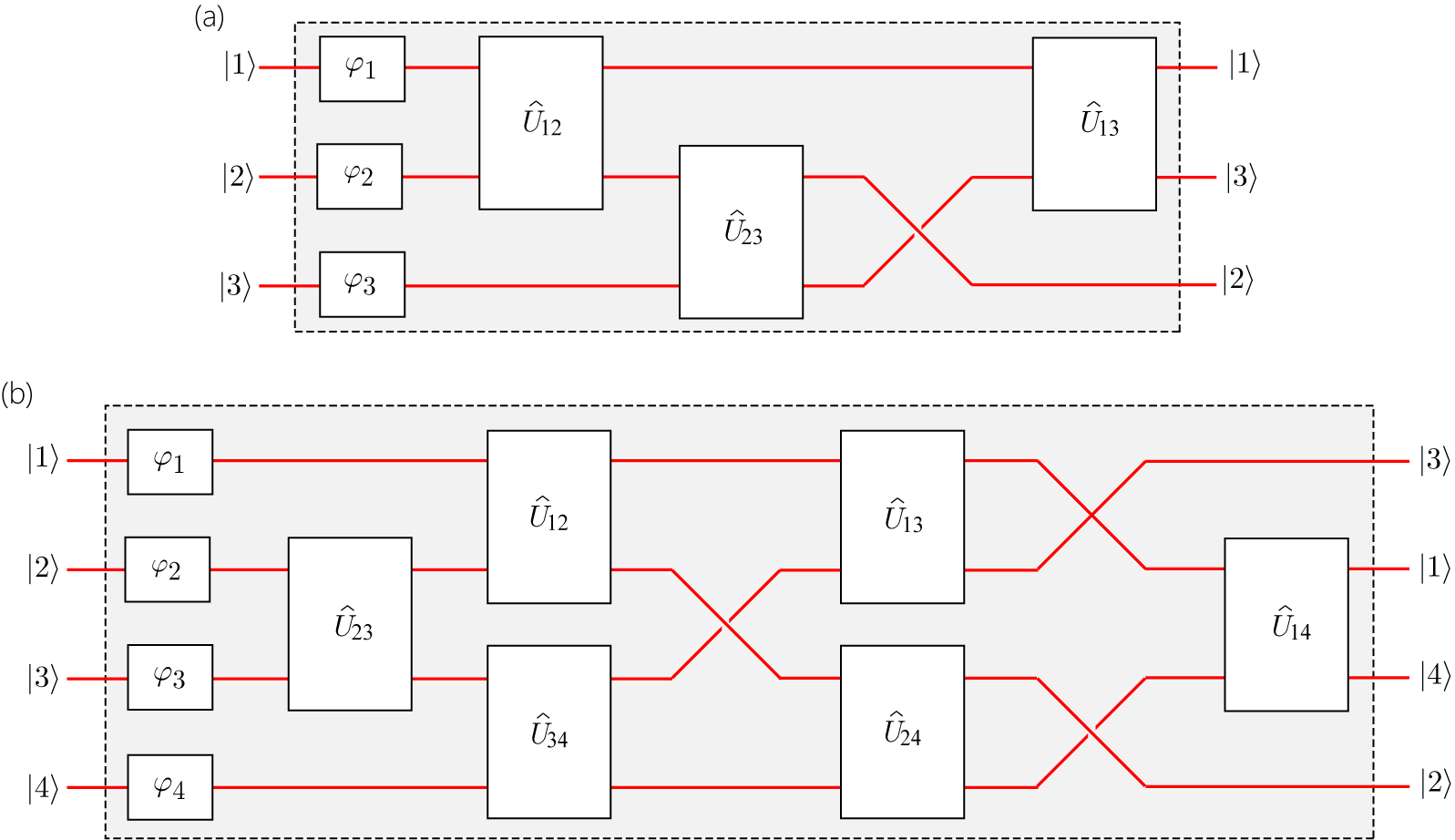}
\caption{(a) An arbitrary $3\times3$ unitary for a three-mode field constructed out of a sequence of three $2\times2$ unitaries $\hat{U}_{12}$, $\hat{U}_{13}$, and $\hat{U}_{23}$, each operating on a pair of modes, in addition to phases $\varphi_{1}$, $\varphi_{2}$, and $\varphi_{3}$ introduced into the three modes. (b) An arbitrary $4\times4$ unitary for a four-mode field constructed out of a sequence of six $2\times2$ unitaries $\hat{U}_{12}$, $\hat{U}_{13}$, $\hat{U}_{14}$, $\hat{U}_{23}$, $\hat{U}_{24}$, and $\hat{U}_{34}$, in addition to phases $\varphi_{1}$, $\varphi_{2}$, $\varphi_{3}$, and $\varphi_{4}$ introduced into the four modes.}
\label{fig:3x3Decomposition}
\end{figure}

The construction of a~$3\times3$ unitary as a concrete example is depicted in Fig.~\ref{fig:3x3Decomposition}(a). Three phase shifts ($\varphi_{1}$, $\varphi_{2}$, and $\varphi_{3}$) are introduced into the modes, and then $\tfrac{1}{2}\times3\times2=3$~restricted unitaries operating on pairs of modes are implemented: $\hat{U}_{12}$, $\hat{U}_{23}$, and $\hat{U}_{13}$, where the indices of each unitary identify the pair of modes on which it operates. One can then write $\hat{U}$ explicitly as a sequence of unitaries:
\begin{equation}
\hat{U}=
\left(\begin{array}{ccc}
1&0&0\\
0&c_{22}&c_{23}\\
0&c_{32}&c_{33}
\end{array}\right)
\left(\begin{array}{ccc}
b_{11}&0&b_{13}\\
0&1&0\\
b_{31}&0&b_{33}
\end{array}\right)\left(\begin{array}{ccc}
a_{11}&a_{12}&0\\
a_{21}&a_{22}&0\\
0&0&1
\end{array}\right)
\left(\begin{array}{ccc}
e^{i\varphi_{1}}&0&0\\
0&e^{i\varphi_{2}}&0\\
0&0&e^{i\varphi_{3}}
\end{array}\right),
\end{equation}
where $\hat{U}_{12}=\left(\begin{array}{cc}a_{11}&a_{12}\\a_{21}&a_{22}\end{array}\right)$, $\hat{U}_{13}=\left(\begin{array}{cc}b_{11}&b_{13}\\b_{31}&b_{33}\end{array}\right)$, and $\hat{U}_{23}=\left(\begin{array}{cc}c_{22}&c_{23}\\c_{32}&c_{33}\end{array}\right)$ are $2\times2$ restricted unitary matrices (Eq.~\ref{eq:UnitaryRestricted}).

A further example for the construction of a general $4\times4$ unitary is illustrated in Fig.~\ref{fig:3x3Decomposition}(b), which comprises four phases ($\varphi_{1}$, $\varphi_{2}$, $\varphi_{3}$, and $\varphi_{4}$) introduced into the modes, and $\tfrac{1}{2}\times4\times3=6$ restricted unitaries ($\hat{U}_{12}$, $\hat{U}_{13}$, $\hat{U}_{14}$, $\hat{U}_{23}$, $\hat{U}_{24}$, and $\hat{U}_{34}$) operating on pairs of modes. The same procedure extends to larger dimensions $N>4$.

Structured coherence manipulated via $2\times2$ \cite{Hashemi26TwoModes}, $3\times3$ \cite{Hashemi26ThreeModes}, and $4\times4$ \cite{Hashemi26FourModes} unitaries has only been implemented very recently on chip. Much further work is needed along these lines.

\subsection{Modal bases associated with other DoFs}

\subsubsection{Temporal modal bases}

\begin{figure}[t!]
\centering
\includegraphics[width=2.75in]{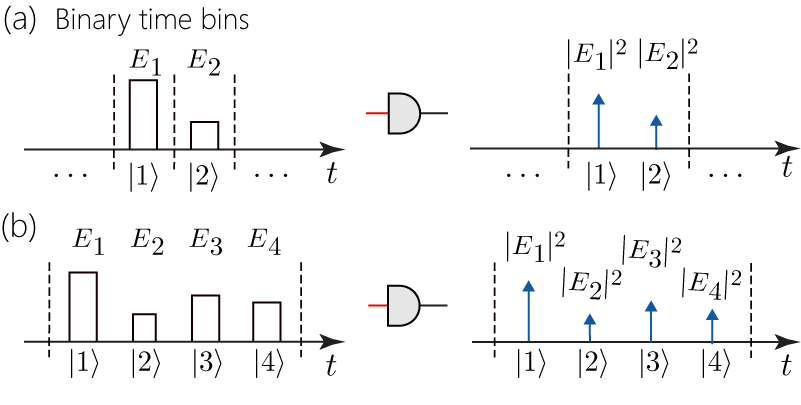}
\caption{Modal bases for the temporal DoF. (a) Binary time-bins. (b) $N$-ary time-bins.}
\label{fig:NonOverlappingTemporalModes}
\end{figure}

We have focused here on the spatial and polarization DoFs, but the matrix formulation for structured coherence is equally applicable to any DoF, including the temporal and spectral DoFs, although it is much less common there. Nevertheless, one scheme that has had significant impact in quantum communications using photons is so-called `time-bins', as depicted in Fig.~\ref{fig:NonOverlappingTemporalModes}(a,b). The example illustrated in Fig.~\ref{fig:NonOverlappingTemporalModes}(a) is the temporal analog of the bimodal spatial field in Fig.~\ref{fig:NonOverlappingSpatialModes}(a). Here we consider a time-window of width $T$ divided into two `bins', each containing an optical pulse of fixed width. The complex amplitudes of these two pulses are $E_{1}$ and $E_{2}$ (with respect to a fixed pulse height). The coherent field can again be written as $|E\rangle=E_{1}|\psi_{1}\rangle+E_2|\psi_{2}\rangle$, where $|\psi_{1}\rangle$ and $|\psi_{2}\rangle$ correspond to fixed amplitude pulses in bins 1 and 2, respectively. Because the pulses in the two bins are temporally non-overlapping, $\langle \psi_{1}|\psi_{2}\rangle=0$, and we normalize the pulse height and width in each bin so that $\langle \psi_{j}|\psi_{j}\rangle=1$ ($j=1,2$). A modal detector would simply be an optical detector with sufficient bandwidth (or response speed) to resolve the pulses in the two bins. The challenge to construct $2\times2$ unitaries $\hat{U}$ that operate on these two bins, which involve fast switches, optical delays, and beam splitters \cite{Xavier25npjQI}. This scheme can be extended to $N$~bins as illustrated in Fig.~\ref{fig:NonOverlappingTemporalModes}(b), so that $|E\rangle=\sum_{j=1}^{N}E_{j}|\psi_{j}\rangle$, where $|E_{j}\rangle$ is a pulse of fixed width and height in bin~$j$ and $E_{j}$ is its complex amplitude, with the normalization $\langle\psi_{j}|\psi_{j}\rangle=1$ and $\sum_{j=1}^{N}|E_{j}|^{2}=1$.

This modal basis can be utilized -- in principle -- with partially coherent light although this has not been realized to date to the best of our knowledge (most realizations have been in the context of quantum communications with photons). Just as in the case of spatial modes, the temporal modes here (the pulses in the time bins) are fixed and stable, and partial coherence arises from random complex amplitudes $E_{j}$ associated with each bin. For two time bins, this results in a $2\times2$ temporal coherence matrix, $\mathbf{G}_{t}=\left(\begin{array}{cc}G_{11}&G_{12}\\G_{21}&G_{22}\end{array}\right)$. As long as unitaries can be implemented on this modal basis, all the results developed for the polarization DoF (Section~\ref{sec:polarizationDoF}) and binary spatial DoF (Section~\ref{sec:SpatialDoF}) can be adapted for binary time bins. For $N>2$, the $N\times N$ unitary can be decomposed into a combination of $2\times2$ unitaries.

\subsubsection{Spectral modal bases}

Similarly to the case of temporal modes, only limited interest has been directed to spectral modal bases. Two examples of discrete modal bases for the spectral DoF are given in Fig.~\ref{fig:NonOverlappingTemporalModes}(c,d), both of which can be classified as non-overlapping modes. In Fig.~\ref{fig:NonOverlappingTemporalModes}(c), a continuous spectrum is binned into discrete spectral windows. This occurs naturally in any spectral analysis device, which inevitably has a finite spectral resolution. This example corresponds to a 1D analog of the spatial case in Fig.~\ref{fig:NonOverlappingSpatialModes}(d). A second example, depicted in Fig.~\ref{fig:NonOverlappingTemporalModes}(d), corresponds to a frequency comb, which are laser fields that are naturally formed of a periodic train of discretized spectral lines. We consider each spectral bin or laser line to be a fixed, stable, and deterministic mode. Structured coherence with the spectral DoF arises from random amplitudes associated with these fixed spectral modes.

Spectral mode detectors are straightforward to implement utilizing gratings or prisms, followed by a detector array. Consequently, one may envision a straightforward extension to spectral bases with large dimension $N$. Nevertheless, the central challenge in utilizing spectral modes in structured coherence is in carrying out spectral transformations (exchange of energy between different wavelengths), which can only be achieved via nonlinear optics. This requirement will likely limit the reach of applications of structured coherence in the spectral domain.

\begin{figure}[t!]
\centering
\includegraphics[width=2.75in]{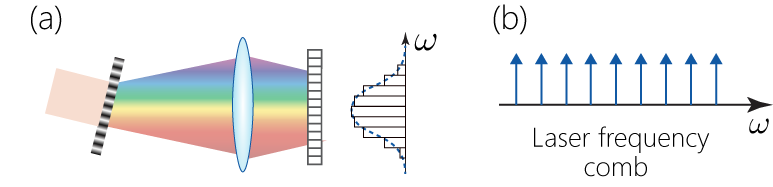}
\caption{Modal bases for the spectral DoF. (a) Spectral bins and (b) laser frequency combs.}
\label{fig:DetectionNonOverlappingTemporalModes}
\end{figure}

\section{Roadmap for structured optical coherence}\label{sec:Roadmap}

The flurry of recent progress in the area of structured coherence indicates several new avenues of research that are ripe for exploration. We list a few of these here to suggest signpots on a roadmap for future developments.

\subsection{Large-dimensional modal bases}

We have examined here binary DoFs in Section~\ref{sec:OneBinaryDoF}, a pair of binary DoFs in Section~\ref{sec:2DoFs}, and an $N$-dimensional DoF in Section~\ref{sec:ExtenstioToN}. More work needs to be done to fully appreciate the properties of $N$-dimensional modal bases with $N\geq3$, where the associated coherence matrix is $N\times N$. Indeed, even $N=3$ presents some challenges. For example, whereas an even-dimensional modal basis can make use of direct products of Pauli matrices, the case of $N=3$ necessitates identifying a new set of matrices to serve as a basis for $3\times3$ Hermitian coherence matrices. For $N=3$, the Gell-Mann matrices utilized in particle physics \cite{GellMann62PR} and in non-paraxial polarization optics \cite{Alonso23AOP} can be exploited to obtain modal Stokes-Gell-Mann parameters as an intermediary to reconstruct the $3\times3$ coherence matrix. Much research is needed to optimize the configurations for generalizing the concept of modal Stokes parameters to large $N$, which we anticipate will make use of assemblies of Pauli and Gell-Mann matrices.

\subsection{Multiple DoFs}

When two DoFs of the optical field are described by modal bases of dimensions $N$ and $M$, the composite modal basis has dimensions $N\times M$, so that the coherence matrix is $(N\times M)\times(N\times M)$. The reduced coherence matrices associated with the two DoFs have dimensions $N\times N$ and $M\times M$. The investigation of structured coherence with large-dimensional modal bases and of fields with two DoFs of large and mismatched dimension is still in its infancy.

One example of this challenge that we anticipate may be of interest is with regards to recently studied spatiotemporally structured optical fields \cite{Shen23JO,Abouraddy25OPN}. For example, space-time wave packets (STWPs) \cite{Yessenov22AOP} are pulsed beams in which the spatial and temporal DoFs are tightly associated. Coherent STWPs have been studied extensively over the past decade and have revealed a host of useful and fascinating properties. However, only a few studies of partially coherent STWPs have been reported to date \cite{Yessenov19Optica,Yessenov19OL}. The prospect of structured spatiotemporal coherence is made particularly intriguing after recently finding that a discrete basis for STWPs can be formed through the Schmidt decomposition of the field with respect to the spatial and temporal DoFs. Finally, only limited work has been done on optical fields in which 3~DoFs are relevant, and no reports have appeared regarding the structured coherence of such fields.

\subsection{Structured coherence in free space and multimode fibers}

In addition to on-chip platforms for the manipulation of large-dimensional coherence matrices, broad swathes of free-space opportunities have gone unexplored to date. Although highly sophisticated field structures have been investigated, they have all been coherent fields. Extending such field structures to structured coherence has yet to be done.

\subsection{On-chip structured coherence}

One of the first goals in this area of structured coherence is to experimentally demonstrate the manipulation of multimoded partially coherent optical fields in photonic integrated circuits, which provide a convenient platform for exploiting large-dimensional coherence matrices that will be useful for advanced communications and cryptography schemes. These require developing efficient layouts for optimizing space and time resources in carrying out specific computational tasks. One immediate task to be optimized is the efficient on-chip reconstruction of a coherence matrix. To date, two strategies have been explored: (a) variational processing (Section~\ref{sec:MeasuringTheDegreeOfCoherence}) \cite{roques2024measuring}; and (2) tomographic reconstruction via measurements of the modal Stokes parameters (Section~\ref{sec:ReconstructingGSingleDoF}) \cite{Abouraddy14OL,Kagalwala15SR}.

\subsection{Applications of structured coherence in optical information processing}

Despite the fundamental interest in structured coherence as a new class of optical fields, we expect that sustained interest will be ultimately determined by success in demonstrating a `coherence advantage': applications in optical information processing in which structured coherence outperforms coherent light. We anticipate that the central feature of structured coherence that may reveal a coherence advantage is the larger number of free parameters involved in identifying an $N\times N$ coherence matrix compared to an $N\times1$ coherent field vector. This feature has already resulted in two distinct results in optical communications: mutual coherence multiplexing to increase the channel density \cite{Nardi22OL} and scattering-free coherence-rank communications \cite{Harling25APLP}. Additional application in optical computing \cite{Dong24Nature} and coherence cryptography \cite{Peng21P,Liu25LPR} are also emerging. These are only a few guideposts for this \textit{terra incognita} of structured optical coherence. Undoubtedly, the next few years will witness progress along these lines, in addition to unanticipated breakthroughs and surprises.

\begin{figure}[t!]
\centering
\includegraphics[width=13.3cm]{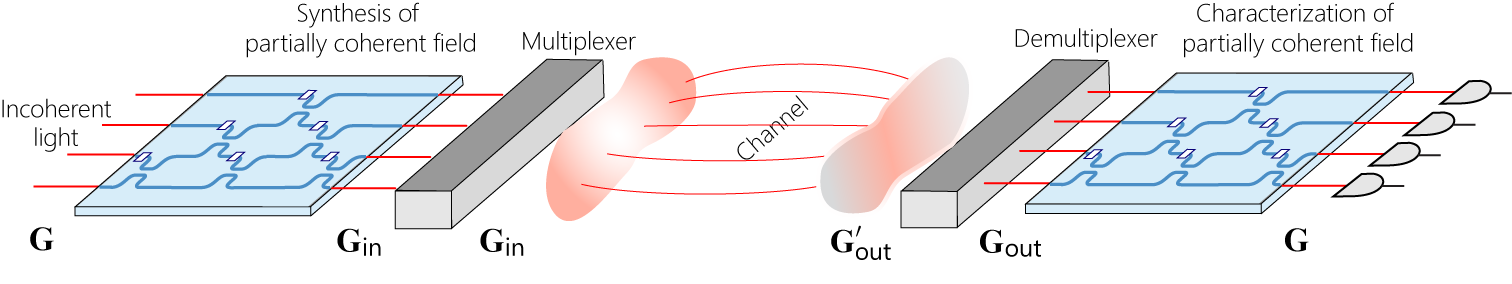}
\caption{Vision for exploiting structured coherence in optical communications and information processing. Generic multimoded incoherent light (on the left) is fed into a photonic integrated circuit where it is pre-processed before a prescribed coherence matrix $\mathbf{G}_{\mathrm{in}}$ being launched through an optical channel. At the end of the channel, light described by the coherence matrix $\mathbf{G}_{\mathrm{out}}$ is coupled into a second photonic integrated circuit where it is post-processed and the coherence matrix reconstructed.}
\label{fig:vision}
\end{figure}

We envision the configuration illustrated in Fig.~\ref{fig:vision} as a paradigm for the synthesis, processing, and detection of structured coherence. Generic incoherent multimoded light ($N$ modes) described by the coherence matrix $\mathbf{G}=\tfrac{1}{N}\hat{\mathbb{I}}_{N}$ is coupled to a photonic integrated circuit, with each mode of the incoherent field coupled to a single-mode on-chip waveguide. Once coupled to the chip, several tasks need to be performed:
\begin{enumerate}
\item Tuning the coherence rank by setting the requisite amplitudes to zero.
\item Adjusting the entropy of the field by varying the values of the non-zero eigenvalues. 
\item Sculpting the coherence matrix by implementing the requisite $N\times N$ unitary $\hat{U}$, comprising a sequence of $2\times2$ unitaries operating on pairs of modes.
\end{enumerate}

Once the target coherence matrix $\mathbf{G}_{\mathrm{in}}$ is synthesized on chip, the field is launched into a physical optical channel, where all the parameters of $\mathbf{G}_{\mathrm{in}}$ (its rank, entropy, or structure) may undergo change. After traversing the optical channel, the field is coupled to a second photonic integrated circuit that may add further processing of the coherence matrix before reconstructing it. This overall construction is likely to be the basis for developing further applications of structured coherence in optical communications and information processing that exploit the coherence advantage.

\section{Conclusion}

To date, freely propagating continuous fields have been the province of partially coherent light. However, progress in programmable on-chip photonic platforms, in addition to the emergence of novel applications of partially coherent light in communications and information processing all point towards the need for a discrete formulation of optical coherence. We have presented here an outline of this formulation restricted to a single binary DoF (two modes) and to dual binary DoFs (four modes). Such a formulation encompasses a range of important concepts that are crucial for such discrete modal bases: the Stokes parameters for the reconstruction of a $2\times2$ coherence matrix associated with any DoF (not necessarily polarization); extraction of the degree of coherence for any binary DoF through diagonalization (or unitarily maximizing the difference between the modal weights) or equalization of the modal weights; tomographic reconstruction of the coherence matrix via dual-DoF composite Stokes parameters; entropy conversion between DoFs; the coherence rank as a classifier of partially coherent fields; and coherence-rank communications across strongly scattering channels. Although these concepts have analogs in the quantum mechanics of a qubit or a qubit pair, there are nevertheless crucial differences that stem from the no-cloning theorem that restricting measurements of quantum systems but not their classical counterparts. Additionally, the flexibility of unitarily coupling and decoupling of different DoFs in classical optics, which is not readily available with two-photon states, makes the concept of coherence-rank relevant to partially coherent fields.

The formulation presented here lays the foundation for extension to larger-dimensional modal bases, which is expected to be a pressing need as programmable on-chip platforms are adopted for the manipulation of partially coherent fields in applications involving information processing.

\vspace{7mm}
\noindent
\textbf{Acknowledgments}

\noindent
U.S. Office of Naval Research (ONR) N00014-20-1-2789.


\bibliography{diffraction}

\end{document}